\documentclass{dmathesis}
\usepackage{etex}

\usepackage{showlabel}
\usepackage{fancyhdr}
\usepackage{epsfig}
\usepackage{cite}
\usepackage{graphicx}
\usepackage{amsmath}
\usepackage{theorem}
\usepackage{amssymb}
\usepackage{latexsym}
\usepackage{epic}

\newcounter{ind}

{\theorembodyfont{\rmfamily}}
{\theorembodyfont{\rmfamily}}
{\theorembodyfont{\rmfamily}}
{\theorembodyfont{\rmfamily}}
{\theorembodyfont{\rmfamily}}
{\theorembodyfont{\rmfamily}}

\usepackage[ruled,vlined]{algorithm2e}
\usepackage[usenames,dvipsnames,table]{xcolor}
\usepackage{algpseudocode}
\usepackage{parskip}
\usepackage{listings}

\usepackage{setspace}
\usepackage[printonlyused,nohyperlinks]{acronym}
\usepackage{subfigure}
\usepackage{url}
	\usepackage{color}
\definecolor{greytext}{gray}{0.5}
\usepackage[format=hang,indention=-1cm,labelfont={bf},margin=1em,]{caption}
\usepackage{verbatim}
\usepackage{framed}
\usepackage{float}
\newfloat{Protocol}{t}{struct}
\usepackage{nomencl}
\usepackage{soul}
\usepackage{multirow}
\usepackage[toc,page]{appendix}
\usepackage{tikz}
\usepackage{pgfplots}

\usepackage{makecell}
\usepackage{amssymb}
\usepackage{picture}
\usepackage{array}
\usepackage{lipsum}
\newcommand\blfootnote[1]{  \begingroup
  \renewcommand\thefootnote{}\footnote{#1}  \addtocounter{footnote}{-1}  \endgroup
}

\newtheorem{theorem}{Theorem}
\newtheorem{claim}[theorem]{Claim}

\makenomenclature 

\makeglossary

\usepackage[bookmarks,bookmarksnumbered,    citebordercolor={0.8 0.8 0.8},filebordercolor={0.8 0.8 0.8},    linkbordercolor={0.8 0.8 0.8},pagebordercolor={0.8 0.8 0.8},    urlbordercolor={0.8 0.8 0.8},    pagebackref,    pdfauthor={AbdelRahman Abdou},
	pdftitle={Internet Location Verification: Challenges and Solutions},
    ]{hyperref}

\usepackage{setspace}
\usepackage[redeflists]{IEEEtrantools}
\acrodefplural{CA}[CAs]{Certification Authorities}
\acrodefplural{MB}[MBs]{Middle Boxes}
\acrodefplural{AS}[ASes]{Autonomous Systems}

\usepackage{enumitem}
\usetikzlibrary{calc}

\usepackage{adjustbox}
\usepackage{array}
\newcolumntype{Z}[2]{    >{\adjustbox{angle=#1,lap=\width-(#2)}\bgroup}    l    <{\egroup}}
\newcommand*\rott{\multicolumn{1}{Z{45}{1em}}}
\begin{document}

\pagenumbering{roman}

\setcounter{page}{1}

\linespread{1.0}

\newpage

\thispagestyle{empty}
\begin{center}
\vfill
 \textcolor{white}.\\
\textcolor{white}.\\
\MakeUppercase{Internet Location Verification: Challenges and Solutions}\\
\textcolor{white}.\\
\textcolor{white}.\\
By\\[0.7\baselineskip]
{\large AbdelRahman Mohamed Abdou}\\[1.4\baselineskip]

\textcolor{white}.\\
A thesis submitted to\\[-0.4\baselineskip]
the Faculty of Graduate Studies and Research\\[-0.4\baselineskip]
in partial fulfillment of\\[-0.4\baselineskip]
the requirements for the degree of\\[1.4\baselineskip]
\textsc{\large Doctor of Philosophy}\\
in\\
Electrical and Computer Engineering\\
\textcolor{white}.\\
{\small Department of Systems and Computer Engineering}\\
\textcolor{white}.\\
\textcolor{white}.\\
\textcolor{white}.\\
\textsc{Carleton University}\\
Ottawa, Ontario, Canada\\[-0.4\baselineskip]
\textcolor{white}.\\
\textcolor{white}.\\
\textcolor{white}.\\
\copyright 2015, AbdelRahman Abdou
\vfill
\end{center}

\newpage
\thispagestyle{empty}
\addcontentsline{toc}{chapter}{\numberline{}Abstract}
\begin{center}
   \textbf{\large Abstract}
\end{center}
\small

\normalsize

This thesis addresses the problem of verifying the geographic locations of Internet clients. First, we demonstrate how current state-of-the-art delay-based geolocation techniques are susceptible to evasion through delay manipulations, which involve both increasing and decreasing the Internet delays that are observed between a client and a remote measuring party. We find that delay-based techniques generally lack appropriate mechanisms to measure delays in an integrity-preserving manner. We then discuss different strategies enabling an adversary to benefit from being able to manipulate the delays. Upon analyzing the effect of these strategies on three representative delay-based techniques, we found that the strategies combined with the ability of full delay manipulation can allow an adversary to (fraudulently) control the location returned by those geolocation techniques accurately.

We then propose Client Presence Verification (CPV) as a delay-based technique to verify an assertion about a client's physical presence in a prescribed geographic region. Three verifiers geographically encapsulating a client's asserted location are used to corroborate that assertion by measuring the delays between themselves and the client. CPV infers geographic distances from these delays and thus, using the smaller of the forward and reverse one-way delay between each verifier and the client is expected to result in a more accurate distance inference than using the conventional round-trip times. Accordingly, we devise a novel protocol for accurate one-way delay measurements between the client and the three verifiers to be used by CPV, taking into account that the client could manipulate the measurements to defeat the verification process.

We evaluate CPV through extensive real-world experiments with legitimate clients (those truly present at where they asserted to be) modeled to use both wired and wireless access networks. Wired evaluation is done using the PlanetLab testbed, during which we examine various factors affecting CPV's efficacy, such as the client's geographical nearness to the verifiers. For wireless evaluation, we leverage the Internet delay information collected for wired clients from PlanetLab, and model additional delays representing the last-mile wireless link. The additional delays were generated following wireless delay distribution models studied in the literature. Again, we examine various factors that affect CPV's efficacy, including the number of devices actively competing for the wireless media in the vicinity of a wireless legitimate CPV client.

Finally, we reinforce CPV against a (hypothetical) middlebox that an adversary specifically customizes to defeat CPV (i.e., assuming an adversary that is aware of how CPV operates). We postulate that public middlebox service providers (e.g., in the form of Virtual Private Networks) would be motivated to defeat CPV if it is to be widely adopted in practice. To that end, we propose to use a Proof-of-Work mechanism that allows CPV to impose constraints, which effectively limit the number of clients (now adversaries) simultaneously colluding with that middlebox; beyond that number, CPV detects the middlebox.

\chapter*{Acknowledgements}
\addcontentsline{toc}{chapter}{\numberline{}Acknowledgements}
\small

\normalsize

Thanks to my supervisors, Dr. Ashraf Matrawy and Dr. Paul Van Oorschot, for their diligent efforts and close mentoring. Both were continuously keen to maintain high quality research outcomes throughout my PhD journey. Their vast knowledge and continuous guidance have helped shape the contributions in this thesis.

I would also like to thank my great wife Hala, who is not only my life partner but has also been a PhD colleague and a supporting friend. She has helped ameliorate many of the figures in this thesis, and was eager to proofread my papers even during her tight schedules. To my parents, Mohamed and Hayam, and my sisters, thanks for your tremendous moral and spiritual support throughout this tough journey. Your regular long distance calls were necessary to maintain my stamina, especially during phases of frustration.

I thank my PhD committee members, Urs Hengartner, Thomas Kunz, Michel Barbeau and Jiying Zhao, for their insightful comments. Thanks also to Pat Morin for his help with standard geometry, and David Lie for providing infrastructure for remote small-scale preliminary testing. I am grateful to the anonymous reviewers, and members of the networking/security community who generously gave their opinions, e.g., during casual discussions in conferences. Those include Michael Freedman, James Muir, Bill Aiello and Andrew Csinger.

Finally, I would like to express my gratitude to members of the Carleton Computer Security Lab (CCSL), especially David Barrera and Furkan Alaca, for hours of proofreading and discussions.

\break
\pagebreak
\pagebreak

\chapter*{Glossary}
\addcontentsline{toc}{chapter}{\numberline{}Glossary}

\glossary{name={$\mathcal{U}$}, description={The universal set}, sort=R} 

\acrodef{UI}{user interface}
\acrodef{UBC}{University of British Columbia}

\begin{acronym}[ANOVA]

\acro{ACK}{Acknowledgement}
\acro{API}{Application Programming Interface}
\acro{AS}{Autonomous System}
\acro{AVOD}{Audio/Video On Demand}
\acro{BGP}{Border Gateway Protocol}
\acro{BSD}{Berkeley Software Distribution}
\acro{CA}{Certification Authority}
\acro{CBG}{Constraint-Based Geolocation}
\acro{CDF}{Cumulative Distribution Function}
\acro{CDN}{Content Distribution Network}
\acro{CDP}{Cisco Discovery Protocol}
\acro{CPV}{Client Presence Verification}
\acro{CTS}{Clear To Send}
\acro{CSMA/CA}{Carrier Sense Multiple Access with Collision Avoidance}
\acro{DB}{Database}
\acro{DCF}{Distributed Coordination Function}
\acro{DIFS}{Distributed (coordination function) Interframe Space}
\acro{DNS}{Domain Name Server}
\acro{DSCP}{Differentiated Services Code Point}
\acro{ECN}{Explicit Congestion Notification}
\acro{FA}{False Accept}
\acro{FR}{False Reject}
\acro{GLS}{Google Location Server}
\acro{GNU}{GNU's not Unix}
\acro{GPS}{Global Positioning System}
\acro{GREN}{Global Research and Educational Network}
\acro{HTTP}{Hypertext Transfer Protocol}
\acro{IANA}{Internet Assigned Numbers Authority}
\acro{ICMP}{Internet Control Message Protocol}
\acro{IHL}{Internet Header Length}
\acro{IP}{Internet Protocol}
\acro{ISP}{Internet Service Provider}
\acro{LBS}{Location-Based Service}
\acro{LBAC}{Location-based Access Control}
\acro{LSP}{Location-Sensitive Provider}
\acro{LIS}{Location Information Server}
\acro{LLDP-MED}{Link Layer Discovery Protocol-Media Endpoint Discovery}
\acro{MAC}{Medium Access Control}
\acro{MB}{Middle Box}
\acro{MitM}{Man in the Middle}
\acro{NAT}{Network Address Translation}
\acro{NCS}{Network Coordination System}
\acro{NFC}{Near Field Communication}
\acro{NIC}{Network Interface Card}
\acro{NRMSD}{Normalized Root-Mean-Square-Deviation}
\acro{NTP}{Network Time Protocol}
\acro{OS}{Operating System}
\acro{OWAMP}{One-Way Active Measurement Protocol}
\acro{OWD}{One-Way Delay}
\acro{P2P}{Peer to Peer}
\acro{PDA}{Personal Digital Assistant}
\acro{PDF}{Probability Distribution Function}
\acro{PGR}{Permitted Geographic Region}
\acro{PID}{Process ID}
\acro{PKI}{Public Key Infrastructure}
\acro{PMF}{Probability Mass Function}
\acro{PoW}{Proof-of-Work}
\acro{QoS}{Quality of Services}
\acro{RBAC}{Role-based Access Control}
\acro{RF}{Radio Frequency}
\acro{RFC}{Request For Comment}
\acro{RFID}{Radio-Frequency Identifier}
\acro{RTP}{Real-time Transport Protocol}
\acro{RTS}{Request to Send}
\newacro{RTS/CTS}{Request to Send and Clear to Send}
\acro{RTT}{Round Trip Time}
\acro{SGR}{Sensitive Geographic Region}
\acro{SIFS}{Short Interframe Space}
\acro{SIP}{Session Initiation Protocol}
\acro{SOI}{Speed of the Internet}
\acro{SSID}{Service Set Identifier}
\acro{TCP}{Transport Control Protocol}
\acro{TIV}{Triangular Inequality Violation}
\acro{ToFU}{Trust on First Use}
\acro{TPM}{Trusted Platform Module}
\acro{TTL}{Time to Live}
\acro{UDP}{User Datagram Protocol}
\acro{VANET}{Vehicular Ad-hoc Networks}
\acro{VoIP}{Voice over IP}
\acro{VPN}{Virtual Private Network}
\acro{W3C}[W3C]{\acroextra{the }World Wide Web Consortium\acroextra{, the standards body for web technologies}}
\acro{WiFi}{Wireless Fidelity}
\acro{WPS}{WiFi Positioning System}
\acro{WSN}{Wireless Sensor Networks}

\end{acronym}

\tableofcontents
\clearpage

\listoffigures 
\listoftables
\pagebreak
\pagenumbering{arabic}
\setcounter{page}{1}

\newcolumntype{L}[1]{>{\raggedright\let\newline\\\arraybackslash\hspace{0pt}}m{#1}}
\newcolumntype{C}[1]{>{\centering\let\newline\\\arraybackslash\hspace{0pt}}m{#1}}
\newcolumntype{R}[1]{>{\raggedleft\let\newline\\\arraybackslash\hspace{0pt}}m{#1}}

\newcommand{\changes}{black}

\onehalfspacing

\chapter{Introduction}
\label{ch:intro}

Remotely proving that an Internet-connected device is geographically present at where it asserts to be remains one of the most challenging problems in today's Internet. This thesis addresses the problem of one party verifying an assertion about the geographic location of a typically remote second \emph{target party} over the Internet. The verifying party is different from and not in physical possession of the target party's device. The target party is a physical device running a process that is able to send and receive Internet data. The goal of the thesis is to devise a mechanism that provides greater assurance about the correctness of an asserted location, compared to current state-of-the art techniques.

More formally, we state the research question as follows:

\begin{center}
\textit{Assume two devices transmitting data between themselves over the Internet. If the geographic location of one of them is asserted, what mechanisms are available to provide assurance of the correctness of this assertion?}
\end{center}

\section{Terminology and Scope}

{\bf Terminology.} Throughout this thesis, the terms \emph{client} and \emph{\ac{LSP}} are such that the client is the target party whose location is important for the appropriate operation of the \ac{LSP}. The term \emph{geolocation} means (physical) location-determination; hence, \emph{geolocating} means determining the location. Additionally, \emph{Internet geolocation} means geolocating an Internet-connected device.

We define the \emph{evasion of geolocation} as deliberately causing a geolocation technique to fail or to return an incorrect location.

{\bf Scope.} While geolocation is of interest, the main focus is \emph{location verification}, meaning that some geolocation mechanism (beyond the scope of this thesis) first asserts a client's location, which is then verified by mechanisms within this thesis.

The thesis focuses on real-time location verification. Non real-time applications, such as in forensic analysis where geolocating IP addresses post-incident is of interest, are out of scope. Solutions devised in this thesis assume a two-way real-time communication between the client and the \ac{LSP}.

\section{Motivation}
Numerous applications can benefit from reliable information about the locations of Internet clients. The following examples are those where evasion incentives may arise, motivating this thesis. Cases where little or no evasion incentive exists, e.g., location-directed ads, are outside the scope of this thesis.

{\bf Fraud prevention.} The geographic location where credit card transactions are taking place could provide higher assurance to the authenticity of these transactions, compared to when the location is not known.
        
{\bf Impersonation prevention.} Impersonation over the Internet, including through password-guessing attacks, can be reduced by restricting a user's login to locations previously associated with the user's account. An impersonating adversary may thus try to evade geolocation in order to place itself fraudulently in that location.

{\bf Policy compliance.} Various legal agreements are location-dependent. For example, cloud providers often promise the sole storage of users' data within user-requested jurisdictions. However, motivations to violate such an agreement may arise due to cheaper overseas operations and maintenance costs. Additionally, video-on-demand providers, e.g., Hulu \cite{hulu}, are often licensed to stream only to restricted geographic regions. Gambling regulations differ across jurisdictions, placing a responsibility on gambling websites to enforce these regulations according to where the gambler is geographically located. Many online retailers are required to charge applicable taxes based on users' locations. In general, the motivation to evade geolocation in the cases under the \emph{Compliance} category is usually for gaining location-dependent benefits.

{\bf Location-based access control.} Sensitive data, such as military documents or patient records, are often allowed to be viewed only from within certain regions. Operations like online bidding, community-related voting, or even ordering home-delivery meals, can be regulated by users' locations, e.g., to reduce spammers.

\section{Deficiency of Existing Geolocation Mechanisms}
\label{intro:sec:Background}

Commonly-used Internet geolocation techniques lack integrity or cross-checking of their results. Such techniques fail to adequately consider a knowledgeable adversary that is motivated to cheat about its location. Most of the geolocation literature focus on achieving higher geolocation accuracy, overlooking adversarial environments.

{\bf Tabulation-based techniques.} These work by having the \ac{LSP} look up the client's IP address in a pre-populated \ac{DB} that maps IP-addresses to locations,~e.g., MaxMind \cite{maxmind}. Studies have found that many of the major tabulation providers are evadable, e.g., by having the client simply hide its true IP address using a \ac{VPN} \cite{MuirPaul}. 

{\bf Self-positioning systems.} This is the class of techniques where an \ac{LSP} requests the client's location information from the client itself. The client's device determines its own location using, e.g., \ac{GPS} \cite{gpsgps}, \ac{WPS} \cite{TGIS:TGIS1152} (see Chapter \ref{ch:background}), cell tower triangulation (in the case of mobile devices) \cite{Trestian:2009:MSC:1644893.1644926}, and communicates it to the \ac{LSP}. No geolocation techniques under this class can be relied upon to geolocate adversaries motivated to forge their locations; the asserted location must be verified for prudent use in location-sensitive services.

{\bf Measurement-based geolocation techniques \cite{laki2011spotter,6197179}.} These exploit the correlation between Internet delays and geographic distances in geolocating clients. Delays are measured between the client and a set of landmarks with known locations, and are mapped to distances according to some predefined (usually calibrated) mapping function. Multilateration is then used to determine the client's location relative to the landmarks. When the client's IP address is used in delay measurements (e.g., using the \textsl{ping} utility), employing a non-local IP address evades those techniques \cite{MuirPaul}, i.e., similar to evading tabulation-based techniques discussed above. Additionally, even when the IP address is not used, delay manipulations can corrupt the geolocation process \cite{Dude}.

\section{Thesis Contributions and Organization}

\subsection{Analyzing the security of delay-based geolocation}
\label{firstthesiscontribution}
Previous literature analyzed the effect of an adversary increasing the measured delays on the accuracy and evadability of measurement-based geolocation techniques \cite{Dude}. Our first contribution is research investigating adversarial evasion capabilities by (1) demonstrating that an adversary could also decrease the measured delays by means of exploiting the lack of integrity in common delay-measurement utilities; (2) demonstrating enhanced adversarial strategies to better utilize delay manipulation for a more accurate misrepresentation of location, thus showing additional vulnerabilities in existing mechanisms; and (3) evaluating the adversarial accuracy in forging its location, now considering the previous two contributions. This is presented in Chapter \ref{ch:attack}.

\subsection{Accurate One-way Delay Estimation}
Due to delay asymmetry \cite{pathak2008measurement}, \acp{OWD} have the potential to improve the performance of delay-dependent applications \cite{4438363}, such as delay-based geolocation. However, delay-based geolocation techniques usually map \acp{RTT} to distances rather than mapping \acp{OWD} because the former is easier to estimate. 

To that end, and as a tool useful in one of our other contributions (see CPV below), we devised the \emph{minimum pairs} protocol (Chapter \ref{ch:owd})---a \ac{OWD}-estimation protocol that requires no more cooperation between the two parties than that required to estimate \acp{RTT}, yet is in many cases more accurate than simply taking half the \ac{RTT} as a \ac{OWD}-estimate. This later conclusion is reached by formally deriving the probability distribution of absolute error for these two alternative protocols, as a function of the delay distribution between network nodes.

The \emph{minimum pairs} protocol is a generic contribution, which we believe to be of independent interest; e.g., it could be used by delay-dependent Internet applications for accurate \ac{OWD}-estimation without the need for the overwhelming cooperation---between the two parties estimating delays---typically required by \ac{OWAMP}-like protocols \cite{rfc4656}.

\subsection{Client Presence Verification (CPV)}
As the main contribution of this thesis, Chapter \ref{ch:cpv} introduces \ac{CPV}---a delay-based location verification technique designed to verify in realtime the presence of an Internet-connected client in a prescribed triangular geographic region. The algorithm employs heuristics to reduce erroneous false rejects/accepts, while retaining reasonable granularity. 

In \ac{CPV}, three verifiers geographically encapsulating the client's asserted location are selected to verify this assertion. They use the \emph{minimum pairs} protocol to estimate \acp{OWD} between themselves and the client, and leverage these delays for evidence supporting the client's presence within the triangle determined by their geographic positions.

\ac{CPV} mitigates common geolocation-evasion tactics explored in the literature \cite{MuirPaul,Dude}, as well as the novel adversarial manipulations discussed above in Section \ref{firstthesiscontribution}. We discuss the integrity of \ac{CPV}'s decisions in the presence of a broad class of adversarial evasion tactics, and argue about the algorithm's defense capabilities against these tactics.

Viability of the \ac{CPV} algorithm is extensively evaluated (from a networking perspective) through real world experiments on PlanetLab \cite{planetlab}. The effect of various factors on the correctness of \ac{CPV} is examined, such as the clients' geographic proximity to the verifiers and the triangle they determine, and the number of delay measurements the verifiers perform. This evaluation is presented in Chapter \ref{ch:wiredecva}.

The PlanetLab nodes employed are connected through a wired access network. To evaluate the use of \ac{CPV} with wireless clients, we use wireless delay distribution models from the literature, and generate delays following these models. Those delays are then added to the delays measured using the wired PlanetLab nodes to model wireless clients. Several factors are considered, including the number of wireless devices in the vicinity of the wireless client. The wireless evaluation is presented in Chapter \ref{ch:wirelessecva}.

\subsection{Hindering Unauthorized Traffic Relaying}
\ac{CPV} raises the bar for an adversary trying to forge its geographic location. It is designed to reject adversaries even when they are using a \ac{MB}, geographically present at their intended location, to hide their IP addresses and relay their traffic from the server. A colluding \ac{MB}, customized to specifically evade \ac{CPV}, may however succeed to mislead \ac{CPV} to accept the \ac{MB}'s location as that of the client. For example, third party \ac{MB} service providers, such as public \acp{VPN} \cite{hotspot}, may well be motivated to customize their infrastructure to evade \ac{CPV} upon its deployment.

We propose to use a \ac{PoW} mechanism, such as client puzzles, to defeat colluding \acp{MB}, hindering their illicit traffic relaying. Our proposal is evaluated using a Markov queuing model, and additionally using simulations. Similar to the \emph{minimum pairs} protocol, this proposal may be of independent interest as a standalone contribution since it can be used to resist unauthorized traffic relaying regardless of the application. In our case, we use it to strengthen \ac{CPV} against colluding \acp{MB}. This contribution is presented in Chapter \ref{ch:puzzles}.

\section{Related Publications}

Detailed explanation to the \ac{CPV} algorithm (Chapter \ref{ch:cpv}), the \emph{minimum pairs} protocol it uses (a proportion of Chapter \ref{ch:owd}), and the algorithm's evaluation in a wired network environment (a proportion of Chapter \ref{ch:wiredecva}) were published as a full paper in the IEEE CNS conference. The paper was nominated for a Best Paper award.
\begin{itemize}[leftmargin=0.5in]
\item A. M. Abdou, A. Matrawy, and P.C. van Oorschot, ``Location Verification on the Internet: Towards Enforcing Location-aware Access Policies Over Internet Clients". In {\it IEEE Communications and Network Security (CNS)}, Oct. 2014.
\end{itemize}

A followup version of the CNS paper, which includes additional analysis to the \ac{CPV} algorithm is accepted for publication in IEEE TDSC.
\begin{itemize}[leftmargin=0.5in]
\item A. M. Abdou, A. Matrawy, and P.C. van Oorschot, ``CPV: Delay-based Location Verification for the Internet". In {\it IEEE Transactions on Dependable and Secure Computing} (to appear; accepted June 14, 2015).
\end{itemize}

In addition to introducing the \emph{minimum pairs} protocol, Chapter \ref{ch:owd} also evaluates the protocol analytically by first deriving the probability distribution of its absolute error, then comparing its accuracy (using the derived distribution) to the \ac{RTT}-halving protocol. The evaluation methodology and the derived model were accepted for publication.
\begin{itemize}[leftmargin=0.5in]
\item A. M. Abdou, A. Matrawy, and P.C. van Oorschot, ``Accurate One-Way Delay Estimation with Reduced Client-Trustworthiness". In {\it IEEE Communications Letters}, vol. 19, no. 5, pp. 735--738, 2015.
\end{itemize}

Finally, Chapter \ref{ch:puzzles} introduces the principle of using a \ac{PoW} mechanism to thwart \acp{MB} (e.g., \acp{VPN} and proxies) from illicitly relaying traffic. The principle and its evaluation, both analytically and using simulations, were published.
\begin{itemize}[leftmargin=0.5in]
\item A. M. Abdou, A. Matrawy, and P.C. van Oorschot, ``Taxing the Queue: Hindering Middleboxes from Unauthorized Large-Scale Traffic Relaying". In {\it IEEE Communications Letters}, vol. 19, no. 1, pp. 42--45, 2015.
\end{itemize}

\chapter{Background and Related Work}
\label{ch:background}

This chapter presents related work on the areas of geolocation and location verification, their limitations and vulnerabilities. Although we only focus on the verification aspect in this thesis, we review Internet geolocation techniques in general to help explore their susceptibility to evasion. Moreover, we analyze measurement-based geolocation techniques as they provide an insight about the nature of delays over the Internet, and the accuracy of mapping delays to distances. A brief survey on location verification in single-hop wireless networks is also presented, as well as state-of-the-art location-proof architectures and their applications.

\section{Internet Geolocation}
\label{backgroundinternetgeo}
An Internet geolocation technique aims to bind a client's identifier (e.g., its IP address) to a geographic location. It either involves Internet delay measurements between the client and a set of reference objects with known locations, and the use of multi-lateration to determine the client's location relative to these objects; or it could be inference-based, where an estimate for the location is inferred from the client's attributes \cite{MuirPaul} and/or behavior \cite{pindrop}. Either way, a \acf{LSP} may ask the client to geolocate itself and inform the \ac{LSP}, or ask a third party geolocation service provider to geolocate the client given an identifier.

Measurement-based techniques can be further categorized into either \emph{delay-based} or \emph{topology-aware} techniques. In what follows, we review proposals in the literature under each category, review inference-based approaches, then discuss other techniques by which a client geolocates itself and informs the \ac{LSP}.

\subsection{Delay-based techniques}
\label{ipgeosection}

Delay-based \ac{IP} geolocation is a class of techniques where the geographic location of the client machine is determined based on the observed network delays between the machine and a set of geographically scattered landmarks with known locations. These techniques assume the client is able to receive and respond to the delay-measurement probes, which in practice, commonly use \ac{ICMP}-based utilities like \textsl{ping} and \textsl{traceroute}.

\paragraph{Factors affecting Internet delays}
Four primary delay components exist between two Internet hosts: propagation, transmission,\footnote{Note that the transmission delay of a packet is measured from the time the first bit of the packet is placed on the transmission media, until the last bit is similarly placed. It is a function of the packet length (in bits) and the media's transmission capacity (in bits per second). In contrast, the propagation delay of a (single) bit is the time required for the bit to propagate through the media from the sender to the receiver. It is a function of the distance spanned in the media (in meters) and the media's transmission speed (in meters per second).} queueing and processing delays at intermediate systems (e.g., routers) \cite{book:kurose}. There are also delays imposed by end-systems protocols, such as the \ac{TCP}'s congestion and flow control mechanisms. The flow control mechanism receives its parameters from the destination. It would decrease the sender's transmission rate if the receiver signals ``slow-down" cues, or increase the rate otherwise. Additionally, other delay components may arise from different aspects such as a low \ac{QoS} provided by the \ac{ISP}, and excessively circuitous routes.\footnote{A network route is said to be \emph{circuitous} when the geographic distance it spans is considerably larger than the (shortest) geographic distance between its source and destination.} Route circuitousness could also be a result of user configuration, as in the case of using a proxy server or an anonymizing browser.

\paragraph{How delay-based techniques work}
Despite a plethora of factors that affect the delays between two nodes over the Internet \cite{Octant,crovella}, numerous studies have established that there is a strong correlation between delays and geographic distances \cite{GeoTrackGeoPing, 1266171, improvacc, Constrainbased, landa2013measuring}. The main characteristic relied on is the propagation delay. Most, if not all, delay-based geolocation techniques mitigate the effect of other delay factors (e.g., queueing due to congestion) by using the minimum of multiple delay measurements (e.g., 10-20 \acp{RTT}) to the client from each landmark. Once delays are measured, the research question addressed by most techniques becomes finding the best function to map them to geographic distances.

An exception is one of the first delay-based techniques: GeoPing \cite{GeoTrackGeoPing}. Instead of mapping delays to distances, GeoPing matches the location of the client to a location where the most similar delay behavior has previously been observed. Assuming $n$ landmarks and $m$ reference nodes with known locations,\footnote{The problem of the geographic placement of such infrastructure to enhance the geolocation accuracy was well studied in the literature \cite{ziviani2003demographic,improvacc}.} the landmarks in Geo-Ping first create a delay vector to each of the $m$ nodes. A delay vector of a node contains $n$ values corresponding to the \acp{RTT} between the landmarks and the node. Each landmark then measures the \ac{RTT} between itself and the client, enabling the landmarks to create a delay vector for that client. The location of the node with the \emph{nearest} delay vector is then returned as the client's calculated location. The authors of GeoPing proposed to calculate nearness between two delay vectors as the $n$-dimensional Euclidean distance between the two vectors \cite{GeoTrackGeoPing}. Ziviani {\it et al.} showed that Manhattan, Canberra, and Chebyshev distances can generate more accurate results when used as alternatives to the Euclidean distance \cite{GeoPing2}. This class of delay-based geolocation returns a location from a discrete space depending on the number of available reference nodes.

The authors of delay-based techniques often contribute a function that can map delays-to-distances accurately. In most proposals \cite{Constrainbased,statgeo,Dong201285}, the function's parameters are landmark-specific, and are calibrated prior to geolocating clients. Calibration occurs by having each landmark measure \acp{RTT} to all other landmarks; \{RTT, distance\} pairs are then used to calibrate the mapping function.\footnote{The ``\emph{distance}" element is the geographic distance between a pair of landmarks.} The fundamental difference between delay-based geolocation techniques in the literature lies in the proposed mapping function. After mapping delays to distances using this function, these distances are used to calculate the client's location using, e.g., multi-lateration. 
Each landmark in \ac{CBG} \cite{Constrainbased} calibrates a linear delay-to-distance function called the \emph{best line}. On a graph where the $x$-axis is the distance (in km) and the $y$-axis is the \ac{RTT} (in milliseconds), the authors of \ac{CBG} define the best line to be the one ``closest to, but below, all data points ($x$, $y$) and has a non-negative intercept"\footnote{The \emph{intercept} is the intersection with the $y$-axis.} \cite{Constrainbased}. After calibration, each landmark measures the \ac{RTT} to the client, and maps it to distance using the best line function. The client's location is then estimated as the centroid of the intersection of circles whose centers are the landmarks and radii are the distances. The authors of \ac{CBG} later proposed a mechanism to estimate and remove delays caused by buffering of the message along the route between landmarks and the client, resulting in a more accurate mapping to distances \cite{geoBUD}.

Dong {\it et al.} \cite{Dong201285} proposed to cluster the \{RTT, distance\} coordinates of the landmarks into $k$ clusters. The coordinates in each cluster are then fitted to a polynomial function, which is then used by the landmark to map delays to distances. Such a segmented polynomial approach makes use of the observation that delay-to-distance ratios vary according to the spanned geographic distance \cite{Dong201285}. 
Youn {\it et al.} \cite{statgeo} proposed to apply kernel density estimation in the calibration phase to approximate the \ac{PDF} of delays with respect to distances, while 
Arif {\it et al.} \cite{maxliklihood} used maximum likelihood estimation \cite{kay1998fundamentals}. A Naive Bayes technique for delay-distance calibration was also considered \cite{LearningBased}.

Laki {\it et al.} \cite{Laki20101490} proposed to calibrate one delay-to-distance mapping function across all landmarks, hypothesizing that the relationship between delays and distances is not landmark-dependent. By doing so, delay measurements from all landmarks were combined together to generate such a global mapping function.

GeoWeight \cite{Arif:2010:GIH:1862199.1862209} is another example geolocation technique that works by calculating a \emph{weight} factor reflecting the client's presence inside a region, for some regions that are determined in realtime. Weights are calculated as the number of overlapping circles in that region. Recall, a \emph{circle} at a landmark is one that has the landmark's location as its center and the estimated distance between the landmark and the client as its radius.

Eriksson {\it et al.} \cite{Eriksson:2012:PLA:2381056.2381058} devised a lightweight geolocation technique that can reduce the number of required probing messages, yet achieve comparable geolocation accuracy. Similar to GeoPing, Eriksson {\it et al.} \cite{Eriksson:2012:PLA:2381056.2381058} rely on delay vectors between a group of passive monitors and landmarks, while leveraging likelihood estimation.

\subsection{Topology-aware techniques}
\label{topologyaware}

Topology-aware geolocation techniques leverage the network topology to generate a richer set of constraints, compared to those of delay-based techniques, to more accurately geolocate clients. Intermediate systems between the client and the landmarks are iteratively geolocated using single-hop delay-based analysis. The increased accuracy comes at the cost of  longer geolocation time and more required resources.

Katz-Bassett {\it et al.} \cite{delayandtopology} proposed to use \textsl{traceroute} measurements from the landmarks to the client in order to identify the network topology. They devised some techniques to refine their topology identification, such as detecting multiple device interfaces and using \ac{DNS} LOC records \cite{rfc1876}. The authors then use the network topology combined with the constraints of the speed of traffic propagation in fiber \cite{speedoflight} to geolocate the client.

Similar proposals involved leveraging negative constraints to enhance the geolocation accuracy \cite{Octant}. In contrast to the regular (positive) constraints, negative constraints exclude regions where the client cannot be present at, i.e., based on the delay measurements.

Others have proposed to leverage large numbers of \emph{passive landmarks} with known locations to further enhance the accuracy \cite{minging}. A passive landmark is usually a public server that responds to \ac{ICMP} queries, e.g., \textsl{ping} or \textsl{traceroute},  but is not under control of the \ac{LSP} or the geolocation service provider---neither can conduct measurements originating from it. After populating a table of thousands of passive landmarks, Wang {\it et al.} \cite{streetlevel} combined delay-based with topology-aware techniques to constrain the region where the client is. Their technique then returns the location of the nearest (delay-wise) passive landmark to the client as the client's location. This last step makes use of the \emph{closest-shortest rule}, which states that shorter delays tend to result from smaller distances \cite{6197179}.

\subsection{Client self-geolocation}
In this geolocation category, the client determines its own location and informs the \ac{LSP}. The client may have determined its location using, e.g., its \ac{GPS}. Another abstract example under this category is having the \ac{LSP} simply asking the (human) user to input its location, e.g., in an \textsl{address} field on the \ac{LSP}'s website \cite{MuirPaul}. Note that, regardless of how the client determines its location, we place all class of techniques by which the client sends location information to the \ac{LSP} under the \emph{self-geolocation} category even if the geolocation method involves, e.g., delay measurements (i.e., similar to those reviewed in Sections \ref{ipgeosection} and \ref{topologyaware}). 

Commonly used over the Internet, the \ac{W3C} geolocation \ac{API} \cite{w3capi} defines an interface that allows the client's web browser to determine and return the client's location to the requesting \ac{LSP}. Browser vendors usually rely on common location-determination technologies, such as \ac{GPS} \cite{gpsgps} or \ac{WPS}\footnote{In \ac{WPS}, a device's location is determined relative to the wireless access points. The device's \ac{NIC} reports a list of visible access points and their signal strengths (reflecting the distance between the \ac{NIC} and the access point) to a ``location provider". Location providers manage lookup tables that map access points to their corresponding geographic coordinates. The location provider calculates the location and informs the browser.} \cite{TGIS:TGIS1152}. Because the client sends its location to the \ac{LSP}, it can submit false information before submitting it \cite{Geolocater}.

\subsection{Inference-based Approaches}
\label{Inferencebasedapproaches}
In this class of approaches, the client's location is inferred from observations of its transmitted data \cite{MuirPaul}. For example, a time zone in a \ac{HTTP} packet header being UTC+12 likely indicates that the client is in New Zealand. If the Chinese language was set in the \texttt{Accept-Language} header, the party is likely to be from China. If the party's domain name ends in \texttt{fr}, it is likely in France. Even the preferred encoding of data gives an insight about the possible locations. 

The client's location could also be inferred from its IP address \cite{Tabulation}. The IP address is used to consult geolocation service providers that maintain lookup tables mapping IP addresses to locations \cite{studyDB}, e.g., MaxMind \cite{maxmind} and HostIP \cite{hostip}. However, such tabulation-based techniques were found unreliable \cite{unreliable}.

\section{Vulnerabilities of Internet Geolocation}
\label{vulnsinbackgroundall}

As this section discusses attempts whereby a client is motivated to forge its own location, the \emph{client} is referred to as the \emph{adversary}.

Asking the adversary to calculate its location and inform the \ac{LSP} of that location enables the adversary to provide misleading information about its location, provided that no additional verification mechanism is employed. In this case, the adversary may not only misrepresent its location, but also accurately control the location where it claims to be at. A verification mechanism could, for example, be to use \ac{TPM} chips \cite{Liu:2012:SAT:2307636.2307670} to trust \ac{GPS}-calculated coordinates. However, location coordinates obtained from a \ac{TPM}-supported \ac{GPS} driver may still be vulnerable to the Cuckoo attack \cite{parno2008bootstrapping}, where an adversary colludes with a remote party having a \ac{TPM}-supported \ac{GPS} to fake the adversary's location.

Against inference-based approaches, the adversary can alter information that indicates its true location \cite{MuirPaul}, misleading the \ac{LSP} into calculating the adversary's presence in the forged location. For example, changing the browser-requested language from $ja$ to $it$ may cause the \ac{LSP} to believe the adversary is in Italy instead of Japan. The adversary's control over the forged location in this case depends on the information used to determine the location.

\subsection{IP-hiding Attacks}

IP geolocation techniques, whether they are measurement- or tabulation-based, are prone to being mislead using \acp{MB} such as proxy servers, \acp{VPN}, anonymizers \cite{dingledine2004tor} or similar IP-hiding technologies. 
We believe that an adversary motivated to misrepresent its location would easily adopt any such technologies, especially given the wide availability of public \ac{VPN}-service providers. A number of these public anonymizers are even available free of charge \cite{hotspot}. As such, a fundamental design goal in any Internet location verification mechanism is to address such a well-known evasion tactic, as we do by the mechanism introduced in Chapter \ref{ch:cpv}.

\acp{MB} tend to alter transport-layer headers and/or react differently to \ac{ICMP} messages, compared to (non-\ac{MB}) end-systems \cite{detal2013revealing}. To detect a \ac{MB}, the provider and the client typically exchange especially-crafted packets and notice unexpected changes on the other end \cite{Honda:2011:SPE:2068816.2068834}. Due to such considerable client cooperation requirement between the two parties, these techniques cannot be implemented by an \ac{LSP} to detect \acp{MB} before geolocating an adversary by its IP addresses.

Attempts to enumerate the IP addresses of \acp{MB} and block them do not ensure their detection due to the dynamic behavior of IP addresses assignment \cite{Xie:2007:DIA:1282380.1282415}, and the risk of falsely blocking IP addresses that are not associated to \acp{MB} \cite{shroud}.

\subsection{Delay-adding Attacks}

Gill {\it et al.} \cite{Dude} analyzed adversarial location-forging abilities when the adversary increases delays to evade a measurement-based geolocation. The authors \cite{Dude} explored the case where the adversary injects delays by not responding to echo-request messages promptly. They found that, although the adversary was able to misrepresent its location, it had little control over the forged location. Additionally, the authors \cite{Dude} found that such adversarial manipulations are better detected when the adversary attempts to fraudulently place itself farther away from its true geographic location.

Gill {\it et al.} \cite{Dude} also tested a more sophisticated adversary that has control over a full network (such as an \ac{AS}-owner or a cloud provider), not just the device it owns. They found that such an adversary can misrepresent its location more accurately against topology-aware techniques, than delay-based ones. This adversary was tested to model a cloud provider \cite{Dude}.

\subsection{Delay-shortening Attacks}
\label{delayshortening}

One of the findings of this thesis is that common \ac{ICMP}-based delay-measurement tools allow an adversary to fully manipulate, i.e., increase and \emph{decrease}, the delays observed by the measuring party, which is made possible due to the lack of integrity in these tools. 

Because those \ac{ICMP}-based tools are commonly used in measurement-based geolocation, full delay manipulation not only allows an adversary to misrepresent its location but also gives the adversary substantial control over the forged location, compared to the evasion tactic proposed by Gill {\it et al.} \cite{Dude}. In such case, susceptibility to evasion stems from the reality that \ac{ICMP}-based delay-measurement tools were not designed for adversarial environments.

We explain this attack in details in Chapter \ref{ch:attack}, where we also propose strategies by which an adversary can increase its control over the location which a geolocation technique perceives to be the adversary's actual location.

\section{Location-verification}
\label{background:locveri}

\subsection{Single-hop wireless networks}
\label{background:wireless}

Verifying the proximity of two devices to each other using delays has been well studied in contexts other than the Internet, such as single-hop wireless networks, e.g., \acp{RFID} and \acp{WSN} \cite{Barbeau}. Brands and Chaum \cite{distboundingp} have proposed a \ac{RF}-based distance bounding protocol that aims at proving an upper bound to the distance between a prover and a verifier. To address the high sensitivity to processing delays and the complexity of achieving highly accurate clock synchronization among the verifiers, Wagner {\it et al.} \cite{wagner} proposed an ultrasound-based approach using a prover and a group of verifiers. Capkun {\it et al.} \cite{1498470} emphasized the importance of having at least three verifiers surrounding a prover to account for delay-adding attacks introduced by the prover or a third party attacker. 
The nature of delays over the Internet differs from those in single-hop wireless networks. Internet delays alleviate some of the challenging problems in the single-hop wireless context (e.g., less sensitivity to processing delays), but introduce new challenges (e.g., stochastic queueing delays due to traffic/route uncertainty \cite{Dong201285}). Thus, proximity verification in single-hop wireless networks is a distinct research problem from the one addressed in this thesis, since our focus is on Internet location verification.

\subsection{Privacy-Preserving Location-Proof Architectures}

Delay-based techniques for single-hop wireless networks (see Section \ref{background:wireless}) are often leveraged in the literature to design \emph{location proof architectures}, also sometimes referred to as a \emph{spatial-temporal attestation service} \cite{SurveyLocAuth}, which enable users to obtain proofs of their presence in a certain location \cite{gonzalez2008guaranteeing}. The proof is often in the form of a certificate, where a \emph{trusted} party in the client's vicinity is available to certify its presence \cite{slowpaper,Lin2014484}.

Saroiu {\it et al.} \cite{locationproofs} proposed a location-proof architecture in which a user gets cell towers or \ac{WiFi} access points to certify that the user is present where it claims to be at. VeriPlace \cite{Luo:2010:PYL:1734583.1734586,veriplace} is another architecture that addresses \emph{wormhole} attacks,\footnote{A wormhole is a relaying attack in wireless networks, where an adversary encapsulates bytes at one location, relays and decapsulates them at another location in the network \cite{1209219}.} while focusing on the users' privacy by employing cryptographic techniques to spread user's identification credentials across different entities. 
Other proposals involved decentralizing the trusted infrastructure \cite{gambs2014props}, or replacing it with Bluetooth-based devices in the vicinity of the client \cite{applaus}, or with \ac{NFC} tags \cite{acsaclbs}. Zerosquare \cite{zerosquare} is a privacy-preserving \emph{location hub}, which allows location-based services to query the users' locations, while regulating access to personal information by separating user information from their location.

The threat of a compromised infrastructure has been addressed as well, where Khan {\it et al.} \cite{whowhenwhere} proposed to use additional (trusted) witnesses in the user's vicinity to verify assertions in the presence of untrusted access points/location managers. The principle of verified location tracking has been explored as well, where a chain of location proofs can represent the history of a user's locations \cite{Khan:2014:OTS:2590296.2590339,stubblebine}.

\paragraph{Relationship to the applications addressed by this thesis} 

Compared to the Internet location-verification problem we address in this thesis, location-proof architectures target a problem with more constraints (e.g., preserving users' privacy and verifying locations with high granularity). Thus, they address a different class of applications than the ones addressed by this thesis.

The applications addressed by location-proof architectures provide the advantage of being privacy-centric; users are assumed to be unwilling to share (or publicly disclose) the credentials identifying them to an \ac{LSP}. Such assumption must hold; otherwise, a user who wants to forge their location may send their credentials to a colluding party to get them bound to a remote location and endorsed by an access point at that location. As stated earlier in Chapter \ref{ch:intro}, we assume no client credentials playing that role in the applications addressed in this thesis.

There are many cases where users could be unwilling to disclose their credentials. For example, doing so may reveal private (location) information such as regular hospital visits. Additionally, sharing identification credentials may threaten losing benefits associated with these credentials \cite{6520850}. Foursquare \cite{foursquare} for example enables coffee shops to reward users, identified by their personal (secret) credentials, when they visit regularly; a user sharing their credentials with a remote colluding party may risk having their rewards lost/stolen.
 
Another fundamental difference between the applications addressed by location-proof architectures and the ones addressed in this thesis lies in the verification granularity and trustworthiness of infrastructure. The granularity addressed by location-proof architectures is very high, e.g., verifying the presence in a hospital ward or inside a coffee shop. Such high granularity is made possible since wireless devices in the user's vicinity are typically relied upon for location verification. To achieve location verification at a global-level, we expect the high granularity of location-proof architectures would come at the cost of large-scale trustworthiness requirements, since a sufficient number of trusted wireless (endorsing) devices must be present to cover geographic regions at a high granularity. In contrast, the location-verification granularity addressed in this thesis is coarser (e.g., state- or country-level) as we explain in Chapter \ref{ch:cpv}, and the required trusted infrastructure is therefore smaller (see Chapters \ref{ch:cpv} and \ref{ch:wiredecva} for details).

\chapter{Accurate Manipulation of Delay-based Internet Geolocation}
\label{ch:attack}

Numerous delay-based Internet geolocation techniques have been proposed in recent years, and are repeatedly positioned as well suited for security-sensitive applications (e.g., location-based access control, credit card verification). Previous literature \cite{Dude} showed that an adversary simply delaying response messages to increase measured \acp{RTT} gains only limited location control in forging its location, and decreasing \acp{RTT} was believed to be infeasible. In contrast, herein we show that indeed an adversary can decrease \acp{RTT} arbitrarily because commonly-used \ac{ICMP}-based utilities are not intended to provide delay-measurement integrity, and explore how an adversary can leverage this to \textit{accurately} manipulate geolocation results. Using several adversarial models, we evaluate (on three delay-based geolocation techniques) how selectively combining this with delay increases can achieve surprisingly high adversarial accuracy in forging location---e.g., modeled adversaries can fraudulently misrepresent their true location by over 15,000 km, some within 100 km of their intended (fraudulent) target location. Thus the new ability to decrease delays, combined with previous delay-increasing tactics, enables significantly greater adversarial location control over previous methods.

\section{Introduction}

The recent proliferation of \acp{LBS} in the Internet has highlighted the requirement for reliable and accurate Internet geolocation tools. Some of these services employ location-based access policies \cite{Bertino:2005:GSA:1063979.1063985}, or restrict operations by clients' geographic locations \cite{trimble2011future}. Examples include media streaming \cite{burnett2013geographically}, online voting/gambling, location-based social networking \cite{acsaclbs}, and fraud prevention \cite{shroud}. Nanjee \cite{nanjee} is one example that provides commercial geolocation services based on active network measurements \cite{streetlevel}. Tabulation-based IP geolocation service providers maintain lookup tables that map IP addresses to locations. Studies have found that many of the major tabulation providers (e.g., MaxMind \cite{maxmind} and HostIP \cite{hostip}) are inaccurate/outdated \cite{studyDB,unreliable} and evadable \cite{MuirPaul}.

IP geolocation techniques that rely on active network (delay) measurements have accuracy advantages over others, e.g., tabulation-based \cite{Tabulation}. They are also resilient to security vulnerabilities that other techniques suffer, such as clients submitting false location information \cite{streetlevel,MuirPaul}. Although IP geolocation is often evadable using \acp{VPN} and similar IP-hiding technologies \cite{MuirPaul}, such technologies can be detected \cite{shroud} and thwarted \cite{commletters}; some \acp{LBS} (e.g., Hulu \cite{hulu}) have recently started employing these practices \cite{huluvpn,netflixvpn2}. For these reasons, delay-based techniques are gaining increasing community support \cite{improvacc,ErikssonCrovella:Infocom13}, specifically advocated \cite{unreliable}, and repeatedly positioned as well suited for security-aware contexts, e.g., ensuring legitimate storage of data in the cloud \cite{gondree2013geolocation}, or locating hidden servers \cite{Castelluccia:2009:GPS:1644893.1644915}.

Since 2001, more than 10 delay-based IP geolocation techniques have been proposed \cite{GeoTrackGeoPing,GeoPing2,Constrainbased,statgeo,minging,maxliklihood,LearningBased,Arif:2010:GIH:1862199.1862209,
laki2011spotter,Dong201285,Eriksson:2012:PLA:2381056.2381058,6197179}. The \ac{RTT} is measured between the client and a set of landmarks with known locations, and the client's location is estimated relative to the landmarks. Delay-based geolocation techniques require some way of measuring delays; because \ac{ICMP} \cite{rfc792} utilities (e.g., \textsl{ping} and \textsl{traceroute}) are ubiquitous and facilitate delay measurements, they are commonly used for that purpose \cite{statgeo,Dong201285}. 
In 2010, Gill {\it et al.}\ \cite{Dude} studied the ability of an adversary to distort geolocation techniques that are based on active delay measurements, and the capability of the geolocating party to detect circumvention. For delay-based techniques, their analysis considered an adversary that can only increase the observed \acp{RTT} by selectively delaying response messages. Their modeled adversary succeeded to misrepresent its location, but was limited to a coarse control over the forged location.
In this chapter, we show that common delay-measuring utilities used by geolocation techniques are subject to (1) modifying and/or (2) predicting packet contents; this enables an adversary to fully manipulate, i.e., increase and decrease, the observed \acp{RTT}. For example, \ac{GNU}'s implementation of \textsl{ping} relies on the \ac{ICMP} echo request/reply protocol, and records the packet-creation time in the \textsc{data} field of the \ac{ICMP} packet \cite{GNUping}. The \ac{RTT} is then calculated by subtracting this timestamp, as read from the echoed packet, from the time the echoed packet was received, expecting the client to return the data unchanged. However, an adversarial client seeking to manipulate the geolocation mechanism can alter the timestamp in the \textsc{data} field before echoing the packet. We show that using this, an intelligent adversary can selectively increase or decrease the observed \ac{RTT} as necessary to beneficially control the calculated location.

Upon being able to manipulate delays, the adversary faces the question: What should the \ac{RTT} between each landmark and the adversarial client be such that the geolocation process calculates the adversary's intended location? We propose different strategies whereby an adversary can utilize information known about the landmarks to answer this question. We model several adversaries adopting different strategies, and compare their accuracy in forging location.

\begin{figure}\centering
\subfigure[Attempted distance on \ac{CBG}= 15,092 km. Dist error = 70.7 km.]
{
\label{example1label}
\includegraphics[scale=0.7]{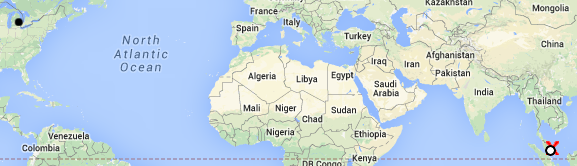}
}
\subfigure[Attempted distance on GeoPing = 8,055 km. Dist error = 71.4 km.]
{
\label{example2label}
\includegraphics[scale=0.7]{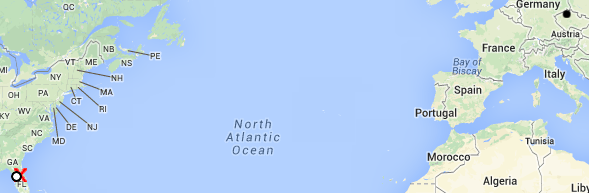}
}
\subfigure[Attempted distance on SegPoly = 6,617 km. Dist error = 75.2 km.]
{
\label{example3label}
\includegraphics[scale=0.7]{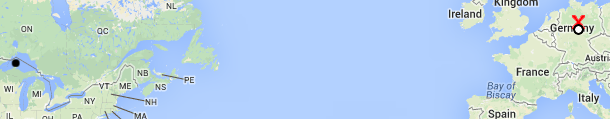}
}
\caption[Adversarial capabilities in forging geographic locations]{Examples of adversarial capabilities after exploiting properties of common delay-measuring utilities. $\bullet$ = true location of adversary; \textcolor{red}{$\times$} = intended location of adversary; \textbf{$\circ$} = locations calculated by (a) \ac{CBG} \cite{Constrainbased}, (b) GeoPing \cite{GeoTrackGeoPing}, and (c) SegPoly \cite{Dong201285}; \emph{attempted dist} is that between $\bullet$ and \textcolor{red}{$\times$}; \emph{dist error} for the adversary is that between \textcolor{red}{$\times$} and \textbf{$\circ$}. Map data: Google, INEGI, Basarsoft.}
\label{forgingexamples}\end{figure}

To study the effectiveness of the modeled adversaries, we implemented three delay-based geolocation techniques, \ac{CBG} \cite{Constrainbased}, GeoPing \cite{GeoTrackGeoPing} and segmented polynomial ({\it SegPoly} for short) \cite{Dong201285}, and evaluated adversarial location-forging accuracy, given the ability to fully manipulate delays. Some adversaries modeled obtained forged locations with \emph{distance errors} (defined as the distance between the adversary's intended location and the one calculated by the geolocation technique) below 100 km; this relatively fine-grained location control was possible even for some who attempted fraudulent relocation more than 15,000 km from their true locations (see Fig.~\ref{forgingexamples}). 

Our work highlights the need for integrity of timing information when relied upon by security-sensitive applications. Contributions:
\begin{enumerate}
\itemsep0em
\item We show how properties of common \ac{ICMP}-based delay-measuring utilities allow an adversary to both increase and decrease measured delays.\item We demonstrate several strategies that enable an adversary, exploiting these properties, to accurately forge the location calculated by delay-based geolocation techniques.
\item We evaluate the manipulation effectiveness to three techniques: \ac{CBG} \cite{Constrainbased}, GeoPing \cite{GeoTrackGeoPing}, and SegPoly \cite{Dong201285}. This demonstrates how powerful an adversary can be upon being able to fully manipulate \acp{RTT}.
\end{enumerate}

The rest of this chapter is organized as follows. Section \ref{ICMPexplain} reviews common delay-measurement utilities. Section \ref{ICMPattack} explains how \acp{RTT} can be fully manipulated, i.e., increased and decreased. The adversarial models are defined in Section \ref{attack:threatmodel}. Section \ref{attackdelbased} analyzes the effect of manipulating \acp{RTT} on delay-based geolocation. Section \ref{impactfull} compares location-forging abilities of different adversarial models. Section \ref{countermeasures} suggests countermeasures. Section \ref{attack:relatedwork} discusses related work, and Section \ref{attack:conclusion} concludes.

\section{Background: RTT Measurement Using Common ICMP-based Utilities}
\label{ICMPexplain}

A \emph{sender} can measure \acp{RTT} between itself and a \emph{receiver} by having the receiver respond to (special) packets of the sender, and timing these responses. Assuming the sender issues these packets to the receiver every $t$ ms, and the first one was created at time $T$, then the sender's system time when packet $i$ was created, for all packets $i\geq0$ is:
\begin{equation}\label{issuedtimes}s_i = T + i\cdot t\end{equation}
If the packets take $\gamma_1$ ms one-way delay from the sender to the receiver, they reach the receiver at times:
\begin{equation}\label{arrive}
m_i = s_i + \gamma_1 = T + i\cdot t + \gamma_1\\
\end{equation}
Assuming the receiver responds promptly, if packets take $\gamma_2$~ms one-way delay from the receiver back to the sender, the responses arrive at times:
\begin{equation*}
r_i	= m_i + \gamma_2 = T + i\cdot t + \gamma_1 + \gamma_2\\
\end{equation*}
The sender calculates the \ac{RTT} for packet $i$ as:
\begin{equation}\label{mainOne}\textsl{RTT}_i = r_i-s_i = \gamma_1 + \gamma_2\end{equation}

To measure \acp{RTT}, network utilities commonly use the \ac{ICMP} protocol \cite{rfc792},\footnote{Some utilities, such as \textsl{tcptraceroute} \cite{tcptraceroute}, rely on \ac{TCP} messages.} as it is implemented by default in most systems' protocol stack. An \ac{ICMP} packet gets wrapped by an IP packet for delivery. Eleven \ac{ICMP} types are specified by \ac{RFC} 792. The type is indicated by the \textsc{type} field of an \ac{ICMP} header. \emph{Echo-request/reply}, types 8 and 0 respectively, and \emph{destination-unreachable}, type 3, are the \ac{ICMP} options commonly used to measure \acp{RTT}. The \ac{RFC} does not specify a mechanism to calculate \acp{RTT} for either types \cite{rfc792}.

\begin{figure}
\centering
\subfigure[]
{
\label{icmpHeaderecho}
\includegraphics[scale=0.52]{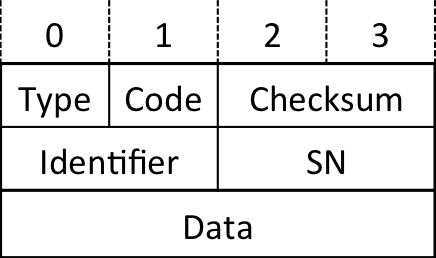}
}\hspace{5pt}
\subfigure[]
{
\label{icmpHeaderdest}
\includegraphics[scale=0.52]{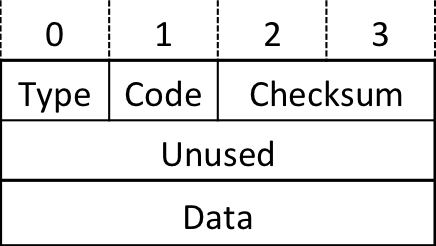}
}\hspace{5pt}
\subfigure[]
{
\label{udpheader}
\includegraphics[scale=0.52]{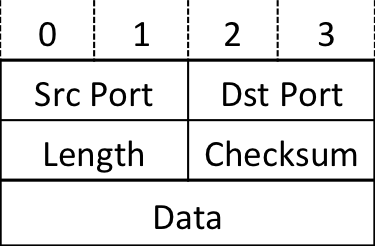}
}\hspace{20pt}
\subfigure[]
{
\label{ipv4header}
\includegraphics[scale=0.5]{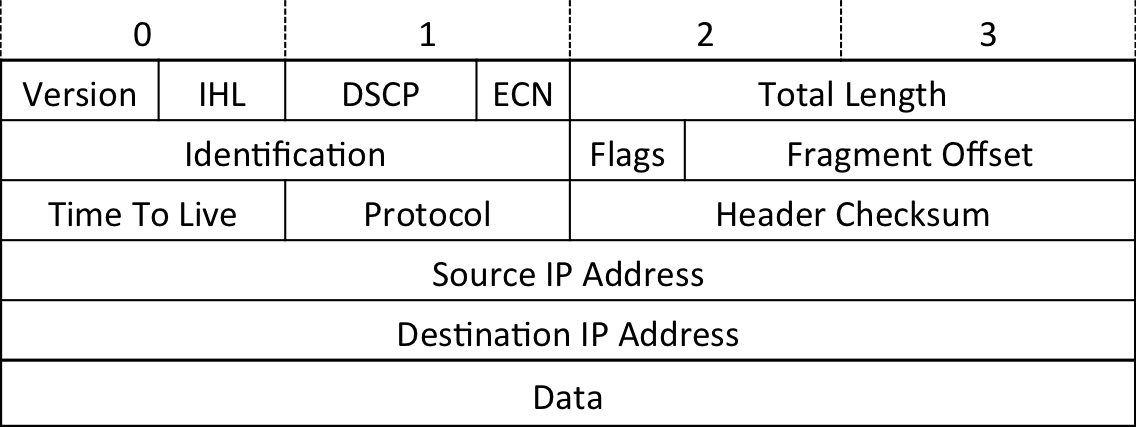}
}
\caption[Headers of protocol data units]{(a) \ac{ICMP} echo-request/reply packet format; (b) \ac{ICMP} destination-unreachable packet format; (c) \ac{UDP} segment format; (d) IPv4 packet format.}
\label{bothIPDUPheaders}
\end{figure}

\subsubsection{Echo-request/reply}Echo-request and echo-reply share the same message format, which is shown in Fig.~\ref{icmpHeaderecho}. To construct an echo-request, the sender sets the \textsc{type} and \textsc{code} fields to 8 and 0 respectively, chooses two 16-bit values for the \textsc{identifier} and \textsc{sequence number} fields, and finally (after filling the \textsc{data}) calculates the checksum and places it in its field. Values in the \textsc{data}, \textsc{identifier} and \textsc{sequence number} fields are up to the implementer, however the latter two may aid in matching requests with replies as specified by the \ac{RFC} \cite{rfc792}. One of the most commonly used network utility that implements the echo-request/reply type of the \ac{ICMP} protocol is \textsl{ping}. We found that many, if not all, \textsl{ping} implementations on Linux, \ac{BSD} and Mac \ac{OS} place the \ac{PID} of the issuing process in the \textsc{identifier} field. The \textsc{sequence number} field usually starts by either 0 or 1, and is incremented by 1 in each subsequent \textsl{ping} message \cite{GNUping,BSDping,Macping}. When the receiver gets an echo-request message, it should change the \textsc{type} field to 0, recalculate the checksum and echo the packet. According to \ac{RFC} 792, ``the data received in the echo message must be returned in the echo-reply message" \cite{rfc792}. However, the \ac{RFC} does not specify a mechanism to ensure the receiver behaves as described. That is, providing integrity checking is not stated as a requirement. And, of course, attackers feel no particular obligation to follow \acp{RFC}.

To calculate the \ac{RTT} using the echo-request/reply options, two common implementations exist: \emph{stateless} and \emph{stateful}. In the stateless implementation, the sender places the timestamp $s_i$ (packet-creation time) in the \textsc{data} field of the \ac{ICMP} packet. 
When the echo-reply is received, the sender observes the receiving time $r_i$, reads $s_i$ from the echoed packet, and uses them to calculate the \ac{RTT} using (\ref{mainOne}). Other examples of such stateless implementation include, but not limited to, \textsl{ping} on FreeBSD \cite{BSDping} and Mac \ac{OS} \cite{Macping}. 

In the stateful echo-request/reply implementation, the sender records $s_i$ in its local memory. The \ac{RTT} is calculated also using (\ref{mainOne}), but reading $s_i$ from the sender's local memory instead of the echo-reply packet. Examples of this stateful implementation that use the echo-request/reply options include \ac{GNU}'s \textsl{traceroute}---\ac{ICMP} option (i.e., \texttt{traceroute -I <host>}) \cite{GNUtrace}, and \textsl{hping3}---\ac{ICMP} option (i.e., \texttt{hping3 -1 <host>}) \cite{hping3}. These stateful utilities commonly fill the \textsc{data} field using a fixed predefined pattern, e.g., all zeros, a list of sequential ASCII characters, or hard-coded strings. 
\subsubsection{Destination Unreachable}To calculate the \ac{RTT} using the destination-unreachable option, the sender creates a \ac{UDP} segment and sends it to the receiver, with a destination port unlikely to be open on the receiver's machine. The sender records $s_i$ in its local memory. As with the \textsc{data} field of the echo-request/reply type explained above, the data of this \ac{UDP} segment is usually filled with fixed predefined patterns \cite{BSDtrace,Mactrace}. If the port was actually closed, the receiver is expected to respond with an \ac{ICMP} destination-unreachable message \cite{rfc1122}; when the sender receives it, the sender records $r_i$, and calculates the \ac{RTT} using (\ref{mainOne}). Utilities implementing this behavior are commonly stateful, $s_i$ is recorded locally, because the receiver is not echoing an exact copy of the sender's packets. \ac{GNU}'s \textsl{traceroute} is an example employing this implementation through its (default) \ac{UDP} probes \cite{GNUtrace}.

The destination-unreachable message format is shown in Fig.~\ref{icmpHeaderdest}. To construct its header, the receiver sets the \textsc{type} and \textsc{code} fields to 3 and 3 respectively, fills the 32-bit \textsc{unused} field with zeros, and finally (after setting the \textsc{data} field) calculates the checksum and places it in its field \cite{rfc792}. To enable the sender match responses with their corresponding processes, the receiver places the IP header and the first 8 payload bytes of the IP packet it received from the sender in the \textsc{data} field of the destination-unreachable message \cite{rfc792}.

\section{Manipulating Latencies}
\label{ICMPattack}

From Section \ref{ICMPexplain}, the cases whereby a sender can measure \acp{RTT} using \ac{ICMP} are:
\begin{itemize}
\itemsep0em
\item Case 1: stateless using echo-request/reply.
\item Case 2: stateful using echo-request/reply.
\item Case 3: stateful using destination-unreachable.
\end{itemize}
Table~\ref{vulns} lists potentially-exploitable properties of common \ac{ICMP}-based network utilities. These properties become vulnerabilities when the measured \acp{RTT} are relied upon by security-sensitive applications; because we investigate the effect of using \ac{ICMP}-based utilities in security-sensitive geolocation purposes, we refer to the properties in Table~\ref{vulns} as \emph{vulnerabilities}. The table also shows which of the \ac{RTT}-measurement cases listed above has which vulnerability. Note that despite having the same effect on \acp{RTT}, the first and second vulnerabilities in Table~\ref{vulns} increase \acp{RTT} in a different way; likewise, the second and third decrease differently. We now discuss how an adversary can increase/decrease the \acp{RTT} when it exploits the corresponding vulnerabilities in each case.

\begin{table}
\centering
\caption[The effect of exploiting properties of common ICMP-based utilities]{Properties of \ac{ICMP}-based utilities, and the effects of exploiting them on the observed \acp{RTT}. A bullet ($\bullet$) in column 3 means Case $i$ has property $j$.}
\scalebox{0.75}{
\begin{tabular}{ r c | c c | C{7pt}C{7pt}C{7pt} | l }
\multicolumn{2}{c|}{\multirow{2}{*}{Property}} & \multicolumn{2}{c|}{Effect} & \multicolumn{3}{c|}{Case} & \multirow{2}{*}{Discovered}\\
&& $\uparrow$ RTT & $\downarrow$ RTT &1&2&3&\\\hline
1&Suspendable responses		&\checkmark	&			&$\bullet$&$\bullet$&$\bullet$&	\cite{Dude}\\
2&Modifiable pkt contents	&\checkmark	&\checkmark	&$\bullet$&&&	herein\\
3&Predictable pkt contents	&			&\checkmark	&&$\bullet$&$\bullet$&	herein\\
\end{tabular}
}
\label{vulns}
\end{table}

Exploiting the first vulnerability in Table~\ref{vulns} enables the adversary increase \acp{RTT} in all three cases, because the adversary needs only hold on to the response messages to increase the \acp{RTT} \cite{Dude}. Decreasing \acp{RTT}, in each case, is achieved as follows.\\

\noindent{\bf Case 1.} The packet-creation time, $s_i$, is recorded in the \ac{ICMP} echo-request in this case. To decrease \acp{RTT}, the adversary exploits the second vulnerability in Table~\ref{vulns}; it increases the value of $s_i$ before including it in the echo-reply. Changing $s_i$ to $s_i+\delta$ decreases the \acp{RTT} the sender observes by $\delta$. Using (\ref{mainOne}), the sender calculates the manipulated \ac{RTT} of packet $i$ as:
\begin{equation}\label{fakeZero}
\text{RTT}'_i = r_i - (s_i + \delta) = \textsl{RTT}_i - \delta
\end{equation}
\acp{RTT} can also be fraudulently increased by $\delta$ ms  by changing $s_i$ to $s_i-\delta$. If the adversary knows the actual \ac{RTT} between itself and the sender, it can mislead the sender into calculating the \ac{RTT} as a specific value of the adversary's choosing, $\tau$, by setting:
\begin{equation}\label{specifictau}\delta=\textsl{RTT}_i-\tau\end{equation}
causing the sender to calculate the manipulated \ac{RTT} as:
\begin{equation}\label{fakerttttt}
\text{RTT}'_i = r_i - (s_i + \delta) = \textsl{RTT}_i - \delta = \tau\\
\end{equation}

\noindent{\bf Case 2.} To decrease \acp{RTT}, the adversary first estimates the \emph{waiting} time, $t$ in (\ref{issuedtimes}), that the sender waits between sending echo-requests. Recall that delay-based geolocation techniques take multiple \ac{RTT} measurements to a client, and use the smallest in geolocation. To estimate $t$, the adversary refrains from responding to the first $n>1$ echo-requests, or drastically delays their responses to ensure none of them will be chosen as the smallest. It then subtracts the receiving time of the echo-request, $m_{i}$ in (\ref{arrive}), from $m_{i+1}$ for all $0\leq i<n-1$ (Section \ref{ICMPexplain}). Because the accuracy of this method depends on the stability of the one-way delay from the sender to the adversary, the adversary averages the waiting time over multiple packets:
\begin{equation}\label{case2first}t=\frac{1}{n-1}\sum_{i=0}^{n-2}(m_{i+1}-m_{i})\end{equation}

The adversary can then estimate the receiving time of the next echo-request packet as:
\begin{equation}m_{i} = m_{i-1}+t\end{equation}
To decrease the \ac{RTT} that the sender observes from packet $i$ by $\delta$ ms, the adversary issues an early (fake) echo-reply at times $m'_i$, instead of $m_i$, such that:
\begin{equation}\label{fakearrive}
m'_i = m_i-\delta = s_i+\gamma_1-\delta
\end{equation}
The sender will then receive replies at times $r'_i$, such that:
\begin{equation}\label{fakerectimmmmmmme}
r'_i	 	= m'_i + \gamma_2 = s_i + \gamma_1 - \delta + \gamma_2
\end{equation}
and hence, calculate the \ac{RTT} of packet $i$ as:
\begin{equation}\label{fakeOne}\text{RTT}'_i = r'_i-s_i = \gamma_1 + \gamma_2 - \delta\end{equation}
If the adversary knows the actual \ac{RTT}, it can use (\ref{specifictau}) to mislead the sender into calculating the \ac{RTT} as $\tau$.

Issuing early \ac{ICMP} echo replies requires the adversary to craft them before receiving their corresponding requests. Exploiting the third vulnerability in Table~\ref{vulns} enables the adversary achieve this because the values in the header of an echo-reply message, Fig.~\ref{icmpHeaderecho}, are highly predictable. When the sender receives an echo-reply (type 0, code 0), it only uses the \textsc{identifier} and \textsc{sequence number} fields to match them with corresponding requests; they are the only two fields an adversary needs to predict, before receiving them in echo-requests. The \textsc{identifier} (commonly being the \ac{PID} of the issuing process) is usually constant across echo-requests issued within the same session, the \textsc{sequence number} is usually 1 plus the previous echo-request \cite{GNUtrace,hping3,fping}. After receiving the first echo-request, which the adversary ignores, it predicts the values of those two fields for subsequent requests.\\

\noindent{\bf Case 3.} Similar to the previous Case, the adversary decreases \acp{RTT} by sending early (fake) destination-unreachable messages. Timing analysis is, thus, similar to that of Case 2. From the destination-unreachable header, Fig.~\ref{icmpHeaderdest}, we see that the \ac{ICMP} header constitutes no difficulties for the adversary to predict; the \textsc{type} and \textsc{code} fields are set to 3 \cite{rfc792}, the \textsc{unused} bytes must be set to 0 \cite{rfc792}, and the \textsc{checksum} is calculated after placing the data. Predicting the \textsc{data} field requires the adversary to predict the sender's IP header and the first 8 bytes of the IP payload. We found that given common implementations of \ac{ICMP}-based utilities, both headers are highly predictable after receiving the first \ac{UDP} segment from the sender.

For the IP header (Fig.~\ref{ipv4header}), the following fields are not expected to change across multiple packets issued within the same session: version, \ac{IHL}, total length, fragmentation bytes (flags+offset), protocol, and source and destination IP addresses. Fragmentation is likely to remain zero because \ac{UDP} segments are typically small in size; otherwise, they may distort the measurement \acp{RTT} due to additional processing and transmission delays of large packets. The protocol number will be set to 17, for \ac{UDP} \cite{rfc768}. The following fields are already prone to changes by intermediate systems (e.g., routers) \cite{rfc791}: \ac{DSCP}, \ac{ECN}, \ac{TTL}, and header checksum. Thus, the sender cannot rely on those fields to match the returned \ac{ICMP} messages to an issuing process (we noticed no utilities relying on them). For the remaining field, IP identification, most systems increment it by 1 in each subsequent IP packet. This summarizes the adversary's ability to predict the contents of the next IP header after receiving at least one.

The first 8 bytes of the IP payload constitute the \ac{UDP} header (Fig.~\ref{udpheader}). On many implementations, including the \textsl{traceroute} utility of \ac{GNU} \cite{GNUtrace}, FreeBSD \cite{BSDtrace}, and Mac \ac{OS} X \cite{Mactrace}, the source and destination port numbers are fixed over a single session, or incremented by one with each subsequent \ac{UDP} segment. Similar to the stateful echo-request utilities (Section \ref{ICMPexplain}), the \textsc{data} field of \ac{UDP} segments is commonly a fixed predefined pattern. However, we found that many utilities overlook the returned values in this field, as well as the returned \ac{UDP} header length and checksum; that is, only the \ac{UDP} source and destination port numbers are used to match destination-unreachable messages with corresponding \ac{UDP} segments.

\section{Adversarial Models}\label{attack:threatmodel}

\subsection{Common Capabilities}

The adversary is a client that tries to misrepresent its own location by manipulating geolocation. The adversary's objective is to have the technique return a location as close as possible to its intended location, rather than its true location. We consider a \ac{LSP} that uses a delay-based geolocation technique that relies on \ac{ICMP} messages to measure delays.\footnote{Note that the adversary can lead the \ac{LSP} to rely on \ac{ICMP}-response messages simply by filtering all \ac{TCP} ports, i.e., no \ac{TCP} response messages are sent on attempted connections to any port.} 

The adversary has full control over its own machine, but no other machines. It cannot influence the delay-distance calibration process of the landmarks (see Section \ref{ipgeosection}, page \pageref{ipgeosection}), nor infer their calibration functions. Note that the adversary is nonetheless a powerful one since, as shown below, the adversary can achieve high accuracies while manipulating geolocation, even lacking knowledge of those parameters. The adversary is able to selectively manipulate the delays between itself and any landmark, as explained in Section \ref{ICMPattack}, and accurately increase or decrease the \acp{RTT} observed by a measuring party. The adversary only knows the geographic locations of the landmarks, but does not know the \ac{RTT} between each landmark and its \emph{intended} location (where the adversary wants to appear to be, in terms of the result computed by the geolocation technique), nor between each landmark and its true location. We make the latter assumption because the landmarks may prevent anyone from \textsl{ping}ing their addresses except, perhaps, themselves.

\subsection{Strategies for Modeling Traffic Speed} 

\begin{table}
\centering
\caption{Capabilities and assumptions of 5 modeled classes of adversaries, their assumed traffic propagation speed, and where they are discussed.}
\scalebox{0.75}{
\begin{tabular}{ r | c c | ccc | c | c | l }
Adv. & \multicolumn{2}{c|}{Able to} & \multicolumn{3}{c|}{Knows} & Traffic & \multirow{2}{*}{Proposed} & \multirow{2}{*}{Section}\\
class & $\uparrow$ RTT & $\downarrow$ RTT &G & T & F & Speed & & \\\hline
\emph{A}&\checkmark	&\checkmark	&\checkmark	&			&			&(1/3){\bf c}	&herein		&\ref{attackdelbased}\\
\emph{B}&\checkmark	&			&\checkmark	&			&			&(2/3){\bf c}	&\cite{Dude}&\ref{impactfull}\\
\emph{C}&\checkmark	&			&\checkmark	&			&			&(1/3){\bf c}	&herein		&\ref{impactfull}\\
\emph{D}&\checkmark	&\checkmark	&\checkmark	&\checkmark	&			&Variable		&herein		&\ref{impactfull}\\
\emph{E}&\checkmark	&			&\checkmark	&			&\checkmark	&Variable		&\cite{Dude}&---\\
\end{tabular}}\\
\footnotesize
G = landmarks' locations; T = adversary-to-landmark \ac{RTT}; F = landmarks' calibrated delay-distance function; {\bf c} = speed of light.
\label{adversarysummary}
\end{table}

Let the adversary's true location be $a$, the set of landmarks be $L$, and the \ac{RTT} (at a given time) between the adversary's true location and each landmark $l\in L$ be $\alpha(a,l)$. To deceive a geolocation process, the adversary manipulates the \acp{RTT}, observed by each landmark $l\in L$, between itself and $l$. To forge its location to $a'$, the adversary ideally deceives each $l\in L$ to measure the \ac{RTT} as one that would be consistent with $\alpha(a', l)$ instead of $\alpha(a,l)$. The challenge for the adversary is that (by our assumption) it does not know both $\alpha(a,l)$ and $\alpha(a',l)$.

If the adversary guesses the speed of traffic propagation, it can estimate the \ac{RTT} (at current time) because it knows the distances between itself and the landmarks. The adversary may use the constant (2/3){\bf c} (i.e., speed of light in fiber \cite{speedoflight}, where {\bf c} is the speed of light in vacuum) as an estimate to the traffic propagation speed \cite{Dude}. However, Katz-Bassett {\it et al.}\ \cite{delayandtopology} found that a speed between (2/9){\bf c} and (4/9){\bf c} better reflects the one-way delay nature of the typically multi-hop Internet routes. We study the adversary's manipulation capabilities when it uses (3/9){\bf c} = (1/3){\bf c} as an approximation to the traffic propagation speed. The adversary's estimated \ac{RTT} between its true/intended location and landmark $l$ is:
\begin{equation}\label{equn10}\beta(a,l) = \frac{2\times dis(a,l)}{(1/3){\bf c}}\end{equation}
and
\begin{equation}\label{equn11}\beta(a',l) = \frac{2\times dis(a',l)}{(1/3){\bf c}}\end{equation}
where $dis(a,l)$ and $dis(a',l)$ are the great circle\footnote{A great circle is one whose center and radius are those of the Earth.} geographic distances \cite{australiageo} between the adversary's true/intended location and landmark $l$. The distance (in the numerator) is doubled because $\beta$ is a round-trip, rather than one-way, delay. To forge its location from $a$ to $a'$, the adversary sets $\delta$ in (\ref{specifictau}) as:
\begin{equation}\label{deltaeqn}\delta=\beta(a,l)-\beta(a',l)\end{equation}
The difference between the adversary's estimated \ac{RTT} and the actual contributes to the adversary's errors in forging location.

{\bf Adversary \emph{A}}. After evaluating this adversary, we compare its efficacy to three other classes of adversaries. We refer to the adversary described above as adversary \emph{A}. Adversaries \emph{B}, \emph{C} and \emph{D} have similar assumptions, except for the factors in Table~\ref{adversarysummary}. Note that in contrast to the \emph{achieved ability} in the second column of the table, the third column presents an \emph{assumed} knowledge.

{\bf Adversary \emph{B}}. Gill {\it et al.}\ \cite{Dude} studied the effect of an adversary increasing the \ac{RTT} observed by a measuring party by delaying response messages. To realize how much an adversary gains by also being able to decrease \acp{RTT} (as explained in Section \ref{ICMPattack}), we implemented the manipulation tactic of Gill {\it et al.} \cite{Dude},\footnote{The results obtained from our implementation closely match those reported by Gill {\it et al.} \cite{Dude}; we believe that any dissimilarities arise from differences in the data sets and the experimental environment.} which is equivalent to adversary \emph{B}, to compare it with adversary \emph{A}.

{\bf Adversary \emph{C}}. Similar to the modeled adversary of Gill {\it et al.} adversary \emph{B} uses (2/3){\bf c} to model traffic speed, whereas \emph{A} uses (1/3){\bf c}. To understand whether \emph{B}'s retrogressions/improvements over \emph{A} were due to its limited delay-manipulation abilities or its parameterization, we involve in the discussion adversary \emph{C} which only differs from \emph{B} in that it uses (1/3){\bf c} as the traffic speed.

{\bf Adversary \emph{D}}. We assume this adversary has the advantage of knowing the \ac{RTT} between itself and each $l\in L$, $\alpha(a,l)$ (e.g., by \textsl{ping}ing each $l\in L$). However, it does not know the \ac{RTT} between the landmarks and its intended location. To estimate it, \emph{D} benefits from its knowledge of $\alpha(a,l)$, and calculates the traffic speed between itself and each $l\in L$ as follows:
\begin{equation}\label{speedDynamic2}\lambda_l = max\left(\frac{2\times dis(a,l)}{\alpha(a,l)},(2/9){\bf c}\right)\end{equation}
The calculated speed $\lambda_l$ reflects $l$'s access network speed; it increases with fast access network, and decreases otherwise. Since $l$'s calibrated delay-to-distance function (see Section \ref{ipgeosection}, page \pageref{ipgeosection}) would have already been affected by the speed of its access network, using $\lambda_l$ increases $l$'s accuracy in calculating the distance between itself and \emph{D}'s intended location, i.e., in favor of the adversary. The lower bound (2/9){\bf c} in (\ref{speedDynamic2}) is applied to avoid the effect of increased circuitousness (indirectness) and highly varying delay-to-distance ratios occurring with short distances over the Internet \cite{subramanian2002geographic}. \emph{D} then estimates the \ac{RTT} that $l$ should observe at the intended location, $a'$ using the speed $\lambda_l$:
\begin{equation}\label{equn11Dynamic}\beta(a',l) = \frac{2\times dis(a',l)}{\lambda_l}\end{equation}
The adversary finally sets $\delta$ in (\ref{specifictau}) as:
\begin{equation}\delta=\alpha(a,l)-\beta(a',l)\end{equation}

{\bf Adversary \emph{E}}. Proposed by Gill {\it et al.}\ \cite{Dude}, adversary \emph{E} was assumed to have access to each landmark's calibration function, and was using this function to model the traffic propagation speed. \label{changeadversaryE}\textcolor{\changes}{It may not be trivial for an adversary to have access to such information in practice. Thus, we only list Adversary E in the table for completeness, but do not explore it further. Note however that having access to each landmark's calibration function makes the adversary stronger. We show later in this chapter that even when lacking knowledge of this information, some of the modeled adversaries are already powerful enough to control the forged location at a country-level granularity.}

\section{Evaluation Results}
\label{attackdelbased}

\begin{figure}\centering
\includegraphics[scale=0.5]{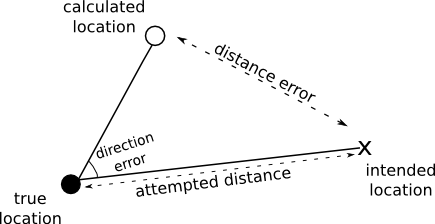}
\caption[Definition of distance and direction errors]{Distance and direction errors. The calculated location is the one returned by the geolocation technique, whereas the intended location is the one the adversary intends to appear at (fraudulently).}
\label{error}\end{figure}

The primary evaluation metrics we use are the adversary's distance error and direction error (Fig.~\ref{error}). The first is the distance between the adversary's intended location and the location calculated by the geolocation technique. We used, again, the great circle distance to calculate this metric. The second metric is the absolute spherical angle (i.e., $\leq$180) between the lines passing through both locations and the adversary' true location. We used spherical trigonometry to calculate this metric, where we adopted 6,371 km as an approximation to the Earth's radius \cite{earthradius}.

\begin{figure*}\centering
\includegraphics[width=0.9\textwidth, height=6cm]{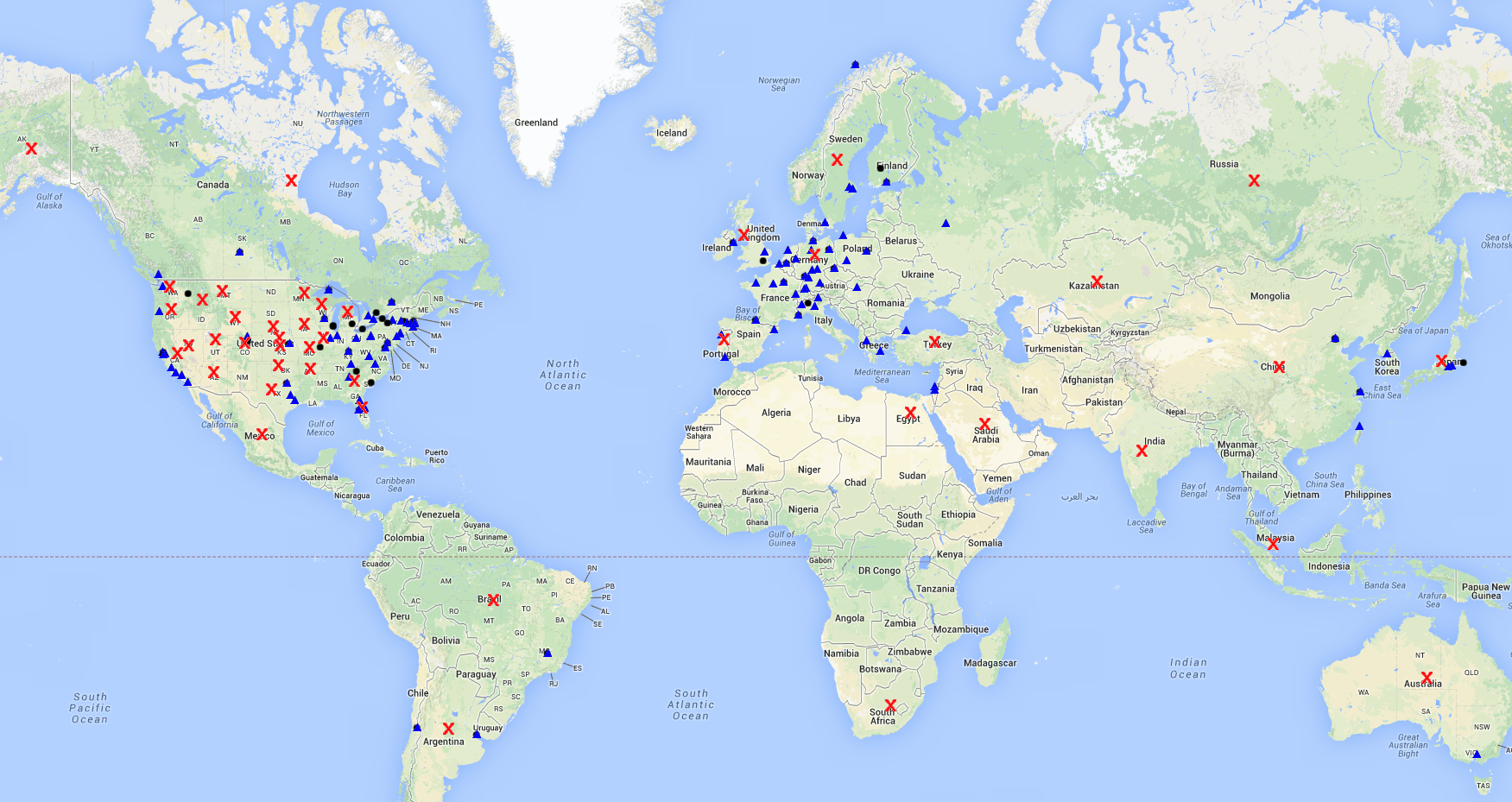}
\caption[Geographic locations of adversaries and landmarks used in the experiments (experimental design)]{Locations of 122 landmarks and 51 modeled adversaries used in our experiments. Each adversary attempted to forge its location to 50 other (intended) locations, for a total of 2,550 modeled attempts to manipulate geolocation. \textcolor{\changes}{$\blacktriangle$} = landmarks; $\bullet$ = true locations of adversaries; \textcolor{red}{$\times$} = intended locations of adversaries. Note: this graph shows experimental design, not results. Map data: Google, INEGI.}
\label{typical}\end{figure*}

We implemented three representative delay-based techniques for evaluating manipulations: GeoPing \cite{GeoTrackGeoPing}, \ac{CBG} \cite{Constrainbased}, and SegPoly \cite{Dong201285}, and believe analogous manipulation effect extends to other techniques. To evaluate the manipulations, we used PlanetLab \cite{planetlab}, where we selected 144 nodes (Fig.~\ref{typical}) to represent 122 landmarks and 51 adversaries (some nodes acted as both). We obtained the delays between these nodes from the iPlane project \cite{Madhyastha:2006:IIP:1298455.1298490}, which were collected on March 27, 2014. Each client made 50 location-forging attempts, marked by $\times$ in Fig.~\ref{typical}, giving a total of 2,550 attempts.

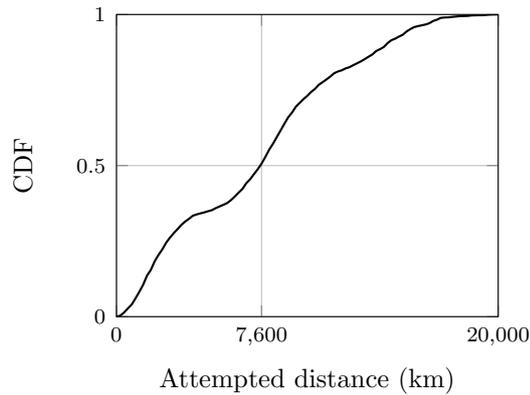
\begin{figure}
\centering
\begin{tikzpicture}
\begin{axis}[
width=2.6in,
height=2.2in,
xlabel=Attempted distance (km),
ylabel=CDF,
xmin=0, xmax=20000,
ymin=0, ymax=1,
xtick={0,7600,20000},
ytick={0,0.5,1},
x tick label style={font=\scriptsize},
grid=major,
y tick label style={font=\scriptsize},
unbounded coords=jump,
label style={font=\footnotesize},
scaled ticks=false, tick label style={/pgf/number format/fixed},
]
\addplot[solid,line width=0.8pt] table[col sep=comma]{csv/Attack/attempted/attempted.csv};
\end{axis}
\end{tikzpicture}
\caption[CDF of the attempted distances (experimental design)]{CDF of the attempted distances. A point ($x$,$y$) means a fraction $y$ of all 2,550 manipulations attempted to move $x$ km or less away from the true location. Note: this graph shows experimental design, not results.}
\label{attempteddist}
\end{figure}

Figure~\ref{attempteddist} shows a \ac{CDF} of the \emph{attempted distances} (see Fig.~\ref{error} for definition); 50\% of all attempts intended to move at least $\sim$7,600 km away from the true locations. Such large distances are not typically state or city level relocation, but rather country or continent level. In fact, we chose the centroids of 20 countries to represent 20 of the 50 intended locations, and the centroids of 30 US states to represent the remaining ones.

\subsection{Manipulation Accuracy}
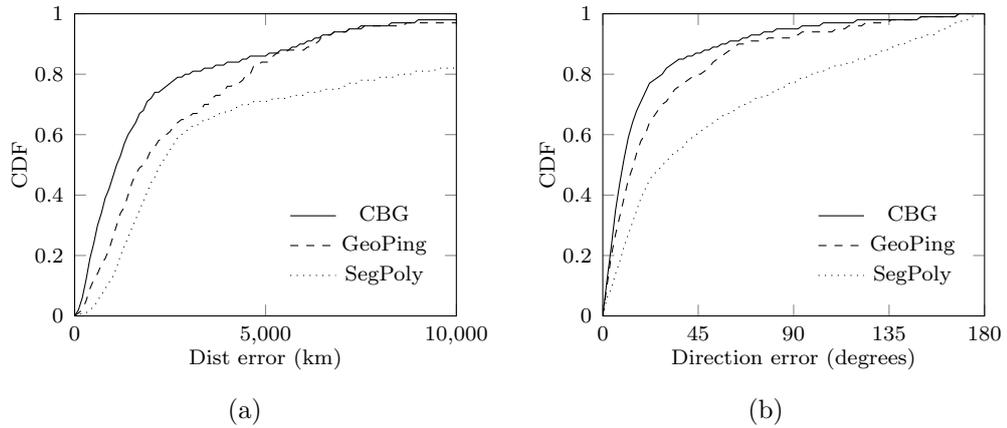
\begin{figure}
\centering
\subfigure[]
{
\label{delayattackerrorkm}
\begin{tikzpicture}
\begin{axis}[
width=2.6in,
height=2.2in,
xlabel=Dist error (km),
ylabel=\ac{CDF},
xmin=0, xmax=10000,
ymin=0, ymax=1,
xtick={0,5000,10000},
x tick label style={font=\scriptsize},
y tick label style={font=\scriptsize},
y label style={font=\scriptsize,at={(0.1,0.5)}},
x label style={font=\scriptsize,at={(0.5,0.06)}},
legend columns=1,
legend style={font=\scriptsize,draw=none,at={(0.75,0.4)},anchor=north,},
unbounded coords=jump,
scaled x ticks=false, tick label style={/pgf/number format/fixed},
]
\addplot[solid] table[col sep=comma]{csv/Attack/Error_Attacker_Two_Distance/CBG3.csv};
\addlegendentry{\ac{CBG}}
\addplot[dashed] table[col sep=comma]{csv/Attack/Error_Attacker_Two_Distance/GeoPing3.csv};
\addlegendentry{GeoPing}
\addplot[dotted] table[col sep=comma]{csv/Attack/Error_Attacker_Two_Distance/SegPoly3.csv};
\addlegendentry{SegPoly}
\end{axis}
\end{tikzpicture}
}
\subfigure[]
{
\label{delayattackerrorangle}
\begin{tikzpicture}
\begin{axis}[
width=2.6in,
height=2.2in,
xlabel=Direction error (degrees),
ylabel=\ac{CDF},
xmin=0, xmax=180,
ymin=0, ymax=1,
xtick={0,45,90,135,180},
x tick label style={font=\scriptsize},
y tick label style={font=\scriptsize},
y label style={font=\scriptsize,at={(0.1,0.5)}},
x label style={font=\scriptsize,at={(0.5,0.06)}},
legend columns=1,
legend style={font=\scriptsize,draw=none,at={(0.75,0.4)},anchor=north,},
unbounded coords=jump,
scaled ticks=false, tick label style={/pgf/number format/fixed},
]
\addplot[solid] table[col sep=comma]{csv/Attack/Error_Attacker_Two_Angle/CBG3.csv};
\addlegendentry{\ac{CBG}}
\addplot[dashed] table[col sep=comma]{csv/Attack/Error_Attacker_Two_Angle/GeoPing3.csv};
\addlegendentry{GeoPing}
\addplot[dotted] table[col sep=comma]{csv/Attack/Error_Attacker_Two_Angle/SegPoly3.csv};
\addlegendentry{SegPoly}
\end{axis}
\end{tikzpicture}
}
\caption[Distance and direction errors of an adversary manipulating vulnerable delay-based geolocation techniques]{CDF of (a) distance errors and (b) direction errors for adversary \emph{A} upon manipulating geolocation. A point ($x$,$y$) means a fraction $y$ of all manipulation attempts resulted in error of $x$ (km or degrees) or less.}
\end{figure}

Figure~\ref{delayattackerrorkm} shows a \ac{CDF} of \emph{A}'s distance errors; one-third of manipulations to \ac{CBG} resulted in errors below 700 km (close to the width of France), and two-thirds below 1,700 km.\footnote{Note that these adversarial errors arise in part due to inherent inaccuracies of the geolocation methods themselves, making the relatively smaller errors more noteworthy.} Both values are less than half the US width; e.g., if Pandora \cite{pandora} used \ac{CBG} to enforce US geographic restriction policies, at least two-thirds of non US-based clients are expected to bypass these restrictions.

The adversary's distance errors were larger while manipulating GeoPing; one-fifth of all manipulations resulted in errors below 850 km, and half had errors below 1,800 km. The difference between \ac{CBG} and GeoPing, however, partly stems from \ac{CBG} being generally more accurate than GeoPing \cite{Constrainbased}. When the adversary can fully manipulate the delays, such higher accuracy may unfortunately help adversaries more accurately control the calculated location. 

For SegPoly, 80\% of manipulation attempts resulted in more than 1,200 km error, which is due to the linear function adversary \emph{A} uses to map distances to delays (distance = delay $\times$ (1/3){\bf c}). The function leads to a large deflection between the distance it wants a landmark to calculate, and the one the landmark actually calculates. Despite using linear mapping against a technique that uses polynomial mapping, 44\% of \emph{A}'s manipulations resulted in less than 2,000 km distance error.

\begin{figure}\centering
\includegraphics[scale=0.18]{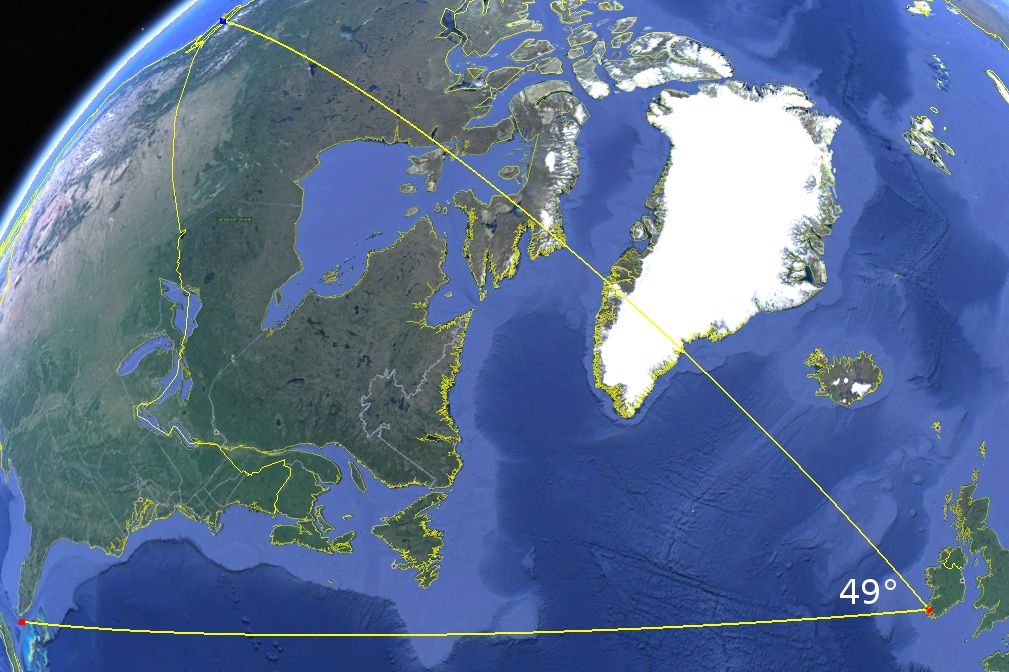}
\caption[The spherical angle at the intersection point, at the west of Europe, of two lines enclosing the contiguous US]{The spherical angle at the intersection point, close to the extreme west of Europe, of two lines enclosing the contiguous US is $\sim$49$^{\circ}$. Map data: Google, SIO, NOAA, U.S. Navy, NGA, GEBCO.}
\label{dierrorexplain}\end{figure}

The \ac{CDF} of the direction error for the three techniques is shown in Fig.~\ref{delayattackerrorangle}; 88\%, 82\%, and 63\% of adversary \emph{A}'s manipulations to \ac{CBG}, GeoPing and SegPoly respectively resulted in direction errors less than 50$^{\circ}$. To interpret this result, one can think of a US bank restricting credit card transactions to the US, e.g., for fraud prevention \cite{shroud,laki2011spotter}. Using \ac{CBG}, and assuming relatively small distance errors, about 88\% of European-based adversaries are expected to succeed to pretend to be in the (contiguous) US. That is because for most adversaries whose true locations are Europe (excluding Iceland) and who intend to be in the US, a direction error below $\sim$50$^{\circ}$ (and distance error below $\sim$5,000 km---the country's width) enables them achieve their objective. Figure~\ref{dierrorexplain} shows the spherical angle at the intersection point, close to the extreme west of Europe, of two lines enclosing the contiguous US. If a European-based adversary, for example, expects to incur a 50$^{\circ}$ direction error clockwise, it can plan to pretend to be in Florida so that its location ends up being calculated as Washington, and vice versa.\footnote{The adversary is not assumed to control whether the direction error is clockwise or counterclockwise; if one fails, it tries the other.} Note that the angle in Fig.~\ref{dierrorexplain} decreases when the two lines intersect further to the east of Europe.

Next we explore the relationship between \emph{A}'s attempted distance and its distance error. The correlation\footnote{The Pearson Correlation Coefficient ranges from -1 to +1, with 0 indicating no correlation, and $\pm$1 indicating extreme +/-ve correlation.} between the two variables are 0.55 for \ac{CBG} and GeoPing, and 0.68 for SegPoly. A powerful adversary should exhibit lower correlation between both variables, meaning that its accuracy does not degrade when its intended location is far away from its true one. The relatively moderate correlation while manipulating \ac{CBG} and GeoPing (0.55) indicates that an adversary able to increase and decrease \acp{RTT} can accurately control extremely remote fraudulent locations. Note that the correlation is positive because manipulations of small attempted distances result in small distance errors. 

Manipulations to SegPoly experienced higher correlation, compared to \ac{CBG} and GeoPing, because of the discrepancy between SegPoly's segmented polynomial mapping function and the linear function adversary \emph{A} uses, which manifests quite clearly as larger delays get mapped to distances.

\subsection{Manipulation Detection}
\label{areabaseddetectability}

\ac{CBG} calculates a client's geographic location as the centroid of a convex region enclosed by the intersection of multiple circles. Gill {\it et al.}\ \cite{Dude} suggested the area of this region could be used to detect manipulations, which involves an adversary increasing the \ac{RTT}, because larger adversary-landmark \acp{RTT} increase the area. We analyze detection abilities of this against an adversary that can also decrease delays. GeoPing generates no intersection regions; we are not immediately aware of a method to precisely detect manipulations against GeoPing.

\begin{figure}
\centering
\begin{tikzpicture}
\begin{axis}[
width=2.6in,
height=2.2in,
xlabel=Area $\times$10$^6$ (km$^2$),
ylabel=\ac{CDF},
xmin=0, xmax=4000000,
ymin=0, ymax=1,
ytick={0,0.2,0.4,0.6,0.8,1},
x tick label style={font=\scriptsize},
y tick label style={font=\scriptsize},
label style={font=\footnotesize},
y label style={font=\scriptsize,at={(0.08,0.5)}},
legend columns=2,
legend style={draw=none,at={(0.5,1.35)},anchor=north,font=\footnotesize},
unbounded coords=jump,
every x tick scale label/.style={xshift=1pt,anchor=south west,inner sep=0pt,color=white}
]
\addplot[solid] table[col sep=comma]{csv/Attack/Area_True/CBG.csv};
\addlegendentry{CBG true}
\addplot[dotted] table[col sep=comma]{csv/Attack/Area_True/SegPoly.csv};
\addlegendentry{SePoly true}
\addplot[dashed] table[col sep=comma]{csv/Attack/Error_Attacker_Two_Area/CBG3.csv};
\addlegendentry{CBG manipulated}
\addplot[loosely dashdotted] table[col sep=comma]{csv/Attack/Error_Attacker_Two_Area/SegPoly3.csv};
\addlegendentry{SegPoly manipulated}
\end{axis}
\end{tikzpicture}
\caption[Attack detectability]{CDF of the intersection-region areas while determining the (51) \emph{true} and the (2,550) \emph{manipulated} locations. A point ($x$,$y$) means a fraction $y$ of all attempts had areas of $x$ km$^2$ or less. Higher (\emph{manipulated}) curves indicate less detectable manipulations.}
\label{areaattack}
\end{figure}
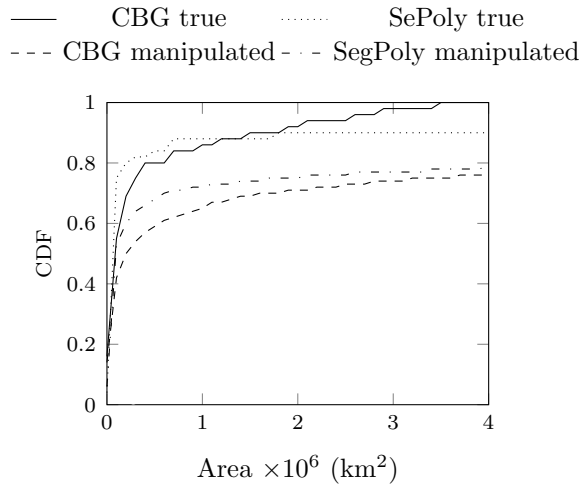

Figure~\ref{areaattack} shows a \ac{CDF} of the intersection-region areas while operating \ac{CBG} and SegPoly to calculate the forged and true locations of adversaries.\footnote{These true locations are calculated from the original delays between the landmarks and the 51 PlanetLab nodes, before changing these delays to model adversaries.} The speed that adversary \emph{A} uses, (1/3){\bf c}, is slow relative to the average traffic propagation speed \cite{delayandtopology}. This results in relatively large \ac{RTT} estimates to the intended location, $\beta(a',l)$ in (\ref{equn11}), increasing the distances the landmarks calculate from mapping those \acp{RTT}. This explains the large areas (low curves in Fig.~\ref{areaattack}) while manipulating geolocation.

However, 92\% of the areas that \ac{CBG} calculated while locating the true nodes were equivalent to 71\% of those while calculating the forged locations, at $x$ = 2$\times$10$^6$ km$^2$. This implies that if the geolocating party decides to reject clients whose intersection-region areas are greater than this value, it falsely rejects 8\% of legitimate clients and falsely accepts 71\% of adversaries. According to these results, detecting manipulations based on the intersection-region areas is not trivial.

\section{Comparing the Adversarial Models}
\label{impactfull}
We compare the adversaries modeled in Table~\ref{adversarysummary} (Section \ref{attack:threatmodel}). The same delay dataset was used across them to establish a comparable experimental set up.

\subsection{Manipulation Accuracy}

Figure~\ref{delayattackerrorkmCOMPARE} compares the distance errors for the four adversaries. About 80\% of \emph{B}'s manipulations to the three techniques resulted in distance errors above $\sim$1,900 km; only 29\%, 48\% and 59\% of \emph{A}'s manipulations to \ac{CBG}, GeoPing, and SegPoly respectively resulted in errors above 1,900 km. The median distance errors for \emph{A} and \emph{C} while manipulating \ac{CBG} were 1,100 km and 3,800 km respectively. The corresponding values for GeoPing were 1,800 km and 3,100 km, and for SegPoly were 2,250 km and 3,600 km. The improvements of adversary \emph{A} over \emph{C} underscore the effectiveness of full delay manipulation on an adversary's location-forging abilities.

Adversary \emph{D} shows a distance error improvement over \emph{A} while manipulating \ac{CBG}, with 66\% of \emph{D}'s manipulations resulting in errors below 1,000 km, versus 46\% of \emph{A}'s manipulations. Surprisingly, \emph{A} showed slight improvement over \emph{D} while manipulating GeoPing. One possible explanation for this could be \emph{D}'s access network; if it is relatively slow, the varying delay-distance mapping decreases the mapped delays between \emph{D} and \emph{all landmarks}. A constant traffic speed protects \emph{A} from the effect of a slow access network. Finally, for SegPoly, almost unnoticeable distance error improvements were made by adversary \emph{D}'s manipulations over \emph{A}.

\newcommand{\x}{1.7in}
\newcommand{\y}{0.13}

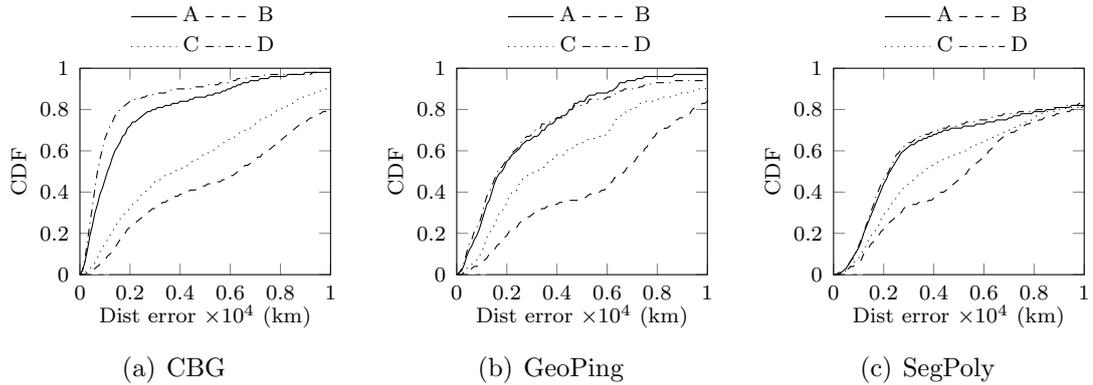
\begin{figure}
\centering
\subfigure[\ac{CBG}]
{
\begin{tikzpicture}
\begin{axis}[
width=1.13*\x,
height=\x,
xlabel=Dist error $\times$10$^4$ (km),
ylabel=\ac{CDF},
xmin=0, xmax=10000,
ymin=0, ymax=1,
x tick label style={font=\scriptsize},
y tick label style={font=\scriptsize},
y label style={font=\scriptsize,at={(\y,0.5)}},
x label style={font=\scriptsize,at={(0.5,0.1)}},
legend columns=2,
legend style={draw=none,at={(0.5,1.35)},anchor=north,font=\scriptsize},
unbounded coords=jump,
every x tick scale label/.style={xshift=1pt,anchor=south west,inner sep=0pt,color=white}
]
\addplot[solid] table[col sep=comma]{csv/Attack/Error_Attacker_Two_Distance/CBG3.csv};
\addlegendentry{A}
\addplot[dashed] table[col sep=comma]{csv/Attack/Error_Attacker_Two_Distance/my_implementation/CBG.csv.gill};
\addlegendentry{B}
\addplot[dotted] table[col sep=comma]{csv/Attack/Error_Attacker_Two_Distance/my_implementation/CBG.csv_C.gill};
\addlegendentry{C}
\addplot[dashdotted] table[col sep=comma]{csv/Attack/Error_Attacker_Two_Distance/CBG_T2.csv};
\addlegendentry{D}
\end{axis}
\end{tikzpicture}
}
\subfigure[GeoPing]
{
\label{delayattackerrorkm_GillGeoPing}
\begin{tikzpicture}
\begin{axis}[
width=1.13*\x,
height=\x,
xlabel=Dist error $\times$10$^4$ (km),
ylabel=\ac{CDF},
xmin=0, xmax=10000,
ymin=0, ymax=1,
x tick label style={font=\scriptsize},
y tick label style={font=\scriptsize},
y label style={font=\scriptsize,at={(\y,0.5)}},
x label style={font=\scriptsize,at={(0.5,0.1)}},
legend columns=2,
legend style={draw=none,at={(0.5,1.35)},anchor=north,font=\scriptsize},
unbounded coords=jump,
every x tick scale label/.style={xshift=1pt,anchor=south west,inner sep=0pt,color=white}
]
\addplot[solid] table[col sep=comma]{csv/Attack/Error_Attacker_Two_Distance/GeoPing3.csv};
\addlegendentry{A}
\addplot[dashed] table[col sep=comma]{csv/Attack/Error_Attacker_Two_Distance/my_implementation/GeoPing.csv.gill};
\addlegendentry{B}
\addplot[dotted] table[col sep=comma]{csv/Attack/Error_Attacker_Two_Distance/my_implementation/GeoPing.csv_C.gill};
\addlegendentry{C}
\addplot[dashdotted] table[col sep=comma]{csv/Attack/Error_Attacker_Two_Distance/GeoPing_T2.csv};
\addlegendentry{D}
\end{axis}
\end{tikzpicture}
}
\subfigure[SegPoly]
{
\begin{tikzpicture}
\begin{axis}[
width=1.13*\x,
height=\x,
xlabel=Dist error $\times$10$^4$ (km),
ylabel=\ac{CDF},
xmin=0, xmax=10000,
ymin=0, ymax=1,
x tick label style={font=\scriptsize},
y tick label style={font=\scriptsize},
y label style={font=\scriptsize,at={(\y,0.5)}},
x label style={font=\scriptsize,at={(0.5,0.1)}},
legend columns=2,
legend style={draw=none,at={(0.5,1.35)},anchor=north,font=\scriptsize},
unbounded coords=jump,
every x tick scale label/.style={xshift=1pt,anchor=south west,inner sep=0pt,color=white}
]
\addplot[solid] table[col sep=comma]{csv/Attack/Error_Attacker_Two_Distance/SegPoly3.csv};
\addlegendentry{A}
\addplot[dashed] table[col sep=comma]{csv/Attack/Error_Attacker_Two_Distance/my_implementation/SegPoly.csv.gill};
\addlegendentry{B}
\addplot[dotted] table[col sep=comma]{csv/Attack/Error_Attacker_Two_Distance/my_implementation/SegPoly.csv_C.gill};
\addlegendentry{C}
\addplot[dashdotted] table[col sep=comma]{csv/Attack/Error_Attacker_Two_Distance/SegPoly_T2.csv};
\addlegendentry{D}
\end{axis}
\end{tikzpicture}
}
\caption[Comparison of distance errors across multiple adversarial models]{Distance errors for the adversaries in Table~\ref{adversarysummary}.}\label{delayattackerrorkmCOMPARE}
\end{figure}

\begin{figure}
\centering
\subfigure[\ac{CBG}]
{
\begin{tikzpicture}
\begin{axis}[
width=1.13*\x,
height=\x,
xlabel=Direction error (deg),
ylabel=\ac{CDF},
xmin=0, xmax=180,
ymin=0, ymax=1,
x tick label style={font=\scriptsize},
y tick label style={font=\scriptsize},
y label style={font=\scriptsize,at={(\y,0.5)}},
x label style={font=\scriptsize,at={(0.5,0.1)}},
legend columns=2,
legend style={draw=none,at={(0.5,1.35)},anchor=north,font=\scriptsize},
unbounded coords=jump,
]
\addplot[solid] table[col sep=comma]{csv/Attack/Error_Attacker_Two_Angle/CBG3.csv};
\addlegendentry{A}
\addplot[dashed] table[col sep=comma]{csv/Attack/Error_Attacker_Two_Angle/my_implementation/CBG.csv.gill};
\addlegendentry{B}
\addplot[dotted] table[col sep=comma]{csv/Attack/Error_Attacker_Two_Angle/my_implementation/CBG.csv_C.gill};
\addlegendentry{C}
\addplot[dashdotted] table[col sep=comma]{csv/Attack/Error_Attacker_Two_Angle/CBG_T2.csv};
\addlegendentry{D}
\end{axis}
\end{tikzpicture}
}
\subfigure[GeoPing]
{
\begin{tikzpicture}
\begin{axis}[
width=1.13*\x,
height=\x,
xlabel=Direction error (deg),
ylabel=\ac{CDF},
xmin=0, xmax=180,
ymin=0, ymax=1,
x tick label style={font=\scriptsize},
y tick label style={font=\scriptsize},
y label style={font=\scriptsize,at={(\y,0.5)}},
x label style={font=\scriptsize,at={(0.5,0.1)}},
legend columns=2,
legend style={draw=none,at={(0.5,1.35)},anchor=north,font=\scriptsize},
unbounded coords=jump,
]
\addplot[solid] table[col sep=comma]{csv/Attack/Error_Attacker_Two_Angle/GeoPing3.csv};
\addlegendentry{A}
\addplot[dashed] table[col sep=comma]{csv/Attack/Error_Attacker_Two_Angle/my_implementation/GeoPing.csv.gill};
\addlegendentry{B}
\addplot[dotted] table[col sep=comma]{csv/Attack/Error_Attacker_Two_Angle/my_implementation/GeoPing.csv_C.gill};
\addlegendentry{C}
\addplot[dashdotted] table[col sep=comma]{csv/Attack/Error_Attacker_Two_Angle/GeoPing_T2.csv};
\addlegendentry{D}
\end{axis}
\end{tikzpicture}
}
\subfigure[SegPoly]
{
\begin{tikzpicture}
\begin{axis}[
width=1.13*\x,
height=\x,
xlabel=Direction error (deg),
ylabel=\ac{CDF},
xmin=0, xmax=180,
ymin=0, ymax=1,
x tick label style={font=\scriptsize},
y tick label style={font=\scriptsize},
y label style={font=\scriptsize,at={(\y,0.5)}},
x label style={font=\scriptsize,at={(0.5,0.1)}},
legend columns=2,
legend style={draw=none,at={(0.5,1.35)},anchor=north,font=\scriptsize},unbounded coords=jump,
]
\addplot[solid] table[col sep=comma]{csv/Attack/Error_Attacker_Two_Angle/SegPoly3.csv};
\addlegendentry{A}
\addplot[dashed] table[col sep=comma]{csv/Attack/Error_Attacker_Two_Angle/my_implementation/SegPoly.csv.gill};
\addlegendentry{B}
\addplot[dotted] table[col sep=comma]{csv/Attack/Error_Attacker_Two_Angle/my_implementation/SegPoly.csv_C.gill};
\addlegendentry{C}
\addplot[dashdotted] table[col sep=comma]{csv/Attack/Error_Attacker_Two_Angle/SegPoly_T2.csv};
\addlegendentry{D}
\end{axis}
\end{tikzpicture}
}
\caption[Comparison of direction errors across multiple adversarial models]{Direction error for the adversaries in Table~\ref{adversarysummary}.}\label{delayattackerrorangleCOMPARE}
\end{figure}
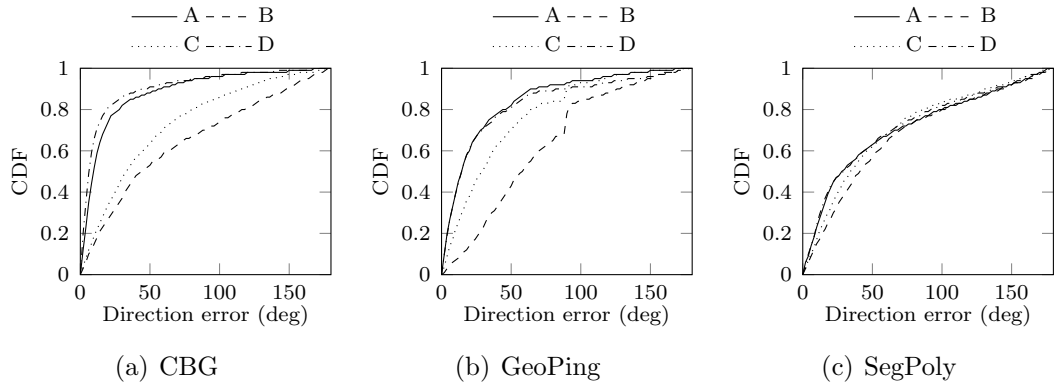

Figure~\ref{delayattackerrorangleCOMPARE} compares the results for the direction errors; 50\%, 38\% and 53\% of adversary \emph{B}'s manipulations to \ac{CBG}, GeoPing and SegPoly respectively resulted in direction errors below 45$^{\circ}$, versus 87\%, 80\% and 60\% of adversary \emph{A}'s manipulations. A lower direction error for the adversary indicates a more accurate (hence, more worrisome) adversary. Similar to the distance errors, adversary \emph{C}'s overall direction error was better than that of \emph{B} but worse than \emph{A}, again highlighting \emph{A}'s devastating abilities. Adversary \emph{D} showed direction error improvements over \emph{A} only while manipulating \ac{CBG}, but no considerable improvements were observed upon manipulating the other two geolocation techniques.

\subsection{Manipulation Detection}

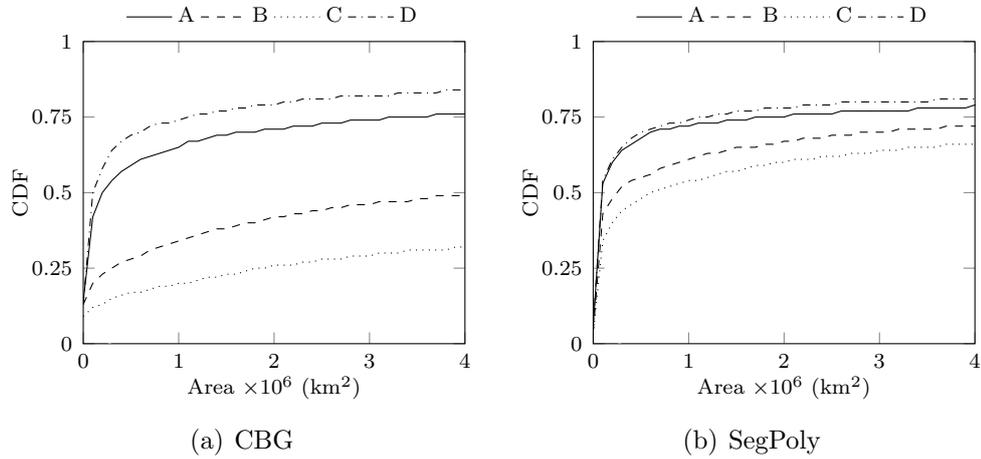
\begin{figure}
\centering
\subfigure[\ac{CBG}]
{
\label{impactcbgarea}
\begin{tikzpicture}
\begin{axis}[
width=2.6in,
height=2.2in,
xlabel=Area $\times$10$^6$ (km$^2$),
ylabel=\ac{CDF},
xmin=0, xmax=4000000,
ymin=0, ymax=1,
xtick={0,1000000,2000000,3000000,4000000},
extra tick style={grid=major},
ytick={0,0.25,0.5,0.75,1},
x tick label style={font=\scriptsize},
y tick label style={font=\scriptsize},
y label style={font=\scriptsize,at={(0.08,0.5)}},
x label style={font=\scriptsize,at={(0.5,0.06)}},
legend columns=-1,
legend style={draw=none,at={(0.5,1.15)},anchor=north,font=\scriptsize},
unbounded coords=jump,
every x tick scale label/.style={xshift=1pt,anchor=south west,inner sep=0pt,color=white}
]
\addplot[solid] table[col sep=comma]{csv/Attack/Error_Attacker_Two_Area/CBG3.csv};
\addlegendentry{A}
\addplot[dashed] table[col sep=comma]{csv/Attack/Error_Attacker_Two_Area/my_implementation/CBG.csv.gill};
\addlegendentry{B}
\addplot[dotted] table[col sep=comma]{csv/Attack/Error_Attacker_Two_Area/my_implementation/CBG.csv_C.gill};
\addlegendentry{C}
\addplot[dashdotted] table[col sep=comma]{csv/Attack/Error_Attacker_Two_Area/CBG_T2.csv};
\addlegendentry{D}
\end{axis}
\end{tikzpicture}
}
\subfigure[SegPoly]
{
\label{impactsegpolyarea}
\begin{tikzpicture}
\begin{axis}[
width=2.6in,
height=2.2in,
xlabel=Area $\times$10$^6$ (km$^2$),
ylabel=\ac{CDF},
xmin=0, xmax=4000000,
ymin=0, ymax=1,
xtick={0,1000000,2000000,3000000,4000000},
ytick={0,0.25,0.5,0.75,1},
x tick label style={font=\scriptsize},
y tick label style={font=\scriptsize},
y label style={font=\scriptsize,at={(0.08,0.5)}},
x label style={font=\scriptsize,at={(0.5,0.06)}},
legend columns=-1,
legend style={draw=none,at={(0.5,1.15)},anchor=north,font=\scriptsize},
unbounded coords=jump,
every x tick scale label/.style={xshift=1pt,anchor=south west,inner sep=0pt,color=white}
]
\addplot[solid] table[col sep=comma]{csv/Attack/Error_Attacker_Two_Area/SegPoly3.csv};
\addlegendentry{A}
\addplot[dashed] table[col sep=comma]{csv/Attack/Error_Attacker_Two_Area/my_implementation/SegPoly.csv.gill};
\addlegendentry{B}
\addplot[dotted] table[col sep=comma]{csv/Attack/Error_Attacker_Two_Area/my_implementation/SegPoly.csv_C.gill};
\addlegendentry{C}
\addplot[dashdotted] table[col sep=comma]{csv/Attack/Error_Attacker_Two_Area/SegPoly_T2.csv};
\addlegendentry{D}
\end{axis}
\end{tikzpicture}
}
\caption[Comparison of attack detectability across multiple adversarial models]{CDF of the intersection-region areas for the adversaries in Table~\ref{adversarysummary}. A point ($x$,$y$) means a fraction $y$ of all attempts resulted in areas of $x$ km$^2$ or less. Higher curves indicate less detectable manipulations.}
\label{areaattackCOMPARE}
\end{figure}

Figure~\ref{areaattackCOMPARE} shows \acp{CDF} of the intersection regions areas; 58\% of \emph{B}'s manipulations to \ac{CBG} resulted in areas above 2$\times$10$^6$ km$^2$, versus only 29\% of \emph{A}'s. Clearly, adversary \emph{A}'s manipulations to \ac{CBG} are harder to detect using the area as the detection factor.

Areas resulting from \emph{B}'s manipulations to SegPoly were significantly smaller compared to its manipulations to \ac{CBG}, and more interestingly, were close to those resulting from \emph{A}'s manipulations to SegPoly (the curves \emph{A} and \emph{B} are close in Fig.~\ref{impactsegpolyarea} than in Fig.~\ref{impactcbgarea}). This is because \emph{B} uses double the speed that \emph{A} uses to model the traffic speed. If both adversaries pretend to be farther from a landmark by a certain distance, \emph{B} increases the \ac{RTT} by half the amount that \emph{A} increases. When the landmark maps those \acp{RTT}, smaller values get mapped to smaller distances, decreasing the area of intersection. Nonetheless, the average distances resulting from adversary \emph{B}'s mapping are not expected to be relatively small since \emph{B} can only increase \acp{RTT}. \emph{B}'s faster traffic speed combined with its ability to only increase delays explains its area similarity with \emph{A}. This argument does not apply to \ac{CBG} because the linear calibration the landmarks use blindly maps larger delays to larger distances.
Adversary \emph{C} had the largest intersection-region areas compared to \emph{A} and \emph{B} because it combines two factors that tend to increase areas: only increasing \acp{RTT} and using a small constant to model traffic speed. Therefore, \emph{C} is the most exposed to being detected based on the area of the intersection region.

Adversary \emph{D} was less detectable than \emph{A}. Half of \emph{D}'s manipulations to \ac{CBG} resulted in intersection-region areas below 0.1$\times$10$^6$ km$^2$, compared to double this number for half of \emph{A}'s manipulations. 

\begin{table*}
\centering
\renewcommand{\arraystretch}{1.2}
\caption[Summary of adversarial capabilities in forging geographic locations]{Median distance (km) \& direction errors (degrees), median areas of intersection regions (km$^2$), and correlation coefficients between the distance errors and the attempted distances for the adversaries in Table~\ref{adversarysummary}; Adversary \emph{B} is similar to that of Gill {\it et al.}\ \cite{Dude}. Smaller values for all fields indicate a more powerful adversary.}
\scalebox{0.7}{
\begin{tabular}{ r | c c c c | C{16pt} C{16pt} C{16pt} C{16pt} | c c c c | c c c c }
Geolocation& \multicolumn{4}{c|}{Dist error (km)} & \multicolumn{4}{c|}{Direction error (deg)} & \multicolumn{4}{c|}{Area $\times$10$^6$ (km$^2$)} & \multicolumn{4}{c}{Correlation}\\
method& \emph{A} & \emph{B} & \emph{C} & \emph{D} & \emph{A} & \emph{B} & \emph{C} & \emph{D} & \emph{A} & \emph{B} & \emph{C} & \emph{D} & \emph{A} & \emph{B} & \emph{C} & \emph{D}\\\hline
\ac{CBG} \cite{Constrainbased}& 1,100 & 6,300 & 3,800 & 700 & 9.5 & 44 & 33 & 6 & 2 & 4.1 & 17.3 &$<$1& 0.54 & 0.89 & 0.73 & 0.33\\
GeoPing \cite{GeoTrackGeoPing} & 1,800 & 6,700 & 3,100 & 1,630 & 14 & 58 & 29 & 14 & -- & -- & -- & -- & 0.55 & 0.86 & 0.64 & 0.57\\
SegPoly \cite{Dong201285} & 2,250 & 5,350 & 3,600 & 2,200 & 28 & 41 & 34 & 29 &$<$1&$<$1&$<$1&$<$1& 0.68 & 0.85 & 0.8 & 0.64\\
\Xcline{1-17}{2\arrayrulewidth}
\end{tabular}
}
\label{tableSummary}
\end{table*}

\subsection{Summary}

Table~\ref{tableSummary} summarizes the differences between the four modeled adversaries. Adversary \emph{A} achieves 83\%, 73\% and 58\% reductions\footnote{Percentage reduction = $\frac{\text{Median error of \emph{B} - Median error of \emph{A}}}{\text{Median error of \emph{B}}}\times100$.} to the median distance errors over \emph{B} while manipulating \ac{CBG}, GeoPing, and SegPoly respectively; and achieves 71\%, 41.94\% and 37.5\% over \emph{C} while manipulating the three techniques. \emph{A}'s improvement over \emph{C} is solely due to its ability to fully manipulate delays (since it is the only difference between them), highlighting the powerful nature of manipulation when an adversary is able to decrease and increase the \acp{RTT}.

Compared to \emph{B}, adversary \emph{C} achieves 40\%, 54\%, and 33\% improvement to the median distance error while manipulating \ac{CBG}, GeoPing, and SegPoly respectively (see Table~\ref{tableSummary}). The only difference between both adversaries is the constant they used to model traffic speed, suggesting that a clever modeling can by itself increase the adversary's location-forging accuracy drastically.

Adversary \emph{D} achieves 36\%, 9.4\%, and 2.2\% improvement to the median distance error over \emph{A} while manipulating the three techniques respectively. Thus, against GeoPing and SegPoly, an adversary's knowledge of the \acp{RTT} between itself and the landmarks did not significantly improve the results.

Adversary \emph{D}'s manipulations to \ac{CBG} resulted in the smallest intersection-region area, thus hiding manipulation attempts on \ac{CBG} is easier when the adversary knows the \acp{RTT} between its true location and the landmarks. However, knowledge of this information did not reveal significant advantages in hiding manipulation attempts on SegPoly since all adversaries resulted in very small areas. These results highlight that the accuracy SegPoly gains using polynomial regression comes at the cost of lower ability to detect manipulations by the constrained region area.
Finally, from Table~\ref{tableSummary}, it is evident that \emph{B} and \emph{C} exhibit the highest correlation between the attempted distance and distance error (i.e., poor performance). Adversaries \emph{A} and \emph{D} relax the correlation, enabling them to fraudulently relocate themselves at extremely remote locations with high accuracy. Thus, the combined ability of increasing and decreasing delays reduces the impact on the distance error when large distances are attempted.

\section{Countermeasures}
\label{countermeasures}

We discuss possible countermeasures that specifically aim at preserving the integrity of delay measurements used by a geolocation technique. We stress that the root cause of the vulnerabilities lies not in the \ac{ICMP} utilities themselves but rather in (improperly) leveraging them to carry out a task (geolocation) for which they were not designed. A high level countermeasure would therefore be to avoid using \ac{ICMP}-based utilities for geolocation. If \ac{ICMP}-based utilities are to be used nonetheless, the following countermeasures could be considered. These measures require only the landmarks conducting geolocation to modify their network stacks.

As discussed in Section \ref{ICMPattack}, the vulnerabilities lie in the adversary's ability to tamper with both the sending ($s$) and receiving ($r$) times of \ac{ICMP}-based network utilities. Every landmark must ensure the integrity of both parameters. Locally recording $s$ enables a landmark to retrieve $s$ from its memory instead of the \textsc{data} field of an echo-reply packet. Obviously, the landmark's own local memory is more reliable than an unprotected packet returned from the receiver/adversary. This precludes the adversary from undetectably tampering with the value of $s$. If a stateless implementation is desired, landmarks may use a Message Authentication Code (MAC) protecting $s$, the \ac{ICMP} identifier, and the \ac{ICMP} sequence number of the echo request; the landmarks can then place the MAC in the \textsc{data} field of the packet along with $s$ (see Fig.~\ref{icmpHeaderecho}). Note that a landmark cannot include the \textsc{type} and \textsc{checksum} fields in the MAC because the receiver must change them in the echo reply, as specified by \ac{RFC} 792 \cite{rfc792}. The landmark can store the non-shared key of the MAC locally. A single key suffices for multiple sessions. Landmarks can then verify the integrity of their own timestamp $s$ retrieved from a received echo reply.

As for the receiving time $r$, recall that a key factor for an adversary to beneficially manipulate it is the adversary's ability to measure the waiting time between a successive pair of echo requests. Consequently, randomizing the waiting times raises the bar for the adversary to accurately predict this time. Such a precaution is simple to implement as it may not necessarily require modifications to the local utilities (e.g., \textsl{ping} and \textsl{traceroute}). However, because the adversary calculates the average waiting time, this precaution does not stop the adversary from undetectably manipulating $r$; it only increases the adversary's error range. Another countermeasure to provide timestamp integrity is to include an element of randomness in the \textsc{data} field of echo requests, i.e., similar to \ac{DNS} cache-poisoning countermeasures \cite{son2010hitchhiker}. For all practical purposes, ample unpredictability should prevent the adversary from successfully issuing fraudulent echo replies, forcing it to wait for echo requests first.

Although the countermeasures presented in this section could be technically simple, we expect little community support to deploy modifications to the widely used generic network utilities for the sole purpose of hardening geolocation.

\section{Related work}
\label{attack:relatedwork}

In 2010, Gill {\it et al.}\ \cite{Dude} studied the effect of delay increases on topology-aware (see Section \ref{topologyaware}, page \pageref{topologyaware}) and delay-based geolocation techniques, choosing one representative technique for each. They modeled two classes of adversaries: simple (controls only its own machine) and sophisticated (controls a full wide area network). The former was able to increase delays (adversaries \emph{B} and \emph{E} herein---Section \ref{attack:threatmodel}) and the latter was able to increase the number of hops to the landmarks. Against delay-based techniques, both adversaries had limited control over the forged location \cite{Dude}.

Muir {\it et al.}\ \cite{MuirPaul} investigated geolocation over the Internet from a security perspective, and enumerated a broad spectrum of tactics for an adversary to manipulate geolocation techniques, including using proxies to hide the IP address, and falsifying location records of public registries (\emph{whois} databases, \ac{DNS} LOC records \cite{rfc1876}, etc). They argued that despite a plethora of proposals to geolocate Internet hosts, none appears to be robust against all classes of adversaries. Our work is complementary as it provides concrete evidence, based on practical evaluations, supporting their assertion with respect to popular implementations of delay-based geolocation techniques.

Goldberg {\it et al.}\ \cite{Goldberg:2008:PMP:1375457.1375480} addressed the problem of path quality monitoring, devising protocols to detect if an adversary sitting in the path between two end systems is manipulating their traffic. Although their research is motivated, in part, by the lack of integrity checking in network-monitoring utilities, their proposed solutions assume the collaboration of the two end systems. In our case, one of the end systems is the adversary itself and therefore collaboration cannot be assumed (hence: none of their solutions fit the problem studied herein).

Delay-based location verification techniques have been proposed \cite{cpv,surveywireless,wirelesscapkun}. However, proposals for single-hop wireless networks \cite{surveywireless,wirelesscapkun} cannot be directly applied to the Internet because of the difference in delay nature between both domains \cite{Dude}.

In \acp{NCS}, such as Vivaldi \cite{dabek2004vivaldi} and Meridian \cite{wong2005meridian}, network nodes are assigned coordinates according to the delays between them. 
\acp{NCS} are generally seen as different from geolocation because the coordinates of a node reflect its \emph{network location} rather than geographic longitude and latitude; thus, no delay-to-distance mapping is required.
Adversarial environments (e.g., to disrupt an \ac{NCS}) were explored \cite{zage2007accuracy,girlich2013bounds}, and proposals for securing \acp{NCS} addressed adversarial delay-increase \cite{Kaafar:2007:SIC:1282380.1282388}.

\section{Conclusion}
\label{attack:conclusion}

Virtually all current implementations of conventional network utilities (e.g., \textsl{ping} and \textsl{traceroute}) fail to check the integrity of the measured \acp{RTT}. Thus, misusing them for delay-based geolocation allows an adversary of moderate abilities to increase \emph{or decrease} the \acp{RTT} observed by the measuring party. Without controlling any devices or network traffic other than its own, the adversary can then manipulate geolocation techniques that are based on active delay measurements to the extent of accurately controlling the calculated (forged) location, e.g., for a resulting error as small as 100 km---providing country-level granularity control.\footnote{Related to geolocating cloud data, Peterson {\it et al.} \cite{peterson2011position} emphasize: ``Of particular interest is establishing data location at a granularity sufficient for placing it within the borders of a particular nation-state."} This may defeat location-aware security systems, e.g., a cloud provider violating service level agreements \cite{gondree2013geolocation}, especially given that such geolocation techniques are increasingly being advocated for use in security-aware contexts \cite{Castelluccia:2009:GPS:1644893.1644915}.

By evaluating several adversarial situations, we have demonstrated that better estimates to the traffic propagation speed can enhance the adversary's accuracy in controlling the forged location. This finding even extends to adversaries only capable of increasing \acp{RTT} \cite{Dude}; e.g., an adversary that uses the constant (1/3){\bf c} as an estimate to the traffic speed, as shown herein, is 40\% more accurate in forging a \ac{CBG}-calculated location \cite{Constrainbased} than the one using (2/3){\bf c} studied in previous literature \cite{Dude}.

We note there are countermeasures, based on well-known and technically simple techniques, which provide integrity to the timing information exploited by the manipulations we discussed in Section \ref{ICMPattack}, and thereby would (if implemented and deployed) preclude the evasion of geolocation that we analyzed in Sections \ref{attackdelbased} and \ref{impactfull}. However, these add overhead to core \ac{ICMP} utilities, and thus may well face deployment resistance since they are unnecessary for core services. Designers of delay-based geolocation usually focus on achieving high location accuracy, but to date have failed to propose integrity-preserving yet deployable delay-measurement algorithms---despite being motivated by security-sensitive applications \cite{Constrainbased,laki2011spotter,maxliklihood,Dong201285}.
The analysis in this chapter provides some useful insights. For example, landmarks in \ac{CBG} \cite{Constrainbased} would ideally allow only themselves to measure \acp{RTT} between each other; in our experiments, an adversary knowing the \acp{RTT} between itself and the landmarks was 36\% more accurate. Additionally, if SegPoly \cite{Dong201285} is used, the areas of the constrained region cannot be relied upon for detecting manipulations since they become considerably smaller. In fact, security-sensitive applications \cite{gondree2013geolocation} should not rely on the constrained region areas for detecting manipulations because, while geolocating adversaries who can fully manipulate delays, the constrained regions become almost indistinguishable from those of legitimate clients.

Our work highlights the importance of ensuring timing integrity in delay-based geolocation. We believe more investigations into these techniques are required, and expect that the search for a geolocation mechanism which is not easily defeated remains challenging. We hope our work raises awareness of the importance of devising such evasion-resilient geolocation mechanisms, and encourages further research in this area.

\chapter[Estimating One-Way Delays with Adversaries]{Accurate One-Way Delay Estimation with Reduced Client-Trustworthiness}
\label{ch:owd}

This chapter proposes a novel protocol that enables a server to estimate \acp{OWD} between itself and a client by cooperating with two other servers, 
requiring neither client-clock synchronization nor client trustworthiness in reporting one-way delays. Due to these benefits, the proposed protocol is of value to, and is used in, the location verification mechanism introduced later in Chapter \ref{ch:cpv}.

We evaluate the protocol by deriving the probability distribution of its absolute error, and compare its accuracy with the well-known round-trip halving protocol. While neither protocol requires client-trustworthiness nor client clock synchronization, the analysis shows that the new protocol is more accurate in many situations.

\newcommand{\intersect}{\text{\ }\pmb{,}\text{\ \ }}
\newcommand{\mycap}{\text{\ }\pmb{\cap}\text{\ }}

\section{Introduction}
\blfootnote{The first part of this chapter, the \emph{minimum pairs} protocol, was published at the 2014 IEEE CNS conference \cite{cpv} (with a full length version accepted for publication in IEEE TDSC \cite{cpvtdsc}). The second part, the evaluation using probability models, was published at the IEEE Communications Letters \cite{abdouaccurate}.}

Delay-dependent applications can benefit from accurate \ac{OWD}-estimation mechanisms \cite{Choi2005819}. For example, measurement-based geolocation (see Chapter \ref{ch:background}) may be performed with a greater precision if accurate \ac{OWD}-estimates were relied upon, rather than \acp{RTT}. The common methods that enable a server to accurately measure one-way delays to/from a client \cite{rfc4656} rely on the client's honest cooperation---the client is assumed to synchronize its clock accurately with the server, calculate and honestly report its view of the delays.\footnote{Although \acp{OWD} are generally measured between peers, we use the server/client terminology to discriminate between the party measuring the delays (server) and the one the delays are measured to/from (client).} 

Because \acp{RTT} are easier to estimate than \acp{OWD}, half the \acp{RTT}---or the \emph{average} ($av$) of the actual forward and reverse \acp{OWD}---are often use as \ac{OWD}-estimates \cite{zeitoun2004level}. However, the asymmetric nature of Internet routes \cite{pathak2008measurement} highlights the potential for \acp{OWD} to improve the efficacy of such delay-dependent applications \cite{4438363}. Compared to half the \acp{RTT}, \acp{OWD} are more likely to exclude noisy delay components caused by, e.g., congested or circuitous routes because the delay is measured in one direction.

A delay-based location verification mechanism requires a combination of both: accurate delay-estimation and minimal client cooperation. The accuracy is required to reduce false client rejects and false adversarial accepts as much as possible, while the minimal cooperation is required to reduce the adversary's attack surface. Using common \ac{OWD}-estimation methods (e.g., \ac{OWAMP} \cite{rfc4656}) allows dishonest clients to forge delay-estimates, and using half the $av$ protocol is expected to result in incorrect decisions.

This chapter introduces a new protocol, \emph{minimum pairs} ($mp$), which allows a server to estimate \acp{OWD} between itself and a client by mainly cooperating with two other servers , while requiring less client cooperation than classical \ac{OWD}-estimation protocols; e.g., neither client-clock synchronization nor client trustworthiness in reporting \acp{OWD} is required by the $mp$ protocol. The required client cooperation is similar to that required by the $av$ protocol (i.e., responding to echo-request messages for measuring \acp{RTT}). These features make $mp$ more suitable for location verification, as we show in Chapter~\ref{ch:cpv}.

The $mp$ protocol is evaluated by deriving the probability distribution of its absolute error. Because the protocol's client-cooperation requirements are similar to that of the $av$, we similarly derive the probability distribution of error for the $av$ protocol, and compare both protocols assuming a Poisson delay-distribution. While neither protocol requires client-trustworthiness nor client clock synchronization, the analysis shows that the $mp$ protocol provides more accurate \ac{OWD}-estimates than $av$ in many situations.

This chapter makes the following contributions:
\begin{itemize}
\itemsep0pt
\item Proposing the \emph{minimum pairs} ($mp$) protocol for accurate \ac{OWD}-estimation, which is to be used later in Chapter \ref{ch:cpv} for location verification.
\item Deriving the \ac{PMF} of the absolute error for the $mp$ and the $av$ protocols as a function of the delay distribution between the client and the servers. 
\item Using the derived probability models to compare the accuracy of both protocols assuming Poisson delay distribution with various representative means. This example comparison can now be drawn since the derived models allow general determination of the more accurate protocol given the probability distribution of delays; Poisson is used as an example.
\end{itemize}

The rest of this chapter is organized as follows. Section \ref{sec:threatmodelmp} explains the threat model. Section \ref{sec:mpprotocoldescription} presents the $mp$ protocol, while Sections \ref{sec:analyzingavprotocol} and \ref{sec:analyzingthemp} derive the \ac{PMF} of absolute errors for the $av$ and $mp$ protocols respectively. Section \ref{AccuracyComparison} provides an example of comparing the accuracy of both protocols assuming Poisson delay distribution with various means. Section \ref{conclusion} concludes.

\section{Threat model}
\label{sec:threatmodelmp}

Recall that in \ac{OWAMP}-like protocols, \acp{OWD} between a server and a client are estimated by having them synchronize their clocks together, and exchange timestamps. The server can calculate the \ac{OWD} (at some moment) only in the direction \emph{client-to-server} by subtracting the timestamp that the client sends from the time the stamp was received; the client does the same procedure for calculating \acp{OWD} in the reverse direction, and informs the server with the calculated \acp{OWD}.

Because the $mp$ protocol is designed to address possibly dishonest clients, it must assume the client is able to:
\begin{enumerate}
\itemsep0em
\item Refrain from appropriately synchronizing its clock with the server;
\item Falsify \acp{OWD} before informing the server about them, during the estimation of \emph{server-to-client} \acp{OWD};
\item Falsify the timestamps before sending them, during the estimation of \emph{client-to-server} \acp{OWD}; or
\item Delay or reject timestamp messages.
\end{enumerate}

As we explain in the next section, the $mp$ protocol neither relies on the client's clock, nor on any information reported by the client. Thus, the first three threats do not affect $mp$. In the next chapter, we show how the location verification mechanism itself handles the fourth threat---delaying or rejecting timestamp messages.

\section{The Minimum Pairs Protocol}
\label{sec:mpprotocoldescription}

The $mp$ protocol is designed to estimate the smaller of the forward and reverse \acp{OWD} at current network conditions. The larger \ac{OWD} can then be estimated as the difference between the smaller and the \ac{RTT}. However, we discard the larger delay between the two parties since the smaller provides a more accurate estimate to the distance between them; the larger delay must have been influenced by route congestion, circuitousness \cite{Octant} (see Section \ref{ipgeosection}, page \pageref{ipgeosection}), or other noisy circumstances that increase delays.

To use $mp$, three servers must cooperate together. These servers will be the ones implementing the location verification algorithm later in Chapter~\ref{ch:cpv}, and will be referred to as \emph{verifiers}. To simplify the discussion, we refer to them as \emph{verifiers} from this point on. We assume each of the three verifier possesses a public-private key pair, and is aware of the public keys of the other two verifiers, possibly through a closed \ac{PKI}.

\paragraph{Notation}
There are three bidirectional edges joining a client with three verifiers, and three bidirectional edges joining the three verifiers, as shown in Fig.~\ref{fig:owd:typical}. Each of the six edges has two \acp{OWD} in opposite directions. Denote $\mathbf{D}^\bullet$ as an ordered list holding six \ac{OWD} estimates at a given time. The estimates correspond to the smaller of the forward and reverse \acp{OWD} (i.e., at current network conditions) at each of the six bidirectional edges in Fig.~\ref{fig:owd:typical}. The superscript $\bullet$ is the protocol used to estimate the delays in $\mathbf{D}^\bullet$.

\subsection{Protocol description}

When requesting a location-sensitive service from an \ac{LSP}, the \ac{LSP} notifies the client of the IP addresses of a set, $V$, of three verifiers, which the client must connect to\footnote{The client and the verifiers may use websockets \cite{rfc6455} to connect to the verifiers, as they are a stable means of delay measurement through the browser \cite{appraising}.} in order to have their location verified. Details about how the verifiers are chosen are discussed in Chapter~\ref{ch:cpv}.

Algorithm \ref{unidirectionalprotocol} explains the $mp$ protocol; see Table~\ref{table:owdnotation} for notation used in the algorithm. Note that the location verification protocol presented in Chapter \ref{ch:cpv} also relies on $av$ as an alternative \ac{OWD}-estimation protocol. For convenience, Algorithm \ref{unidirectionalprotocol} also calculates \ac{OWD}-estimates following the $av$ protocol.
\begin{figure}\centering\includegraphics[scale=0.9]{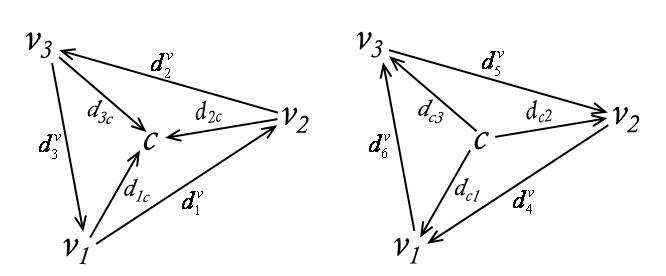}
\caption[Notation of OWDs between a client and three verifiers]{Notation of OWDs between client $c$ and verifiers $v_1$, $v_2$ and $v_3$.}
\label{fig:owd:typical}\end{figure}
\begin{table}
\centering
\caption[Notation used in Chapter \ref{ch:owd}]{Notation}
\begin{tabular}{cl}
Notation	&	Description\\\hline
$S_a(m)$	&	denotes message $m$ digitally signed by entity $a$.\\
$A\xrightarrow{m}B$		&	$A$ sends message $m$ to $B$.\\
$t_a$		&	the most recent timestamp according to verifier $a$'s clock.\\
$e_{ij}$ (line \ref{estimatedTime}) & corresponds to $d_{ic}+d_{cj}$ (see Fig.~\ref{fig:owd:typical}).\\\Xcline{1-2}{2\arrayrulewidth}
\end{tabular}
\label{table:owdnotation}
\end{table}

\begin{algorithm}
\DontPrintSemicolon
\KwIn{The set of the three verifiers, $V$ (see Fig.~\ref{fig:owd:typical}).}
\KwOut{$\mathbf{D}^{mp}$ and $\mathbf{D}^{av}$}
\Begin
{
    \nl\label{constructmessage}\ForEach{$v_i$ {\bf in} $V$}
	{
		\nl$v_i$ retrieves its current system time $b := t_i$\;
		\nl\label{sendclient}$v_i\xrightarrow{b,S_{i}(b)}$ client\;
		\nl\ForEach{$v_j$ {\bf in} $V$}
        {
			\nl\label{forwardingmsg}client $\xrightarrow{b,S_{i}(b)}v_j$\;
			\nl$v_j$ records the message-receiving time $r := t_j$\;
			\nl\label{valsig}$v_j$ validates $S_{i}(b)$\;
			\nl\If{invalid signature}
			{\nl Abort ``possible client cheating attempt"}
			\nl\label{estimatedTime}$e_{ij} := r - b$\;					}
	}
	\nl\For{$i:=1$ {\bf to} $6$}
	{
		\nl\label{selfowd} The verifiers in $V$ measure $d^v_{i}$ (see Fig.~\ref{fig:owd:typical})\;
	}

	\;\tcc{Calculating $\mathbf{D}^{mp}$}	
	\nl\label{simul}$m := \{min(e_{12},\ e_{21}),\ min(e_{23},\ e_{32}),\ min(e_{31},\ e_{13}) \}$\;
	\nl\For{$i:=1$ {\bf to} $3$}
	{
		\nl $j=((i+1)\text{ mod }3)+1$\;
		\nl $k=(i\text{ mod }3)+1$\;
		\nl\label{simul1}$x_i := (m_i + m_j - m_k)/2$\;
		\nl\label{simul2}$y_i := min(d^v_{i},d^v_{i+3})$\;
		\nl\label{simul3}Append $x_i$ and $y_i$ to $\mathbf{D}^{mp}$\;
	}
	
	\;\tcc{Calculating $\mathbf{D}^{av}$}
	\nl\label{avm1}\For{$i:=1$ {\bf to} $3$}
	{
		\nl\label{avmaa}$x_i := e_{ii}/2$\;
		\nl\label{avmbb}$y_i := (d^v_i+d^v_{i+3})/2$\;
		\nl\label{avm3}Append $x_i$ and $y_i$ to $\mathbf{D}^{av}$\;
	}

	\;\nl\Return{$\mathbf{D}^{mp}$ \emph{and} $\mathbf{D}^{av}$}\;
}
\caption{The $mp$ protocol. See notation inline.}
\label{unidirectionalprotocol}
\end{algorithm}

\paragraph{Algorithm Explanation}
The three verifiers take turns to send the client digitally signed timestamps of their most recent system time (line \ref{sendclient}). Once received, the client is required to forward this message to the three verifiers.\footnote{This behavior can be implemented in the browser through javascript.} 
When all three verifiers are done their turns, they will have nine values of delays corresponding to $d_{ic}+d_{cj}$ for all $1\leq i,j \leq 3$. The $mp$ protocol estimates the smaller of $d_{ic}$ and $d_{ci}$ independently, for all $1\leq i\leq 3$, as follows. First, for all $1\leq i,j \leq 3$ and $i\neq j$, the larger of $d_{ic}+d_{cj}$ and $d_{jc}+d_{ci}$ is discarded (line \ref{simul}) because the smaller sums are likely to correspond to the smaller OWDs. Second, the three remaining sums are equated to the corresponding smaller OWDs, and estimates to the smaller delays are obtained by solving simultaneously for $x_1, x_2, x_3$:
\[x_i + x_j = min(d_{ic}+d_{cj},\ d_{jc}+d_{ci})\ \ \ \forall\ 1\leq i < j \leq 3\]
where $x_i$ is the estimate to the smaller of $d_{ic}$ and $d_{ci}$. \label{changesimult}\textcolor{\changes}{To work out these equations, let $m_1$ be the minimum between $d_{1c}+d_{c2}$ and $d_{2c}+d_{c1}$; similarly, $m_2$ is the minimum between $d_{2c}+d_{c3}$ and $d_{3c}+d_{c2}$; and $m_3$ is the minimum between $d_{3c}+d_{c1}$ and $d_{1c}+d_{c3}$. The simultaneous equations are:}\\
\textcolor{\changes}{$x_1 + x_2 = m_1$}\\[-5pt]
\textcolor{\changes}{$x_2 + x_3 = m_2$}\\[-5pt]
\textcolor{\changes}{$x_3 + x_1 = m_3$}\\
\textcolor{\changes}{Solving these equations yields:}\\
\textcolor{\changes}{$x_1 = (m_1 + m_3 - m_2)/2$}\\[-5pt]
\textcolor{\changes}{$x_2 = (m_2 + m_1 - m_3)/2$}\\[-5pt]
\textcolor{\changes}{$x_3 = (m_3 + m_2 - m_1)/2$}\\
This is demonstrated in lines \ref{simul} to \ref{simul1} of Algorithm \ref{unidirectionalprotocol}.

Discarding the larger delays (line \ref{simul}) provides a fundamental advantage to $mp$ over $av$, as it helps reduce the unfavourable effect of delay spikes occurring in one direction but not the other. Compared to $av$, the probability of $mp$ to exclude delay spikes is higher.
In line \ref{selfowd}, estimating the smaller \acp{OWD} of the edges between the verifiers (i.e., $d_i^v$ in Fig.~\ref{fig:owd:typical}) is simpler, since the verifiers trust each other; for example, the \ac{OWAMP} \cite{rfc4656} tool can be used. Again, the verifiers discard the larger of the forward and reverse \acp{OWD} for each of the three edges between them (line \ref{simul2}). Finally, the set $\mathbf{D}^{mp}$ holds the six smaller \ac{OWD} estimates (line \ref{simul3}).

\subsection{Clock synchronization among the verifiers}

In $mp$, the verifiers may choose to synchronize their clocks to the nearest millisecond to increase the accuracy of \ac{OWD} estimates \cite{crovella,4542794}, or use techniques that do not require accurate synchronization \cite{6785479,6177249}. For example, Gurewitz {\it et al.} \cite{gurewitz2006one} proposed a technique that estimates \acp{OWD} in the absence of accurate clock synchronization between network nodes. Strong cooperation between these nodes is, however, required. The nodes conduct many \ac{OWD} measurements among themselves using the poorly synchronized clock, and use those preliminary estimates to derive constraints of an objective function. The function uses optimization techniques, and reaches a per-link \ac{OWD} estimate that minimizes the error with respect to the provided constraints. 

While this class of techniques addresses imperfect clock synchronization, the $mp$ protocol addresses client untrustworthiness. Therefore, such a class of techniques can be used among the verifiers if accurate clock synchronization cannot be achieved. However, due to its strong cooperation and trustworthiness requirement, it cannot be used with potentially dishonest clients.

\section{Analyzing the Average Protocol ($av$)}
\label{sec:analyzingavprotocol}
In this section, the \ac{PMF} of absolute error is derived for the $av$ protocol. The \emph{absolute error} is the absolute difference between the smaller of the forward and reverse \acp{OWD} and the \ac{OWD} estimated by the protocol. Let $f_x(d)$ be the \ac{PMF} of the delay of edge $d$, for each of the six bidirectional edges in Fig.~\ref{fig:owd:typical}.

Throughout this section (and Section~\ref{sec:analyzingthemp}), we focus on the \acp{OWD} between the client and verifier $v_1$ in Fig.~\ref{fig:owd:typical}. Similar analysis applies to the other two bidirectional edges.

\subsection{Absolute error of $av$}
The \emph{av} protocol estimates the smaller \ac{OWD} between $v_1$ and $c$ as:
\begin{equation}
\label{tavmain}
t^{av} = \frac{\text{RTT}}{2} = \frac{d_{1c} + d_{c1}}{2}\end{equation}
The absolute error of the \emph{av} protocol is:\begin{equation*}
\varepsilon^{av}	= \left| t^{av} - min(d_{1c},d_{c1}) \right|
\end{equation*}
The magnitude of the error thus depends on the difference between $d_{1c}$ and $d_{c1}$. Table~\ref{cases_av} lists the three cases. Denoting by $\varepsilon^{av}_i$ the error in Case $i$, then:
\begin{equation*}
\varepsilon^{av}_1	= \left| \frac{d_{1c} + d_{c1}}{2} - d_{1c} \right|= \frac{d_{c1} - d_{1c}}{2}
\end{equation*}
We can drop the \emph{``absolute"} sign ($||$) because in Case 1, $d_{1c}<d_{c1}$. The error for the remaining two cases is given in Table~\ref{cases_av}.
\newcolumntype{L}[1]{>{\raggedright\let\newline\\\arraybackslash\hspace{0pt}}m{#1}}
\newcolumntype{C}[1]{>{\centering\let\newline\\\arraybackslash\hspace{0pt}}m{#1}}
\newcolumntype{R}[1]{>{\raggedleft\let\newline\\\arraybackslash\hspace{0pt}}m{#1}}
\definecolor{Gray1}{gray}{0.98}
\definecolor{Gray2}{gray}{0.95}
\definecolor{Gray3}{gray}{0.75}
\begin{table}
\centering
\caption{Cases relating $d_{1c}$ with $d_{c1}$, the calculated delay ($t^{av}$) in each case, and the error ($\varepsilon^{av}$) of the \emph{av} protocol.}

\begin{tabular}{r | c | c | c  r }\\
\multirow{2}{*}{Case ($i$)}	&Condition&\multirow{2}{*}{$t^{av}_i$}	& \multirow{2}{*}{$\varepsilon^{av}_i$}&\\[4pt]
&$d_{1c}$ [{\tiny relation}] $d_{c1}$	&&&\\\hline
\rowcolor{Gray2}1		&$<$	&$\displaystyle (d_{1c}+d_{c1})/2$	&$\displaystyle (d_{c1}-d_{1c})/2$&\\[8pt]
\rowcolor{Gray1}2		&$=$	&$\displaystyle (d_{1c}+d_{c1})/2$	&0&\\[8pt]
\rowcolor{Gray2}3		&$>$	&$\displaystyle (d_{1c}+d_{c1})/2$	&$\displaystyle (d_{1c}-d_{c1})/2$&\\[8pt]\hline
\end{tabular}\\
\label{cases_av}
\end{table}

\subsection{PMF of error for $av$}
The \ac{PMF} of $\varepsilon^{av}_i$ depends on the probability of occurrence of Case $i$. Thus, for all $x\geq 0$:
\begin{equation}
\label{main_av}
\begin{split}P\{\varepsilon^{av} = x\}	&= \sum_{i=1}^{3}P\{\text{Case } i\}\cdot P\{\varepsilon^{av}_i = x\text{\ \ }|\text{\ \ }\text{Case } i\}\\
							&= \sum_{i=1}^{3}P\{\text{Case } i\}\cdot \frac{P\{\varepsilon^{av}_i = x\intersect\text{Case } i\}}{P\{\text{Case } i\}}\\
						&= \sum_{i=1}^{3}P\{\varepsilon^{av}_i = x\intersect\text{Case } i\}
\end{split}
\end{equation}
where the \emph{``comma"} indicates the intersection of the two events.
Expanding the term at $i=1$ yields:
\begin{equation}
\label{firsterm}
\begin{split}
P\{\varepsilon_1^{av}=x\intersect\text{Case 1}\}
&= P\left\{\frac{d_{c1}-d_{1c}}{2}=x\intersect d_{1c}<d_{c1}\right\}\\
&= P\{d_{c1}=2x+d_{1c}\intersect d_{1c}<d_{c1}\}\\
&= P\{d_{c1}=2x+d_{1c}\intersect d_{1c}<2x+d_{1c}\}\\
&= P\{d_{c1}=2x+d_{1c}\intersect x>0\}\\
&= \left(\displaystyle\sum_{i=0}^{\infty}P\{d_{1c}=i\}\cdot P\{d_{c1} = 2x + i\}\right)\cdot P\{x>0\}\\
&=
\begin{cases}
\displaystyle\sum_{i=0}^{\infty}f_i(d_{1c})\cdot f_{2x+i}(d_{c1}),	&x>0\\[12pt]
0,			&\text{otherwise}
\end{cases}
\end{split}
\end{equation}
Since $\varepsilon_2^{av}=0$ (see Table~\ref{cases_av}), therefore,
\begin{equation*}
\begin{split}
P\{\varepsilon_2^{av}=x\intersect \text{Case 2}\}&=P\{x=0\intersect d_{1c}=d_{c1}\}\\
&=
\begin{cases}
\displaystyle P\{d_{1c}=d_{c1}\},	&x=0\\[4pt]
0,			&\text{otherwise}
\end{cases}
\end{split}
\end{equation*}
where:
\begin{equation*}
P\{d_{1c}=d_{c1}\} = \sum_{i=0}^{\infty}f_i(d_{1c})\cdot f_i(d_{c1})
\end{equation*}
The term for $i=3$ in (\ref{main_av}), $P\{\varepsilon_3^{av}=x\intersect \text{Case 3}\}$, can be expanded analogous to Case 1. We thus rewrite (\ref{main_av}) as:
\begin{equation}
\label{pmfavgfinal}
\begin{split}
P\{\varepsilon^{av} = x\}
&= 
\begin{cases}
	\displaystyle P\{d_{1c}=d_{c1}\},											& x=0\\[4pt]
	\displaystyle P\{\varepsilon^{av}_1 = x\intersect \text{Case 1}\}+P\{\varepsilon^{av}_3 = x\intersect \text{Case 3}\},	& x>0
\end{cases}\\
&=
\begin{cases}
	\displaystyle \sum_{i=0}^{\infty}f_i(d_{1c})\cdot f_i(d_{c1}),			& x=0\\[4pt]
	\displaystyle \sum_{i=0}^{\infty}f_i(d_{1c})\cdot f_{2x+i}(d_{c1}) + \sum_{i=0}^{\infty}f_i(d_{c1})\cdot f_{2x+i}(d_{1c}),	& x>0
\end{cases}\\
\end{split}
\end{equation}

\section{Analyzing the Minimum Pairs Protocol ($mp$)}
\label{sec:analyzingthemp}

In this section, the \ac{PMF} of absolute error is derived for the $mp$ protocol. Again, we focus our analysis on the \acp{OWD} between the client and $v_1$. Throughout the section, the notation $d_{ij}^{+}$ is used to denote $d_{ic}+d_{cj}$; likewise, $d_{ij}^{-}$ denotes $d_{ic}-d_{cj}$.

\subsection{Absolute error of $mp$}

\newcommand*\circled[1]{\tikz[baseline=(char.base)]{\node[shape=circle,draw,inner sep=2pt] (char) {#1};}}
\begin{table*}
\centering
\caption{Cases relating $d_{ij}^{+}$ with $d_{ji}^{+}$, the calculated delay in each case ($t^{mp}_i$), and the absolute error ($\varepsilon^{mp}$) of the \emph{mp} protocol. In each Case, a circled condition is implied by the other two.}
\scalebox{0.67}{
\begin{tabular}{r|ccc|c|C{60pt}|cc}\\
\multirow{2}{*}{Case ($i$)}&\multicolumn{3}{c|}{Conditions}&\multirow{2}{*}{Order}&\multirow{2}{*}{$t^{mp}_i$}& \multicolumn{2}{c}{$\varepsilon^{mp}_{i,j}$}\\[4pt]
&$d_{31}^{+}$ [{\tiny relation}] $d_{13}^{+}$&$d_{21}^{+}$ [{\tiny relation}] $d_{12}^{+}$&$d_{32}^{+}$ [{\tiny relation}] $d_{23}^{+}$&&&$d_{1c}\leq d_{c1}$&$d_{1c}> d_{c1}$\\[4pt]\hline

\rowcolor{Gray2}1&\circled{$<$}	&$\leq$				&$<$	&$d_{33}^{-}<d_{22}^{-}\leq d_{11}^{-}$&$\displaystyle d_{c1}+d_{22}^{-}/2$&$\displaystyle \left|d_{22}^{-}/2-d_{11}^{-}\right|$&$\displaystyle \left|d_{22}^{-}/2\right|$\\[8pt]
\rowcolor{Gray2}2&$<$			&\circled{$<$}		&$\geq$	&$d_{22}^{-}\leq d_{33}^{-}<d_{11}^{-}$&$\displaystyle d_{c1}+d_{33}^{-}/2$&$\displaystyle \left|d_{33}^{-}/2-d_{11}^{-}\right|$&$\displaystyle \left|d_{33}^{-}/2\right|$\\[8pt]

\rowcolor{Gray1}3&$\leq$		&$>$				&\circled{$<$}	&$d_{33}^{-}\leq d_{11}^{-}<d_{22}^{-}$&$\displaystyle d_{11}^{+}/2$&$\displaystyle -d_{11}^{-}/2$&$\displaystyle d_{11}^{-}/2$\\[8pt]
\rowcolor{Gray1}4&$=$			&$=$				&\circled{$=$}	&All three are equal&$\displaystyle d_{11}^{+}/2$&$\displaystyle -d_{11}^{-}/2$&$\displaystyle d_{11}^{-}/2$\\[8pt]
\rowcolor{Gray1}5&$\geq$		&$<$				&\circled{$>$}	&$d_{22}^{-}<d_{11}^{-}\leq d_{33}^{-}$&$\displaystyle d_{11}^{+}/2$&$\displaystyle -d_{11}^{-}/2$&$\displaystyle d_{11}^{-}/2$\\[8pt]

\rowcolor{Gray2}6&$>$			&\circled{$>$}		&$\leq$	&$d_{11}^{-}<d_{33}^{-}\leq d_{22}^{-}$&$\displaystyle d_{1c}-d_{33}^{-}/2$&$\displaystyle \left|-d_{33}^{-}/2\right|$&$\displaystyle \left|d_{11}^{-}-d_{33}^{-}/2\right|$\\[8pt]
\rowcolor{Gray2}7&\circled{$>$}	&$\geq$				&$>$	&$d_{11}^{-}\leq d_{22}^{-}<d_{33}^{-}$&$\displaystyle d_{1c}-d_{22}^{-}/2$&$\displaystyle \left|-d_{22}^{-}/2\right|$&$\displaystyle \left|d_{11}^{-}-d_{22}^{-}/2\right|$\\[8pt]\hline

&$d_{33}^{-}$ [{\tiny relation}] $d_{11}^{-}$&$d_{22}^{-}$ [{\tiny relation}] $d_{11}^{-}$&$d_{33}^{-}$ [{\tiny relation}] $d_{22}^{-}$&&\\[4pt]
&\multicolumn{3}{c|}{Rearranged Conditions}&&&&\\[4pt]

\end{tabular}
}\\
\label{cases}
\end{table*}

In Algorithm~\ref{unidirectionalprotocol}, lines \ref{simul1} to \ref{simul3} define three simultaneous equations that estimate the smaller \ac{OWD} ($t^{mp}$). Although the \emph{mp} protocol does not enable the verifiers to calculate $d_{ii}^{-}$ for all $i\in\{1,2,3\}$, it enables them to sort these differences. For example, assume in line~\ref{simul1} that $d_{2c}+d_{c1} \leq d_{1c}+d_{c2}$. Rearranging yields $d_{22}^{-} \leq d_{11}^{-}$. Also assuming in line~\ref{simul2} that $d_{3c}+d_{c2}<d_{2c}+d_{c3}$ (equivalent to $d_{33}^{-} < d_{22}^{-}$), the verifiers can deduce that $d_{33}^{-}<d_{22}^{-}\leq d_{11}^{-}$.

The order of $d_{11}^{-}$, $d_{22}^{-}$ and $d_{33}^{-}$ identifies the cases in Table~\ref{cases}; possible outcomes of the $min()$ function in lines \ref{simul1} to \ref{simul3} are indicated at the header of the \emph{``Conditions"} column, with their rearrangements indicated at the bottom. Two conditions imply the third; the implied condition is circled in Table~\ref{cases}.

The smaller between $d_{1c}$ and $d_{c1}$ is indicated by the $t^{mp}_i$ column in Table~\ref{cases}. In Case 1 for example, where $d_{31}^{+}<d_{13}^{+}$, $d_{21}^{+}\leq d_{12}^{+}$, and $d_{32}^{+}<d_{23}^{+}$, the simultaneous equations of lines \ref{simul1} to \ref{simul3} will be $\beta_1+\beta_2=d_{21}^{+}$, $\beta_2+\beta_3=d_{32}^{+}$, and $\beta_3+\beta_1=d_{31}^{+}$. In Algorithm~\ref{unidirectionalprotocol}, $\beta_1$ is returned as the estimate to the smaller between $d_{1c}$ and $d_{c1}$, which evaluates to:
\begin{equation*}
\begin{split}
t_1^{mp} = \beta_1 &= \frac{d_{21}^{+}+d_{31}^{+}-d_{32}^{+}}{2}\\
		&= \frac{d_{2c}+d_{c1}+d_{3c}+d_{c1}-(d_{3c}+d_{c2})}{2}\\
		&= \frac{d_{2c}-d_{c2}+2d_{c1}}{2}= d_{c1}+\frac{d_{22}^{-}}{2}
\end{split}
\end{equation*}
Similarly, $t_i^{mp}$ can be calculated for the remaining cases.

The returned \ac{OWD} estimate ($t^{mp}$) can indicate whether there were large delay asymmetries between each verifier and the client. For example, if $t^{mp}<0$, then the difference between the forward and reverse delays of some links between the client and the verifiers is relatively large.

\subsection{Comparison between $t^{mp}$ and $t^{av}$}
As is now shown, in none of the seven cases will the $mp$ protocol return a larger estimate to the smaller \ac{OWD} than that of the $av$ protocol; that is, the inequality $t_i^{mp}\leq t^{av}$ holds for all $i\in\{1..7\}$. In Case 1, we have (Table~\ref{cases}):
\begin{equation}
\label{tmpcase1}
t^{mp}_1 = d_{c1}+\frac{d_{22}^{-}}{2}\end{equation}
Since $d_{22}^{-}\leq d_{11}^{-}$ in this case (second rearranged condition, bottom of the \emph{``Conditions"} column in Table~\ref{cases}), therefore:
\begin{equation*}
t^{mp}_1 \leq d_{c1}+\frac{d_{11}^{-}}{2}
\end{equation*}
Simplifying yields
\begin{equation*}
t^{mp}_1 \leq  \frac{d_{1c}+d_{c1}}{2} = t^{av}\text{\ \ \ \ \ \ from (\ref{tavmain})}
\end{equation*}
Analogous analysis applies to Cases 2, 6 and 7, which we omit for conciseness. The equation $t_i^{mp}=t^{av}$ already holds for $i\in\{3,4,5\}$ (see Table~\ref{cases}). Thus, the $mp$ protocol never returns an estimate, to the smaller between the forward and reverse \acp{OWD}, that is larger than that of the $av$ protocol.

\subsection{PMF of error for $mp$}
The \ac{PMF} of error depends on the probability of occurrence of each case in Table~\ref{cases}, and the probabilities of $d_{1c}\leq d_{c1}$ and $d_{1c}> d_{c1}$ in each case. We index those two additional conditions using the variable $j\in\{1,2\}$ respectively. For example, to calculate the error in Case 1 given additional condition 2 (which is $d_{1c}> d_{c1}$):
\begin{equation*}
\varepsilon_{1,2}^{mp}	=|t_1^{mp}-min(d_{1c},d_{c1})|=\left|d_{c1}+\frac{d_{22}^{-}}{2}-d_{c1}\right|=\left|\frac{d_{22}^{-}}{2}\right|
\end{equation*}

The probability that the error is equal to $x$ is the probability that any of the expressions listed under the $\varepsilon^{mp}_{i,j}$ column in Table~\ref{cases} evaluates to $x$, for all $x\geq 0$. The \ac{PMF} of the absolute error can, thus, be expressed as:
\begin{equation}
\label{gloablformp}
\begin{split}P\{\varepsilon^{mp} = |x|\}	&= \sum_{i=1}^{7}\sum_{j=1}^{2}P\{X_{i,j}\}\cdot P\{\varepsilon^{mp}_{i,j}=|x|\text{\ \ }|\text{\ \ }X_{i,j}\}\\
						&= \sum_{i=1}^{7}\sum_{j=1}^{2}P\{X_{i,j}\}\cdot \frac{P\{\varepsilon^{mp}_{i,j}=|x|\intersect X_{i,j}\}}{P\{X_{i,j}\}}\\
						&= \sum_{i=1}^{7}\sum_{j=1}^{2}P\{\varepsilon^{mp}_{i,j}=|x|\intersect X_{i,j}\}\\
\end{split}
\end{equation}
where $X_{i,j}$ is the intersection of all three conditions under the \emph{``Conditions"} column of Case $i$ with additional condition $j$. Because the error, $\varepsilon^{mp}_{i,j}$, in each of those 14 cases is the absolute difference, then:
\begin{equation}\label{sixsix}
P\{\varepsilon^{mp}_{i,j} = |x|\intersect X_{i,j}\}=
\begin{cases}
	\displaystyle P\{\varepsilon^{mp}_{i,j} = 0\intersect X_{i,j}\},				& x=0\\[4pt]
	\displaystyle P\{\varepsilon^{mp}_{i,j} = x\intersect X_{i,j}\}+P\{\varepsilon^{mp}_{i,j} = -x\intersect X_{i,j}\},	& \text{otherwise}
\end{cases}
\end{equation}
At $i=1$ and $j=1$, the event $X_{1,1}$ is (from Table~\ref{cases}):
\begin{equation*}
X_{1,1} = (d_{31}^{+}<d_{13}^{+})\mycap (d_{21}^{+} \leq d_{12}^{+})\mycap (d_{32}^{+}<d_{23}^{+})\mycap (d_{1c}\leq d_{c1})
\end{equation*}
The condition $d_{31}^{+}<d_{13}^{+}$ can be removed because it is implied by the other two conditions in Case 1, Table~\ref{cases}. Therefore:
\begin{equation*}
\begin{split}
X_{1,1} 
&= (d_{21}^{+} \leq d_{12}^{+})\mycap (d_{32}^{+}<d_{23}^{+})\mycap (d_{1c}\leq d_{c1})\\
&= (d_{22}^{-} \leq d_{11}^{-})\mycap (d_{33}^{-}<d_{22}^{-})\mycap (d_{11}^{-}\leq 0)
\end{split}
\end{equation*}
By substitution, we have\vspace{5pt}\\
$P\{\varepsilon^{mp}_{1,1}=x\intersect X_{1,1}\}$\nopagebreak\vspace{-2pt}
\begin{equation*}
\begin{split}
&=P\{\frac{d_{22}^{-}}{2}-d_{11}^{-}=x\intersect d_{22}^{-}\leq d_{11}^{-}\intersect d_{33}^{-}<d_{22}^{-}\intersect d_{11}^{-}\leq 0\}\\
&=\sum_{i=-\infty}^{0}P\{d_{22}^{-}=2(i+x)\intersect d_{22}^{-}\leq i\intersect d_{33}^{-}<d_{22}^{-}\intersect d_{11}^{-}=i\}\\
&=\sum_{i=-\infty}^{0}\left(P\{d_{11}^{-}=i\}\cdot P\{d_{22}^{-}=2(i+x)\intersect d_{22}^{-}\leq i\intersect d_{33}^{-}<d_{22}^{-}\}\right)\\ &=\sum_{i=-\infty}^{0}\left(g_i(d_{1c},d_{c1})\cdot\sum_{j=-\infty}^{i}P\{j=2i+2x\intersect d_{22}^{-}=j\intersect d_{33}^{-}<j\}\right)\\ 
&=\sum_{i=-\infty}^{0}\left(g_i(d_{1c},d_{c1})\cdot\sum_{j=-\infty}^{i}\left(P\{j=2i+2x\}\cdot P\{d_{22}^{-}=j\}\cdot P\{d_{33}^{-}<j\}\right)\right)\\ 
&=\sum_{i=-\infty}^{0}\left(g_i(d_{1c},d_{c1})\cdot\sum_{j=-\infty}^{i}\left(P\{j=2i+2x\}\cdot g_j(d_{2c},d_{c2})\cdot \sum_{k=-\infty}^{j-1}P\{d_{33}^{-}=k\}\right)\right)\\ 
&=\sum_{i=-\infty}^{0}\left(g_i(d_{1c},d_{c1})\cdot \sum_{j=-\infty}^{i}\left(P\{j=2i+2x\}\cdot g_j(d_{2c},d_{c2})\cdot \sum_{k=-\infty}^{j-1}g_k(d_{3c},d_{c3})\right) \right)\\ 
\end{split}
\end{equation*}
where the function $g_x(Y,Z)$ is the probability $P\{Y-Z = x\}$ for two independent discrete random variables $Y$ and $Z$. It is calculated as follows:
\begin{equation*}
\begin{split}
&g_x(Y,Z) = P\{Y-Z = x\} = P\{Y = x + Z\}  \\
&= \sum_{i=-\infty}^{\infty}P\{Z=i\}\cdot P\{Y = x + i\}  = \sum_{i=\infty}^{\infty}f_i(Z)\cdot f_{x+i}(Y)
\end{split}
\end{equation*}
This concludes an example expansion to one of the terms in (\ref{sixsix}). Analogous expansion could be made for the remaining terms, which we omit for conciseness.

\section{Examples of Accuracy Comparison}
\label{AccuracyComparison}

\begin{table}
\centering
\caption{Means of the Poisson distributions of the delays for each edge in Fig.~\ref{fig:owd:typical}, and their corresponding chart in Fig.~\ref{CDFs}.}
\begin{tabular}{rl | cc c cc c cc  }
\multicolumn{2}{c|}{\multirow{2}{*}{Scenario}}		& \multicolumn{8}{c}{Mean (ms)}\\
&&$d_{1c}$&$d_{c1}$&&$d_{2c}$&$d_{c2}$&&$d_{3c}$&$d_{c3}$\\\hline
\multirow{6}{*}{Fig. 2}&(a)&30&30&&30&30&&30&30\\
&(b)&30&7&&8&25&&5&5\\
&(c)&2&20&&5&50&&7&80\\
&(d)&35&5&&45&70&&2&15\\
&(e)&10&10&&30&12&&30&60\\
&(f)&10&10&&30&3&&20&5\\\Xcline{1-10}{2\arrayrulewidth}
\end{tabular}
\label{means}
\end{table}

It has been established that Internet delays follow a Gamma distribution with varying parametrization \cite{Mukherjee92onthe,bovy2002analysis}. We model the \acp{OWD} of the six edges of Fig.~\ref{fig:owd:typical} as independent and discrete random variables that follow Poisson distributions,\footnote{Note that this is not the packet arrival times.} and take on integer values (e.g., delays in milliseconds). Poisson is used because it is a discrete distribution that is a special case of Gamma. Table~\ref{means} lists the distribution means in six example scenarios. The scenarios were chosen to analyze the effect of delay asymmetry between the client and the verifiers. Figure~\ref{CDFs} plots the Cumulative Distribution Functions (CDFs) of the absolute errors for each scenario in Table~\ref{means}, using (\ref{main_av}) and (\ref{gloablformp}) for the $av$ and the $mp$ protocols respectively.

Scenario (a) (Table~\ref{means}) addresses delay symmetry in all six edges.\footnote{\label{changeclarifympav}\textcolor{\changes}{Note that the numbers in Table~\ref{means} do not represent the delays on each edge. The delays are rather modeled as a random variable following Poisson distributions with the means listed in the table.}} Figure~\ref{awelwa7ed} shows that $mp$ is more accurate than $av$ in this scenario, with a 54\% chance of producing an absolute error $<$1.5 ms, versus 35\% for $av$. 

Scenario (b) addresses the effect of delay symmetry between the client and one verifier. In this scenario, we deduce that $mp$ will operate in Case 2 most of the time (from the \emph{``Order"} column in Table~\ref{cases}), and thus $\varepsilon^{mp}=\varepsilon_{2,2}^{mp}$ as it is highly probable that $d_{1c}>d_{c1}$. Because $d_{3c}$ and $d_{c3}$ have equal means (5 ms), the error $\varepsilon_{2,2}^{mp} = |d_{33}^{-}/2|$ becomes relatively small, as shown in Fig.~\ref{tanywa7ed}. The $mp$ protocol has a 90\% chance of resulting in $<$2.5 ms absolute error, versus 0.1\% for the $av$, making it significantly more accurate in this scenario.

Scenarios (c) and (d) explore delay asymmetry in all six edges. Despite the huge asymmetries in (c), $mp$ has a $\sim$25\% chance to result in $<$2.5 ms absolute error, versus $\sim$0.2\% for $av$. The smaller delay variations of scenario (d), compared to (c), caused $mp$ to be substantially more accurate (Fig.~\ref{rabe3wa7ed}).

Scenarios (e) and (f) analyze the effect of delay symmetry between $d_{1c}$ and $d_{c1}$, and asymmetry in the other two links. In Fig.~\ref{khameswa7ed}, where the two graph lines coincide, the accuracy of $mp$ is similar to that of $av$ because, with higher probability, $mp$ operates in Case 3 of Table~\ref{cases} (the resulting \ac{OWD}-estimates are similar to $av$). In (f), delay asymmetry between the client and \{$v_2$, $v_3$\} mislead $mp$, but do not affect the average of $d_{1c}$ and $d_{c1}$. Because $d_{1c}$ and $d_{c1}$ are highly symmetric (see Table~\ref{means}), $av$ is more accurate.

\newcommand{\widthowd}{2.1in}
\newcommand{\heightowd}{\widthowd}

\begin{figure}
\centering
\subfigure[]{
\label{awelwa7ed}
\begin{tikzpicture}
\begin{axis}[
width=\widthowd,
height=\heightowd,
xlabel=absolute error (ms),
ylabel=CDF,
xmin=0, xmax=10,
ymin=0, ymax=1,
ytick ={0,1},
xtick ={0,5,10},
extra x ticks={1.5},
extra x tick style={grid=major},
extra y ticks={0.35,0.54},
extra y tick style={grid=major,yticklabels={,,}},
tick label style={font=\scriptsize},
x label style={font=\scriptsize},
y label style={font=\scriptsize,at={(0.24,0.5)}},
legend columns=-1,
legend style={draw=none,at={(0.52,1.15)},anchor=north,}
]
\addplot[] table[col sep=comma]{csv/OWD/CDF/owd-a.csv};
\addlegendentry{\emph{mp}}
\addplot[dashed] table[col sep=comma]{csv/OWD/CDF/rtt-a.csv};
\addlegendentry{\emph{av}}
\end{axis}
\end{tikzpicture}
}
\subfigure[]{
\label{tanywa7ed}
\begin{tikzpicture}
\begin{axis}[
width=\widthowd,
height=\heightowd,
xlabel=absolute error (ms),
ylabel=CDF,
xmin=0, xmax=20,
ymin=0, ymax=1,
ytick ={0,1},
xtick ={0,10,20},
tick label style={font=\scriptsize},
x label style={font=\scriptsize},
y label style={font=\scriptsize,at={(0.24,0.5)}},
legend columns=-1,
legend style={draw=none,at={(0.52,1.15)},anchor=north,}
]
\addplot[] table[col sep=comma]{csv/OWD/CDF/owd-b.csv};
\addlegendentry{ \emph{mp}}
\addplot[dashed] table[col sep=comma]{csv/OWD/CDF/rtt-b.csv};
\addlegendentry{\scriptsize \emph{av}}
\end{axis}
\end{tikzpicture}
}
\subfigure[]{
\label{taletwa7ed}
\begin{tikzpicture}
\begin{axis}[
width=\widthowd,
height=\heightowd,
xlabel=absolute error (ms),
ylabel=CDF,
xmin=0, xmax=20,
ymin=0, ymax=1,
ytick ={0,1},
xtick ={0,10,20},
tick label style={font=\scriptsize},
x label style={font=\scriptsize},
y label style={font=\scriptsize,at={(0.24,0.5)}},
legend columns=-1,
legend style={draw=none,at={(0.52,1.15)},anchor=north,}
]
\addplot[] table[col sep=comma]{csv/OWD/CDF/owd-c.csv};
\addlegendentry{ \emph{mp}}
\addplot[dashed] table[col sep=comma]{csv/OWD/CDF/rtt-c.csv};
\addlegendentry{\scriptsize \emph{av}}
\end{axis}
\end{tikzpicture}
}\hspace{20 pt}
\subfigure[]{
\begin{tikzpicture}
\label{rabe3wa7ed}
\begin{axis}[
width=\widthowd,
height=\heightowd,
xlabel=absolute error (ms),
ylabel=CDF,
xmin=0, xmax=30,
ymin=0, ymax=1,
ytick ={0,1},
xtick ={0,15,30},
tick label style={font=\scriptsize},
x label style={font=\scriptsize},
y label style={font=\scriptsize,at={(0.24,0.5)}},
legend columns=-1,
legend style={draw=none,at={(0.52,1.15)},anchor=north,}
]
\addplot[] table[col sep=comma]{csv/OWD/CDF/owd-d.csv};
\addlegendentry{ \emph{mp}}
\addplot[dashed] table[col sep=comma]{csv/OWD/CDF/rtt-d.csv};
\addlegendentry{\scriptsize \emph{av}}
\end{axis}
\end{tikzpicture}
}
\subfigure[]{
\begin{tikzpicture}
\label{khameswa7ed}
\begin{axis}[
width=\widthowd,
height=\heightowd,
xlabel=absolute error (ms),
ylabel=CDF,
xmin=0, xmax=10,
ymin=0, ymax=1,
ytick ={0,1},
xtick ={0,5,10},
tick label style={font=\scriptsize},
x label style={font=\scriptsize},
y label style={font=\scriptsize,at={(0.24,0.5)}},
legend columns=-1,
legend style={draw=none,at={(0.52,1.15)},anchor=north,}
]
\addplot[] table[col sep=comma]{csv/OWD/CDF/owd-e.csv};
\addlegendentry{ \emph{mp}}
\addplot[dashed] table[col sep=comma]{csv/OWD/CDF/rtt-e.csv};
\addlegendentry{\scriptsize \emph{av}}
\end{axis}
\end{tikzpicture}
}
\subfigure[]{
\begin{tikzpicture}
\label{sadeswa7ed}
\begin{axis}[
width=\widthowd,
height=\heightowd,
xlabel=absolute error (ms),
ylabel=CDF,
xmin=0, xmax=20,
ymin=0, ymax=1,
ytick ={0,1},
xtick ={0,10,20},
tick label style={font=\scriptsize},
x label style={font=\scriptsize},
y label style={font=\scriptsize,at={(0.24,0.5)}},
legend columns=-1,
legend style={draw=none,at={(0.52,1.15)},anchor=north,}
]
\addplot[] table[col sep=comma]{csv/OWD/CDF/owd-f.csv};
\addlegendentry{ \emph{mp}}
\addplot[dashed] table[col sep=comma]{csv/OWD/CDF/rtt-f.csv};
\addlegendentry{\scriptsize \emph{av}}
\end{axis}
\end{tikzpicture}
}\caption[Absolute errors between the estimated and the actual OWD]{Absolute errors between the estimated and the actual OWD, assuming Poisson delay distributions (see Table~\ref{means} for means) for the edges in Fig.~\ref{fig:owd:typical}.}
\label{CDFs}
\end{figure}
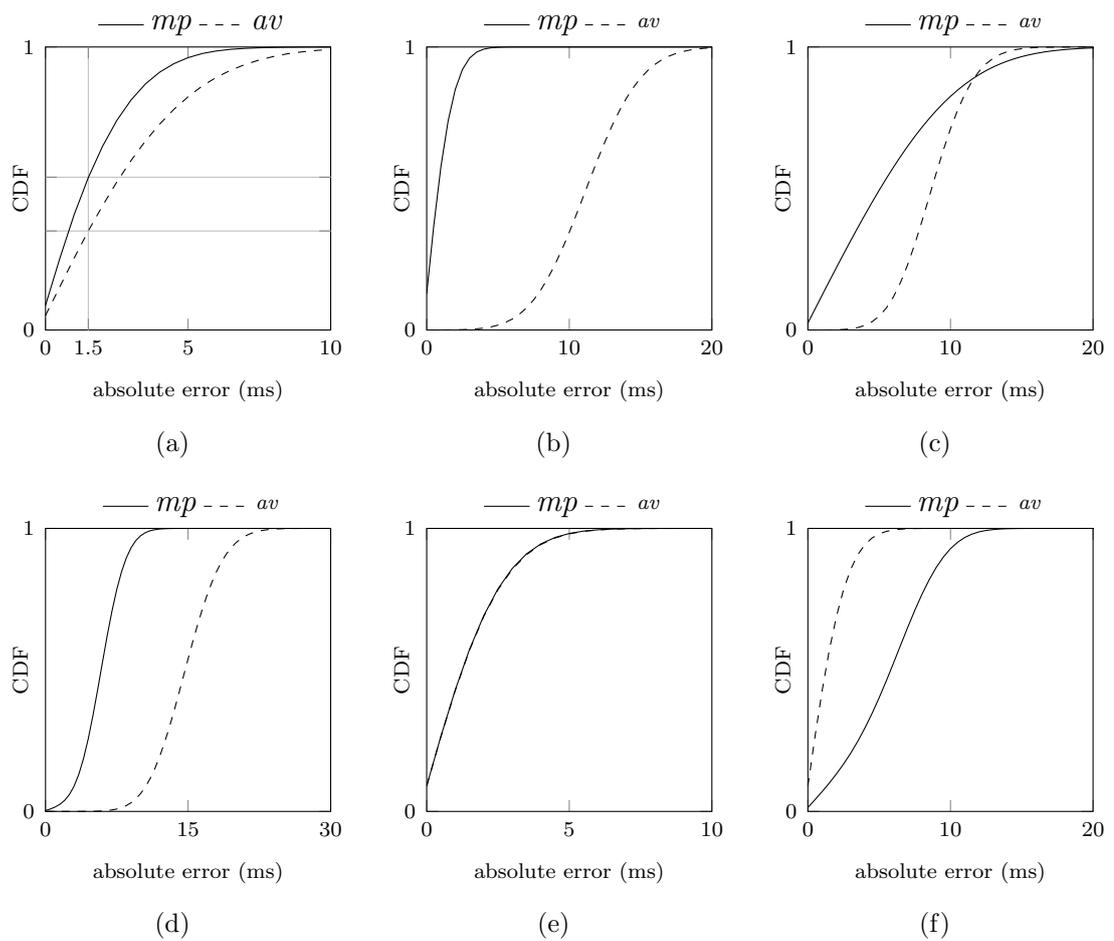

\section{Related Work}
\label{ch:owd:relatedwork}

\label{changeowdlit}
\textcolor{\changes}{
Most research in the area of accurate \ac{OWD} estimation is primarily to achieve accurate clock synchronization \cite{surowd}, e.g., by predicting delay jitters \cite{hua2013analysis}. Estimation errors and the accuracy of clock synchronization are two metrics generally used to evaluate an \ac{OWD}-estimation technique. Commonly, there is a tradeoff between the two metrics. The \ac{OWAMP} tool \cite{rfc4656} is a popular example that relies on clock synchronization to accurately estimate \acp{OWD}. In the lack of synchronized clocks, the typical method is to measure the \ac{RTT}, and use its half as an estimate to the \ac{OWD} \cite{zeitoun2004level}.}

\textcolor{\changes}{
Other methods leveraged the accuracy of \ac{GPS} clocks to enhance \ac{OWD} estimation \cite{6993464}. Additionally, since network queuing delays constitute the most unpredictable delay component, researchers have worked towards devising techniques that enable a sender and a receiver of a \ac{VoIP} application estimate one-way queuing delays without requiring perfect clock synchronization \cite{5168507}. Despite addressing imperfect clock synchronization, all these proposals assume honest cooperation between both parties, and thus cannot be used in hostile environments.}

\section{Conclusion}
\label{conclusion}

This chapter proposed a novel \ac{OWD}-estimation protocol that combines accuracy and reduced-cooperation advantages over current state-of-the art techniques that provide one advantage but not the other. The protocol was formally analyzed by deriving the probability distribution of its absolute error, and comparing it with that of $av$. The comparison establishes that the $mp$ protocol is in many cases more accurate in estimating \acp{OWD} than the commonly-used $av$ protocol. This is achieved with the added bonus of the $mp$'s reduced client-cooperation requirements, making it suitable for adversarial environments, but comes at the cost of requiring extra infrastructure (the verifiers).

The probability distribution models derived herein for the $mp$ protocol did not consider errors due to imperfect clock synchronization among the verifiers because such errors can be mitigated as shown in the literature \cite{6177249}.

We highlight that the degree of delay asymmetry between the verifiers and the client is a key element affecting the accuracy of both protocols. The \acp{PMF} derived herein are thus useful to an application deciding between both protocols. This follows from the properties of the \acp{PMF} derived herein: (1) they allow determination of which protocol is more accurate in estimating \acp{OWD} given the delay environment, and (2) they are generic---they evaluate the probability mass of error given any discrete delay distribution (Poisson was used herein). \label{changegammavspoisson} \textcolor{\changes}{Note however that, despite being generic, the \acp{PMF} derived herein must be used with discrete delay distributions. We did not pursue the avenue of \acp{PDF} that can be used with continuous delay distributions models, and therefore cannot advise on whether any technical difficulties would be encountered. However, it would appear that analogous steps would provide a corresponding analysis for the case of continuous distributions. This is left as future work.}

\chapter{CPV: Delay-based Location Verification for the Internet}
\label{ch:cpv}

The number of location-aware services over the Internet continues growing. Some of these require the client's geographic location for security-sensitive applications. Examples include location-aware authentication \cite{5738981,otppwd2}, location-aware access policies, fraud prevention, complying with media licensing \cite{HBOcracking}, and regulating online gambling/voting. An adversary can evade existing geolocation techniques, e.g., by faking \ac{GPS} coordinates or employing a non-local IP address through proxy and virtual private networks. This chapter presents \ac{CPV}, a delay-based technique designed to verify an assertion about a device's presence inside a prescribed geographic region. \ac{CPV} does not identify devices by their IP addresses. Rather, the device's location is corroborated in a novel way by leveraging geometric properties of triangles, which prevents an adversary from manipulating measured delays. To achieve high accuracy, \ac{CPV} mitigates Internet path asymmetry using the \ac{OWD}-estimation protocol introduced in Chapter \ref{ch:owd}, and leverages delay-related information for evidence supporting/refuting the asserted location. We explain the threat model, detail the \ac{CPV} algorithm, and discuss its security benefits.

\section{Introduction}
\blfootnote{The content of this chapter is accepted for publication at IEEE TDSC \cite{cpvtdsc}.}

Over the Internet, \acp{LSP} are those that customize their content/services based on the geographic locations of their \emph{clients} (the software that communicates with the \ac{LSP}, typically a web-browser). Some \acp{LSP} restrict their services to certain geographic regions, such as media streaming \cite{burnett2013geographically} (e.g., hulu.com); others limit certain operations to a specific location, such as online voting (e.g., placespeak.com), online gambling (e.g., ballytech.com), location-based social networking \cite{acsaclbs} (e.g., foursquare.com), or fraud prevention (e.g., optimalpayments.com). \acp{LSP} may also use location information as an additional authentication factor to thwart impersonation and password-guessing attacks (e.g., facebook.com). Privacy laws differ by jurisdiction, which allows/bans content based on region \cite{trimble2011future}. The nature of the provided services may motivate clients to forge their location to gain unauthorized access.

Existing geolocation technologies, commonly used in practice, are susceptible to evasion \cite{MuirPaul}, as discussed in Section \ref{vulnsinbackgroundall} (page \ref{vulnsinbackgroundall}). Tabulation-based techniques, where a geolocation service provider maintains tables that map IP addresses to locations---e.g., MaxMind \cite{maxmind}, can be evaded through IP address-masking technologies \cite{shroud} such as proxy servers and anonymizers \cite{dingledine2004tor}. Geolocation that is based on active delay measurements \cite{laki2011spotter,Arif:2010:GIH:1862199.1862209} is prone to an adversary corrupting the delay-measuring process \cite{Dude}. A location verification technique is therefore required to provide greater assurance of the veracity of the specified location.
 
Various solutions have been proposed to verify location claims in wireless networks \cite{wagner,1498470}. However, solutions in this domain cannot be directly adopted by multi-hop networks, e.g., the Internet, due to delay characteristics of different domains. For example, Internet delays are stochastic \cite{Dong201285}, whereas in single-hop wireless networks, delays can be estimated from the distance the signal spans and the speed of its propagation.

Verifying the location of Internet clients is a challenging problem \cite{MuirPaul}. A practical approach must address critical challenges such as handling of IP address-masking, and ensuring the correctness of location information submitted by the client. We present and evaluate \ac{CPV}, a delay-based technique designed to verify a client's geographic location. Experimental results show that \ac{CPV} provides a high level of assurance that a correct (i.e., honest) location assertion is verified to a granularity equivalent to a circle of radius $\sim$400$km$. \ac{CPV} is designed to resist known geolocation-circumvention tactics as it (1) does not rely on the client's IP address, (2) does not rely on client-submitted information, and (3) is designed such that manipulating the delays is not in the dishonest client's favor, e.g., \ac{CPV} precludes the attacks of Chapter \ref{ch:attack}, as well as those of Gill {\it et al.} \cite{Dude}.

A common challenge faced by delay-based geolocation techniques is to find an accurate delay-to-distance mapping function, and thus factors affecting the correctness of this mapping have been well studied in the literature \cite{improvacc,landa2013measuring}. \ac{CPV} undertakes a set of measures to mitigate the effect of these factors. For example, it mitigates path asymmetry \cite{pathak2008measurement} by relying on \ac{OWD}-estimates, instead of \acp{RTT}, to/from a potentially dishonest client, using the \emph{minimum pairs} protocol introduced in Chapter \ref{ch:owd}. Additionally, \ac{CPV} mitigates network instability \cite{6558501} by iterating the \ac{OWD}-estimation process.

In Chapter \ref{ch:wiredecva}, the effect of several factors on the correctness of \ac{CPV} is analyzed by evaluating its \ac{FR} and \ac{FA} rates using PlanetLab \cite{planetlab}, where all modeled clients are assumed to use wired access networks. Further in Chapter \ref{ch:wirelessecva}, the correctness of \ac{CPV} is analyzed when only legitimate clients are using wireless access networks.

The rest of this chapter is organized as follows. Section \ref{cpv:background} provides a summary of the literature on delay behavior over the Internet, and its relationship to geographic distances. The threat model is discussed in Section \ref{threatmodel}, and \ac{CPV} is explained in Section \ref{locationverification}. A security discussion is presented in Section \ref{SecurityAnalysis}. Section \ref{cpv:conclusion} concludes.

\section{Background}
\label{cpv:background}

Delay characterization between Internet hosts plays a prominent role in numerous applications such as distributed web-caching, server placement in \acp{CDN}, clock synchronization, overlay \ac{P2P} networks, Internet geolocation, application-layer mutlicast, and timeout estimations in \ac{TCP}. Due to the importance of understanding the impacts of delays between Internet hosts on delay-dependent applications, factors affecting these delays have been well studied \cite{improvacc,wang2007towards,Octant,landa2013measuring} including the spanned geographic distances, routing policies, etc.

Delay-based IP geolocation includes a broad class of techniques aiming to calculate the geographic location of a client based on the delays observed between the client and a set of landmarks with known locations \cite{Constrainbased}. Most techniques apply regression analysis to find a function that best models the relationship between the measured delays and geographic distances \cite{laki2011spotter,Dong201285}. Multilateration is then used on the distances mapped between the landmarks and the client to constrain the region where the client is located. Recent techniques incur a median error of as low as a few kilometres \cite{laki2011spotter}. To infer distances from delays, the speed at which packets are transmitted over the Internet has been approximated by Katz-Bassett {\it et al.} \cite{delayandtopology} to 4/9 the speed of light in vacuum, a ratio called the \ac{SOI} \cite{delayandtopology}. However, the actual speed is affected by several factors such as time of the day, region and characteristics of the underlying network. Based on 19 million \ac{RTT} measurements in the Internet, Landa {\it et al.} \cite{landa2013measuring} found that the knowledge of the geographic distance between two nodes, their /8 IP prefixes, and their countries can help scope down delay-estimation errors to within $\sim$22$ms$.

\acfp{NCS} \cite{dabek2004vivaldi} model a network as a geometric space by assigning coordinates to each node in the network. The coordinates denote a node's position relative to other nodes in the \emph{network delay space}, i.e., according to its delay to/from them. One essential advantage of \acp{NCS} is the ability to locate a node's network position relative to \emph{almost all} other nodes without overwhelming the network with storms of delay sampling \cite{donnet2010survey}. \acp{NCS} are vulnerable to an adversary falsifying its coordinates \cite{girlich2013bounds}.

The aforementioned delay studies provide solid evidence of a strong correlation between Internet delays and geographic distances \cite{yook2002modeling}, which is commonly speculated to stem from improved global network connectivity \cite{Constrainbased}. \ac{CPV} leverages these results to address location verification.

\section{Threat Model}
\label{threatmodel}

We now explain the threat model addressed by \ac{CPV}. Note that this threat model is different from the adversarial models explained in Section \ref{attack:threatmodel} (page \pageref{attack:threatmodel}); those in Chapter \ref{ch:attack} explain how various adversarial capabilities can manipulate delay-based geolocation. 

The adversary is a human user that programs its client software to evade a geolocation process, to intentionally misrepresent its location. The adversary is in physical possession of the client device (e.g., laptop or smartphone), which is connected to the Internet and thereby to the \ac{LSP}. The adversary has full control over its client device; it can install/uninstall any software.

We consider within scope an adversary that uses public proxies, \acp{VPN} and/or anonymizers to hide its IP address or to hide any other identifying information that may reveal its true location. The adversary is also capable of manipulating delays, as explained in Chapter \ref{ch:attack}.

\ac{CPV} is designed to verify the output of a geolocation technique. The adversary must thus be able to mislead that technique first to forge its location. We assume, for simplicity, that the geolocation step prior to the operation of \ac{CPV} is an unverified location assertion; \ac{CPV} is then to verify this assertion. By considering this case, whereby the adversary can simply assert a location (e.g., the \ac{LSP} asks its users to simply input their location), the adversary is powerful enough to evade any basic geolocation technique.\footnote{Some geolocation techniques are harder to evade than others. See Chapter \ref{ch:attack}.}

We define the {\it target location} as the location the adversary attempts to appear at. The following two use cases explain adversarial motivation to forge location, both of which are within the threat model.

\noindent \textbf{Impersonation. }
To mitigate online impersonation of users' accounts, typically done through password-guessing attacks, logins can be restricted to location(s) (e.g., country) associated with the legitimate user's account. To impersonate a user, the adversary needs to not only guess the user's password, but also the user's associated location, and place itself fraudulently in that location. In this case, the adversary's target location changes widely according to the account being attacked.

\noindent \textbf{Violation of geographic-restriction policies. }
When an \ac{LSP} customizes its services/content based on the location of its users, such as location-sensitive multimedia providers (e.g., Pandora \cite{pandora} and Hulu \cite{hulu}), adversaries may be motivated to evade geolocation to gain location-dependent benefits. This threat is harder to defend against than the previous one, since the adversary's target location is fixed (i.e., the adversary does not have to keep modifying its geolocation evasion mechanism to appear at different parts of the world), and immediately known to the adversary.

\section{CPV: Client Presence Verification}
\label{locationverification}

\begin{figure}
\centering
\includegraphics[scale=0.4]{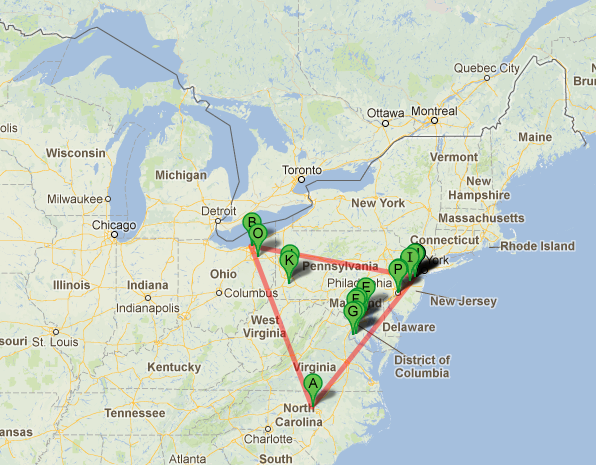}
\caption[An example of a triangle and several inside clients]{An example of 13 clients inside a triangle projected by verifiers in Duke University, Case Western Reserve University and Rutgers.}
\label{experimentsamplefig}
\end{figure}

\ac{CPV} builds on the established result that Internet delays and geographic distances have strong positive correlation \cite{subramanian2002geographic} (see Section \ref{cpv:background}). In \ac{CPV}, when a client asserts its presence in a geographic location, delays are measured between the client and three \emph{verifiers}\footnote{In practice, verifiers could be dedicated servers maintained by an independent party providing location verification as a service.} encompassing the asserted location. These delays are then processed to provide assurance that the client is truly present (geographically) inside the triangle determined by the three verifiers. The size of that triangle is the verification granularity. Figure~\ref{experimentsamplefig} shows an example triangle and several inside clients.

To reduce falsely rejecting legitimate (honest) clients and falsely accepting adversaries, factors affecting the delay-distance correlation (e.g., route circuitousness, queuing delays and congestion) must be addressed. The forward and reverse paths between any two hosts over the Internet are often affected by those factors differently, resulting in delay asymmetry \cite{pathak2008measurement}. The less affected path is likely to be the faster one (i.e., with a smaller \ac{OWD}), and thus better represents the distance between the two hosts. Relying on the smaller \ac{OWD} between the client and the verifiers rather than the \acp{RTT} is, thus, expected to improve \ac{CPV}'s accuracy in judging location assertions. 
\ac{CPV} uses the \emph{minimum pairs} protocol for \ac{OWD}-estimation (see Chapter \ref{ch:owd}). Accurate \ac{OWD}-estimation is one measure utilized by \ac{CPV} for accurate delay-to-distance mapping. By the end of this section, a summary is provided on how \ac{CPV} manages the delay-measurement process to reduce the factors affecting this mapping, without jeopardizing the integrity of the location verification process.

After mitigating these factors, \ac{CPV} uses a simple function to map delays to distances, and verifies assertions based on these distances (see Section \ref{CPVdescription} below).

\subsection{Operational Requirements}

\ac{CPV} requires geographically-distributed verifiers whose locations are consistent with the \ac{LSP}'s \acp{PGR}. \acp{PGR} are the regions in which clients are permitted to receive services/content or carry out location-specific operation (e.g., login or vote). The client must not control any of the verifiers involved in corroborating its assertion. To successfully enforce the \ac{LSP}'s location-aware policies, the verifiers must:
\begin{enumerate}
\itemsep0em
\item be publicly reachable over the Internet; and
\item the convex hull of the verifiers must encapsulate the \ac{LSP}'s \acp{PGR}.
\end{enumerate}

\subsection{Notation and definitions}
\label{notationsection}

The set of verifiers available to the \ac{LSP} is denoted $\mathbb{V}$. For any triangle, $\bigtriangleup$, the set of the three verifiers determining $\bigtriangleup$ is denoted $V_\bigtriangleup\subset\mathbb{V}$. For any geographic location $l=\{\text{latitude, longitude}\}$, $E_l$ is the set of triangles enclosing $l$, such that all $\bigtriangleup_l\in E_l$ are near equilateral in the network delay-space (see Section \ref{cpv:background}), and do not cross the \ac{PGR} border.

A client and three verifiers make four triangles. The function $valid(\mathbf{D})$ checks for \acp{TIV} in the four triangles whose side lengths are mapped from the six \acp{OWD} in $\mathbf{D}$. It returns true only if, for each of the four triangles, the sum of each two sides is greater than the third. The function $area\_v(\mathbf{D})$ calculates the area of the triangle determined by the three verifiers; the side lengths of that triangle are mapped from the three \acp{OWD} in $\mathbf{D}$ that belong to the edges between the verifiers. The function $area\_c(\mathbf{D})$ similarly calculates the areas of the three triangles determined by each pair of verifiers and the client, and returns the summation of those areas.

\subsection{CPV description}
\label{CPVdescription}

\ac{CPV}'s verification process begins with an asserted client location as input, $l=\{lat,lon\}$. The \ac{LSP} chooses a triangle $\bigtriangleup_{l}\in E_l$, and informs the client of the IP addresses of the verifiers in $V_{\bigtriangleup_l}$. The client connects to the verifiers and the verification process, Algorithm \ref{verifierprotocol}, begins.

\begin{algorithm}
\DontPrintSemicolon
\KwIn{Number of iterations, $n_{\bigtriangleup_l}$; tolerance of area inequality, {\large$\epsilon_{\bigtriangleup_l}$}; and acceptance threshold {\large$\tau_{\bigtriangleup_l}$}.}
\KwOut{Accept/Reject client's location assertion}
\Begin
{
	\nl \textsl{pass} $:= 0$\;
	\nl \label{goloop}\For{$i:=1$ \emph{\KwTo} $n_{\bigtriangleup_l}$}
	{
		\nl $\mathbf{D}_i := \phi$\;
		\nl \label{getvalues}Estimate, in real time, the one-way delays for $\mathbf{D}^{mp}$ and $\mathbf{D}^{av}$ using \ \ \textcolor{white}{.}\ \ \ \  \ \ \textcolor{white}{.}\ \ \ \ Algorithm \ref{unidirectionalprotocol} (see Chapter \ref{ch:owd}).\;
		\nl \label{minimumpairs}\lIf{valid($\mathbf{D}^{mp}$)}
		{$\mathbf{D}_i := \mathbf{D}^{mp}$}
		\nl \label{average}\lElseIf{valid($\mathbf{D}^{av}$)}
		{$\mathbf{D}_i := \mathbf{D}^{av}$}\;

		\nl\label{newAddition}\If{$\mathbf{D}_i\neq\phi$}
		{
			\nl \label{areaDfirst}$\delta_i:=area\_c(\mathbf{D}_i) - area\_v(\mathbf{D}_i)$\;
			\nl \label{bothtrue}\If{$\delta_i$ $\leq$ {\large$\epsilon_{\bigtriangleup_l}$} {\bf and} $acceptable(\mathbf{D}_i)$}
			{
				\nl\label{passplusplus}\textsl{pass} $:=$ \textsl{pass} $ + 1$\;
			}
		}
	}
	\nl \label{accratio}$\Gamma := $ \textsl{pass}$ / n_{\bigtriangleup_l}$\;
	\nl \label{laststep}\If{$\Gamma < $ {\large$\tau_{\bigtriangleup_l}$}}{\nl Reject client's location assertion}
	\nl\Else{\nl Accept client's location assertion}
}
\caption{Executed by the verifiers in $V_{\bigtriangleup_l}$ when a client asserting to be at location $l$ connects to them. See inline for the definition of the function $acceptable()$; similarly, see Section \ref{notationsection} for the definitions of the functions $valid()$, $area\_c()$ and $area\_v()$.}
\label{verifierprotocol}
\end{algorithm}

First (in line \ref{getvalues}), the verifiers estimate the \emph{smaller} of the forward and reverse \acp{OWD} at the six edges between the verifiers and the client using two protocols: {\it minimum pairs} ($mp$) and {\it average} ($av$), as explained in Chapter \ref{ch:owd}. The six \acp{OWD} are then mapped to distances according to the simple mapping function $f(x)=x$, i.e., $x\ ms$ is equal to $x\ km$. The resulting distances are never used in an absolute form; they are only processed relative to each other. This design provides the advantage of resilience to factors that affect the network comprising the client and the three verifiers, e.g., a network congestion that affects the delays of the six edges altogether. 

\label{changeclarifyinput}
\ac{OWD} estimation is done iteratively (line \ref{goloop}), where the input parameter $n_{\bigtriangleup_l}$ specifies the number of iterations to be performed, to account for possible delay instability \cite{Zhang:2001:CIP:505202.505228}. The \emph{confidence ratio}, $\Gamma$ (line \ref{accratio}), represents the verifiers' confidence of the truthfulness of the asserted location. It is calculated as the proportion of iterations where the values of $area\_c(\mathbf{D}^{mp})$ and $area\_v(\mathbf{D}^{mp})$ (see Section \ref{notationsection} for notation) match within a suitable error tolerance, $\epsilon$. From a geometric perspective, we have the following claim (see Appendix \ref{app:proofs} for proofs):
\begin{claim}
\label{claimarea}
Let $P$ be a point in the Cartesian plane, and let $\bigtriangleup XYZ$ be the triangle determined by the points $X$, $Y$ and $Z$. If $P$ is strictly outside $\bigtriangleup XYZ$, then the sum of the areas of $\bigtriangleup XYP$, $\bigtriangleup XPZ$ and $\bigtriangleup PYZ$ is greater than the area of $\bigtriangleup XYZ$.
\end{claim}

\acp{TIV} are evident in the Internet \cite{tiv}. Because \ac{CPV} relies on triangular areas in verifying location assertions, \acp{TIV} can thwart \ac{CPV}'s successful operation. Additionally, an adversary can increase the estimated \acp{OWD} of the $mp$ protocol, flattening some triangles and resulting in \acp{TIV}. Thus, the verifiers become less confident about the truthfulness of the asserted location as more \acp{TIV} occur, which is a security precaution to reduce potential false accepts. This can be seen in line \ref{bothtrue}, where $\mathbf{D}_i$ must hold a valid set of delays (from lines \ref{minimumpairs} or \ref{average}) for $\Gamma$ (line \ref{accratio}) to increase.

Iterating the delay-estimation process helps reduce the number of benign \acp{TIV} \cite{wang2007towards}, hence reducing the number of \acp{FR}. Additionally, more than one delay-estimation protocol (namely, both $mp$ and $av$) further lessens the effect of \acp{TIV}; $av$ is used as a fallback if the estimates in $\mathbf{D}^{mp}$ result in TIVs \cite{wang2007towards}. In lines \ref{minimumpairs} and \ref{average}, $\mathbf{D}^{mp}$ is checked first because it is more resilient to delay spikes, as discussed in Chapter \ref{ch:owd}.

\label{changeiterations1}\textcolor{\changes}{On the other hand, such iterative delay-estimation approach may affect the usability of CPV, as it increases the time required by CPV to reach a decision. Some applications may require a decision before providing the location-sensitive service to users, such as online credit card transactions. However, in other applications, the verification algorithm may run in the background (i.e., continuously and concurrent to the location-sensitive application), such as media streaming. As such, despite its potential usability drawbacks, the impact of the number of iterations on the usability of CPV depends essentially on the application.}

The error tolerance, $\epsilon$ (line \ref{bothtrue}), accounts for route circuitousness \cite{Octant}, congested routes, or other factors that contribute to inaccuracies in the delay-distance mapping over the Internet. If an adversary's true location is so far from the asserted location that one of the \emph{inner} triangles (those having the client as one of their vertices) becomes obtuse, the triangle becomes flattened and its area decreases. An unnecessarily large error tolerance may thus falsely accept this adversary.

To mitigate this effect, we include the $acceptable(\mathbf{D})$ function (line \ref{bothtrue}), which checks that the \ac{OWD} between verifier $v$ and the client is not larger than the \acp{OWD} between $v$ and the other two verifiers. The function returns true only if the previous statement is true for the three delay-mapped distances in $\mathbf{D}$ that are between the client and the verifiers. From a geometric perspective, using the notation $\overline{AB}$ for the length of line segment $AB$, we have the following claim (see Appendix \ref{app:proofs} for proofs):
\begin{claim}
\label{claimexcessive1}
Let $W$ be a point in the Cartesian plane, and let $\bigtriangleup XYZ$ be the triangle determined by the points $X$, $Y$ and $Z$ such that $\overline{XZ}\leq\overline{XY}$. If $\overline{XW}>\overline{XY}$, then $W$ is strictly outside of $\bigtriangleup XYZ$.
\end{claim}

{\bf Calibration of input parameters.} 
\label{changecalibration}
\textcolor{\changes}{Calibration of input parameters. To set the three input parameters of Algorithm \ref{verifierprotocol} for each $\bigtriangleup$, the three verifiers in $V_\bigtriangleup$ can operate CPV to verify the geographic presence/absence of network nodes that are known (as a ground-truth) to be inside/outside $\bigtriangleup$ (e.g, using other verifiers in V). Based on the delays between the verifiers and these nodes, the input parameters should be set such that CPV accepts inside nodes, and rejects outside ones. For example, in line \ref{accratio} (Algorithm \ref{verifierprotocol}), if $\Gamma\geq$ 0.6 for all such nodes, then $\tau_{\bigtriangleup_l}$ should be set to 0.6.}

{\bf Summary.} 
\ac{CPV}'s measures to reduce factors negatively affecting delay-to-distance mapping can be summarized as follows:
\begin{enumerate}
\itemsep0em
\item Two protocols are used to estimate \acp{OWD} instead of one to reduce the effect of \acp{TIV}.
\item Active delay measurement is used with each client, which reflects the most recent delay status in the region \cite{Zhang:2001:CIP:505202.505228}.
\item No universal delay-to-distance mapping is used. Rather, mapping is done relative to other delays in the region.
\item Delay-estimation is conducted iteratively to more accurately converge to the actual delays at current network conditions \cite{hsu1947complete}.
\item The three verifiers are chosen within a geographical proximity of the asserted location to
	\begin{enumerate}
	\itemsep0em
	\item reflect regional delays \cite{improvacc,Dong201285};
	\item span fewer Autonomous Systems, which reduces route circuitousness \cite{subramanian2002geographic};
	\item reduce the number of \acp{TIV} \cite{wang2007towards}; and
	\item exhibit stronger positive correlation between delays and distances \cite{landa2013measuring}.
	\end{enumerate}
\end{enumerate}

\section{Security Discussion}
\label{SecurityAnalysis}

\subsection{Classical Geolocation Attacks}
\label{geoattacks}
\begin{figure}
\centering
\subfigure[]
{
\label{lielayout}
\includegraphics[scale=0.7]{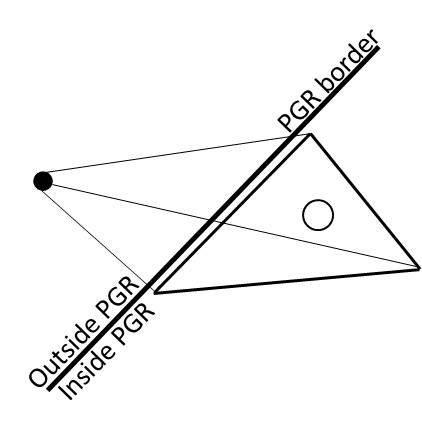}
}
\hspace{20pt}
\subfigure[]
{
\label{proxylayout}
\centering
\includegraphics[scale=0.7]{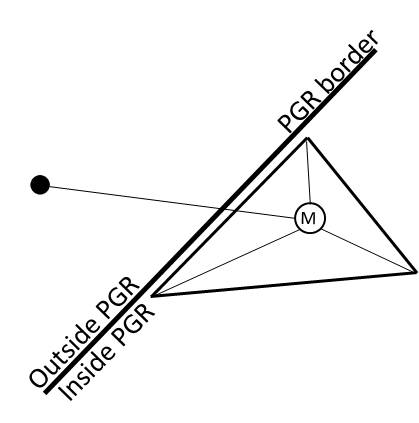}
}
\label{lielayoutproxylayout}
\caption[An adversary asserting a false location]{An adversary asserting a false location (a) without using a middlebox, and (b) using a middlebox at the asserted location. $\bullet$=true location; {\large$\circ$}=asserted location; $M$=middlebox; \ac{PGR}=Permitted Geographic Region.}
\end{figure}

\noindent{\bf Submitting false information. }Although this may mislead simple geolocation techniques \cite{MuirPaul}, it does not defeat \ac{CPV} because the verification process (Algorithm \ref{verifierprotocol}) is independent of any information submitted by the client. Chapters \ref{ch:wiredecva} and \ref{ch:wirelessecva} analyze \ac{CPV}'s efficacy in detecting false location assertions (Fig.~\ref{lielayout}) due to area mismatch or large client-verifier delays.

\noindent{\bf Using middleboxes. }Some IP geolocation techniques can be circumvented if a client's IP address is concealed using generic \acp{MB} such as proxies, anonymizers, or \acp{VPN} \cite{MuirPaul}. These do not threaten the integrity of the verification process of \ac{CPV} because delay measurements are conducted over the client's application layer. \acp{MB} that blindly relay application-layer traffic (Fig.~\ref{proxylayout}) will also relay the timestamps (see Section \ref{locationverification}) to the client \cite{shroud}. Chapter \ref{ch:puzzles} shows how a \ac{MB} specifically designed to defeat \ac{CPV} by searching application-layer traffic for timestamps could be mitigated using a \ac{PoW} mechanism.

\noindent{\bf Manipulating delays to increase calculated distances. }Delay-adding attacks \cite{Dude} can be attempted on \ac{CPV} when the adversary inserts a delay before forwarding timestamps. Assuming verifier $i$ sent a timestamp, the adversary failing to forward it promptly to verifier $j$ enlarges $d_{ic}$ and $d_{cj}$ fraudulently, increasing the value of $d_{ic}+d_{cj}$ (see Fig.~\ref{fig:owd:typical} in Chapter~\ref{ch:owd} for notation). Because the $mp$ protocol estimates the smaller \ac{OWD} at each edge by solving simultaneous equations, selectively delaying timestamps can result in delay estimates that are smaller than the actual delay. For example, solving simultaneously the equations $a+b=7$, $a+c=8$ and $b+c=9$ gives $a=3$, $b=4$, and $c=5$. Whereas $a+b=7$, $a+c=8$, and $b+c=13$ results in $a=1$, $b=6$ and $c=7$. Thus, \emph{increasing} $b+c$ resulted in a \emph{smaller} value for $a$.

However, the adversary cannot reduce the \emph{summation} of $d_{ic}$ and $d_{cj}$ as this requires speeding up the traffic propagation between the adversary and the verifiers \cite{Dude}. From a geometric perspective, increasing the summation of any pair of edges does not help an adversary outside a triangle to forge its location making it inside. Formally, using the notation {\footnotesize $\overline{AB}$} for the length of line segment $AB$, we have the following claim (see Appendix \ref{app:proofs} for proofs):
\begin{claim}
\label{claimlines}
Let $P$ be a point in the Cartesian plane, and let $\bigtriangleup XYZ$ be the triangle determined by the points $X$, $Y$ and $Z$. If $P$ is strictly outside $\bigtriangleup XYZ$, then increasing the sums {\footnotesize $\overline{XP}+\overline{PZ}$, $\overline{XP}+\overline{PY}$} or {\footnotesize $\overline{YP}+\overline{PZ}$} without reducing at least one of the other sums cannot place $P$ inside $\bigtriangleup XYZ$.
\end{claim}

\noindent{\bf Manipulating delays to cause \acp{TIV}. }As shown in Algorithm \ref{verifierprotocol}, \ac{CPV} holds the number of \acp{TIV} against the client (the condition $\mathbf{D}_i\neq\phi$ in line \ref{bothtrue} means $\mathbf{D}_i$ must not violate the triangle inequality to increment \textsl{pass}). In conclusion, manipulating delays does not help the adversary, but rather signals the adversary's evasion attempts.

\subsection{Attempts to Evade CPV}
\label{UnsuccessfulAttacks}
To study potential vulnerabilities in \ac{CPV}, we review steps where the verifiers interact with the client.

\noindent{\bf Connecting to the verifiers.} Assuming the adversary's target location (location it is trying to appear at) is $l$, connecting to a set of verifiers $V_{\bigtriangleup_{l'}}\neq V_{\bigtriangleup_{l}}$ does not help the adversary in pretending to be at $l$ as those verifiers cannot verify the adversary's presence inside $\bigtriangleup_{l}$.

\noindent{\bf Forwarding the timestamp.} Because the verifiers sign the timestamps, the adversary can neither forge nor inject fake ones. Delaying a timestamp is discussed in Section \ref{geoattacks}.

\subsection{Poor Verifier Deployment and PGR Proximity}
\label{PGRProximity}

\begin{figure}
\centering
\subfigure[]
{
    \label{borderProximitySimple}
    \includegraphics[scale=0.5]{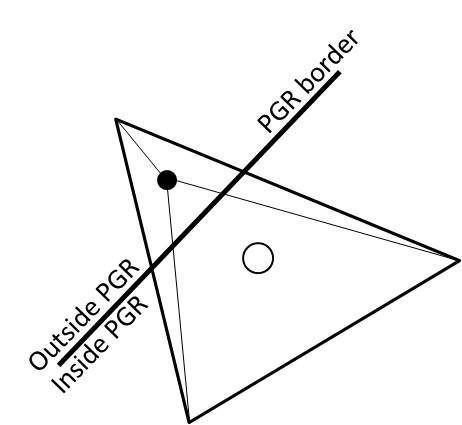}
}
\hspace{20pt}
\subfigure[]
{
    \label{enclosure}
    \includegraphics[scale=0.5]{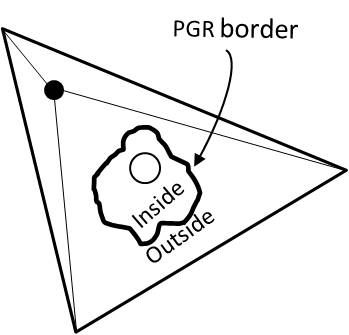}
}
\hspace{20pt}
\subfigure[]
{
    \label{borderattack}
    \includegraphics[scale=0.5]{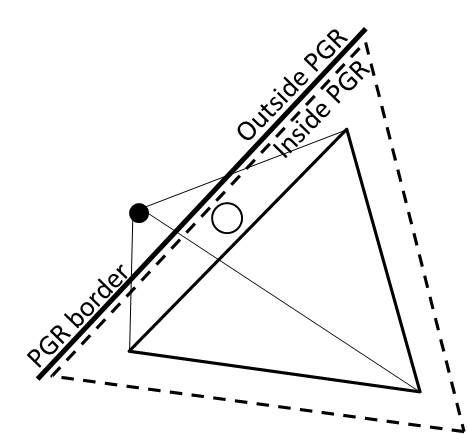}
}
\caption[Examples of inappropriate and insufficient deployment of verifiers]{(a) and (b) inappropriately deployed verifiers; (c) insufficiently deployed verifiers. $\bullet$=true location; {\large$\circ$}=asserted location; \ac{PGR}=Permitted Geographic Region.}
\label{closeborder1}
\end{figure}

Adversaries bordering the \ac{PGR} may be able to exploit inappropriate or insufficient verifier deployment. Figures \ref{borderProximitySimple} and \ref{enclosure} show examples of inappropriately deployed verifiers with respect to the \ac{PGR}, where a triangle crosses the \ac{PGR} border or encloses the \ac{PGR} inside itself. As shown, a close adversary could be outside the \ac{PGR} but inside those triangles. Verifying the presence inside the triangle does not ensure presence inside the \ac{PGR} in those cases. Figure~\ref{borderattack} shows potential vulnerability due to insufficient verifiers/triangles: not all regions inside the \ac{PGR} are covered with triangles. The verifiers determining the shown (solid) triangle should not overly relax $\epsilon_\bigtriangleup$ to account for the uncovered region (relaxing $\epsilon_\bigtriangleup$ is depicted by the \emph{dashed} triangle in Fig.~\ref{borderattack}). Otherwise, the verifiers falsely accept an adversary close to the \ac{PGR} asserting to be at the uncovered region of the \ac{PGR}, as shown in Fig.~\ref{borderattack}.

\begin{figure}
\centering
\subfigure[$\times$=possible true locations.]
{
\label{bordercrossing}
\includegraphics[scale=0.55]{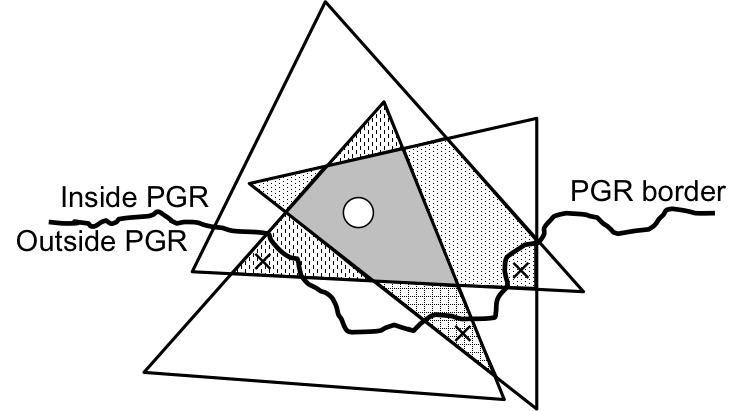}
}
\hspace{20pt}
\subfigure[M=middlebox;]
{
\label{borderattackproxy}
\includegraphics[scale=0.55]{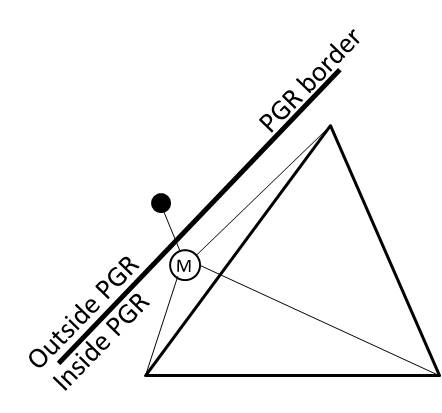}
}
\caption[Possible defenses againstinappropriate and insufficient deployment of verifiers]{Defenses against a bordering adversary that exploits inappropriate or insufficient verifier deployment. $\bullet$=true location; {\large$\circ$}=asserted location; PGR=Permitted Geographic Region.}
\label{closeborder2}
\end{figure}

{\bf Possible countermeasures. } To address \ac{PGR} border crossing, additional overlapping triangles could be used to enclose the asserted location as long as a single triangle, or the intersection of multiple triangles, crosses the \ac{PGR} border. The intersection region of the triangles must (1) not cross the \ac{PGR} border and (2) enclose the asserted location, as shown in Fig.~\ref{bordercrossing}. Client presence inside the \ac{PGR} is then verified only if the verifiers of each triangle accept the assertion. For example, in Fig.~\ref{bordercrossing}, if the client's (adversary's) true location was at any of the areas marked with $\times$, two triangles may falsely accept the assertion. Two triangles are insufficient in that case because the \ac{PGR} border crosses the overlapping areas of each two of the three triangles. Verifying the presence inside all three suffices to verify the correctness of the assertion. 

As for insufficient deployment of verifiers, whenever an assertion is made in a region not covered by any triangle, the \ac{LSP} (location-sensitive provider) could use a measurement-based IP geolocation technique instead of relying on client-dependent geolocation (such as \ac{GPS}). A bordering adversary must then evade this technique prior to bypassing \ac{CPV}. It would then be challenging for the adversary to precisely target a location not covered by any triangle only through delay manipulation \cite{Dude}. In such a case, using a measurement-based IP geolocation technique motivates the adversary to use a \ac{MB} inside the uncovered region of the \ac{PGR} (Fig.~\ref{borderattackproxy}). However, \acp{MB} tend to increase delays \cite{shroud}, which helps the verifiers detect the adversary's false assertion.

\section{Conclusion}
\label{cpv:conclusion}

\ac{CPV} is a delay-based technique which, to the best of our knowledge, is the first to verify a client's location over the Internet without assuming the client's possession of a secret personal identifier (see Section \ref{background:locveri}). \ac{CPV} mitigates delay spikes injected by the Internet as it iterates the delay-measuring process, and corroborates the client's location based on the \emph{smaller} \ac{OWD}, as estimated using the \emph{minimum pairs} protocol (Chapter \ref{ch:owd}). In \ac{CPV}, delays are estimated between a client and three verifiers, which enclose the client's unverified location within their convex hull. The verifiers estimate the delays over the client's application layer to overcome IP hiding tactics, typically carried out using \acfp{MB}. For clients using web-browsers, \ac{CPV} requires no extra client-side software; the client's browsing experience is retained as the verification process could run in the browser. These advantages highlight \ac{CPV}'s potential for practical adoption.

In the following chapter, \ac{CPV} is evaluated using detailed experiments in a real-world environment when legitimate clients are using wired access networks. Further, in Chapter \ref{ch:wirelessecva}, experimental logs collected from the wired testing are modified to represent last mile delays of a client using a wireless access network, and \ac{CPV} is reevaluated under these conditions.

\chapter{Evaluating CPV in Wired Networks}
\label{ch:wiredecva}
\blfootnote{The content of this chapter was published at the 2014 IEEE CNS conference \cite{cpv}  (with a full length version accepted for publication in IEEE TDSC \cite{cpvtdsc}).}

In this chapter, \ac{CPV} is evaluated in wired networks through detailed experiments on PlanetLab \cite{planetlab}, exploring various factors that affect its efficacy, including the granularity of the verified location, and the verification time. The evaluation of \ac{CPV} in wireless netowrks is presented in Chapter~\ref{ch:wirelessecva}.

We use the rates of \acfp{FR} and \acfp{FA} as the assessment metrics. If a client asserts to be at location $l$, an \ac{FR} occurs when this client is actually present somewhere inside $\bigtriangleup_l$, and is judged by the verifiers in $V_{\bigtriangleup_l}$ as absent from $\bigtriangleup_l$. By contrast, an \ac{FA} occurs when that client is actually absent from $\bigtriangleup_l$, and is judged by the verifiers in $V_{\bigtriangleup_l}$ as present in $\bigtriangleup_l$. 
\begin{figure}\centering\includegraphics[scale=0.45]{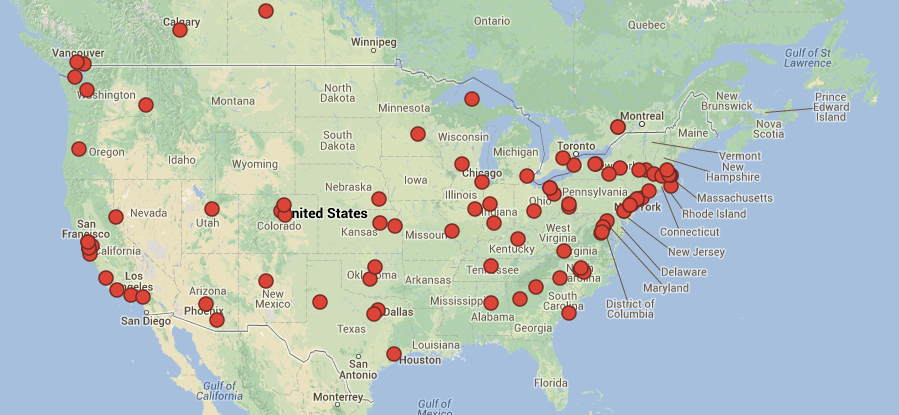}
\caption[PlanetLab nodes used in evaluating CPV]{Locations of the 80 PlanetLab nodes used in our experiments. Map data: Google, INEGI.}
\label{nodelocations}\end{figure}

We use 80 PlanetLab \cite{planetlab} nodes in USA and Canada (Fig.~\ref{nodelocations}), and identified 34 different sized triangles satisfying the requirements stated in Section \ref{locationverification}. The triangles were chosen with internal angles ranging 50-70 degrees so as to be near-equilateral in the network delay-space, as specified in Section \ref{notationsection}, page \pageref{notationsection}.
Triangular areas ranged from $\sim$32,000 km$^2$, almost the size of Maryland state, to $\sim$500,000 km$^2$, almost the size of Spain.

We assume that the \acf{PGR} (see Chapter \ref{ch:cpv}) is a triangular-shaped region that perfectly coincides with the dimensions of the triangle. One triangle was considered at a time. For each triangle, all nodes---except the three determining the triangle---acted as clients; all clients had provided assertion to be at the centroid of that triangle. Combining clients of all triangles, \emph{legitimates}\footnote{We use the word \emph{legitimates} (i.e., as a noun) to refer to legitimate clients.} (clients actually inside) totalled 146 and \emph{adversaries} (clients actually outside) totalled 2,301 for a total of 2,447 experiments. The verifiers determining each triangle were verifying assertions of all clients concurrently. The verifiers used \ac{NTP} \cite{rfc5905} to synchronize their clocks. Knowing the ground truth of legitimates and adversaries with respect to each triangle, our objective is to identify the optimal values for the tolerance of the area inequality ($\epsilon_\bigtriangleup$) and the acceptance threshold ($\tau_\bigtriangleup$) for each of the 34 triangles, and quantify the \acp{FR} and \acp{FA} at these values.

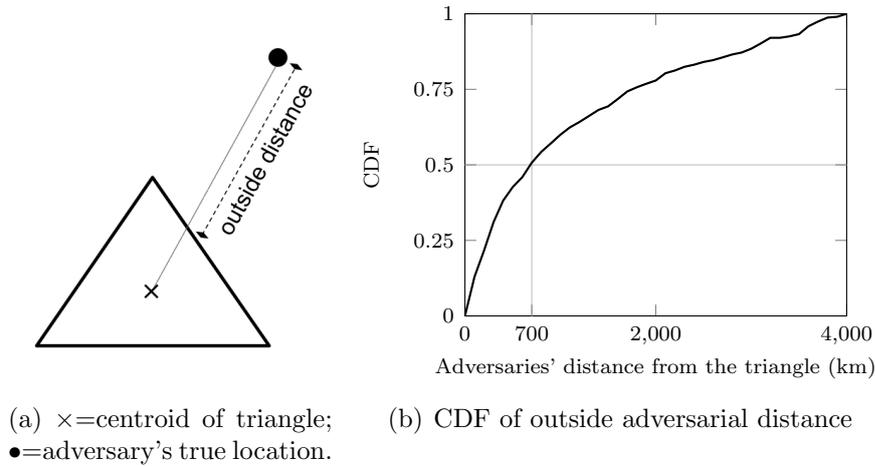
\begin{figure}
\centering
\subfigure[$\times$=centroid of triangle; $\bullet$=adversary's true location.]
{
\label{outsidedistancedefinition}
\includegraphics[scale=0.55]{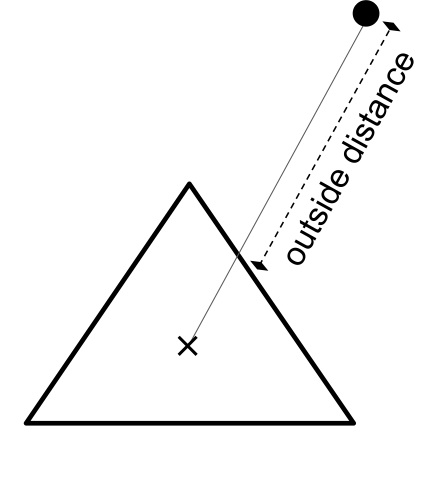}
}
\subfigure[CDF of outside adversarial distance]
{
\label{dis_away_cdf}
\begin{tikzpicture}
\begin{axis}[
width=2.6in,
height=2.2in,
xlabel={Adversaries' distance from the triangle (km)},
ylabel=CDF,
xmin=0, xmax=4000,
ymin=0, ymax=1,
xtick={0,2000,4000},
ytick={0,0.25,0.75,1},
extra x ticks={700},
extra y ticks={0.5},
extra tick style={grid=major},
x tick label style={font=\scriptsize},
x label style={font=\scriptsize,at={(0.5,0.03)}},
y tick label style={font=\scriptsize},
y label style={font=\scriptsize},
legend columns=-1,
legend style={draw=none,at={(0.52,1.25)},anchor=north,}
]
\addplot[solid,line width=0.8pt]table[col sep=comma]{csv/Wired/dis_adv.csv};
\end{axis}
\end{tikzpicture}
}
\caption[Adversaries' distances from the triangles' closest side (experimental design)]{Adversaries' distances from the triangles' closest side. (a) How the external distance is calculated with respect to a triangle; (b) A point $(x,y)$ means the proportion $y$ of adversaries were $x$ km away from the closest side. Note: this graph shows experimental design, not results.}
\end{figure}

To see how far adversaries were from the triangles in the experiments, we define the adversaries' \emph{outside distance} with respect to each triangle in our experiments as the distance between the adversary's true location and the point of intersection between lines $A$ and $B$; line $A$ is the one passing through the adversary's true location and the triangle's centroid; line $B$ is the triangle's closest side to the adversary (see Fig.~\ref{outsidedistancedefinition}). Figure~\ref{dis_away_cdf} shows a CDF of the 2,301 adversaries' outside distance. Half the adversaries were less than 700 km away from the triangle's closest side (i.e., the triangle encapsulating their fraudulently asserted location), and no adversary was farther than 4,000 km away. For reference, the width of the United States is approximately 4,000 km. The argument is that if \ac{CPV} rejects relatively nearby adversaries, it will reject more distant ones.

{\bf Implementation details.} \label{changeimplement}\textcolor{\changes}{
Both the CPV server and the client were implemented as Java applications. The server's code was run on the three PlanetLab nodes chosen as CPV servers at each experiment; the client code's was run on the remaining nodes representing the CPV clients. Each CPV client was informed with the server's IP addresses and port numbers. When all three CPV servers are started and waiting for clients to connect, all clients were started in parallel and the verification process begins across all clients simultaneously. Note that in practice, the CPV algorithm requires no specific client-side software because the client side can be implemented using javascript and websockets.}

Experiments were run over the course of a month (April 2013) and at different times of the day. The number of iterations, $n_\bigtriangleup$ (Algorithm \ref{verifierprotocol} on page~\pageref{verifierprotocol} ), was fixed at $n_\bigtriangleup=600$ for all $\bigtriangleup$ in the 34-triangle set to study the factors affecting \ac{CPV} over a relatively long period of time (a total of $\sim$13.3 million delay measurements were taken between all nodes). Fewer iterations might be sufficient to judge a client, as we show in Section \ref{timeSufficient} below.

{\bf Limitations of PlanetLab. } \label{changePlanetLablimitations}\textcolor{\changes}{Despite being generally used as an experimental testbed representing the global Internet, PlanetLab measurements should not absolutely be deemed as so \cite{plproblem}. Many of PlanetLab nodes are connected through the \ac{GREN}, e.g., Internet 2 \cite{Internet2} and CANARIE \cite{CANARIE}, in which traffic could be fully routed within the network. Accordingly, all experiments conducted in this thesis are subject to PlanetLab's network settings \cite{6838709}.}

The rest of this chapter is organized as follows. Section \ref{anExampleSection} details an example from the experiments, which involves three clients: one legitimate and two adversaries. Sections \ref{tivsection}, \ref{areadiscsection}, and \ref{wired:confidenceratio} respectively analyze the rates of \acp{TIV}, examines the use of the triangular areas as \ac{CPV}'s primary assertion-verification metric, and analyzes \ac{CPV}'s confidence ratio associated with all experimented clients. In Section \ref{Adjacency}, the effect of legitimate clients' closeness to the triangles' sides is examined, and in Section \ref{timeSufficient} the appropriate number of iterations is analyzed. Section \ref{wired:comparison} analyzes hypothetical modifications to \ac{CPV}, where the \ac{OWD}-estimation process is modified and \ac{CPV}'s efficacy is reevaluated. Finally, Section \ref{wired:conclusion} concludes.

\section{An Example}
\label{anExampleSection}

We detail the results of one of the triangles in our 34-triangle set, and three of the clients being verified by that triangle. One of the clients was legitimate, the other two were adversaries. Figure~\ref{geoMapExample} shows the geographic location of the triangle and the three clients, labelled $D$, $E$ and $F$. The area difference, $\delta_i$ (line \ref{areaDfirst} of Algorithm \ref{verifierprotocol}) for all $1\leq i\leq 600$, is plotted for the three clients in Fig.~\ref{geoMapExampleArea}.

{\bf Number of \acfp{TIV}. }Some iterations have no corresponding values for the area difference (visible in high resolution). Those are the ones where $valid(\mathbf{D}^{mp})$ and $valid(\mathbf{D}^{av})$ (lines \ref{minimumpairs} and \ref{average} of Algorithm \ref{verifierprotocol}) returned false, i.e., the mapped distances resulted in at least one \ac{TIV} of the four triangles determined by the three verifiers and the client. Of all 600 iterations, the number of iterations where both functions returned false for $D$, $E$ and $F$ are 114, 11 and 0 respectively. The number of \acp{TIV} is high for $D$ likely due to its relatively close position to two of the three triangle's sides (versus one side as with $E$).

{\bf Area difference ($\delta$). }From Fig.~\ref{geoMapExampleArea}, the median of $\delta_i$, $\tilde{\delta}$, for clients $D$, $E$ and $F$ is 30 km$^2$, 66 km$^2$ and 209 km$^2$ respectively. The median corresponding to $F$ is substantially larger than that of $D$ and $E$ because $F$ is relatively far away from the triangle. The smallest recorded area difference for $F$ is $\delta_{325}$ = 102 km$^2$. Therefore, any value for $\epsilon_\bigtriangleup$ in the range $\epsilon_\bigtriangleup<102$ keeps the variable \textsl{pass}$=0$ (line \ref{passplusplus}, Algorithm \ref{verifierprotocol}) for all iterations, resulting in $\Gamma$ = 0. Consequently, at $\epsilon_\bigtriangleup<102$, any value for $\tau_\bigtriangleup$ (the acceptance threshold, Section \ref{locationverification}) in the range $\tau_\bigtriangleup>0$ rejects $F$'s assertion. Client $E$ was less than 50 km away from the triangle's nearest side $AC$, thus the average area difference of $E$ is close to that of $D$. However, at $\epsilon_\bigtriangleup=45$, there is a visible distinction between both nodes---there existed a value for $\epsilon_\bigtriangleup$ (i.e., 45 km$^2$) that enabled the verifiers to correctly judge the assertions of both clients, $D$ and $E$, despite being geographically collocated.

\definecolor{g1}{rgb}{1,0.4,0.4}
\definecolor{g2}{rgb}{0.4,1,0.4}
\definecolor{g3}{rgb}{0.4,0.4,1}
\begin{figure}
\centering
\subfigure[The area of the shown triangle is $\sim$230,000 km$^2$. Clients $E$ and $F$ are outside, whereas $D$ is inside. Map data: Google, INEGI.]
{
\label{geoMapExample}
\includegraphics[scale=0.6]{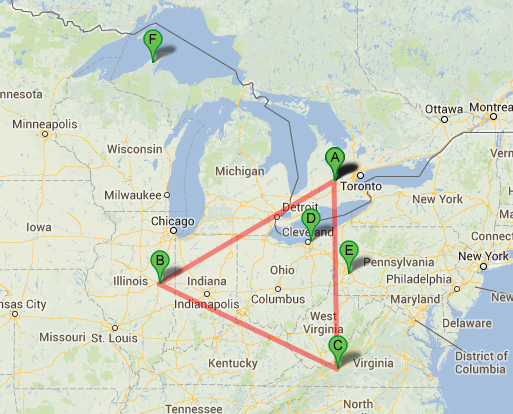}}

\subfigure[]
{
\label{geoMapExampleArea}
\begin{tikzpicture}
\begin{axis}[
width=2.6in,
height=2.2in,
xlabel=Iteration number ($i$),
ylabel=$\delta_i$ (km$^2$),
xmin=1, xmax=600,
ymin=-50, ymax=400,
extra y ticks = {45},
xtick={1,300,600},
x tick label style={font=\scriptsize},
y tick label style={font=\scriptsize},
y label style={at={(0.01,0.5)}},
legend columns=-1,
legend style={draw=none,at={(0.5,1.17)},anchor=north,},
unbounded coords=jump,
]
\addplot[color=g3] table[col sep=comma]{csv/Wired/example/minpInside.csv};
\addlegendentry{$D$}
\addplot[color=g2] table[col sep=comma]{csv/Wired/example/minpOutsideNear.csv};
\addlegendentry{$E$}
\addplot[color=g1] table[col sep=comma]{csv/Wired/example/minpOutsideFar.csv};
\addlegendentry{$F$}
\addplot[dashed,color=black,line width=1pt]plot coordinates {(0,45)(600,45)};
\end{axis}
\end{tikzpicture}
}
\subfigure[At $\epsilon_\bigtriangleup$ = 45 km$^2$]
{
\label{geoMapExampleAcceptRatio}
\begin{tikzpicture}
\begin{axis}[
width=2.6in,
height=2.2in,
xlabel=Iteration number ($i$),
ylabel=$\Gamma$ (after $i$ iterations),
xmin=1, xmax=600,
ymin=-0.1, ymax=1,
xtick={1,300,600},
x tick label style={font=\scriptsize},
y tick label style={font=\scriptsize},
y label style={at={(0.01,0.5)}},
legend columns=-1,
legend style={draw=none,at={(0.5,1.17)},anchor=north,}
]
\addplot[color=blue] table[col sep=comma]{csv/Wired/example/aRInside.csv};
\addlegendentry{$D$}
\addplot[color=green] table[col sep=comma]{csv/Wired/example/aROutsideNear.csv};
\addlegendentry{$E$}
\addplot[color=red] table[col sep=comma]{csv/Wired/example/aROutsideFar.csv};
\addlegendentry{$F$}
\end{axis}
\end{tikzpicture}
}
\caption[CPV verifying location assertions of one legitimate client and two adversaries]{An example from our experiments showing a triangle and three clients (best viewed in color).}
\label{areaDifferenceBig}
\end{figure}
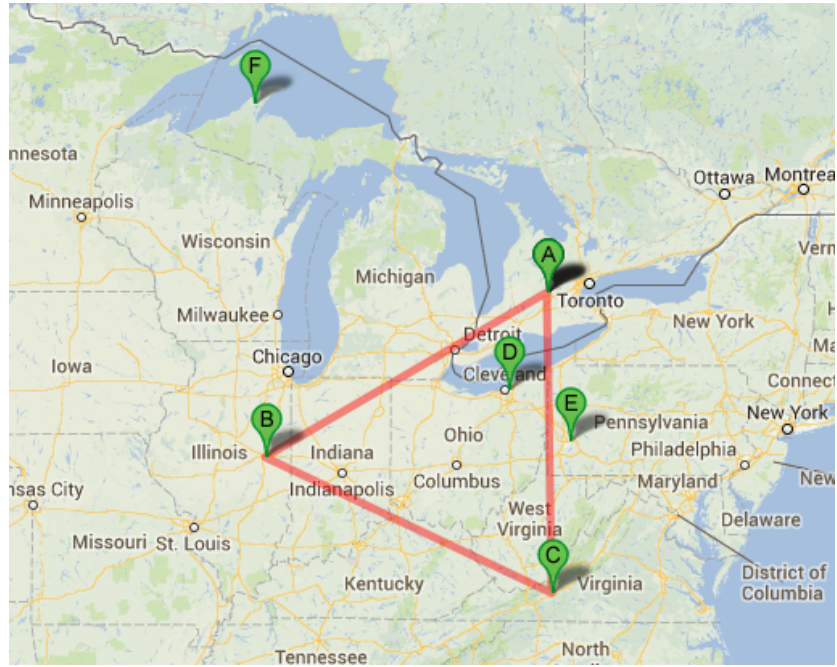
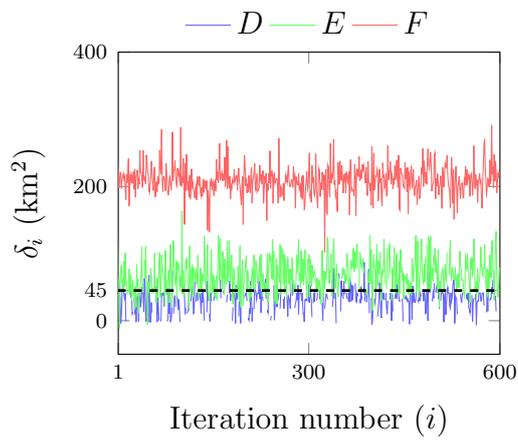
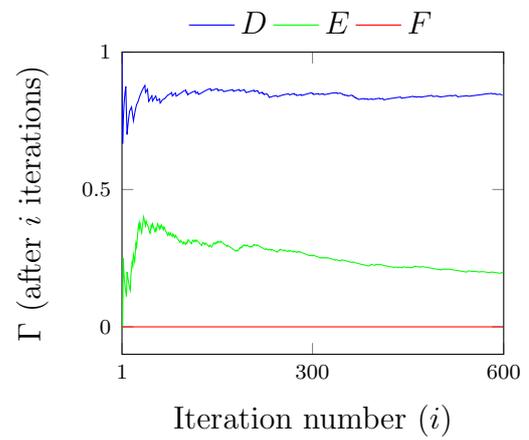

{\bf Confidence ratio ($\Gamma$). }In Algorithm \ref{verifierprotocol}, $\Gamma$ is calculated when all $n$ iterations are performed. Figure~\ref{geoMapExampleAcceptRatio} plots $\Gamma$ (at $\epsilon_\bigtriangleup=45$ km$^2$), assuming it was calculated at each iteration. Despite the relatively close values of $\delta_i$ between $D$ and $E$ in Fig.~\ref{geoMapExampleArea}, their $\Gamma$ greatly differs. At $i=100$, $\Gamma$ is 0.86 and 0.3 for $D$ and $E$ respectively. Therefore, after 100 iterations, any $\tau_\bigtriangleup$ in the range $0.3<\tau_\bigtriangleup\leq0.86$ enables the verifiers to decide that $D$ is a legitimate and $E$ is an adversary. When all 600 iterations are performed, $\Gamma$ becomes 0.84 and 0.2 for $D$ and $E$ respectively, showing no significant change from the $100^{th}$ iteration. 
{\bf Summary. }Table~\ref{examplesummary} summarizes the results of this example. The following three sections analyze each of the three variables (rows) in the table for all 2,447 experiments. The respective section is reported in the table.

\begin{table}
\centering
\caption[Summary of results for three example clients from the experiments]{Results for clients $D$, $E$, and $F$. The ``Section" column shows the section where each variable (row) is analyzed further for all experiments.}
\begin{tabular}{ r | ccc | l }
\multirow{2}{*}{Variable} & \multicolumn{3}{c|}{Client}&\multirow{2}{*}{Section}\\
&\emph{D}&\emph{E}&\emph{F}&\\\hline
Number of TIVs								&114&11&0&\ref{tivsection}\\
$\tilde{\delta}$ (km$^2$)		&30&66&209&\ref{areadiscsection}\\
$\Gamma$ (0 to 1)		&0.84&0.2&0&\ref{wired:confidenceratio}\\\Xcline{1-5}{2\arrayrulewidth}
\end{tabular}
\label{examplesummary}
\end{table}

\section{Triangle Inequality Violations}
\label{tivsection}

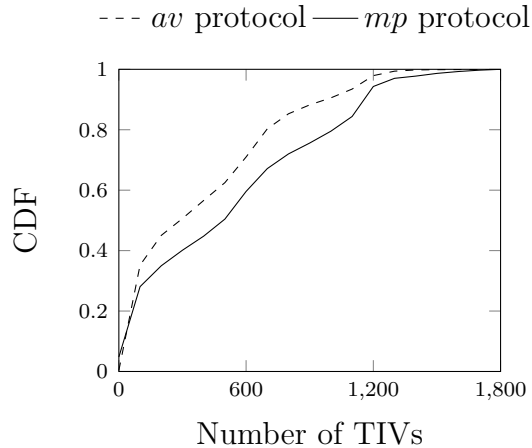
\begin{figure}
\centering
\begin{tikzpicture}
\begin{axis}[
width=2.6in,
height=2.2in,
xlabel={Number of TIVs},
ylabel=CDF,
xmin=0, xmax=1800,
ymin=0, ymax=1,
xtick={0,600,1200,1800},
x tick label style={font=\scriptsize},
y tick label style={font=\scriptsize},
legend columns=-1,
legend style={draw=none,at={(0.52,1.25)},anchor=north,}
]
\addplot[dashed]table[col sep=comma]{csv/Wired/TIV/CDFrttTIV.csv};
\addlegendentry{$av$ protocol}
\addplot[]table[col sep=comma]{csv/Wired/TIV/CDFminpTIV.csv};
\addlegendentry{$mp$ protocol}
\end{axis}
\end{tikzpicture}
\caption[Number of TIVs involving the client]{Number of TIVs involving the client. A point $(x,y)$ means the proportion $y$ of clients suffered $x$ or fewer TIVs.}
\label{tivchart}
\end{figure}

For each client, four delay-based triangles are calculated at each iteration, three of which have the client as one of the triangle's vertices for a total of 3$\times$600 = 1,800 triangles involving the client. Figure~\ref{tivchart} shows a CDF of the number of \acp{TIV}, resulting from either $mp$-estimated or $av$-estimated delays, for each client (legitimate or adversary). Note that Algorithm \ref{verifierprotocol} does not call $valid(\mathbf{D}^{av})$ if $valid(\mathbf{D}^{mp})$ is true (line \ref{minimumpairs}).\footnote{Recall form Section \ref{notationsection} on page \pageref{notationsection}, the function $valid(\mathbf{D})$ checks for \acp{TIV} in the four triangles whose side lengths are mapped from the six \acp{OWD} in $\mathbf{D}$.} We thus counted the number of \acp{TIV} for $av$ by running a modified version of Algorithm \ref{verifierprotocol} (page \ref{verifierprotocol}), where line \ref{minimumpairs} is removed (and the \emph{else} at the beginning of line \ref{average}).

For the triangles described by $mp$-estimated delays, very few clients (5\%) suffered no \acp{TIV}, and 86\% suffered at least 10 (of 1,800 possible) \acp{TIV}. While these results confirm that \acp{TIV} occur frequently in the Internet ({\it cf.} \cite{tivrecent}), they emphasize the importance of iterative delay-measurement to mitigate \acp{TIV}. For example, half the clients suffered fewer than 28\% (or 500) \acp{TIV} in total, enabling \ac{CPV} to use the remaining 1,300 valid triangles to verify location assertions.

The case was slightly different using $av$-estimated delays; almost all clients suffered at least one \ac{TIV} and 93\% suffered at least 10 of the possible 1,800 \acp{TIV}. However, $av$ was overall better in avoiding \acp{TIV} than $mp$. Half the clients suffered fewer than 300 \acp{TIV} (versus 500 for $mp$). Because $av$ estimates the \ac{OWD} of a triangle's side as the average of both directions, it tends to reduce the discrepancy between the three sides, leading to fewer \acp{TIV} than $mp$.

\section{The ``Area" as a Discrimination Metric}
\label{areadiscsection}

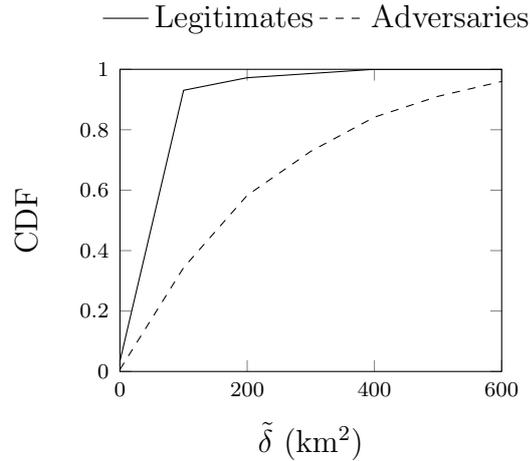
\begin{figure}
\centering
\begin{tikzpicture}
\begin{axis}[
width=2.6in,
height=2.2in,
xlabel=$\tilde{\delta}$ (km$^2$),
ylabel=CDF,
xmin=0, xmax=600,
ymin=0, ymax=1,
tick label style={font=\scriptsize},
x tick label style={font=\scriptsize},
y tick label style={font=\scriptsize},
legend columns=-1,
legend style={draw=none,at={(0.52,1.25)},anchor=north,}
]
\addplot[] table[col sep=comma]{csv/Wired/areaDifference/combinedInsideAreaDifference.csv};
\addlegendentry{Legitimates}
\addplot[dashed] table[col sep=comma]{csv/Wired/areaDifference/combinedOutsideAreaDifference.csv};
\addlegendentry{Adversaries}
\end{axis}
\end{tikzpicture}
\caption[Triangular areas: legitimate clients versus adversaries]{Median area difference ($\tilde{\delta}$) for 146 legitimates, and 2,301 adversaries. A point $(x,y)$ means $\tilde{\delta}$ was less than or equal to $x$ km$^2$ for the proportion $y$ of clients.}
\label{medianareadifference}
\end{figure}

We analyze the effectiveness of using the areas of triangles (those determined by the verifiers and the client---see Chapter \ref{ch:cpv}) as a metric to distinguish legitimates from adversaries. Figure~\ref{medianareadifference} shows a CDF of the median area difference, $\tilde{\delta}$, for all 146 legitimates and 2,301 adversaries. These area differences are either calculated from the $mp$ or the $av$ protocols (see Algorithm \ref{verifierprotocol}). Note that, from Fig.~\ref{dis_away_cdf} (Section \ref{ch:wiredecva}), about one-third of all adversaries were within 400 km of the triangle's sides (e.g., $E$ and $F$ in Fig.~\ref{geoMapExample} were within 50 km and 850 km of the triangle's side respectively).

The results in Fig.~\ref{medianareadifference} show that 93\% of all legitimates had $\tilde{\delta}<$100 km$^2$, whereas two-thirds of all adversaries had more than that value. The results affirm that, although the experiments involved numerous adversaries that are close to the sides of the triangles encompassing their asserted location, triangular areas distinguished between them. In conclusion, the triangular area served as a successful discrimination metric to distinguish between legitimates and adversaries.

\section{The Confidence Ratio}
\label{wired:confidenceratio}

\begin{figure}
\centering

\subfigure[]
{
\label{cdfatfixedepsilon}
\begin{tikzpicture}
\begin{axis}[
width=2.7in,
height=2.2in,
xlabel=$\Gamma$,
ylabel=CDF,
xmin=0, xmax=1,
ymin=0, ymax=1,
x tick label style={font=\scriptsize},
y tick label style={font=\scriptsize},
legend columns=-1,
legend style={draw=none,at={(0.4,1.25)},anchor=north,}
]
\addplot[] table[col sep=comma]{csv/Wired/cr/CDF_CS_legits.0.csv};
\addlegendentry{Legitimates}
\addplot[dashed] table[col sep=comma]{csv/Wired/cr/CDF_CS_adver.0.csv};
\addlegendentry{Adversaries}
\end{axis}
\end{tikzpicture}
}
\subfigure[]
{
\label{barParameters}
\begin{tikzpicture}
\begin{axis}[
width=2.7in,
height=2.2in,
ybar stacked,
bar width=1pt,
ymin=0, ymax=300,
xticklabels={0,,,,,,,,,,10,,,,,,,,,,20,,,,,,,,,,30},
xtick=data,
xlabel=$\bigtriangleup$,
ylabel=$\epsilon_\bigtriangleup$ (km$^2$),
x tick label style={font=\scriptsize},
y tick label style={font=\scriptsize},
y tick label style={font=\scriptsize}
]
\addplot [fill=blue,draw=black] table[col sep=comma]{csv/Wired/cr/bar.epsilon.0.csv};
\end{axis}
\end{tikzpicture}
}
\caption[CPV's confidence of the assertion truthfulness: legitimate clients versus adversaries]{(a) Confidence ratios ($\Gamma$) for 146 legitimates, and 2,301 adversaries. A point $(x,y)$ means $\Gamma$ was less than or equal to $x$ for the proportion $y$ of clients. (b) Values of $\epsilon_\bigtriangleup$, for each $\bigtriangleup$ in the 34-triangles set.}
\end{figure}
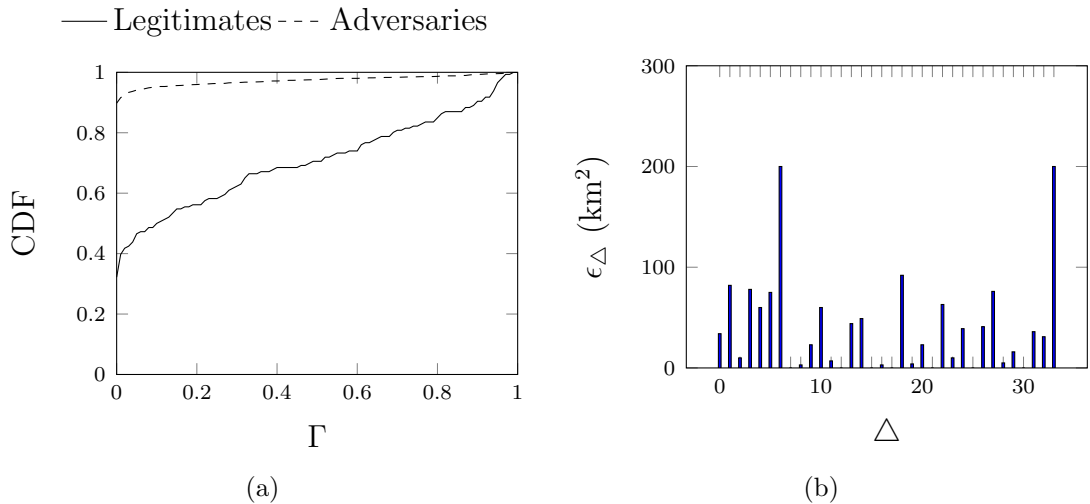

Figure~\ref{cdfatfixedepsilon} shows the CDF of $\Gamma$ for legitimates and adversaries; the values of $\Gamma$ associated with 90\% of all adversaries was 0, i.e., certain values for \ac{CPV}'s input parameters led the algorithm to be 100\% confident about the absence of those adversaries from the triangles encompassing their asserted location. The case was different with legitimates, where only 30\% had a $\Gamma$ value above 0.5, and half had a value above 0.1. Thus in our experiments, \ac{CPV} detected falsified location assertions easier than realizing the correctness of true (honest) assertions. The values of $\epsilon_\bigtriangleup$ that result in this $\Gamma$ distribution are shown in Fig.~\ref{barParameters}.

\label{changebroadcasters}\textcolor{\changes}{For FRs and FAs, tolerating one over the other depends on the application using CPV. For example, media broadcasters aiming to assert their legal compliance with license agreements would likely tolerate FAs more than FRs. On the other hand, FRs might be more tolerable for a sensitive banking transactions than FAs. Tuning \ac{CPV}'s input parameters enables applications to control which false decision should the algorithm tolerate more.}

\section{Proximity to Triangle's Sides}
\label{Adjacency}

\begin{figure}
\centering
\subfigure[$away(\bigtriangleup,g)=70/700=0.1$]
{
\label{insidedistancedefinition}
\includegraphics[scale=0.45]{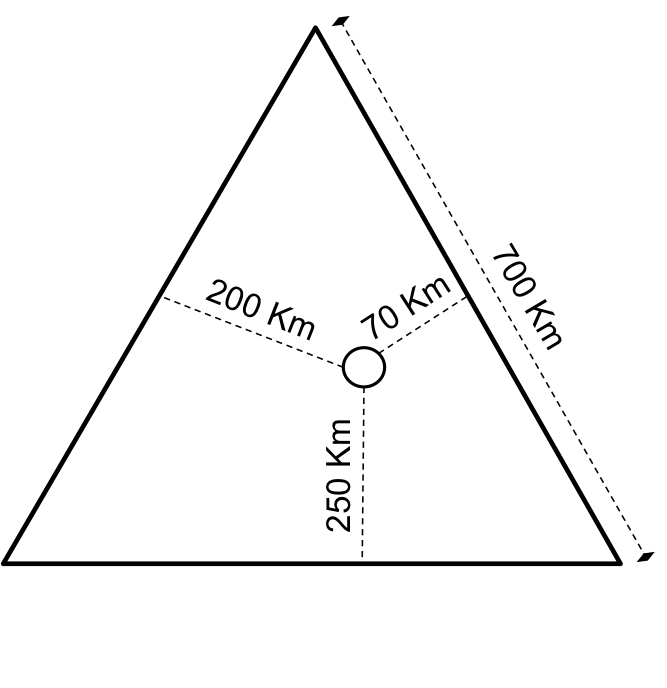}
}
\subfigure[CDF of $away(\bigtriangleup,g)$]
{
\label{dis_legitaway_cdf}
\begin{tikzpicture}
\begin{axis}[
width=2.6in,
height=2.2in,
xlabel={$away(\bigtriangleup,g)$},
ylabel=CDF,
xmin=0, xmax=0.3,
ymin=0, ymax=1,
xtick={0,0.2,0.3},
ytick={0,0.25,0.5,0.75,1},
extra x ticks={0.1},
extra y ticks={0.66},
extra tick style={grid=major},
tick label style={font=\scriptsize},
legend columns=-1,
legend style={draw=none,at={(0.52,1.25)},anchor=north,}
]
\addplot[solid,line width=0.8pt]table[col sep=comma]{csv/Wired/dis_legit.csv};
\end{axis}
\end{tikzpicture}
}
\caption[Legitimates' distances from the triangles' closest side (experimental design)]{Legitimates' distances from the triangles' closest side. (a) Calculation of $away(\bigtriangleup,g)$; (b) A point $(x,y)$ means the proportion $y$ of adversaries were $x$ km away from the closest side. Note: this graph shows experimental design, not results.}
\end{figure}
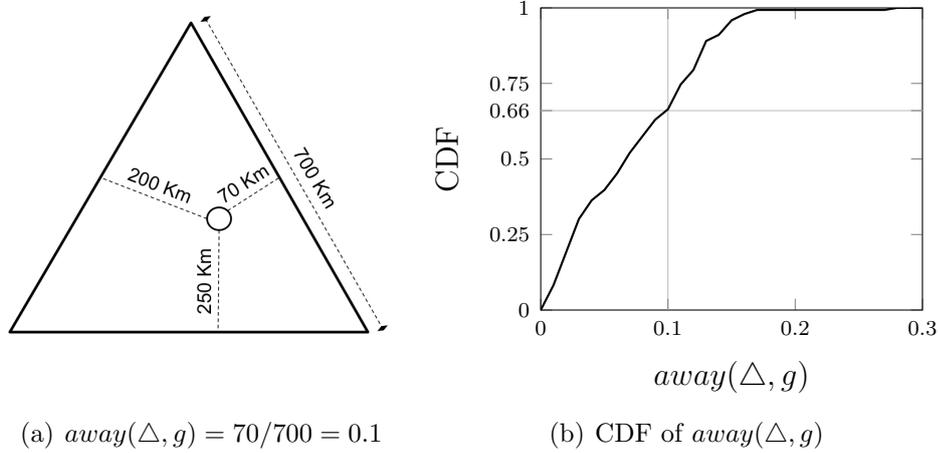

This subsection analyzes the effect of a legitimate's proximity to the sides of its enclosing triangle. Let $away(\bigtriangleup,g)$ be the ratio of the distance between a point $g$ inside $\bigtriangleup$ and side $z_{\bigtriangleup}^g$ to the length of $z_{\bigtriangleup}^g$, where $z_{\bigtriangleup}^g$ is the closest side to $g$ (see Fig.~\ref{insidedistancedefinition} for an example). If $away(\bigtriangleup,g)=0$, then $g$ lies on one of the three sides of $\bigtriangleup$. We evaluate how \ac{CPV}'s efficacy changes (as expected it improves) as we test with fewer legitimates close to the sides (i.e., with relatively small values of $away()$). Figure~\ref{dis_legitaway_cdf} shows a CDF of $away(\bigtriangleup,g)$ for all 146 legitimate clients in the experiments with respect to each $\bigtriangleup$ in the 34 triangle set. The location $g$ of two-thirds of legitimate clients was such that $away(\bigtriangleup,g)\leq0.1$.

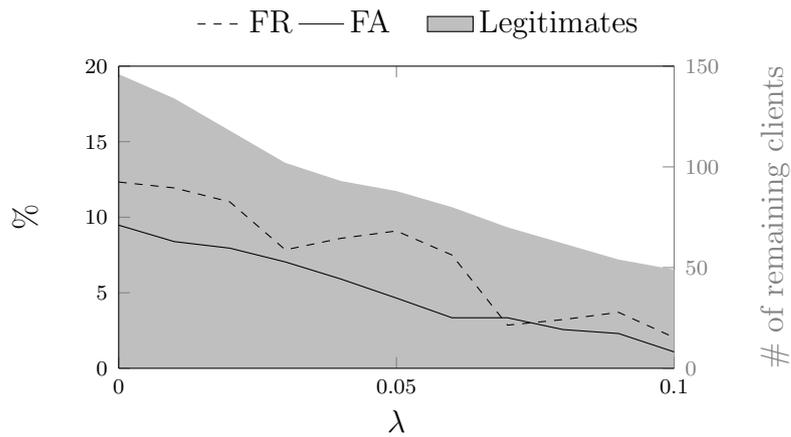
\begin{figure}
\centering
\begin{tikzpicture}
\begin{axis}[
width=3.5in,
height=2.2in,
ymin=0, ymax=150,
xmin=0, xmax=0.1,
ylabel={\# of remaining clients},
axis y line*=right,
axis x line=none, 
y axis line style={gray},
y tick label style={font=\scriptsize,color=gray},
y label style={color=gray,at={(1.35,0.5)}},
area legend,
legend style={draw=none,at={(0.75,1.22)},anchor=north,}
]
\addplot[fill=lightgray,draw=none] table[col sep=comma]{csv/Wired/lambda/remaininginside.csv}\closedcycle;
\addlegendentry{Legitimates}
\end{axis}
\begin{axis}[
width=3.5in,
height=2.2in,
xlabel=$\lambda$,
ylabel=\%,
xmin=0, xmax=0.1,
xtick = {0,0.05,0.1},
ymin=0, ymax=20,
scaled ticks=false,
axis y line*=left,
tick label style={/pgf/number format/fixed},
x tick label style={font=\scriptsize},
x label style={at={(0.5,0.03)}},
y tick label style={font=\scriptsize},
legend columns=-1,
legend style={draw=none,at={(0.32,1.22)},anchor=north,}
]
\addplot[dashed] table[col sep=comma]{csv/Wired/lambda/fr.csv};
\addlegendentry{FR}
\addplot[] table[col sep=comma]{csv/Wired/lambda/fa.csv};
\addlegendentry{FA}
\end{axis}
\end{tikzpicture}
\caption[The effect of legitimate clients' closeness to the triangular sides on CPV]{FRs and FAs when legitimates at location $g=\{x,y\}$ are excluded from the experiments, such that $away(\bigtriangleup,g)<\lambda$. The shaded region is the number of remaining legitimates.}
\label{cborderwithslowcborderwithoutslow}
\end{figure}

Figure~\ref{cborderwithslowcborderwithoutslow} shows the number of \acp{FR} and \acp{FA} after excluding legitimates at locations $g$, such that $away(\bigtriangleup,g)<\lambda$ for all $0\leq\lambda\leq0.1$. The number of remaining legitimates is shown on the same chart as the $y$-axis on the righthand side.\footnote{Most of the PlanetLab nodes used in our experiments are located within cities, which explains the relatively large number of nodes close to triangles' sides.} All adversaries in our experiments were included in the plot regardless of their triangle proximity. As more legitimates are excluded, the effect of the remaining ones on the \acp{FR} increases. When the remaining clients suffer relatively high network delays, the \acp{FR} oscillate as shown in the plot. Of the chosen PlanetLab nodes, we noticed three nodes suffering exceptionally high delays for unknown reasons. Their distance from the triangle's closest side was such that $0.002\leq away()\leq0.28$. Those nodes contribute to the oscillation intensity occurring in Fig.~\ref{cborderwithslowcborderwithoutslow} as $\lambda$ increases, and become very hard to partition from adversaries as more legitimates get excluded. At $\lambda=0.1$, the \acp{FR} were 2\% versus 12.3\% at $\lambda=0$. This improvement emphasizes the importance of appropriate triangle choice with respect to the asserted location. For an asserted location $l$, it is recommended that $\bigtriangleup_l$ be chosen such that $away(\bigtriangleup_l,l)\geq0.1$. 
Although the number of adversaries included in the experiments was unchanged over the spectrum of $\lambda$ in Fig.~\ref{cborderwithslowcborderwithoutslow}, \acp{FA} improve as $\lambda$ increases; the \acp{FA} were 9\% at $\lambda=0$, and dropped to 1.1\% at $\lambda=0.1$. Such improvement stems from the ability to find smaller $\epsilon$ values that do not falsely reject legitimates---now far from the triangle's sides, i.e., at $\lambda=0.1$. Smaller $\epsilon$ values reduce \acp{FA}.

\section{Number of Iterations}
\label{timeSufficient}

In this section, we study the effect of the number of \ac{CPV} iterations, $n$, on the efficacy of the verification process. Note that large number of iterations comes at the cost of an increased \ac{CPV} runtime, during which the client is waiting to get its location verified before receiving services.

\begin{figure}
\centering
\begin{tikzpicture}
\begin{axis}[
width=3.7in,
height=2.2in,
xlabel=$n$,
ylabel=\%,
xmin=1, xmax=600,
ymin=-5, ymax=60,
xmode = log,
xtick={1,10,100,1000},
log base 10 number format code/.code={$\pgfmathparse{10^(#1)}\pgfmathprintnumber{\pgfmathresult}$},
x tick label style={font=\scriptsize},
x label style={font=\scriptsize,at={(0.5,0.03)}},
y tick label style={font=\scriptsize},
y label style={font=\scriptsize},
legend style={draw=none,at={(1.25,0.95)},anchor=north,}
]
\addplot[dashed]table[col sep=comma]{csv/Wired/iterations/fr_p1.csv};
\addlegendentry{FR at $\lambda=0.1$}
\addplot[]table[col sep=comma]{csv/Wired/iterations/fa_p1.csv};
\addlegendentry{FA at $\lambda=0.1$}
\addplot[dotted,thick]table[col sep=comma]{csv/Wired/iterations/fr.csv};
\addlegendentry{FR at $\lambda=0$}
\addplot[dash pattern=on 3pt off 2pt on \the\pgflinewidth off 1pt on \the\pgflinewidth off 1pt on \the\pgflinewidth off 2pt]table[col sep=comma]{csv/Wired/iterations/fa.csv};
\addlegendentry{FA at $\lambda=0$}
\end{axis}
\end{tikzpicture}
\caption[The effect of the number of iterations on CPV]{FRs and FAs when $n$ iterations in Algorithm~\ref{verifierprotocol} (page \pageref{verifierprotocol}) are performed.}
\label{varynumberiteartions}
\end{figure}

Figure~\ref{varynumberiteartions} shows the change in \acp{FR} and \acp{FA} with $n$ (log$_{10}$ scale). \acp{FR} and \acp{FA} generally decrease as more iterations are performed, at $\lambda=0.1$ and $\lambda=0$. The results for $\lambda=0.1$ are quite sensible: \acp{FR} and \acp{FA} decrease almost monotonically when more iterations are performed. With two iterations, at $\lambda=0.1$, the \acp{FA} dropped to $\sim$9\% from over 50\% when only one iteration was performed. Fewer than 10 iterations did not enable the verifiers to identify legitimates appropriately as the \acp{FR} were between 6-22\%, i.e., no values for $\epsilon_\bigtriangleup$ and $\tau_\bigtriangleup$ existed to partition legitimates and adversaries. However, between 10 and 20 iterations, \acp{FR} and \acp{FA}, at $\lambda=0.1$, remained at $\sim$2\% and $\sim$1\% respectively. 
At $\lambda=0$, \acp{FA} dropped from $\sim$56\% when one iteration was performed, to $\sim$10\% when 9 iterations were performed. It then oscillated between $\sim$10\% and $\sim$6\% when fewer than 100 iterations are performed, climbing steadily to $\sim$8\% for the rest of the iterations. This rise happened simultaneously with an improvement in the \acp{FR} (at $\lambda=0$). As more iterations are performed, it becomes more feasible to find $\epsilon_\bigtriangleup$ values that partition legitimates from adversaries. To accommodate legitimates that are very close to the triangles' sides, large values of $\epsilon_\bigtriangleup$ were required, which resulted in falsely accepting more adversaries. This explains the rise in \acp{FA} as more iterations were performed, at $\lambda=0$. Over the entire range of $n$, the \acp{FR} at $\lambda=0$ decreased from $\sim$34\% at $n=1$ to $\sim$12\% at $n=600$. Even when legitimates are highly adjacent to their enclosing triangles' sides, large number of iterations can improve the ability of finding $\epsilon$ and $\tau$ values that better partition legitimates from adversaries. This highlights the importance of the iterative delay-measurement of \ac{CPV} (see Algorithm~\ref{verifierprotocol} on page \pageref{verifierprotocol}), especially when the chosen verifiers determine a triangle whose sides are close to the asserted location.

\textcolor{\changes}{In the conducted experiments, each iteration took six seconds because each verifier sent a probing packet every 2 seconds. Such duration could be modified according to the application's requirements. For example, increasing the duration of each iteration (i.e., increasing the delay between each subsequent probing message) diversifies the network conditions during which the delays are measured. This comes at the cost of increased verification time thus, affecting CPV's usability. In general, a 30 ms delay between network probing/monitoring packets should be sufficient to avoid packet interference \cite{holleczek2006statistical}. Finding an optimal balance between both ends of the spectrum is left for future investigation.}\label{changeiterations2}

\section{\emph{Minimum pairs} versus \emph{Average} protocol}
\label{wired:comparison}

\definecolor{Gray}{gray}{0.85}
\newcolumntype{G}[1]{>{\columncolor{Gray}{\centering\let\newline\\\arraybackslash\hspace{0pt}}}m{#1}}
\newcolumntype{B}{>{\bfseries}C}
\newcolumntype{D}[1]{>{\columncolor{Gray}{\centering\let\newline\\\arraybackslash\hspace{0pt}}\bfseries}m{#1}}
\newcolumntype{g}{>{\columncolor{Gray}}c}
\newcommand{\width}{1.2}
\newcommand{\wid}{2}
\definecolor{LightCyan}{rgb}{0.88,1,1}
\definecolor{LightGreen}{rgb}{0.88,1,0.88}
\definecolor{LightGray}{gray}{0.95}

\begin{table*}
\centering
\renewcommand{\arraystretch}{1.2}
\caption[Hypothetical modifications to CPV's OWD-estimation process: $av$ only versus $mp$ only versus both protocols]{Results of modified versions of CPV. The shaded column is the unmodified version---see Algorithm \ref{verifierprotocol} on page \pageref{verifierprotocol}.}
\scalebox{0.7}{
\begin{tabular}{ r|cc || C{\width cm}C{\width cm}B{\wid cm} | C{\width cm}C{\width cm}B{\wid cm} | G{\width cm}G{\width cm}D{\wid cm} }

\multirow{2}{*}{Case}&\multirow{2}{*}{$\lambda$}&\multirow{2}{*}{$n$}& \multicolumn{3}{c|}{$av$ only} & \multicolumn{3}{c|}{$mp$ only}& \multicolumn{3}{g}{CPV ($mp$ and $av$)}\\

&&		&FR\% & FA\% & FR+FA 		&FR\% & FA\% & FR+FA		& FR\% & FA\%& FR+FA	\\\hline
1&0&10	&45	&4.4&49			&39	&3.8&43			&35	&3.9&\ \ \ 39	\\
2&0&100	&25	&5.3&30			&26	&4.9&31			&21	&5.1&\ \ \ 26	\\
3&0&600	&14	&7.1&21			&17	&6.5&24			&13	&7.3&\ \ \ 20	\\\hline

4&0.1&10		&24	&1.7&26			&10	&2.3&12			&4.1&2.1&\ \ \ 6.2		\\
5&0.1&100	&10&0.7&11			&2.0&1.0&3.0		&2.0&1.1&\ \ \ 3.1		\\
6&0.1&600	&2.0&1.7&3.7		&2.0&1.0&3.0		&2.0&1.0&\ \ \ 3.0		\\\Xcline{1-12}{2\arrayrulewidth}

\end{tabular}
}\\
{\scriptsize $\lambda$ = legitimates-exclusion threshold (see Section \ref{Adjacency}); $n$ = number of iterations (see Algorithm \ref{verifierprotocol});\\[-4pt]
$av$ = the ``average" protocol; $mp$ = the ``minimum pairs" protocol.}
\label{tableSummary}
\end{table*}

Table \ref{tableSummary} summarizes PlanetLab results of different \ac{CPV} evaluation scenarios. The columns represent modified versions of \ac{CPV}, i.e., different from the behavior given in Algorithm \ref{verifierprotocol}. In line \ref{getvalues} of Algorithm \ref{verifierprotocol}, two \ac{OWD}-estimation protocols are used ($mp$ and $av$) to alleviate the effect of \acp{TIV}. Table \ref{tableSummary} lists the results when only the $av$ protocol is used (``$av$ only" column), when only $mp$ is used (``$mp$ only" column), and when both are used (``\ac{CPV}" column). The results are shown for various combinations of the exclusion threshold, $\lambda$ (see Section \ref{Adjacency}), and the number of iterations, $n$. The table shows the \acp{FR}, the \acp{FA}, and their sum in each respective case.

From Table~\ref{tableSummary}, the summation of \acp{FR} and \acp{FA} when both \ac{OWD}-estimation protocols are used (right-most column under ``\ac{CPV}") is smaller in four out of six of the cases (table rows) compared to the summation when each protocol is used solely, e.g., 39 is less than 43 and 49 in the first case. Thus, the use of both \ac{OWD}-estimation protocols tends to enhance the accuracy of the location verification process.

Using the $mp$ protocol solely gave better results than $av$ solely in four out of six cases. The $av$ protocol was better at $\lambda=0$ and $n\geq100$. Recall from Section \ref{tivsection} that the $mp$ protocol results in more \acp{TIV}. Since \ac{CPV} counts the number of TIVs against the client, more TIVs tend to increase \acp{FR}, as shown by the results under the ``$mp$ only" column in Table~\ref{tableSummary}. At $\lambda=0$ and $n\geq100$, there were 26\% and 17\% \acp{FR} using the $mp$ protocol, versus 25 and 14\% using $av$. In conclusion, \ac{CPV} works best when utilizing both delay-estimation protocols to mitigate the unfavorable effect of \acp{TIV}.

\section{Conclusion}
\label{wired:conclusion}

Three remarks can be made in conclusion from the evaluation conducted in this chapter.
\begin{enumerate}

\item Reducing the factors that negatively affect the delay-to-distance mapping process (such as \acp{TIV} \cite{tivrecent}) improves the accuracy of the location verification process. \ac{CPV} leverages several heuristics to reduce such factors, e.g., iterating the delay-measurement process and using multiple delay-estimation protocols. The results in Sections \ref{tivsection}, \ref{timeSufficient} and \ref{wired:comparison} provide evidence that \ac{CPV}'s accuracy improves upon applying these heuristics. 

\item Comparing the areas of triangles projected using the delays between three verifiers and a client enables the verifiers to realize if the client is geographically encapsulated by the triangle determined by the verifiers's locations. Section \ref{areadiscsection} provide evidence supporting this conjecture.

\item The adjacency of a legitimate client to the sides of the triangle enclosing their geographic location can dramatically affect the correctness of \ac{CPV}'s verification. From the analysis in Section \ref{Adjacency}, clients that were away of the triangle's closest side at least 10\% of the length of that side were likely to get their assertions correctly accepted.

\end{enumerate}

In summary, the evaluation conducted in this chapter using a real world experimental testbed with wired-connected clients shows that certain \ac{CPV} parameterization enabled the algorithm to \ac{FR} and \ac{FA} rates of 2\% and 1\% respectively. However, to achieve these results in practice, a sufficient number of verifiers must be available to find the appropriate triangles, ones whose sides are far enough from the asserted location (see Section \ref{Adjacency}).

\label{changetrianglesize}
\textcolor{\changes}{From a geographic perspective, all triangles used in the experiments conducted in this chapter had side lengths ranging from $\sim$260 km to $\sim$1,100 km; the reported results pertain to this range. Due to the increased route circuitousness (see Section \ref{ipgeosection}, page \pageref{ipgeosection}) that happens with short distances over the Internet \cite{6197179}, extremely small triangle sizes are expected to result in higher FR/FA rates. The rate by which the results worsens as triangles become smaller is left for future exploration.}

\textcolor{\changes}{Since the triangle size is the verification granularity, larger triangles may become less practical from the application's perspective. However, some applications may only need coarse verification granularity, e.g., to preserve user's privacy; larger triangles in that case may be beneficial.}

In the next chapter, \ac{CPV} will be similarly evaluated, but with legitimate clients modeled to use 802.11 (wireless) access networks.

\chapter{Evaluation with Wireless CPV Clients}
\label{ch:wirelessecva}

The nature of delays in wireless and wired networks is different. This chapter evaluates CPV when legitimate clients (those inside the triangles) are connected through \label{changewireless2}\textcolor{\changes}{WiFi access networks. In the rest of this thesis, we refer to those clients simply as \emph{wireless clients}. }A wireless client is assumed to be one hop away from its access point, which serves as the client's gateway to the Internet. Beyond the gateway, all hops until the verifiers are assumed to be wired. \label{changesatellite}\textcolor{\changes}{That is, none of the verifiers are assumed to use a wireless access network, e.g., satellite, which is a reasonable assumption since the location verification service provider is assumed to own/control the verifier infrastructure.}

In Chapter \ref{ch:wiredecva}, CPV was evaluated with clients connected through wired access networks. Using the PlanetLab testbed, evaluation was performed by having sets of three verifiers (running on PlanetLab nodes) measure OWDs to/from legitimate clients and adversaries using the $mp$ and $av$ protocols (Chapter \ref{ch:owd}). The measured OWDs where logged, and the CPV algorithm (Chapter \ref{ch:cpv}) was run locally on the collected logs. Knowing the ground truth of inside and outside clients (i.e., legitimates and adversaries), CPV's false reject/accept rates were quantified.

To evaluate CPV in wireless networks, we use the OWDs collected in Chapter \ref{ch:wiredecva} between the client and the verifiers, and add an additional delay component to each delay value to model wireless transmission. The added component represents the single-hop delay between the wireless client and its access point, and is modeled as a random variable that follows wireless latency-distributions studied in the literature \cite{1249764}.

Assume two clients, a legitimate and an adversary, both having their location assertions verified by CPV. Their access networks follow one of the four combinations shown in Table~\ref{wirelesscomb}. The table also shows in which chapter the combination is explored.
\begin{table}
\centering
\caption{Combinations of access networks for a legitimate client and an adversary}
\begin{tabular}{ccc}
Legitimate client&Adversary&Chapter\\\hline
Wired&Wired&\ref{ch:wiredecva}\\
Wired&Wireless&--\\
Wireless&Wired&\ref{ch:wirelessecva}\\
Wireless&Wireless&--\\
\end{tabular}
\label{wirelesscomb}
\end{table}
Wireless adversaries are not modeled in this thesis. The reason is that wireless networks tend to, among other effects, increase delays and the delay variance, which in CPV increase the likelihood of rejecting assertions. Therefore, by modeling wireless legitimates and wired adversaries, we test CPV in the most demanding (to the defender) situation among the four possible combinations in Table~\ref{wirelesscomb}.

This evaluation methodology addresses the effect of delays in wireless networks, while retaining the advantages of PlanetLab, e.g., real-world network delays, logical and geographical network topology, exterior gateway routing policies, congestion behavior. In addition, by using the data logs collected from the wired evaluation phase (Chapter \ref{ch:wiredecva}), we unify all experimental parameters across wireless and wired testing. Root causes of improvement/retrogression can then be more reliably identified.

This chapter aims to study the impact of the varying wireless delays on CPV, by specifically exploring the following three questions:
\begin{enumerate}
\itemsep0pt
\item {\bf Assuming $k$ wireless devices actively competing for the wireless media with the legitimate client, how does $k$ affect CPV? } Here, the number of wireless legitimate clients is varied, and CPV's efficacy is analyzed. We test by modeling clients using IEEE 802.11b as a representative access technology.
\item {\bf For a given triangle verifying assertions of wireless legitimates and a wired adversary, what is the minimum distance the adversary should be away from the triangle's nearest side so that CPV correctly rejects it? } To answer this question, we test CPV when varying the width of the adversary-free region outside the triangle. We do this by progressively excluding nearby adversaries from the experiments and reevaluating CPV.
\item {\bf How many CPV iterations should the verifiers perform in order to essentially eliminate the effect of the additional wireless delays?} As explained in Chapter \ref{ch:cpv}, the verifiers in CPV estimate the delays iteratively. We derive the number of iterations required to essentially eliminate the effect of the wireless networks, as a function of the number of wireless devices $k$ and the acceptance threshold $\tau$ (see Chapter \ref{ch:cpv}).
\end{enumerate}

{\bf Chapter Roadmap. } Section \ref{sec:wireless:background} provides background on the mechanisms by which 802.11 networks manage access to the shared medium. Section \ref{modlit} reviews recent literature that models delays of single-hop wireless networks. The reviewed models are then used to evaluate CPV in Section \ref{evaCPVwireless}. Section \ref{wireless:num:iter} analyzes the effect of the number of iterations on the efficacy of CPV when legitimate clients are using wireless access networks.

\section{Background on 802.11}
\label{sec:wireless:background}

\ac{DCF} is the technique used in IEEE 802.11 (wireless) networks \cite{IEEE802} to manage access to the shared wireless media \cite{book:kurose}. It employs the \ac{CSMA/CA} method.

In \ac{DCF}, when the \ac{MAC} layer of a device has a data frame to send, it checks if the medium is busy and starts transmission if it is free for a length of time called the \ac{DIFS} \cite{book:kurose}. If the medium is busy, the device backs off for $X$ time slots, where $X$ is a number chosen uniformly at random in the range $[0,W_{\text{min}}]$. The countdown of the back-off timer is paused whenever a transmission (i.e., from other devices) is sensed. The device transmits only if the media was found vacant for a period equal to \ac{DIFS} after the back-off time reaches zero. Otherwise, the device backs off for another uniformly-chosen random number of time slots in the range $[0,2\cdot W_\text{min}]$. The process is repeated as long as the medium is sensed to be busy anytime during the countdown, with the back-off interval doubling on each repetition until it reaches a maximum of $W_{\text{max}} = 2^m\cdot W_\text{min}$, for some predefined value $m$.

Upon successful reception, the receiver sends an \ac{ACK}. Transmitting the \ac{ACK} follows the \ac{DCF} procedure described above. If an \ac{ACK} is not received, the sender of the original data frame attempts several further retransmissions following the \ac{DCF} procedure, and eventually gives up if those fail.

If two wireless devices, $A$ and $B$, using one access point are not in the transmission ranges of each other, they are said to be \emph{hidden terminals}. $A$ and $B$ may thus fail to sense each others' transmission, in which case simultaneously transmitting may cause collision at the access point. To address the hidden terminal problem \cite{fullmer1997solutions}, \ac{RTS/CTS} frames are optionally used. If $A$ is the device with data to send, it first sends an \ac{RTS} control frame to the access point. This frame indicates the time $A$ needs to send its data frame and receive the \ac{ACK}. The access point responds by broadcasting a \ac{CTS} containing such timing information, which would also be received by $B$. $B$ then refrains from using the medium for the specified period of time. The analysis included throughout this chapter considers the case whereby \ac{RTS/CTS} frames are used.

\section{Wireless Delay Models in the Literature}
\label{modlit}

This section reviews two wireless delays models in the literature, both assume a single-hop wireless network with one access point and $k$ wireless devices. The $k$ devices are \emph{saturated}, i.e., always have frames to send. The channel is assumed ideal, meaning that the only source of frame corruption is collision.

Note that the focus of this section is not to compare the two wireless delay models, nor not to evaluate their accuracies. We rather review these models to use them in evaluating CPV later in Sections \ref{evaCPVwireless} and \ref{wireless:num:iter} below.

\subsection{Average back-off time at a stage}
Carvalho and Garcia-Luna-Aceves \cite{1249764} derived the average time a device spends backing off. Recall from Section \ref{sec:wireless:background} that a device backs-off for $X= \mathcal{U}\{0, 2^m\cdot W_\text{min}\}$ time slots. Thus, the expected backing-off time, $\alpha$, is the time spent while counting down $X$ time slots plus the time where the countdown is paused during a sensed transmission \cite{1249764}:
\begin{equation}
\label{alphaprob}
\alpha = \sigma p_i + t_cp_c + t_sp_s
\end{equation}
The constant $\sigma$ is the length of the time slot (in $\mu sec$); $p_i$ is the probability the channel is idle (i.e., the subscript is not an index, it denotes ``idle") during a time slot; and $p_c$ and $p_s$ are the probabilities of collision and successful transmission respectively during a time slot. $t_s$ and $t_c$ are the number of time units a device spends while pausing the countdown during a successful transmission and during a transmission with collision respectively. Bianchi {\it et al.} \cite{bianchi2000performance} expressed these durations as follows:
\begin{equation}
\label{tsucccess}
t_s = \frac{l(\text{RTS})+l(\text{CTS})+l(H)+l(P)+l(\text{ACK})}{\text{rate}} + (3\cdot\text{SIFS}+\text{DIFS}) + 4\delta
\end{equation}
\begin{equation}
\label{tcollision}
t_c = \frac{l(\text{RTS})}{\text{rate}} + \text{DIFS} + \delta
\end{equation}
where the function $l(.)$ indicates the frame (or packet) length in bits; RTS/CTS are the Ready/Clear To Send frames (see Section \ref{sec:wireless:background}); $\delta$ is the propagation delay (in $\mu sec$); SIFS is a technology-specific amount of time (in $\mu sec$); $H$, $P$ and ACK are the header, data packet, and acknowledgement packets respectively; and \emph{rate} is the media's transmission rate in Mbps.

Using a 2-dimensional discrete-time Markov process, Bianchi {\it et al.} derived the probability, $\psi$, that a transmission occurs (successful or with collision) at a time slot as:
\begin{equation}
\label{psinonlinear}
\psi = \frac{2(1-2p)}{(1-2p)(W_\text{min}+1)+pW_\text{min}(1-(2p)^m)}
\end{equation}
where $p$ is the probability of collision occurring at a time slot. Note that $p$ is different from $p_i$, $p_c$ and $p_s$ in (\ref{alphaprob}). Bianchi {\it et al.} \cite{bianchi2000performance} then assumed that a packet collides with a constant and independent probability regardless of the number of retransmissions it suffers. Assuming $k$ devices in the network, if one device transmits, the only case that results in no collision is when none of the $k-1$ other devices transmit, i.e., the probability of no collision is $(1-\psi)^{k-1}$. Therefore, $p$ can be expressed in terms of $\psi$ as \cite{bianchi2000performance}:
\begin{equation}
\label{probp}
p = 1-(1-\psi)^{k-1}
\end{equation}
Thus, the relationship between $p$ and $\psi$ is non-linear. Carvalho and Garcia-Luna-Aceves \cite{1249764} linearized this model in order to use $\psi$ to derive the expected total back-off time (see Section \ref{expectedtotalbackoff} below).

Using $\psi$ and assuming $k$ devices, the probability ($P_\text{tr}$) that at least one of the $k$ devices is transmitting, and the probability ($P_\text{suc}$) that a transmission for any of the $k$ devices is successful are calculated as follows \cite{567423,bianchi2000performance}:
\begin{equation*}
P_\text{tr} = 1-(1-\psi)^k
\end{equation*}
\begin{equation*}
P_\text{suc} = \frac{k\psi(1-\psi)^{k-1}}{P_\text{tr}}
\end{equation*}

The probabilities $p_i$, $p_c$ and $p_s$ in (\ref{alphaprob}) are therefore calculated as $p_i = 1 - P_\text{tr}$, $p_c = P_\text{tr}(1 - P_\text{suc})$, and $p_s = P_\text{tr}P_\text{suc}$ \cite{1249764}.

\subsection{Expected total back-off time}
\label{expectedtotalbackoff}
Carvalho and Garcia-Luna-Aceves \cite{1249764} give an approximate solution to the nonlinear relation between $\psi$ in (\ref{psinonlinear}) and $p$ in (\ref{probp}), and reduce $\psi$ to:
\begin{equation}
\label{psilinear}
\psi = \frac{2W_\text{min}}{(W_\text{min}+1)^2}(1-p)
\end{equation}
Using (\ref{psilinear}), the authors derived $p$ independent of $\psi$ as \cite{1249764}:
\begin{equation*}
p = \frac{2W_\text{min} (k-1)}{(W_\text{min}+ 1)^2 + 2W_\text{min}(k-1)}
\end{equation*}

Carvalho and Garcia-Luna-Aceves \cite{1249764} then used this approximation to obtain $\alpha$ in terms of $\sigma$, $k$, $W_\text{min}$, $t_s$ and $t_c$, as explained above. Finally, they derived the expected time a device backs off $\overline{T}_B$ as \cite{1249764}:
\begin{equation}
\label{tbarbb}
\overline{T}_B = \frac{\alpha(W_\text{min}F - 1)}{2q} + \left(\frac{1-q}{q}\right)t_c
\end{equation}
where 
\begin{equation*}
F = \frac{q-2^m(1-q)^{m+1}}{1-2(1-q)}
\end{equation*}
and $q = 1-p$ represents the probability of no collision.

\subsection{Mean delay and jitter, the model of Carvalho {\it et al.}}
\label{Carvalhomodel}
Carvalho and Garcia-Luna-Aceves \cite{1249764} expressed the expected delay $E[T]$ of a frame as the expected time a device backs off $\overline{T}_B$ in (\ref{tbarbb}) plus the frame transmission time $t_s$ in (\ref{tsucccess}):
\begin{equation}
\label{mean_2003}
E[T] = \overline{T}_B + t_s
\end{equation}

The variance of $T$ was derived as:
\begin{equation*}
\text{Var}[T] = \left[\frac{\alpha(W_\text{min}\gamma-1)}{2}+t_c\right]^2 \frac{1-q}{q^2}
\end{equation*}
where
\begin{equation*}
\gamma = \frac{(2q^2-4q+1-mq(2q-1))(2-2q)^m+2q^2}{(2q-1)^2}
\end{equation*}

Thus the jitter (or the standard deviation) is:
\begin{equation}
\label{std_2003}
\text{Std}[T] = \sqrt{\displaystyle\text{Var}(T)}
\end{equation}

\subsection{CDF of delays}
The model of Carvalho and Garcia-Luna-Aceves \cite{1249764} only provides information about the mean and jitter of the delays given some number of wireless devices $k$. We assume that delays will follow a Gaussian distribution with mean and variance derived as in (\ref{mean_2003}) and (\ref{std_2003}) respectively. However, since the distribution (which would be the delays in that case) goes from $-\infty$ to $\infty$, the model can result in negative delay values. Thus, we assume a truncated Gaussian \cite{book:prob} in the range $[0, \infty]$.

The mean of the Gaussian distribution truncated from $a$ to $b$ is given by \cite{book:prob}:
\begin{equation*}
\text{GausMean}_{\mu,\sigma}(a,b) = \mu-\sigma\cdot Z(\alpha,\beta)
\end{equation*}
where $\mu$ and $\sigma$ are respectively the mean and standard deviation of the parent (non-truncated) Gaussian distribution; $\alpha=(a-\mu)/\sigma$ and $\beta=(b-\mu)/\sigma$; and the function $Z(.)$ is defined as:
\begin{equation*}
Z(\alpha,\beta) = \frac{\phi(\beta)-\phi(\alpha)}{\Phi(\beta)-\Phi(\alpha)}
\end{equation*}
The functions $\phi(.)$ and $\Phi(.)$ are respectively the PDF and the CDF of the standard (i.e., with $\mu=0$ ms and $\sigma=1$ ms) Gaussian distribution.

The standard deviation of the Gaussian distribution truncated from $a$ to $b$ is \cite{book:prob}:
\begin{equation*}
\text{GausStd}_{\mu,\sigma}(a,b) = \sqrt{\displaystyle\sigma^2\cdot\left(  1 - \frac{\beta\cdot\phi(\beta)-\alpha\cdot\phi(\alpha)}{\Phi(\beta)-\Phi(\alpha)} -  Z^2(\alpha,\beta) \right)}
\end{equation*}

To obtain a CDF of the wireless delays that has a mean and standard deviation as in (\ref{mean_2003}) and (\ref{std_2003}), we need to solve simultaneously for $\mu$ and $\sigma$:
\begin{equation}
\label{simul1mean}
\text{GausMean}_{\mu,\sigma}(0,\infty) = E[T]
\end{equation}
and
\begin{equation}
\label{simul1std}
\text{GausStd}_{\mu,\sigma}(0,\infty) = \text{Std}[T]
\end{equation}
Those are two equations in two unknowns, which can be solved using numerical methods. Finally, using $\mu$ and $\sigma$, the CDF of the Gaussian distribution truncated from $a$ to $b$ is \cite{book:prob}:
\begin{equation}
\label{GausCDFtrunc}
\text{GausCDF}_{\mu,\sigma}(x;a,b) = \frac{\Phi(\zeta)-\Phi(\alpha)}{\Phi(\beta)-\Phi(\alpha)}
\end{equation}
where $\zeta=(x-\mu)/\sigma$. Table~\ref{paramsparams} shows the mean and standard deviations calculated using (\ref{mean_2003}) and (\ref{std_2003}) for various values of $k$, and the corresponding $\mu$ and $\sigma$ of the parent (non-truncated) Gaussian distribution calculated by solving (\ref{simul1mean}) and (\ref{simul1std}) simultaneously.

\begin{figure}
\centering

\begin{tikzpicture}
\begin{axis}
[
no markers,
xmin=0,xmax=100,
axis lines*=left, 
xlabel=delay (ms),
ylabel=CDF,
height=5cm,
width=12cm,
ymin=0,ymax=1,
label style={font=\footnotesize},
tick label style={font=\footnotesize,},
enlargelimits=false, clip=true, axis on top,
grid = major,
scaled ticks=false, tick label style={/pgf/number format/fixed},
legend columns=1,
legend style={font=\footnotesize,draw=none,at={(1.15,1.05)},anchor=north,},
]
\addplot[solid] table[col sep=comma]{csv/Wireless/CDF_2003_trunc/cdf.k2.csv};
\addlegendentry{$k=2$}
\addplot[red,dashed] table[col sep=comma]{csv/Wireless/CDF_2003_trunc/cdf.k5.csv};
\addlegendentry{$k=5$}
\addplot[blue,thick,dotted] table[col sep=comma]{csv/Wireless/CDF_2003_trunc/cdf.k10.csv};
\addlegendentry{$k=10$}
\addplot[green,dashdotted] table[col sep=comma]{csv/Wireless/CDF_2003_trunc/cdf.k20.csv};
\addlegendentry{$k=20$}
\addplot[brown,thick,densely dashdotted] table[col sep=comma]{csv/Wireless/CDF_2003_trunc/cdf.k30.csv};
\addlegendentry{$k=30$}
\end{axis}
\end{tikzpicture}

\caption{Truncated Gaussian CDFs of single-hop wireless delays that a frame endures when there are $k$ saturated wireless devices in the network.}
\label{figCDF2003}
\end{figure}
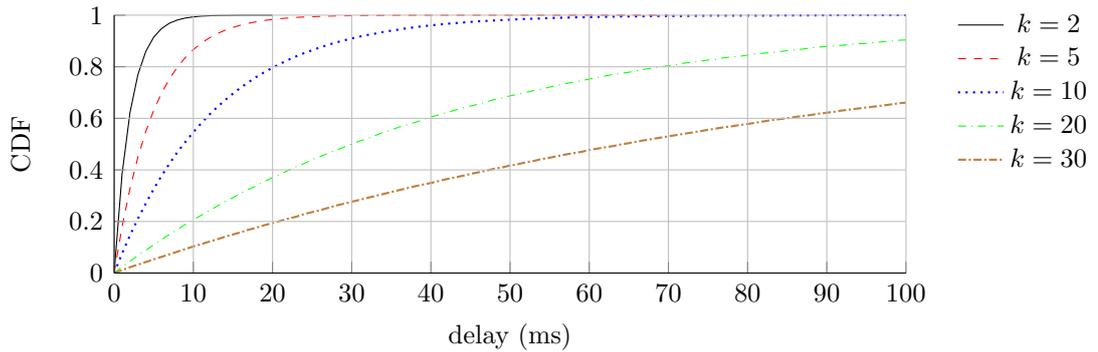

Figure~\ref{figCDF2003} plots the delay distribution, $\text{GausCDF}_{\mu,\sigma}(x;0,\infty)$, using (\ref{GausCDFtrunc}) for various values of $k$. Unsurprisingly, the chart shows that the wireless delays generally increase with $k$. These delay distributions are used in Sections \ref{evaCPVwireless} and \ref{wireless:num:iter} to evaluate CPV in wireless networks.

\begin{table}
\centering
\caption{Mean $\mu$, and standard deviation $\sigma$, of the single-hop wireless delays when $k$ devices are simultaneously competing with the media.}
\begin{tabular}{r|c|rrrrrrrrrrr}
\multirow{2}{*}{
Parameters (ms)}&\multirow{2}{*}{Eqn.}&\multicolumn{7}{c}{$k$}\\\cline{3-9}
					&						&2&3&4&5&10&20&30\\\hline
			
$E[T]$		&(\ref{mean_2003})&2&3&4&5&12&40&87\\
Std$[T]$	&(\ref{std_2003})&0.6&1.6&2.9&4.7&21&89&186\\\hline

$\mu$		&--&-110&-159&-208&-246&-691&-2419&-5156\\
$\sigma$	&--&15&22&29&36&95&328&700\\\hline
\end{tabular}
\label{paramsparams}
\end{table}

The model of Carvalho and Garcia-Luna-Aceves provides an upper bound on the average delay a frame is expected to suffer \cite{1249764}; when they compared their model to simulations, delays from the simulations were always smaller. One reason for the simulation delays being smaller is that there is a non-zero probability that a frame backs off indefinitely \cite{1249764}. However, the DCF standard \cite{IEEE802} specifies that the MAC layer must discard the frame if transmission failed after R back off trials, for some predefined value of R. Transmission retrials from upper layers may then take care of the discarded frames.

\subsection{CDF of delays, the model of Raptis {\it et al.}}
\label{RaptisModel}

Similar to Carvalho and Garcia-Luna-Aceves \cite{1249764}, Raptis {\it et al.} \cite{Raptis2009} used the basis of Binachi \cite{bianchi2000performance} to derive a CDF (and jitter) for the single-hop 802.11 access delays. However, Raptis {\it et al.} \cite{Raptis2009} took into consideration the reality that the frame being transmitted will be discarded after failing transmission in $R$ back-off stages. The authors \cite{Raptis2009} began by deriving the expected delay that a frame suffers after a failed transmission at stage $j$ ($0\leq j\leq R$) as:
\begin{equation}
\label{euunderscorej}
U_j = (j+1)\cdot t_c + \alpha\cdot\sum_{i=0}^{j} \frac{W_i-1}{2}
\end{equation}
where $t_c$ and $\alpha$ are analogous to those in (\ref{tcollision}) and (\ref{alphaprob}) respectively, and
\begin{equation}
W_i = 
\begin{cases}
2^i \cdot W_\text{min},	& \text{if } 0\leq i\leq m\\
2^m \cdot W_\text{min},	& m< i\leq R
\end{cases}
\end{equation}

To derive the CDF of delays, Raptis {\it et al.} \cite{Raptis2009} first calculated the probability that a frame is successfully transmitted at stage $j$ as:
\begin{equation}
\label{Qjjj}
Q_j = \frac{p^j(1-p)}{1-p^{R+1}}
\end{equation}
Since at any stage $j$, selecting any back-off value in the range $0\leq i< W_j$ is equiprobable, then the probability of transmitting a frame at stage $j$ after backing off for $i$ stages is (independent of $i$):
\begin{equation}
\label{Pjjj}
P_j = Q_j\cdot \frac{1}{W_j}
\end{equation}

\begin{figure}
\centering
\begin{tikzpicture}
\begin{axis}[
axis lines*=left, 
width=12cm,
height=2.2in,
xlabel=Delays (ms),
ylabel=CDF,
xmin=0, xmax=100,
ymin=0, ymax=1,
x tick label style={font=\scriptsize},
grid=major,
y tick label style={font=\scriptsize},
unbounded coords=jump,
xlabel style={font=\scriptsize},
ylabel style={font=\scriptsize},
scaled ticks=false, tick label style={/pgf/number format/fixed},
enlargelimits=false, clip=true, axis on top,
legend columns=1,
legend style={font=\footnotesize,draw=none,at={(1.15,1.05)},anchor=north,},
]
\addplot[solid] table[col sep=comma]{csv/Wireless/CDF_2009/cdf.k2.csv};
\addlegendentry{$k=2$}
\addplot[red,dashed] table[col sep=comma]{csv/Wireless/CDF_2009/cdf.k5.csv};
\addlegendentry{$k=5$}
\addplot[blue,thick,dotted] table[col sep=comma]{csv/Wireless/CDF_2009/cdf.k10.csv};
\addlegendentry{$k=10$}
\addplot[green,dashdotted] table[col sep=comma]{csv/Wireless/CDF_2009/cdf.k20.csv};
\addlegendentry{$k=20$}
\addplot[brown,thick,densely dashdotted] table[col sep=comma]{csv/Wireless/CDF_2009/cdf.k30.csv};
\addlegendentry{$k=30$}
\end{axis}
\end{tikzpicture}
\caption[CDF of single-hop wireless delays that a frame endures when there are $k$ saturated wireless devices in the network]{CDF of single-hop wireless delays that a frame endures when there are $k$ saturated wireless devices in the network \cite{Raptis2009}.}
\label{figCDF2009}
\end{figure}
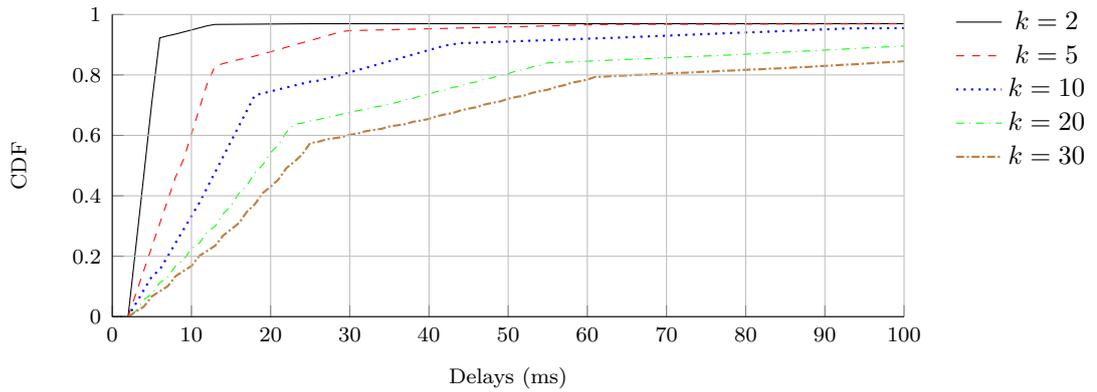

Using (\ref{Pjjj}), Raptis {\it et al.} \cite{Raptis2009} derive the CDF of delays as follows. Let $\Omega$ be a finite set of delays, such that $\Omega_{j,i}$ is the delay a frame suffers before it gets successfully transmitted at stage $j$, given that $i$ back-off slots were selected at stage $j$. For any randomly-chosen delay value $D$, the probability that $D\leq d$ for all $0\leq d\leq \infty$ is given by \cite{Raptis2009}:
\begin{equation}
\label{RaptisCDF}
P\{D\leq d\} = \sum_{j=0}^{R}\sum_{i=0}^{W_j-1} P_{j,i}(d)
\end{equation}
where
\begin{equation}
\label{pjiofd}
P_{j,i}(d) = 
\begin{cases}
P_j						,	& \text{if } \Omega_{j,i}\leq d\\
0	 					,	& \text{otherwise}
\end{cases}
\end{equation}
Using (\ref{RaptisCDF}), Fig.~\ref{figCDF2009} plots the wireless delay CDFs of Raptis {\it et al.} \cite{Raptis2009} at various values of $k$. Once again, the model shows that delays generally increase with $k$, which is unsurprising. However the distributions derived by Raptis {\it et al.} \cite{Raptis2009} (Fig.~\ref{figCDF2009}) are not exactly similar to those derived by Carvalho and Garcia-Luna-Aceves \cite{1249764} (Fig.~\ref{figCDF2003}). Differences between both models are discussed in Section \ref{Differencesmodels} below.

\subsubsection{Jitter}
Similar to Carvalho and Garcia-Luna-Aceves \cite{1249764}, Raptis {\it et al.} \cite{Raptis2009} also derived an expression for the delay jitter in a single-hop wireless network with $k$ devices. To do that, the authors \cite{Raptis2009} first derived the expected total delay that a frame suffers before being successfully transmitted at stage $j$ as:
\begin{equation}
\label{edunderscorej}
\omega_j = U_j - t_c + t_s
\end{equation}
Then, using (\ref{edunderscorej}) and (\ref{Qjjj}), the expected delay, $E[T]$, a frame suffers before being successfully transmitted is \cite{Raptis2009}:
\begin{equation}
\label{EYYy}
E[T] = \sum_{j=0}^{R} (\omega_j \cdot Q_j)
\end{equation}
And the expected value for the square of a delay, $T^2$, is \cite{Raptis2009}:
\begin{equation}
\label{YYsquared}
E[T^2] = \sum_{j=0}^{R} \left(P_j \cdot \sum_{i=0}^{W_j-1} (E[\Omega_{j,i}])^2\right)
\end{equation}
where $E[\Omega_{j,i}]$ is the average of $\{\Omega_{0,0},..\ \Omega_{j,i}\}$, and is calculated as \cite{Raptis2009}:
\begin{equation}
E[\Omega_{j,i}] = t_s + i\cdot \alpha + U_{j-1}
\end{equation}

Finally, in contrast to the delay jitter of Carvalho and Garcia-Luna-Aceves \cite{1249764} in (\ref{std_2003}), the jitter of Raptis {\it et al.} \cite{Raptis2009} is calculated using (\ref{YYsquared}) and (\ref{EYYy}) as:
\begin{equation}
\label{std_2009}
\text{Std}[T] = \sqrt{\displaystyle E[T^2] - (E[T])^2}
\end{equation}

In Sections \ref{evaCPVwireless} and \ref{wireless:num:iter}, we use the CDFs in (\ref{GausCDFtrunc}) and (\ref{RaptisCDF}) to evaluate CPV.

\subsection{Differences between the models}
\label{Differencesmodels}

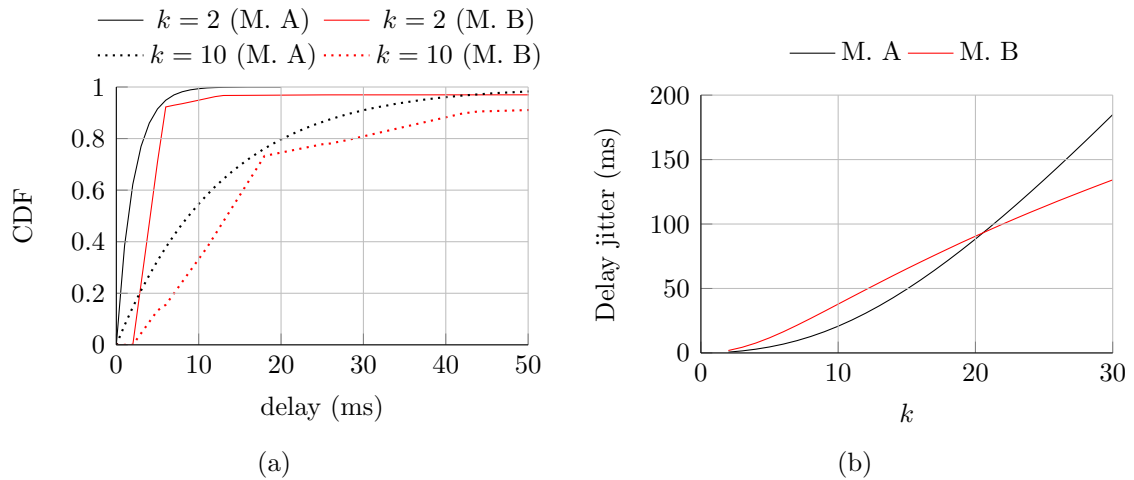
\begin{figure}
\subfigure[]
{
\label{figCDFcomparewirelesslita}
\begin{tikzpicture}
\begin{axis}
[
no markers,
xmin=0,xmax=50,
axis lines*=left, 
xlabel=delay (ms),
ylabel=CDF,
height=5cm,
width=7cm,
ymin=0,ymax=1,
label style={font=\footnotesize},
tick label style={font=\footnotesize,},
enlargelimits=false, clip=true, axis on top,
grid = major,
scaled ticks=false, tick label style={/pgf/number format/fixed},
legend columns=2,
legend style={font=\footnotesize,draw=none,at={(0.5,1.35)},anchor=north,},
]
\addplot[solid] table[col sep=comma]{csv/Wireless/CDF_2003_trunc/cdf.k2.csv};
\addlegendentry{$k=2$ (M. A)}
\addplot[solid,red] table[col sep=comma]{csv/Wireless/CDF_2009/cdf.k2.csv};
\addlegendentry{$k=2$ (M. B)}
\addplot[thick,dotted] table[col sep=comma]{csv/Wireless/CDF_2003_trunc/cdf.k10.csv};
\addlegendentry{$k=10$ (M. A)}
\addplot[red,thick,dotted] table[col sep=comma]{csv/Wireless/CDF_2009/cdf.k10.csv};
\addlegendentry{$k=10$ (M. B)}
\end{axis}
\end{tikzpicture}
}
\subfigure[]
{
\label{figCDFcomparewirelesslitb}
\begin{tikzpicture}
\begin{axis}
[
no markers,
xmin=0,xmax=30,
axis lines*=left, 
xlabel=$k$,
ylabel=Delay jitter (ms),
height=5cm,
width=7cm,
ymin=0,ymax=200,
label style={font=\footnotesize},
tick label style={font=\footnotesize,},
enlargelimits=false, clip=true, axis on top,
grid = major,
scaled ticks=false, tick label style={/pgf/number format/fixed},
legend columns=2,
legend style={font=\footnotesize,draw=none,at={(0.5,1.25)},anchor=north,},
]
\addplot[solid] table[col sep=comma]{csv/Wireless/Jitter/2003.csv};
\addlegendentry{M. A}
\addplot[red] table[col sep=comma]{csv/Wireless/Jitter/2009.csv};
\addlegendentry{M. B}
\end{axis}
\end{tikzpicture}
}
\caption[Comparison between the model of Carvalho {\it et al.} and that of Raptis {\it et al.}]{Comparison of the reviewed models. M. A means using the model of Carvalho {\it et al.} \cite{1249764}; M. B means using the model of Raptis {\it et al.} \cite{Raptis2009}. (a) Truncated Gaussian delay distribution with parameters derived from the model of Carvalho {\it et al.} \cite{1249764}, and the distribution derived by Raptis {\it et al.} \cite{Raptis2009} at $k=2$ and $k=10$. (b) The jitter follows that derived by the authors \cite{1249764,Raptis2009}.}
\label{figCDFcomparewirelesslit}
\end{figure}

Figure~\ref{figCDFcomparewirelesslita} plots the truncated Gaussian distribution with the parameters obtained from the model of Carvalho and Garcia-Luna-Aceves \cite{1249764} modeling single-hop wireless delays, and the distribution derived by Raptis {\it et al.} \cite{Raptis2009} at $k=2$ and $k=10$. The distributions are not drastically different. Their dissimilarities might however stem from differences in their assumptions, e.g., Raptis {\it et al.} assumes the frame is discarded after failing transmissions in $R$ stages, while Carvalho {\it et al.} does not make this assumption.

Figure~\ref{figCDFcomparewirelesslitb} shows the difference in the jitter between both models, obtained using (\ref{std_2003}) and (\ref{std_2009}) respectively. At first glance, the individual values of the two curves over the region up to $k = 20$ are reasonably similar, but the model of Raptis {\it et al.} appears almost linear, while that of Carvalho {\it et al.} gives values lower in the region up to $k = 20$, but rising much faster starting for values shortly beyond $k = 20$.

In the rest of this chapter, both models are used to analyze CPV in wireless networks, with a truncated Gaussian distribution assumed for the parameters of Carvalho and Garcia-Luna-Aceves.

\subsection{Summary of reviewed literature on wireless models}
All the models reviewed herein, in Section \ref{modlit}, consider a wireless network with a single access point and no hidden terminals, typically addressing a small (e.g., home) network. In public places (e.g., coffee shops or hotel rooms), this may not be the case. However, the models already incorporate the additional delays due to the RTS/CTS mechanism of the 802.11 DCF and thus, we believe the existence of hidden terminals is unlikely to result in significant difference in delays.

Another assumption made in the reviewed literature is that the physical media is error-free; in other words, failed transmissions are only caused due to collision. The reviewed literature have compared their analytical models using simulations, which highlighted almost negligible effect of these assumptions in practice \cite{bianchi2000performance,1249764,Raptis2009}.

The reviewed literature assumes all $k$ devices are saturated (i.e., always have packets to send). However, $k$ devices are typically expected to alternate between phases of transmission, reception and idle activity. We believe this assumption tends to cause the delays resulting from the derived models to be larger than those in practice.

\section{Evaluating CPV in 802.11 Networks}
\label{evaCPVwireless}

We evaluate CPV with wireless clients using the delay models discussed in Section \ref{modlit}. All results reported in this chapter follow CPV's recommendation of $\lambda = 0.1$ (see Chapter \ref{ch:wiredecva}). The area tolerance $\epsilon_\bigtriangleup$ and the acceptance threshold $\tau_\bigtriangleup$ are calibrated per triangle. Similar to Chapter \ref{ch:wiredecva}, the objective is to quantify the FRs and FAs at some values of $\epsilon_\bigtriangleup$ and $\tau_\bigtriangleup$ that allow CPV to adequately distinguish legitimates from adversaries.

To analyze CPV with wireless clients, we varied the number of legitimate clients modeled to use wireless access networks.\footnote{Characteristics of such a network are explained in Section \ref{wire_eval_assu} below.} The number of wireless legitimate clients in each $\bigtriangleup$ affects the calibration of CPV's input parameters ($\epsilon_\bigtriangleup$ and $\tau_\bigtriangleup$), and is thus expected to affect the overall results. Each wireless network was modeled to have $k$ actively-transmitting wireless devices, with one of those $k$ being CPV's legitimate client.

\begin{figure}
\centering
\includegraphics[scale=0.34]{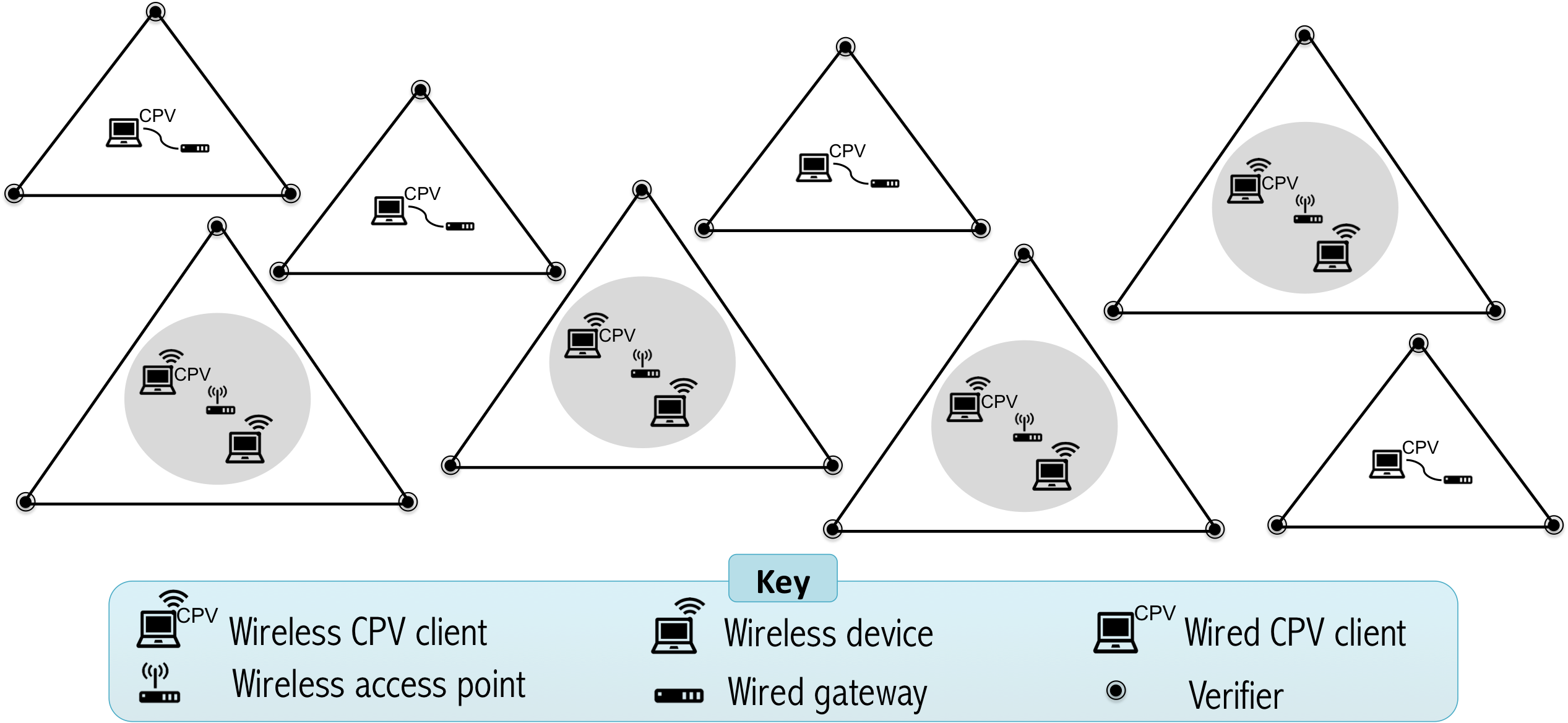}
\caption{An example of eight CPV clients, half of which are using a wireless access network that has $k=2$ devices.}
\label{proportionvsk}
\end{figure}

Recall from Chapter \ref{ch:wiredecva} that at $\lambda = 0.1$, our PlanetLab experiments had 49 legitimate clients. Thus, we can model a maximum of 49 distinct wireless access networks, with $k\geq 2$ wireless devices in each. For example, if a proportion of $\sim$0.2 of all 49 legitimate clients was using a wireless access network with $k=4$, this means there are 10 distinct wireless access networks modeled at different geographic regions, and each network has 4 wireless devices (constant across all 10 networks). Fig.~\ref{proportionvsk} shows an example of eight legitimate CPV clients; a proportion equal to 0.5 of them is using a wireless access network that has $k=2$ devices.

\subsection{Evalution assumptions (wireless access)}
\label{wire_eval_assu}
Each wireless legitimate client is assumed to be competing for the wireless media with $k-1$ other wireless devices. All $k$ devices (i.e., including the legitimate client whose assertion is being verified by CPV) use the same wireless access point, which is one hop away. We assume no hidden terminals (recall Section \ref{sec:wireless:background})---the transmission of any device is sensed by all others.

All $k$ devices are using an 802.11b access network over Direct-Sequence Spread Spectrum (DSSS) on the physical layer with a 11Mbps data rate. Characteristics of DSSS are shown in Table~\ref{DSSSTable}. Following the reviewed models in Section \ref{modlit}, all $k$ devices are assumed \emph{saturated} (i.e., the packet queues of all $k$ device are never empty), and are transmitting at the same time according to a Constant Bit Rate (CBR) with a packet size equal to 8148 bits.

\begin{table}
\centering
\caption{DSSS characteristics}
\begin{tabular}{ll}
{\bf Item}&{\bf Value}\\\hline
$W_{min}$	&32 time slots\\
$W_{max}$	&1024 time slots\\
Retransmission limit (R)					&6 stages\\\hline

Physical header (PHY)& 192bits at 1 Mbit/s\\
MAC header			&224 bits at 11 Mbit/s\\
ACK length			&112 bits at 11 Mbit/s + PHY\\
RTS length			&160 bits at 1 Mbit/s + PHY\\
CTS length			&112 bits at 1 Mbit/s + PHY\\\hline

Propagation delay ($\delta$)	&1 $\mu sec$\\
Slot time ($\sigma$)	&20 $\mu sec$\\
SIFS		&10 $\mu sec$\\
DIFS		&50 $\mu sec$\\\Xcline{1-2}{2\arrayrulewidth}
\end{tabular}
\label{DSSSTable}
\end{table}

Finally, because an element of randomness (i.e., the delay component resembling a wireless network) is now introduced to the results, experimentation scenarios were run 10 times and the average result is reported.

\subsection{Effect of number of wireless devices ($k$) on CPV}

\label{changeconfidence} 
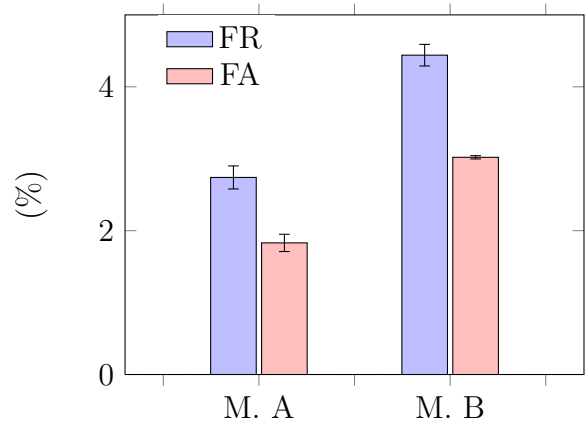
\begin{figure}
\centering
\begin{tikzpicture}
\begin{axis}[
width=3in,
height=2.5in,
xtick={1,2,3,4,5},
xticklabels={,M. A,,M. B,},
ybar,
ymin=0,ymax=5,
bar width = 0.6cm,
area legend,
legend style={at={(0.2,1.0)},anchor=north,draw=none},
ylabel=(\%)
]
\addplot[fill=blue!25,draw=black,point meta=y,every node near coord/.style={inner ysep=5pt},error bars/.cd,y dir=both,y explicit] 
table [y error=error] {
x   y           error    label
1   0    0 1
2   2.74   0.16 2
3   0    0 3
4   4.44    0.15 4
5   0    0 5
};
\addlegendentry{FR}
\addplot[fill=red!25,draw=black,point meta=y,every node near coord/.style={inner ysep=5pt},error bars/.cd,y dir=both,y explicit,] 
table [y error=error] {
x   y           error    label
1   0    0 1
2   1.83   0.12 2
3   0    0 3
4  3.02    0.023 4
5   0    0 5
};
\addlegendentry{FA}
\end{axis}
\end{tikzpicture}
\caption[Statistical confidence of CPV results in wireless networks]{Statistical confidence of CPV results in wireless networks. M. A means using the model of Carvalho {\it et al.} \cite{1249764}; M. B means using the model of Raptis {\it et al.} \cite{Raptis2009}.}
\label{res0wireless}
\end{figure}

\textcolor{\changes}{Figure~\ref{res0wireless} shows the mean FRs and FAs of 100 runs resulting from using the models of Carvalho {\it et al.} \cite{1249764} and Raptis {\it et al.} \cite{Raptis2009}. All 49 legitimate clients were using a wireless access network, and there was a total of $k=5$ devices in the network of each wireless CPV client. The number of CPV iterations (see Chapter \ref{ch:cpv}) was fixed at $n_\bigtriangleup = 600$ for all $\bigtriangleup$. FRs and FAs for both models lied between $\sim$1.8\% and $\sim$4.5\%.}

\textcolor{\changes}{Because FRs and FAs are estimated empirically from 100 runs, we calculate the error margin of these estimates for a 90\% confidence level. To calculate the error margin, we first calculate the critical value as follows \cite{miller1965probability}:}
\begin{equation*}
\alpha = 1 - \frac{\text{confidence level}}{100} = 1 - 0.9 = 0.1
\end{equation*}
\begin{equation*}
\text{Critical Probability } (p*) = 1 - \frac{\alpha}{2} = 1 - \frac{0.1}{2} = 0.95
\end{equation*}
\begin{equation*}
\text{Degree of Freedom } (df) = n - 1 = 100 - 1 = 99
\end{equation*}
\textcolor{\changes}{From the statistics tables \cite{miller1965probability}, at $df = 99$ and $p* = 0.95$, the critical value is 1.66.}

\textcolor{\changes}{Next, we calculate the standard error (SE). For the FRs obtained using the model of Carvalho {\it et al.} \cite{1249764}:}
\begin{equation}
\text{SE (FRs)} = \frac{\text{Std}}{\sqrt{n}} = \frac{0.97}{\sqrt{100}} = 0.097.
\end{equation}
\textcolor{\changes}{Table \ref{MargineofError} shows the SE for the rest of the results. Finally, the Margin of Error (ME) at 90\% confidence level is calculated as:}
\begin{equation}
\text{ME (FRs)} = \text{critical value} \times \text{SE} = 1.66 \times 0.097 = 0.16.
\end{equation}
\textcolor{\changes}{The ME at 90\% confidence level for the rest of the results is reported in Table \ref{MargineofError}.}
\begin{table}
\centering
\caption{SE and Margin of Error (ME) at 90\% confidence level for the rest of the results}
\begin{tabular}{ccccc}
Model	& Parameter &Std	 & SE & ME at 90\% CI \\\hline
\multirow{2}{*}{Carvalho {\it et al.} \cite{1249764}}
& FRs &  0.97 &  0.097 & $\pm$0.16\\
& FAs &  0.74 &  0.074 & $\pm$0.12\\
\multirow{2}{*}{Raptis {\it et al.} \cite{Raptis2009}}
& FRs &  0.92 &  0.092 & $\pm$0.15\\
& FAs &  0.14 &  0.014 & $\pm$0.02\\
\Xcline{1-5}{2\arrayrulewidth}
\end{tabular}\\[7pt]
Std = Standard deviation; SE = Standard error; ME = Margin of Error.
\label{MargineofError}
\end{table}
\textcolor{\changes}{The MEs at 90\% confidence level are depicted using vertical lines atop the bars in Figure~\ref{res0wireless} for the mean FRs and FAs. None of the MEs exceeds $\pm$0.16\%, highlighting that the means estimated from the sample runs are relatively precise.}

\textcolor{\changes}{Note that ideally, the statistical confidence of any results reported thereafter could be measured, although it is not planned in any of the experiments conducted in the remains of this thesis.}

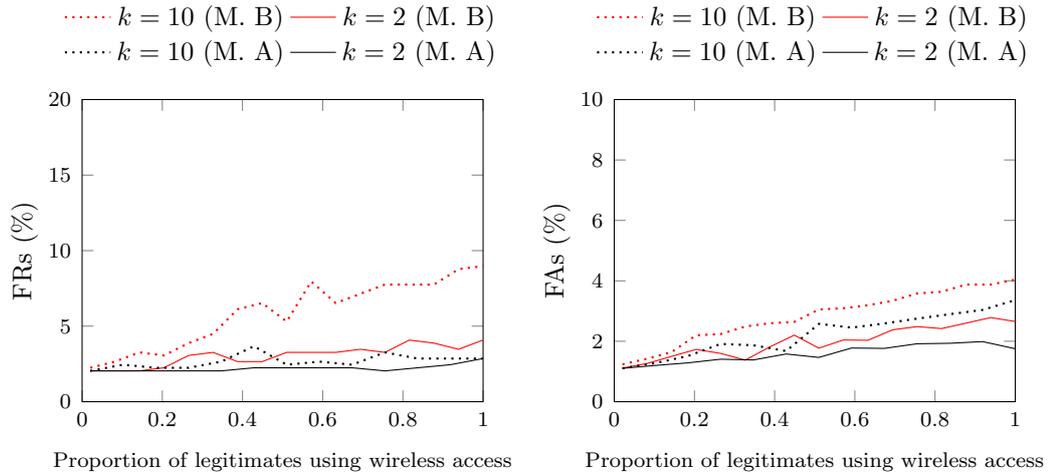
\begin{figure}
\centering
\begin{tikzpicture}
\begin{axis}[
width=2.7in,
height=2.2in,
xlabel=Proportion of legitimates using wireless access,
ylabel=FRs (\%),
xmin=0, xmax=1,
ymin=0, ymax=20,
x tick label style={font=\scriptsize},
y tick label style={font=\scriptsize},
unbounded coords=jump,
xlabel style={font=\scriptsize},
ylabel style={at={(0.09,0.5)},font=\footnotesize},
scaled ticks=false, tick label style={/pgf/number format/fixed},
legend columns=2,
legend style={font=\footnotesize,draw=none,at={(0.5,1.35)},anchor=north,},
]
\addplot[red,thick,dotted] table[col sep=comma]{csv/Wireless/i600_2009/fr_10.csv};
\addlegendentry{$k=10$ (M. B)}
\addplot[red] table[col sep=comma]{csv/Wireless/i600_2009/fr_2.csv};
\addlegendentry{$k=2$ (M. B)}
\addplot[thick,dotted] table[col sep=comma]{csv/Wireless/i600_2003_trunc/fr_10.csv};
\addlegendentry{$k=10$ (M. A)}
\addplot[] table[col sep=comma]{csv/Wireless/i600_2003_trunc/fr_2.csv};
\addlegendentry{$k=2$ (M. A)}
\end{axis}
\end{tikzpicture}
\begin{tikzpicture}
\begin{axis}[
width=2.7in,
height=2.2in,
xlabel=Proportion of legitimates using wireless access,
ylabel=FAs (\%),
xmin=0, xmax=1,
ymin=0, ymax=10,
x tick label style={font=\scriptsize},
y tick label style={font=\scriptsize},
unbounded coords=jump,
xlabel style={font=\scriptsize},
ylabel style={at={(0.09,0.5)},font=\footnotesize},
scaled ticks=false, tick label style={/pgf/number format/fixed},
legend columns=2,
legend style={font=\footnotesize,draw=none,at={(0.5,1.35)},anchor=north,},
]
\addplot[red,thick,dotted] table[col sep=comma]{csv/Wireless/i600_2009/fa_10.csv};
\addlegendentry{$k=10$ (M. B)}
\addplot[red] table[col sep=comma]{csv/Wireless/i600_2009/fa_2.csv};
\addlegendentry{$k=2$ (M. B)}
\addplot[thick,dotted] table[col sep=comma]{csv/Wireless/i600_2003_trunc/fa_10.csv};
\addlegendentry{$k=10$ (M. A)}
\addplot[] table[col sep=comma]{csv/Wireless/i600_2003_trunc/fa_2.csv};
\addlegendentry{$k=2$ (M. A)}
\end{axis}
\end{tikzpicture}
\caption[CPV results in wireless access networks with a fixed number of CPV iterations]{FRs and FAs when a proportion of the 49 legitimate clients (i.e., PlanetLab nodes inside triangles) use a wireless access network that has $k$ wireless devices. $n_\bigtriangleup = 600$ CPV iterations for all $\bigtriangleup$. M. A means using the model of Carvalho {\it et al.} \cite{1249764}; M. B means using the model of Raptis {\it et al.} \cite{Raptis2009}.}
\label{res1wireless}
\end{figure}

Figure~\ref{res1wireless} shows the FRs and FAs when $k=2$ and $k=10$. Again, the number of CPV iterations was fixed at $n_\bigtriangleup = 600$ for all $\bigtriangleup$. Using the model of Carvalho and Garcia-Luna-Aceves \cite{1249764}, there was degradation in CPV's efficacy with an increased $k$, but such degradation was not severe. For example, when all 49 legitimate clients were using a wireless access network (i.e., at $x=1$ in Fig.~\ref{res1wireless}), the sum FR+FA went from $\sim$4.61\% at $k=2$ to $\sim$6.22\% at $k=10$. We believe these results stem from the non-zero probability that the wireless delay is (relatively) negligible, e.g., 3 ms. At $k=10$, the truncated Gaussian distribution in Fig.~\ref{figCDF2003} indicates that there is a $\sim$20\% chance the transmitted frame (holding the verifiers' timestamps) suffers $< $3 ms delay, i.e., if one iteration was performed. As more iterations are performed, the chances that one or more iterations result in such negligible delay increase. Because CPV requires only a proportion $\tau$ of the performed iterations to pass the triangular area checks (which is more likely to happen with smaller delays between the verifiers and the client, as discussed in Chapters \ref{ch:cpv} and \ref{ch:wiredecva}), it still accepts a client when a proportion of $1-\tau$ of all iterations result in large delays and area mismatch. The required number of iterations is derived in terms of $k$ and the acceptance threshold $\tau$ in Section \ref{wireless:num:iter} below.

Using the model of Raptis {\it et al.} \cite{Raptis2009}, and assuming that half the legitimate clients are wireless, the sum FR+FA went from 5.1\% at $k=2$ to 8.3\% at $k=10$. Those numbers are to be compared to 3.1\% (2.0\% + 1.1\%) when none of the legitimate clients are using a wireless access network. In conclusion, under this model, when a wireless CPV legitimate client competes for the media with another device (i.e., $k=2$), it has double the chances of being falsely rejected compared to a wired legitimate client.

\begin{figure}
\centering
\subfigure[Using Carvalho and Garcia-Luna-Aceves's model \cite{1249764}]
{
\begin{tikzpicture}
\begin{axis}[
mesh/ordering=y varies,
xticklabel style={/pgf/number format/fixed},
width=2.5in,
height=2.5in,
xlabel=$n$,
xtick={60,200,400,600},
xmin=60,xmax=600,
ylabel=$k$,
y label style={rotate=45},
ymin=2,ymax=30,
zlabel=FR+FA (\%),
z label style={rotate=90},
minor z tick num=1,
zmin=0,zmax=50,
grid=both,
x tick label style={font=\scriptsize},
x label style={at={(0.35,-0.08)},align=center,rotate=-9,font=\footnotesize},
y tick label style={font=\scriptsize},
y label style={at={(0.94,0.01)},font=\footnotesize},
z tick label style={font=\scriptsize},
z label style={at={(-0.15,0.45)},rotate=270,font=\footnotesize},
legend style={font=\footnotesize,at={(1,0.86)},draw=none,opacity=0.9},
extra tick style={major grid style=red,tick label style={color=red}}
]
\addplot3[opacity=1,surf,shader=faceted,forget plot]table {csv/Wireless/3D/sum_2003_starting60_trunc.csv};
\end{axis}
\end{tikzpicture}
}
\hspace{8pt}
\subfigure[Using Raptis {\it et al.}'s model \cite{Raptis2009}]
{
\begin{tikzpicture}
\begin{axis}[
mesh/ordering=y varies,
xticklabel style={/pgf/number format/fixed},
width=2.5in,
height=2.5in,
xlabel=$n$,
xtick={60,200,400,600},
xmin=60,xmax=600,
ylabel=$k$,
y label style={rotate=45},
ymin=2,ymax=30,
zlabel=FR+FA (\%),
z label style={rotate=90},
minor z tick num=1,
zmin=0,zmax=50,
grid=both,
x tick label style={font=\scriptsize},
x label style={at={(0.35,-0.08)},align=center,rotate=-9,font=\footnotesize},
y tick label style={font=\scriptsize},
y label style={at={(0.94,0.01)},font=\footnotesize},
z tick label style={font=\scriptsize},
z label style={at={(-0.15,0.45)},rotate=270,font=\footnotesize},
legend style={font=\footnotesize,at={(1,0.86)},draw=none,opacity=0.9},
extra tick style={major grid style=red,tick label style={color=red}}
]
\addplot3[opacity=1,surf,shader=faceted,forget plot]table {csv/Wireless/3D/sum_2009_starting60.csv};
\end{axis}
\end{tikzpicture}
}
\caption[CPV results in wireless access networks (varying number of CPV iterations)]{FR+FA when half of the evaluated legitimate clients were using a wireless access network with $k$ devices.}
\label{3dFRFAwireless}
\end{figure}
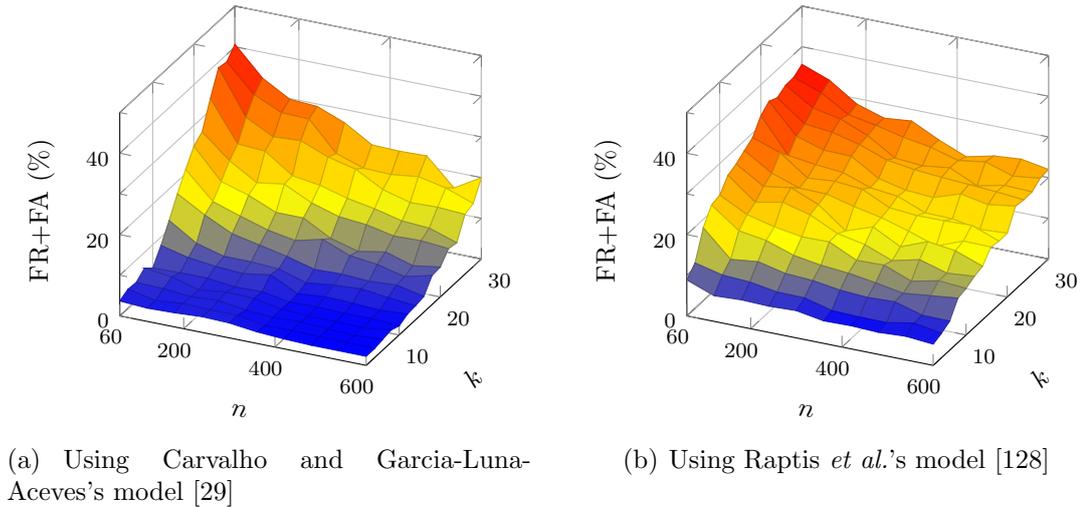

Figure~\ref{3dFRFAwireless} shows the summation of FRs and FAs with respect to the number of iterations $n$ (i.e., $n_\bigtriangleup$ for all $\bigtriangleup$), and the number of wireless devices, $k$, in each wireless network when 25 of the 49 legitimate CPV clients are using a wireless access network.\footnote{Recall that the number of wireless legitimate clients being verified by triangle $\bigtriangleup$ affects the calibration of $\epsilon_\bigtriangleup$ and $\tau_\bigtriangleup$, which is how those 25 wireless clients are expected to influence CPV's decisions on others.} Using the model of Carvalho and Garcia-Luna-Aceves \cite{1249764}, the effect of $k$ on the results begins to manifest starting around $k=1$. For example, at $k=2$ the sum FR+FA is almost constant regardless of the performed number of CPV iterations, $n$. In contrast, at $k=30$, the impact of $n$ on the sum FR+FA is large. In conclusion, increasing the number of CPV iterations has large impact only when more than $k=15$ devices are present in each wireless network.

The case is different using the wireless models of Raptis {\it et al.} \cite{Raptis2009}, where $k$ has a significant impact on the results, for all values of $k$. For example, at $k=6$, the sum FR+FA decreases from $\sim$18\% at $n=60$ to $\sim$7\% at $n=600$; and at $k=30$, FR+FA decreases from $\sim$36\% at $n=60$ to $\sim$22\% at $n=600$. These results highlight the potential for a larger number of iterations to mitigate the effect of the wireless delays on CPV.

Both models agree that CPV's efficacy decreases as $k$ increases, suggesting that CPV may perform poorly in public places where numerous devices are actively competing for the media.

\subsection{Minimum adversarial distance from the triangle}

\begin{figure}
\centering
\begin{tikzpicture}
\begin{axis}[
width=2.7in,
height=2.2in,
xlabel=$k$,
ylabel=Minimum distance (km),
xmin=0, xmax=34,
ymin=0, ymax=2200,
x tick label style={font=\scriptsize},
y tick label style={font=\scriptsize},
unbounded coords=jump,
ylabel style={font=\footnotesize},
scaled ticks=false, tick label style={/pgf/number format/fixed},
legend columns=2,
legend style={font=\footnotesize,draw=none,at={(0.5,1.15)},anchor=north,},
]
\addplot[mark=*,red,error bars/.cd, y dir=both, y explicit,] table[x index=0, y index=1, y error index=2,col sep=comma]{csv/Wireless/DistAdv/2009.csv};
\addlegendentry{M. B}
\addplot[mark=x,error bars/.cd, y dir=both, y explicit,] table[x index=0, y index=1, y error index=2,col sep=comma]{csv/Wireless/DistAdv/2003_trunc.csv};
\addlegendentry{M. A}
\end{axis}
\end{tikzpicture}
\caption[Adversarial external distance from the triangle required to maintain CPV results similar to a setting of all-wired legitimate clients.]{The minimum distance, between the (outside) adversary and the triangle, that enables CPV to maintain similar efficacy compared to when all clients are using a wired access network. Results are obtained when 25 of all 49 legitimate clients are using a wireless access network, and when $n_\bigtriangleup=600$ CPV iterations, for all $\bigtriangleup$. The error bars indicate the smallest and largest $y$ (minimum distance) obtained from 10 runs, and the marker is their average. M. A means using the model of Carvalho {\it et al.} \cite{1249764}; M. B means using the model of Raptis {\it et al.} \cite{Raptis2009}.}
\label{mindistancewireless}
\end{figure}
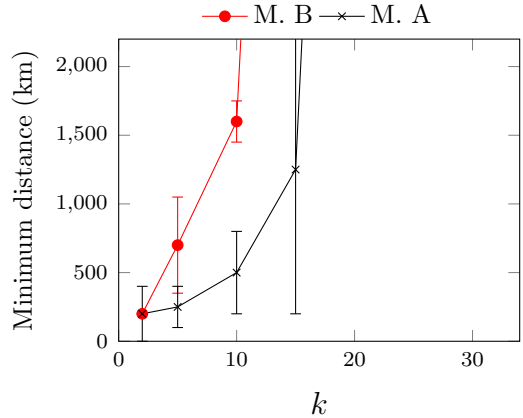

Figure~\ref{mindistancewireless} shows the minimum distance, between an (outside-triangle) adversary and the triangle encapsulating the adversary's asserted location, that enables CPV to maintain similar efficacy compared to when all clients are using a wired access network. Recall from Chapter \ref{ch:wiredecva}, FR+FA when all legitimates were wired-connected is $\sim$3\% at $\lambda=0.1$. Results are obtained when 25 of all 49 legitimate clients are using a wireless access network, and when $n_\bigtriangleup=600$ iterations for all $\bigtriangleup$.

Using the model of Carvalho and Garcia-Luna-Aceves \cite{1249764}, and at $k=5$, the sum FR+FA$\approx$3\% when (outside-triangle) adversaries were at least $\sim$250 km away from the triangles' sides. At $k=15$, the minimum distance adversary-free distance outside the triangle that maintains FR+FA$\approx$3\% becomes $1,250$ km. 

With the model of Raptis {\it et al.} \cite{Raptis2009}, the minimum adversarial distance is 700 km at $k=5$ (see Fig.~\ref{mindistancewireless}) and $\sim$1,600 at $k=10$. In conclusion, the minimum distance clearly increases with $k$ in both models, suggesting that as more saturated devices exist in the network of CPV's legitimate wireless clients, the likelihood of accepting (outside) adversaries close the triangles' sides increases.

\newcommand*{\permcomb}[4][0mu]{{{}^{#3}\mkern#1#2_{#4}}}
\newcommand*{\perm}[1][-3mu]{\permcomb[#1]{P}}
\newcommand*{\comb}[1][-1mu]{\permcomb[#1]{C}}

\section{Required Number of CPV Iterations}
\label{wireless:num:iter}

This section addresses the following question. Assume that the number of wireless devices in the client's access network, $k$, is known to the verifiers; how many CPV iterations (see Chapter \ref{ch:cpv}) should they perform such that with very high probability the legitimate client gets accepted? It is important to answer this question because, as the results of the previous section show, increasing the number of CPV iterations reduces the impact of the wireless delays on the efficacy of CPV. It is thus important to know what the appropriate number should be in order to mitigate such impact.

To answer this question, let $t$ be a small delay value (i.e., due to the wireless access network) that when added to the (Internet) end-to-end delays of a legitimate client that CPV would typically accept, will not cause CPV to falsely reject this client (i.e., due to the increased delay). Using the wireless delay models in Section \ref{modlit}, we can obtain the probability $p_k(t) = P_k\{D<t\}$ that a transmitted frame (carrying the verifiers' signed timestamps) experiences less than $t$ ms additional delay while sharing the wireless media with $k-1$ other actively participating devices. 

If two CPV iterations are performed, the probability that the frames experience $< t$ ms delay in one of them (either the first or the second) is:
\begin{equation}
\begin{split}
\varrho_1(t,k,2)	&= p_k(t)\cdot (1-p_k(t)) + (1-p_k(t))\cdot p_k(t)\\
					&= 2\cdot p_k(t)\cdot (1-p_k(t))
\end{split}
\end{equation}
Note that this equation is similar to the (basic) probability of getting a number $x$ once from a dice that is rolled twice, such that $x<3$ (i.e., the probability of getting either 1 or 2). This probability would be: either getting $x$ from the first roll but not the second, or from the second roll but not the first; the number of dice rolls is analogous to the number of CPV iterations.

For three iterations:
\begin{equation}
\begin{split}
\varrho_1(t,k,3)	&= 3\cdot p_k(t)\cdot (1-p_k(t))^2
\end{split}
\end{equation}
In general, the probability that a transmitted frame experiences $< t$ ms in exactly one of $n$ iterations is given by:
\begin{equation}
\varrho_1(t,k,n) = n \cdot p_k(t)\cdot (1-p_k(t))^{n-1}
\end{equation}

Considering more than one iteration, the probability $\varrho_2$ that the transmitted frames (holding the timestamps) experience $< t$ ms in exactly two of $n$ iterations is given by:
\begin{equation}
\varrho_2(t,k,n) = \left(\frac{n(n-1)}{2}\right) \cdot p_k(t)^2\cdot (1-p_k(t))^{n-2}
\end{equation}
That is because there are $n(n-1)/2$ ways of choosing two of $n$ iterations. In general, there are $\comb{n}{r}$ ways of choosing $r$ of $n$ iterations, where:
\begin{equation}
\comb{n}{r} = \frac{n!}{r!(n-r)!}
\end{equation}
Accordingly, the probability that the transmitted frames experience $< t$ ms in exactly $r$ of $n$ iterations is given by:
\begin{equation}
\varrho_r(t,k,n) = \comb{n}{r} \cdot p_k(t)^r\cdot (1-p_k(t))^{n-r}
\end{equation}
And thus, the probability that the wireless delay is $< t$ ms in at least $r$ of $n$ iterations is given by:
\begin{equation}
\label{numiterationsrequired}
\rho_r(t,k,n) = \sum_{i=r}^{n} \varrho_i(t,k,n)
\end{equation}

Calculating this probability is fundamental to the operation of CPV. For example, let the number of iterations that CPV performs be $n=600$, and let CPV be calibrated such that it requires at least 30 of those 600 iterations to pass the triangular area check (explained in Chapter \ref{ch:cpv}). Assuming that $t=3$, then using (\ref{numiterationsrequired}) we can calculate the probability, $\rho_{30}(3,k,600)$, that the timestamps exchanged between the verifiers and the client are delayed (additionally by the wireless access network) $< 3$ ms in at least 30 of the 600 iterations. This probability will thus serve as an upper bound probability of that client being correctly accepted. It is ``upper bound" because if $\rho_{30}(3,k,600)=1$, the client may still get falsely rejected due to other non-wireless related factors (see Chapter \ref{ch:wiredecva}). Equation (\ref{numiterationsrequired}) is used below to derive a function calculating the number of CPV iterations required to mitigate the negative effect of wireless delays.

Note that $p_k(t)$ is calculated using the CDFs in (\ref{GausCDFtrunc}) and (\ref{RaptisCDF}). For example, for the model of Carvalho {\it et al.}, we have:
\begin{equation}
p_k(t) = \text{GausCDF}_{\mu,\sigma}(t;0,\infty)
\end{equation}
where $\mu$ and $\sigma$ are functions of $k$ as discussed in Section \ref{modlit}. Example values for $p_k(3)$ are listed in Table~\ref{probsless3} for various values of $k$.

\begin{figure}
\centering
\begin{tikzpicture}
\begin{axis}[
axis lines*=left, 
width=2.7in,
height=2.2in,
xlabel=$n$,
ylabel={$\rho_5 (3,k,n)$},
xmin=4, xmax=512,
ymin=0, ymax=1,
x tick label style={font=\scriptsize},
grid=major,
y tick label style={font=\scriptsize},
unbounded coords=jump,
scaled ticks=false, tick label style={/pgf/number format/fixed},
enlargelimits=false, clip=true, axis on top,
legend columns=2,
legend style={font=\footnotesize,draw=none,at={(0.5,1.35)},anchor=north,},
xmode = log,
log basis x={2},
log base 2 number format code/.code={$\pgfmathparse{2^(#1)}\pgfmathprintnumber{\pgfmathresult}$},
]
\addplot[solid] table[col sep=comma]{csv/Wireless/iterations/r5_2003_trunc/itr.3ms.k2};
\addlegendentry{$k=2$ (M. A)}
\addplot[dotted,thick] table[col sep=comma]{csv/Wireless/iterations/r5_2003_trunc/itr.3ms.k10};
\addlegendentry{$k=10$ (M. A)}
\addplot[red,solid] table[col sep=comma]{csv/Wireless/iterations/r5_2009/itr.3ms.k2};
\addlegendentry{$k=2$ (M. B)}
\addplot[red,dotted,thick] table[col sep=comma]{csv/Wireless/iterations/r5_2009/itr.3ms.k10};
\addlegendentry{$k=10$ (M. B)}
\end{axis}
\end{tikzpicture}
\begin{tikzpicture}
\begin{axis}[
axis lines*=left, 
width=2.7in,
height=2.2in,
xlabel=$n$,
ylabel={$\rho_{20} (3,k,n)$},
xmin=16, xmax=2048,
ymin=0, ymax=1,
x tick label style={font=\scriptsize},
grid=major,
y tick label style={font=\scriptsize},
unbounded coords=jump,
scaled ticks=false, tick label style={/pgf/number format/fixed},
enlargelimits=false, clip=true, axis on top,
legend columns=2,
legend style={font=\footnotesize,draw=none,at={(0.5,1.35)},anchor=north,},
xmode = log,
log basis x={2},
log base 2 number format code/.code={$\pgfmathparse{2^(#1)}\pgfmathprintnumber{\pgfmathresult}$},
]
\addplot[solid] table[col sep=comma]{csv/Wireless/iterations/r20_2003_trunc/itr.3ms.k2};
\addlegendentry{$k=2$ (M. A)}
\addplot[dotted,thick] table[col sep=comma]{csv/Wireless/iterations/r20_2003_trunc/itr.3ms.k10};
\addlegendentry{$k=10$ (M. A)}
\addplot[red,solid] table[col sep=comma]{csv/Wireless/iterations/r20_2009/itr.3ms.k2};
\addlegendentry{$k=2$ (M. B)}
\addplot[red,dotted,thick] table[col sep=comma]{csv/Wireless/iterations/r20_2009/itr.3ms.k10};
\addlegendentry{$k=10$ (M. B)}
\end{axis}
\end{tikzpicture}
\caption[The probability that the wireless delays are $< t=3$ ms in at least 5 and 20 of $n$ iterations]{The probability that a transmitted frame experiences $< t=3$ ms of wireless delay in at least 5 and 20 of $n$ iterations, when $k$ wireless devices are sharing the access network. See Table~\ref{probsless3} (or similarly Figures~\ref{figCDF2003} and \ref{figCDF2009} at $x=3$ ms) for the values of $p_k(t)$. M. A means using the model of Carvalho {\it et al.} \cite{1249764}; M. B means using the model of Raptis {\it et al.} \cite{Raptis2009}.}
\label{cdfexamplerho}
\end{figure}
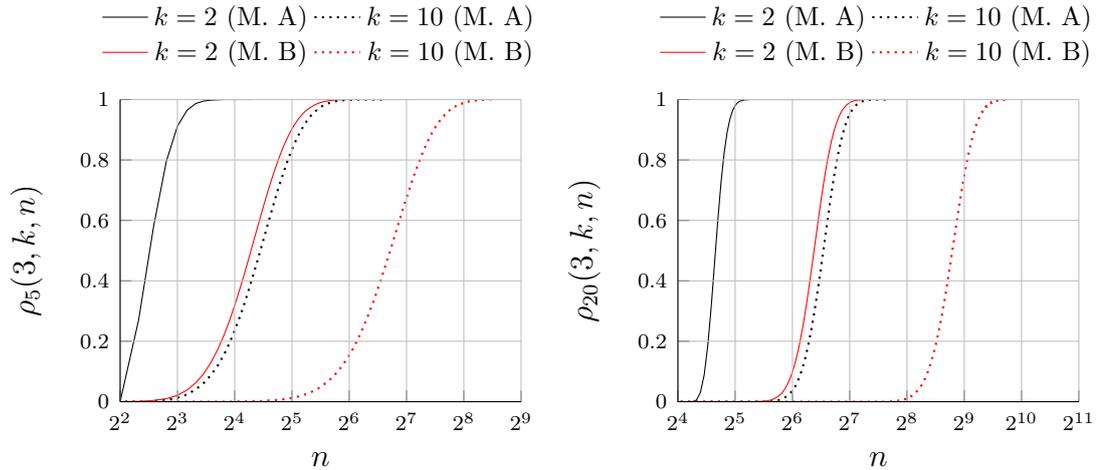

Figure~\ref{cdfexamplerho} shows a plot of $\rho_5(3,k,n)$ and $\rho_{20}(3,k,n)$ against $n$ at $k=2$ and $k=10$. The charts show that at $k=2$, the verifiers need to perform 11 (or 45) iterations using the model of Carvalho and Garcia-Luna-Aceves \cite{1249764} (or that of Raptis {\it et al.} \cite{Raptis2009}) to be almost certain (i.e., with probability $\rho_5(3,2,n) \geq$0.99) that the transmitted frames will endure $<3$ ms delay in at least 5 iterations. To achieve $<3$ ms wireless delay in 20 or more iterations, and at $k=10$, the verifiers will need to perform $\sim$150 and $\sim$700 iterations respectively using the models of Carvalho {\it et al.} and Raptis {\it et al.} to satisfy $\rho_{20}(3,10,n) \geq$0.99.

\begin{table}
\centering
\caption{The probability $p_k(3)$ that an additional delay of $<$ 3 ms is incurred by the wireless network at different values of $k$.}
\begin{tabular}{c|c|cccccc}
&\multirow{2}{*}{Model}		&\multicolumn{6}{c}{$k$}\\\cline{3-8}
&								&	2	&	5	&	10	&	20	&	25	&	30\\\hline
\multirow{2}{*}{$p_k(3)$}
&\cite{1249764}		& 0.77		&	0.45	&	0.21	&	0.07	&	0.04	&	0.03\\
&\cite{Raptis2009}	& 0.24		&	0.08	&	0.04	&	0.02	&	0.02	&	0.02\\\hline
\end{tabular}
\label{probsless3}
\end{table}

CPV requires a proportion $0\leq \tau_\bigtriangleup\leq 1$, for each $\bigtriangleup$, to pass the triangular area-match checking in order to accept a client.\footnote{Recall that the CPV algorithm handles triangular inequality violations (TIVs) and area-mismatches similarly, both are treated as area-mismatch.} By policy, if $n\cdot\tau_\bigtriangleup$ of the $n$ iterations pass the area checks, the client gets accepted. To mitigate the effect (on CPV's decisions) of wireless delays with probability $\geq$0.99, the verifiers need to perform $n$ iterations that satisfy:
\begin{equation}
\label{numiterationsrequiredfinal}
\rho_{n\tau_\bigtriangleup}(t,k,n)\geq 0.99
\end{equation}

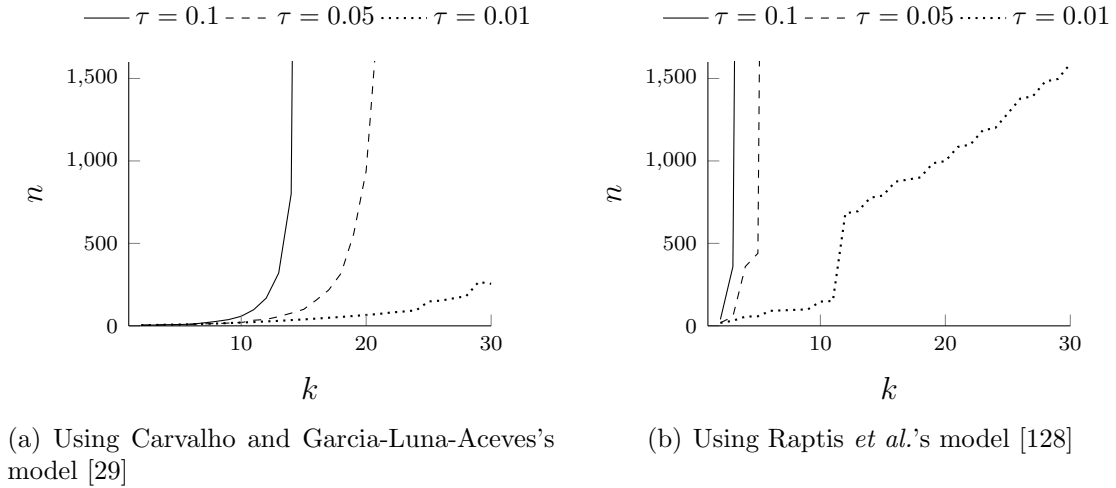
\begin{figure}
\centering
\subfigure[Using Carvalho and Garcia-Luna-Aceves's model \cite{1249764}]
{
\begin{tikzpicture}
\begin{axis}[
axis lines*=left, 
width=2.5in,
height=2in,
xlabel=$k$,
ylabel=$n$,
xmin=1, xmax=30,
ymin=0, ymax=1600,
x tick label style={font=\scriptsize},
y tick label style={font=\scriptsize},
unbounded coords=jump,
scaled ticks=false, tick label style={/pgf/number format/fixed},
enlargelimits=false, clip=true, axis on top,
legend columns=-1,
legend style={font=\footnotesize,draw=none,at={(0.5,1.25)},anchor=north,},
]
\addplot[solid] table[col sep=comma]{csv/Wireless/iterations/nAt_2003_trunc/tau0.1};
\addlegendentry{$\tau=0.1$}
\addplot[dashed] table[col sep=comma]{csv/Wireless/iterations/nAt_2003_trunc/tau0.05};
\addlegendentry{$\tau=0.05$}
\addplot[dotted,thick] table[col sep=comma]{csv/Wireless/iterations/nAt_2003_trunc/tau0.01};
\addlegendentry{$\tau=0.01$}
\end{axis}
\end{tikzpicture}
}
\hspace{4pt}
\subfigure[Using Raptis {\it et al.}'s model \cite{Raptis2009}]
{
\begin{tikzpicture}
\begin{axis}[
axis lines*=left, 
width=2.5in,
height=2in,
xlabel=$k$,
ylabel=$n$,
xmin=1, xmax=30,
ymin=0, ymax=1600,
x tick label style={font=\scriptsize},
y tick label style={font=\scriptsize},
unbounded coords=jump,
scaled ticks=false, tick label style={/pgf/number format/fixed},
enlargelimits=false, clip=true, axis on top,
legend columns=-1,
legend style={font=\footnotesize,draw=none,at={(0.5,1.25)},anchor=north,},
]
\addplot[solid] table[col sep=comma]{csv/Wireless/iterations/nAt_2009/tau0.1};
\addlegendentry{$\tau=0.1$}
\addplot[dashed] table[col sep=comma]{csv/Wireless/iterations/nAt_2009/tau0.05};
\addlegendentry{$\tau=0.05$}
\addplot[dotted,thick] table[col sep=comma]{csv/Wireless/iterations/nAt_2009/tau0.01};
\addlegendentry{$\tau=0.01$}
\end{axis}
\end{tikzpicture}
}
\caption{Required number of iterations to essentially eliminate the effect of wireless network delays at different values of $\tau$.}
\label{kagainstn}
\end{figure}

Using linear iterative root finding \cite{traub1982interative}, we solved (\ref{numiterationsrequiredfinal}) for $n$ at various values of $k$. A plot of both variables is shown in Fig.~\ref{kagainstn} for different values of $\tau$. Once again, the differences between the wireless delay models in the reviewed literature manifest in our analysis. For example, using the model of Carvalho and Garcia-Luna-Aceves \cite{1249764}, if $\tau=0.05$, then only 8 iterations are required to mitigate the effect of the wireless delays on CPV, versus 440 iterations using the model of Raptis {\it et al.} \cite{Raptis2009}. At $k=30$ wireless devices, and $\tau=0.01$, the required number of iterations is $\sim$250 and $\sim$1590 respectively.

\section{Conclusion}
\label{wirelssconc}

In this chapter, the efficacy of CPV was evaluated when legitimate clients are using wireless access networks with varying $k$---the number of saturated wireless devices competing for the media in each network. Evaluation was performed using wireless delay distributions derived in the literature.

The results show that wireless networks are likely to impact the correctness of CPV's decisions. The significance of that impact depends fundamentally on $k$. For example, the summation of FRs and FAs jumped from 3\% (see Chapter \ref{ch:wiredecva}) when all legitimate clients are using wired access networks to $\sim$4.5\% at $k=2$ and to $\sim$7\% at $k=10$ (those numbers are the averages obtained upon using both of the reviewed models).

Moreover, we found that CPV is more likely to falsely accept adversaries close to the triangles' sides when there are wireless legitimate (inside) clients. For example, when $k=10$, some adversaries within $\sim$1,000 km of the triangles' sides were falsely accepted, some of which were correctly rejected when no wireless legitimate clients were considered (Chapter \ref{ch:wiredecva}). Adversaries that are farther away than this distance are unlikely to (illicitly) benefit from the existence of wireless legitimate clients.

Finally, the analysis conducted in this chapter shows that increasing the number of CPV iterations can mitigate the negative effect of wireless delays on CPV. As such, we derived the number of iterations required to achieve that mitigation. Using the derived expressions, we found that the required number of iterations rapidly increases with $k$. When CPV is calibrated to be more tolerant to high delays between the client and the verifiers (i.e., at smaller values of $\tau$), the rate for which the required number of iterations increases with $k$ slows down. These results highlight the importance of conducting the appropriate number of iterations, especially when CPV is verifying locations of wireless legitimate clients.

In general, the results in this chapter suggest that the impact of wireless networks on delay-based Internet applications should be given more attention, e.g., most delay-based geolocation techniques in the literature are not evaluated with wireless networks. \label{changewireless}\textcolor{\changes}{Investigating the behavior of CPV with clients using other wireless access technologies (e.g., LTE networks) is left for future investigation.}

\chapter{Hindering Middleboxes from Unauthorized Traffic Relaying}
\label{ch:puzzles}

When employed by online content providers, access-control policies can be evaded whenever clients collude with a \acf{MB} that meets the policies. A colluding \ac{MB}, commonly being the gateway of a \ac{VPN}, typically contacts the content provider on behalf of the clients it colludes with, and relays the provider's outbound traffic to those clients. 
To address this problem, we propose a solution to hinder colluding \acp{MB} from unauthorized relaying of traffic to a large number of clients. To the best of our knowledge, this is the first work to address this problem. Our solution increases the cost of collusion by leveraging client puzzles in a novel way, and uses network properties to help the content provider detect if its outbound traffic is being further relayed beyond a transport-layer connection. Our evaluation shows that using client puzzles places an upper bound on the number of clients a \ac{MB} can collude with in parallel. The upper bound follows a hyperbolic decay with the rate of creation of puzzles and the time required to solve a puzzle---both factors are influenced by the content provider, but grows almost linearly with the \ac{MB}'s computational resources.

\section{Introduction}
\blfootnote{The content in this chapter was published at the IEEE Communications Letters \cite{commletters}.}

Online content providers, such as Hulu \cite{hulu}, often have access-control policies, which either customize or prevent content-delivery to certain classes of clients. By \emph{client}, we mean the software used to communicate with the content provider, e.g., a web browser. For instance, an access policy may only allow access to clients within 300 $km$ of where the site is hosted (e.g, for data sovereignty \cite{peterson2011position}), or to those with certain IP addresses \cite{ipblacklisting}. Another policy may ban clients at a specific geographic location \cite{Bertino:2005:GSA:1063979.1063985,bbcchina} (see Chapters \ref{ch:attack} and \ref{ch:cpv}), or clients whose devices have certain system fingerprints (operating system, user-agent, etc) \cite{nikiforakis2013cookieless}. A content provider (or \emph{provider} for short) may also classify clients by their access networks \cite{Wei20083205}, or their network distance from the server (in terms of hop counts, network latency, etc) \cite{Jin:2003:HFE:948109.948116}.

When access policies are in effect, the motivation to bypass them may arise. A client that does not meet the access policies may try to bypass them using a \ac{MB} that meets those policies. \acp{MB} are commonly transport-layer proxy servers, gateways of \acp{VPN} or anonymizing networks. The \ac{MB} requests the provider's content and grants the client access to it by simply relaying the provider's outbound traffic. Many \acp{MB} claim to own thousands of IP addresses, which makes blocking them by enumerating their IP addresses almost infeasible. To detect an intercepting \ac{MB}, a provider can collaborate with a \emph{cooperative} client \cite{detal2013revealing}. However, this is infeasible within our threat model as we address a client that aims to bypass the provider's access policies; i.e., the client is the provider's adversary. Solutions that aim to prevent \acp{MB} from intercepting a connection (such as Secure Socket Layer \cite{rescorla2001ssl}) fail to prevent those \acp{MB} from relaying traffic because the client would be ready to share cryptographic credentials, such as encryption keys, with the \ac{MB} to deceive the provider. 

We propose to use client puzzles \cite{juels1999client} to increase the cost of collusion per client on the \ac{MB}. Our solution leverages network properties (average latency between network hosts) which, together with the puzzles, impose a \emph{limit} on the number of simultaneous clients an \ac{MB} can collude with. Exceeding the limit divulges the \ac{MB}'s relaying actions to the provider. This chapter makes the following contributions:
\begin{itemize}
\itemsep0em
\item Proposing and studying a solution that uses client puzzles to limit unauthorized traffic relaying (Section \ref{clientpuzzlessss}).
\item Using a Markovian queueing model to evaluate our solution, and to find the upper limit of the number of clients the \ac{MB} can collude with at a time (Section \ref{queueana}).
\item Evaluating the rates of false rejects and false accepts through simulations.
\end{itemize}

\section{Proposed Approach}
\label{clientpuzzlessss}

Our objective is to enable a provider detect if a \emph{content recipient}\footnote{We use this term to refer to the machine intended by the provider as the final content destination.} is a \emph{legitimate client} (i.e., connected to the provider without an \ac{MB} and not relaying the provider's traffic anywhere else) or an \ac{MB}. To achieve this objective, we use client puzzles \cite{juels1999client} to increase the computation required by the \ac{MB} per client; thus, increasing the \ac{RTT} the provider observes. The success of detecting an \ac{MB} is dependent on the number of simultaneous clients receiving the relayed traffic from the \ac{MB}. As the number increases, the detection success increases. If the number of clients reaches a certain threshold (Section \ref{queueana}), the provider realizes that the \ac{MB} is relaying its traffic. The provider is assumed to be able to:
\begin{itemize}
\itemsep0em
\item Estimate the average \ac{RTT} from itself to a content recipient \cite{landa2013measuring,Octant}.
\item Estimate the mean time to solve a puzzle with certain difficulty across different client machines spanning a range of computational power (demonstrated in \cite{juels1999client}).
\end{itemize}

For each connection made to provider $w$ from content recipient $d$, $w$ estimates $N_w(d)$, which is the average network \ac{RTT} from itself to $d$. The provider $w$ then periodically creates non-parallelizable puzzles \cite{tritilanunt2007toward}, and sends them to $d$. To solve a puzzle, $d$ must allocate a portion of its resources for some time depending on the puzzle difficulty set by $w$. The resource demanded by the puzzle depends on the type chosen by $w$, which could be processing- \cite{juels1999client} or memory-type \cite{doshi2006efficient} puzzles. We assume $w$ uses processing-type puzzles throughout this chapter. However, any type can be chosen as long as $w$ is able to estimate the client's puzzle-solving time to some degree of certainty (second assumption above). Upon solving a puzzle, $d$ is required to return the solution to $w$, which verifies it and bans $d$ if the solution was incorrect. Verification happens in constant time independent of the puzzle difficulty \cite{juels1999client}.

Denoting $t_c$ as the mean time to solve a puzzle across various clients, $w$ expects to see a \ac{RTT} of:
\begin{equation}\label{tcexpected}\text{RTT}_e = N_w(d) + t_c\end{equation}
When $w$ receives a solution, it calculates the actual round-trip time, $\text{RTT}_a$, from the puzzle-arrival time and compares it with $\text{RTT}_e$. If $\text{RTT}_a\leq\text{RTT}_e$, the provider assumes that $d$ is not an \ac{MB}. Otherwise, it suspects that $d$ is an \ac{MB} because the existence of an \ac{MB} between the provider and a client is likely to increase $\text{RTT}_a$---an explanation follows.

If $d$ is an \ac{MB}, it has two options: either relaying all of $w$'s outbound traffic including the puzzles to client $c$, so that $c$ solves them; or extracting the puzzles from the traffic and solving them on behalf of $c$. Relaying the puzzles to $c$ costs an additional network \ac{RTT}, $N_{\text{MB}}(c)$, between the \ac{MB} and $c$. An analogous effect occurs if the puzzles were outsourced to a remote party. The actual \ac{RTT} then becomes:

\begin{equation}\label{relayingalltrafic}\text{RTT}_a = N_w(d) + N_{\text{MB}}(c) + t_c\end{equation}
We do not expect $w$ to be able to estimate $N_{\text{MB}}(c)$. To satisfy $\text{RTT}_a\leq\text{RTT}_e$, the \ac{MB} and $c$ have to satisfy $N_{\text{MB}}(c) + t_c \leq t_c$, which happens when $N_{\text{MB}}(c)=0$; that is, the colluding client and the \ac{MB} are one physical machine, or very close to each other. We believe it is not a cost effective (scalable) attack for an \ac{MB} to be close to a meaningful number of clients. Assuming proper estimations to $t_c$ and $N_w(d)$ (i.e., $\text{RTT}_e$), it would be challenging for the \ac{MB} to relay the puzzles to $c$, and satisfy $\text{RTT}_a\leq\text{RTT}_e$. We study the effect of inappropriate estimation of $\text{RTT}_e$ in Section \ref{simresults} below.

To avoid the additional $N_{\text{MB}}(c)$, the \ac{MB} will be inclined to choose the second option: solve the puzzles on behalf of the clients. An additional queueing time, $q$, is expected to contribute to $\text{RTT}_a$ because the \ac{MB} will solve many puzzles, which correspond to the number of clients it simultaneously colludes with. The actual \ac{RTT} would then be:
\begin{equation}\text{RTT}_a = N_w(d) + q + t_{\text{MB}}\end{equation}
where $t_{\text{MB}}$ is the \ac{MB} puzzle-solving time. Recall, the content recipient $d$ is the \ac{MB}. Again, we do not expect $w$ to be able to estimate $t_{\text{MB}}$. To maintain $\text{RTT}_a\leq\text{RTT}_e$, the \ac{MB}'s computational resources must satisfy:
\begin{equation}\label{doubleu}W \leq t_c\end{equation}
where $W = q + t_{\text{MB}}$, which is the average time a puzzle spends at the \ac{MB} from the moment it arrives unsolved to the \ac{MB} until it departs the \ac{MB} solved. The queueing time $q$ is affected by: the rate at which $w$ sends puzzles to each client connection; the number of clients simultaneously colluding with the \ac{MB}; the \ac{MB}'s processing capabilities; and the puzzles' difficulty. The last two factors also affect $t_{\text{MB}}$. Although this option seems more appealing to the \ac{MB} than the previous one, it forces the \ac{MB} to limit the number of simultaneous clients to avoid being caught by the provider.

If an \ac{MB} chooses to combine both options, solving some puzzles by itself and relaying others, the provider will likely observe larger \ac{RTT} for the relayed puzzles and hence reject the client. The provider may allow some proportion, $\rho$, of \acp{RTT} to be larger than the expected \ac{RTT} before rejecting a client to account for delay spikes. In such case, the benefit of relaying some puzzles will be limited by the provider's parametrization, which upper bounds the proportion of puzzles the \ac{MB} can relay, without getting its clients rejected, by $\rho$.

\section{Evaluation and Analysis}
\label{queueana}

In this section, we derive $W$ (Section \ref{clientpuzzlessss}) as a function of the parameters affecting it. \label{changeevaluation}\textcolor{\changes}{We choose an analytical evaluation method rather than an empirical one to calculate the theoretical maximum number of clients a \ac{MB} can simultaneously collude with (i.e., relay content to) to maintain $W$ that satisfies equation (\ref{doubleu}).} 

We use the notation in Table~\ref{table:puzzles:notation}. Note that of all the variables in the table, a provider needs only estimate $k$ and $g$, which is left for future investigation.
\begin{table}
\centering
\caption[Notation used in Chapter \ref{ch:puzzles}]{Notation}
\begin{tabular}{cL{12cm}}
Notation	&	Description\\\hline
$\delta$	&	the number of clients simultaneously \emph{colluding with} (i.e., being relayed the provider's content from) the \ac{MB}.\\
&\\
$t$		&	($t_c$ in Section \ref{clientpuzzlessss}) the mean of an exponential distribution representing the time required to solve a single puzzle across different client machines, measured in seconds/puzzle. The provider is required to estimate this mean according to the chosen puzzle difficulty.\\
&\\
$r$		&	the rate the provider generates puzzles to each client connection, measured in puzzles/second.\\
&\\
$b$	&	the proportion of a client's time available to solve puzzles;\footnote{We assume a legitimate client uses all of its available computational resources to solve each puzzle it receives promptly.} $b = r t$. If $b=1$, the average client spends all of its time solving puzzles.\\
&\\
$k$		&	the number of distinct puzzles the \ac{MB} can solve simultaneously. It is possibly influenced by the number of available processing cores to the \ac{MB}.\\
&\\
$g$	&	the factor by which an \ac{MB} processing core is faster than the average client. It is possibly influenced by the cores' clock rate.\\\Xcline{1-2}{2\arrayrulewidth}
\end{tabular}
\label{table:puzzles:notation}
\end{table}

We focus only on the \ac{MB}'s processing power ($k$ and $g$) as needed to solve processing-type puzzles, and exclude from consideration resources (e.g., bandwidth, I/O, memory, etc) needed for the \ac{MB} to relay content to clients. The motivation for this is to allow focus on how the puzzle rate and difficulty constrain the \ac{MB}; i.e., this is the limiting factor. It follows that if the \ac{MB} has sufficient resources to solve the puzzles sent to it, then we assume it will have sufficient additional resources to relay content to an arbitrary number of clients. We assume the \ac{MB} does not store a local copy of the traffic it receives from the provider; it initiates a connection to the provider with each client connection request.

We use the $M/M/k$ queueing model \cite{grossfundamentals} to represent the queueing system at the \ac{MB}, where we assume the puzzle arrival is modelled by a Poisson process, and the puzzle-solving time is exponentially distributed. This model considers $k$ serving units, which in our case is the number of puzzles the \ac{MB} is able to solve in parallel. The waiting time of this model is \cite{grossfundamentals}:
\begin{equation}\label{actualdoubleu}W = \frac{1}{\mu} + \left( \frac{(k\rho)^k}{k!(1-\rho)} + \sum_{i=0}^{k-1} \frac{(k\rho)^i}{i!} \right)^{-1}\left( \frac{\rho(k\rho)^k}{\lambda (1-\rho)^2 k!} \right) \end{equation}
where
\begin{equation}\label{rhorho}\rho = \frac{\lambda}{k\mu} \end{equation}
In the queueing terminology, $\lambda$ is the customer arrival rate to the system and $\mu$ is the customer departure rate from each of the $k$ serving units ($\mu=\frac{1}{\text{service time/customer}}$), both measured in customers/time unit. Customers arriving and departing the system resemble, in our case, unsolved puzzles arriving and solved puzzles departing the \ac{MB}. Customer-service time at each serving unit resembles puzzle-solving time at each of the \ac{MB}'s cores.

To realize the maximum $\delta$ that satisfies (\ref{doubleu}), we first need to represent $W$ as a function of $\delta$. We use the waiting time of (\ref{actualdoubleu}), and express $\lambda$ and $\mu$ in terms of $\delta$, $r$, $t$ and $g$. Because the provider sends puzzles at a rate of $r$ puzzles/second to each client connection, the puzzle arrival rate at the \ac{MB} is $\lambda = nr$ puzzles/second. The rate of solving puzzles at each of the $k$ cores is $g$ times faster than that of a client; hence, $\mu = g/t$.
Substituting in (\ref{rhorho}), we get:
\begin{equation}\label{rhoooo}\rho = \frac{n r t}{k g} = \frac{n b}{k g} \end{equation}
Note that the \ac{MB} can prevent its queue from growing indefinitely by maintaining $\lambda<k\mu$ \cite{grossfundamentals}, which occurs if it keeps the number of simultaneous clients $\delta<k g/b$. However, only satisfying this inequality can still disclose the \ac{MB}'s relaying actions to the provider, as it does not ensure satisfying (\ref{doubleu}). By substituting $\rho$ obtained as in (\ref{rhoooo}) for that in (\ref{actualdoubleu}), we express $W$ in terms of $\delta$, $t$, $r$, $k$ and $g$. Inequality (\ref{doubleu}) (which can be rewritten as $W/t-1\leq0$) then becomes:
\begin{equation}\label{finalineq}\frac{1}{g}+\left( \frac{(\frac{nb}{g})^k}{k!(1-\frac{nb}{kg})} + \sum_{i=0}^{k-1} \frac{(\frac{nb}{g})^i}{i!} \right)^{-1}\left( \frac{\frac{nb}{kg}(\frac{nb}{g})^k}{nb (1-\frac{nb}{kg})^2 k!} \right)-1\leq0\end{equation}
Using linear iterative root finding \cite{traub1982interative}, we can find the maximum integer value of $\delta$ that satisfies (\ref{finalineq}).

To study the behavior of $\delta$ with respect to $b$, $k$ and $g$, we consider a range of values for each of those parameters in the intervals $[2^{-6},1]$, $[1,80]$ and $[1,4]$ respectively. \label{changegpuzzles} \textcolor{\changes}{Note that, as of this writing, the fastest clock frequency being manufactured in the industry is the IBM zEC12, which has a frequency of 5.5 GHz} \cite{zEC12}\textcolor{\changes}{. On the other hand, processor speeds of smartphones (i.e., representing slow clients) are generally in the range of 1.2 to 1.9 GHz. As such, the processor speed of a \ac{MB} is unlikely to exceed four times that of a regular client, hence the upper bound of the selected interval of $g$. Note that it is still possible for a client to be using a machine slower than 1.2 GHz.\footnote{Note---it is likely that such slower devices will already be precluded from enjoying many now-common services that require more powerful processors (e.g., streaming media content, running a browser that is new enough to support web-sockets for CPV, etc).} As such, values of $g>4$ could still be of interest to evaluate. However, as we show later below, the selected range does not affect the conclusions drawn about the effect of the puzzles to hinder traffic relaying.}

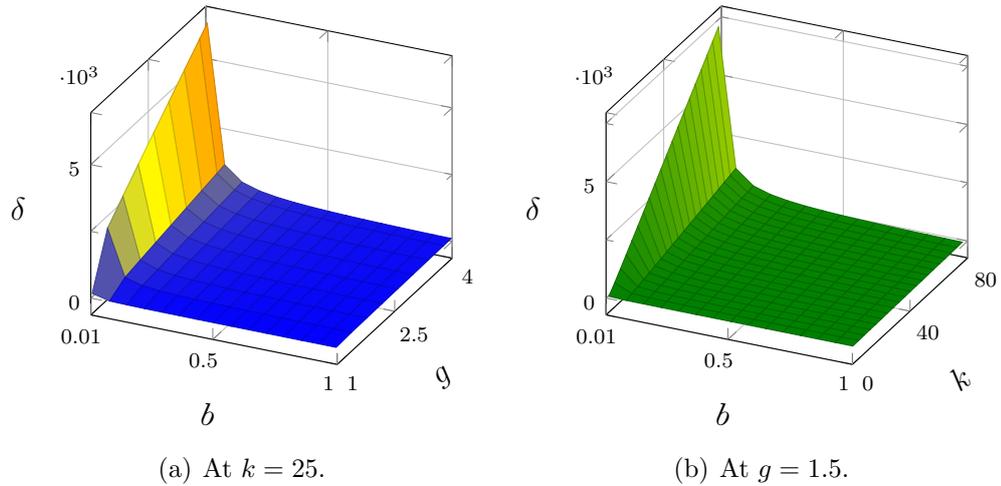
\begin{figure}
\centering
\subfigure[At $k=25$.]
{
\label{variationsatcequal15}
\begin{tikzpicture}
\begin{axis}[
mesh/ordering=y varies,
xticklabel style={/pgf/number format/fixed},
width=2.5in,
height=2.5in,
xlabel=$b$,
xtick={0.01,0.5,1},
xmin=0.01,xmax=1,
minor x tick num=1,
ylabel=$g$,
y label style={rotate=37},
ytick={1,2.5,4},
zlabel=$\delta$,
minor z tick num=1,
scaled z ticks = base 10:-3,
grid=both,
x tick label style={font=\scriptsize},
x label style={at={(0.35,-0.08)},align=center,rotate=-9},
y tick label style={font=\scriptsize},
y label style={at={(0.94,0.01)}},
z tick label style={font=\scriptsize},
z label style={at={(-0.15,0.45)},rotate=270,},
legend style={draw=none}
]
\addplot3[surf,shader=faceted]table {csv/Puzzles/3D/at_c_equal_25.dat};
\end{axis}
\end{tikzpicture}
}
\subfigure[At $g=1.5$.]
{
\label{variationsatgequal2}
\begin{tikzpicture}
\begin{axis}[
mesh/ordering=y varies,
xticklabel style={/pgf/number format/fixed},
width=2.5in,
height=2.5in,
xlabel=$b$,
xtick={0.01,0.5,1},
xmin=0.01,xmax=1,
minor x tick num=1,
ylabel=$k$,
y label style={rotate=37},
ytick={0,40,80},
ymin=0,ymax=80,
zlabel=$\delta$,
minor z tick num=1,
scaled z ticks = base 10:-3,
grid=both,
x tick label style={font=\scriptsize},
x label style={at={(0.35,-0.08)},align=center,rotate=-9},
y tick label style={font=\scriptsize},
y label style={at={(0.94,0.01)}},
z tick label style={font=\scriptsize},
z label style={at={(-0.15,0.45)},rotate=270},
legend style={draw=none}]
]
\addplot3[surf,shader=faceted,colormap/greenyellow]table {csv/Puzzles/3D/at_g_equal_1p5.dat};
\end{axis}
\end{tikzpicture}
}
\caption[Maximum theoretical number of clients that can simultaneously collude with the MB without being detected by the provider.]{Maximum theoretical number of clients that can simultaneously collude with the MB without being detected by the provider. The lines on the surfaces are equally spaced on the $b$, $g$ and $k$ axes. See Table~\ref{table:puzzles:notation} for notation.}
\label{variationslowend}
\end{figure}

Figure~\ref{variationsatcequal15} shows the change of $\delta$ at $k=25$, and Fig.~\ref{variationsatgequal2} at $g=1.5$. We ignore $\delta$ when $b>1$ because the provider should never set $b$ in that range. Otherwise, unsolved puzzles start to accumulate at legitimate clients, increasing the \ac{RTT} due to additional queueing delay, and falsely rejecting these clients.

From (\ref{finalineq}), we can see that $\delta$ and $b$ always occur multiplied together, hence by replacing all occurrences of $\delta b$ with $\gamma$, we can express $\delta$ in terms of $\gamma$ and $b$ as $\delta=\gamma/b$. That is, \emph{$\delta$ follows a hyperbolic decay with $b$ (for all $b>0$) with a scale factor of $\gamma$}. The maximum value of $\gamma$ that makes $W$ satisfy (\ref{doubleu}) grows with $k$ and $g$. For example, in Fig.~\ref{variationsatgequal2}---where $g=1.5$---every integer value, $\kappa$, on the $k$ axis defines the scale factor, $\gamma=f(1.5,\kappa)$, of a hyperbolic decay of $\delta$ with respect to $b$ at $\kappa$.

The results plotted in Fig.~\ref{variationslowend} show that $\delta$ follows an almost linear growth with $g$ and $k$, versus a hyperbolic decay with $b$. The provider influences $b$ through $t$ and $r$, the \ac{MB} controls $k$ and influences $g$ by investing in hardware. This puts the \ac{MB} in a critical situation as the provider has a more significant impact on $\delta$ than the \ac{MB} has. These results illustrate the potential of puzzles in limiting the number of colluding clients.

\subsection{Simulation Results}
\label{simresults}

The analytical evaluation showed how client puzzles affect the number of clients the \ac{MB} could support in case the \ac{MB} decides to solve the puzzles on behalf of the clients it colludes with. We now study the case where the \ac{MB} decides to forward the puzzles to those colluding clients. We use the network simulator (ns-2) \cite{breslau2000advances} to evaluate the rate of \acfp{FR}, where a legitimate client is rejected by the provider; and \acfp{FA}, where a client colluding with the \ac{MB} is accepted. Because wireless access networks have unique latency-estimation issues (see Chapter \ref{ch:wirelessecva}), they are beyond the scope of the evaluation performed in this chapter.

We assume the provider will endure some error while estimating $\text{RTT}_e$ in (\ref{tcexpected}). This error scales $\text{RTT}_e$ by a factor $\beta$, such that:
\begin{equation}\label{rewg3q4tg}\text{RTT}_e = \beta\times\text{RTT}_a\end{equation}
See (\ref{relayingalltrafic}) for $\text{RTT}_a$. 
\acp{FR} tend to increase when $\beta < 1$, \acp{FA} tend to increase when $\beta > 1$.

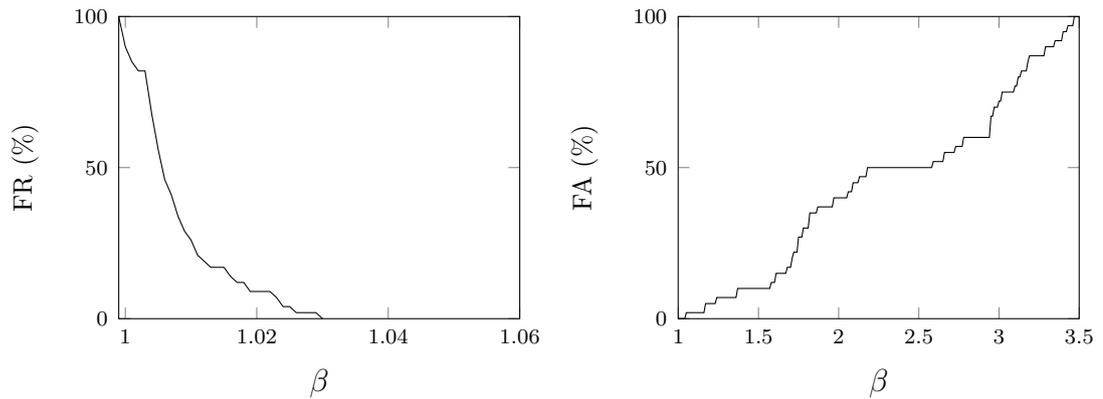
\begin{figure}
\centering
\begin{tikzpicture}
\begin{axis}[
width=2.7in,
height=2.2in,
xlabel=$\beta$,
ylabel=\ac{FR} (\%),
xmin=0.999, xmax=1.06,
ymin=0, ymax=100,
ytick={0,50,100},
legend columns=-1,
legend style={draw=none,at={(0.52,1.2)},anchor=north,},
x tick label style={font=\scriptsize},
y tick label style={font=\scriptsize},
unbounded coords=jump,
y label style={font=\footnotesize},
scaled ticks=false, tick label style={/pgf/number format/fixed},
]
\addplot[solid] table[col sep=comma]{csv/Puzzles/fr.csv};
\end{axis}
\end{tikzpicture}
\begin{tikzpicture}
\begin{axis}[
width=2.7in,
height=2.2in,
xlabel=$\beta$,
ylabel=\ac{FA} (\%),
xmin=0.999, xmax=3.5,
ymin=0, ymax=100,
ytick={0,50,100},
legend columns=-1,
legend style={draw=none,at={(0.52,1.2)},anchor=north,},
x tick label style={font=\scriptsize},
y tick label style={font=\scriptsize},
unbounded coords=jump,
y label style={font=\footnotesize},
scaled ticks=false, tick label style={/pgf/number format/fixed},
]
\addplot[solid] table[col sep=comma]{csv/Puzzles/fa.csv};
\end{axis}
\end{tikzpicture}
\caption[FRs and FAs of proposed approach]{FR and FA obtained from simulations; $\beta$ represents the error of the provider's RTT estimation.}
\label{2dfrfa}
\end{figure}

Our simulation scenarios involved several runs with 100 nodes and random connectivity patterns. Nodes distribution and link latencies were designed to resemble networks distributed over a large geographic region. One node was set to be the provider, another was set to be the \ac{MB}, while other nodes simulated clients. Some clients were connected directly to the provider (legitimate clients), others (colluding clients) were connected through the \ac{MB}. \acp{FR} and \acp{FA} are shown in Fig.~\ref{2dfrfa}. For the runs we conducted, the error scale in the range $1.03<\beta<1.1$ yields $0\%$ \acp{FR} and $2\%$ \acp{FA}. We believe these results show promising potential for the solution we propose herein.

\section{Further Considerations}
\label{PracticalRecommendations}

How many puzzles per second should the provider send to a client, and what should their difficulty be? Figure~\ref{variationslowend} showed a tradeoff between allowing more clients to collude with an \ac{MB}, and overwhelming legitimate clients. To deal with this tradeoff, providers may set $b$ to the value that satisfies a central tendency of $\delta$, such as the mean $\bar{\delta}$, over desired intervals of $b$, $k$ and $g$. 

\begin{figure}
\centering
\begin{tikzpicture}
\begin{axis}[
mesh/ordering=y varies,
xticklabel style={/pgf/number format/fixed},
width=2.5in,
height=2.5in,
xlabel=$b$,
xtick={0.5},
extra x ticks = {0.07},
xmin=0.01,xmax=1,
ylabel=$k$,
y label style={rotate=37},
ytick={0,40,80},
ztick={5000},
ymin=0,ymax=80,
zlabel=$\delta_f$,
minor z tick num=1,
scaled z ticks = base 10:-3,
grid=both,
x tick label style={font=\scriptsize},
x label style={at={(0.35,-0.08)},align=center,rotate=-9},
y tick label style={font=\scriptsize},
y label style={at={(0.94,0.01)}},
z tick label style={font=\scriptsize},
z label style={at={(-0.15,0.45)},rotate=270},
legend style={font=\footnotesize,at={(1,0.86)},draw=none,opacity=0.9},
extra tick style={major grid style=red,tick label style={color=red}}
]
\addplot3[opacity=0.5,surf,shader=faceted,colormap/greenyellow,forget plot]table {csv/Puzzles/3D_Fitted/at_g_equal_1p5_briefed.dat};
\addplot3 [color=blue,line width=0.8pt] coordinates {(0.07,1,8.9) (0.07,80,8.9*80)};
\addlegendentry{$\bar{\delta}_f$}
\end{axis}
\end{tikzpicture}
\caption[Surface fitting of FRs]{Fitted surface at $g=1.5$ represented by (\ref{fittedN}). The values of the constants in equation (\ref{fittedN}) are: $A=5.64$, $B=-58.13$, $C=3.9$ and $D=4.37$. \ac{NRMSD} over the displayed $b$ and $k$ intervals is 0.04 (or 4\%). The blue line represents the mean $\bar{\delta}_f$.}
\label{fittedsurfaces}
\end{figure}
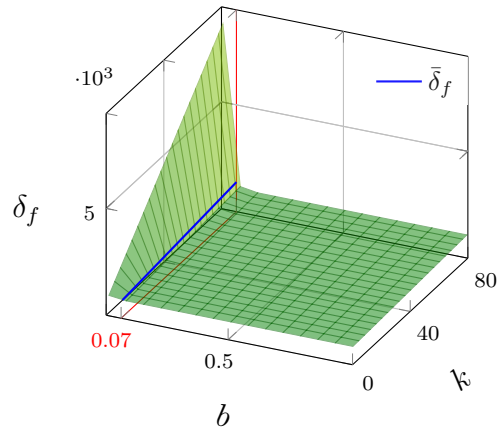

One way to calculate $\bar{\delta}$ is to, first, approximate a function that mimics the behavior of $\delta$. This can be done using curve fitting \cite{philips2011zunzun}. For example, at $g=1.5$ and $2^{-6}\leq b\leq 1$, $\delta_f$ can mimic the behavior of $\delta$, such that:
\begin{equation}\label{fittedN}n_f = k(A e^{b B + C} + D)\end{equation}
where $A$, $B$, $C$ and $D$ are constants---their values are shown on Fig.~\ref{fittedsurfaces}. The mean, $\bar{\delta}_f$, in terms of $k$ is:
\begin{equation}\bar{\delta}_f = \frac{1}{1-2^{-6}}\int_{2^{-6}}^1 \! k(A e^{b B + C} + D) \, \mathrm{d}b = 8.9k\end{equation}
Substituting $\bar{\delta}_f$ for $\delta_f$ in (\ref{fittedN}), and solving for $b$, we get:
\begin{equation}b = \frac{1}{B}\left(\ln\frac{8.9k-Dk}{kA}-C\right) = 0.07 \end{equation}
That is, considering the abstraction given in Section \ref{queueana} and our queueing model, when $b$ is restricted to the range $2^{-6}\leq b\leq 1$ and $g=1.5$, the mean of $\delta_f$ occurs at $b=0.07$. Beyond this value of $b$, puzzles will overwhelm legitimate clients without significant drop in the number of colluding clients $\delta$ whereas below, $\delta$ rapidly increases with little reduction in the puzzle workload on legitimate clients. This highlights selection of an example value of $b$ which may be of practical interest.

To set $b$, the provider adjusts $r$ and $t$ such that their product $b$ results in the desired value. Because the network \ac{RTT} is typically measured in $ms$ \cite{crovella}, a puzzle that takes a relatively long time (e.g., 1 sec) to solve on an average client machine may overshadow the network \ac{RTT}. Providers need to consider that when setting the puzzle difficulty, as it affects $t$.

Finally, providers may consider varying the puzzles' difficulty randomly, and discarding the observed \ac{RTT} of puzzles that are harder than certain undisclosed threshold to avoid having their solving time overshadow the network \ac{RTT}. This may penalize an \ac{MB} significantly as it will not be able to distinguish \emph{time-sensitive} puzzles (those where the provider will account for their \ac{RTT}) from others, and will have to solve them in order of arrival. Having a number of relatively difficult puzzles in the \ac{MB}'s queue will raise the waiting time of all others behind them, making it easier for the provider to capture the highly-delayed responses of timed puzzles, thus, detecting the \ac{MB}.

\section{Conclusion}
\label{conclusion}
This chapter addressed the problem of unauthorized relaying of a content provider's traffic, commonly performed by a \ac{MB} to enable colluding clients to bypass access-control policies set by the provider. We proposed to use client puzzles and delay estimation to enable providers hinder such unauthorized relaying of traffic, and used a queueing-model to evaluate our solution. 

The evaluation shows that in the presence of the proposed solution, the maximum number of clients that can simultaneously collude with the \ac{MB} without being detected by the provider follows a hyperbolic decay with the rate of creation of puzzles and the time required to solve them. Both of these factors are influenced by the content provider. Additionally, the number of colluding clients follows an almost-linear growth with the \ac{MB}'s computational resources, which is rather influenced by the \ac{MB}. This enables a provider using the proposed solution to have a higher control on the maximum number of simultaneously colluding clients than the \ac{MB} itself. However the rate of puzzle creation overwhelms legitimate clients too, deriving the need to find an appropriate balance between overwhelming legitimate clients and limiting the \ac{MB}'s collusion. We discussed how this balance could be obtained given the situation.

\chapter{Conclusion}

As location-oriented service/content providers are emerging over the Internet, verifying the geographic locations of Internet clients is becoming increasingly crucial. A plethora of security applications---such as fraud detection, location-based authentication, and online voting---can benefit from a realtime location-verification tool.

Measurement-based Internet geolocation approaches highlight a strong correlation between the Internet's delays and geographic distances, and provide a strong evidence of the ability to utilize these delays to locate clients, given appropriate delay processing. Despite the achieved accuracy of recent techniques, the process of refining the measured delays could be exploited by an \emph{adversary} motivated to forge its location.  Accordingly, any \emph{secure} delay-based geolocation approach has to consider both \emph{menaces}: the adversary and the Internet-added delay uncertainty.

\section{Satisfying Thesis Objectives}

In this thesis, we first investigate the reliability of current state-of-the-art delay-based geolocation techniques in the presence of an adversarial client motivated to deliberately misrepresent its own geographic location (Chapter \ref{ch:attack}). Our findings illustrate that such techniques are not ready for use in hostile environments yet as they fail to employ an integrity-preserving delay-measurement process, which is the fundamental component relied upon by all such techniques. The difficulty to fix current status quo stems from the challenges of getting community support to modify the default implementation of \ac{ICMP}-based utilities in the network stack, and disseminate the modifications for the sole purpose of hardening geolocation.

We then proceed to devise \ac{CPV} (Chapter \ref{ch:cpv}), a delay-based algorithm designed to provide a higher level of assurance about the correctness of a device's location, compared to the assurance provided by current state-of-the art geolocation techniques. To reduce potential false rejects/accepts, we support \ac{CPV} by a novel \ac{OWD}-estimation protocol that requires similar amount of client cooperation as in estimating \acp{RTT}, yet achieves higher accuracy in many cases (Chapter \ref{ch:owd}). To identify these cases, we derived the probability distribution of absolute error for both protocols as a function of the underlying delay distribution. \ac{CPV} has been extensively evaluated in wired (Chapter \ref{ch:wiredecva}) and wireless networks (Chapter \ref{ch:wirelessecva}), and the results show its potential to be adopted in practice.

We show how the \ac{CPV} algorithm can be further reinforced against a customized \ac{MB}, which is specifically designed to defeat \ac{CPV} by exchanging the algorithm's control messages with the verifiers on behalf of the adversary (Chapter \ref{ch:puzzles}). By attaching a cryptographic puzzle to these control messages and verifying the solution each time the messages are echoed, we force the \ac{MB} to solve all the puzzles destined to all the adversaries it colludes with. We proved how this technique enables \ac{CPV} to place a ceiling on the number of adversaries the \ac{MB} can collude with in parallel, without being detected by \ac{CPV}.

\subsection*{The bigger picture}
Table~\ref{table:contributionsummary} shows solutions designed to ensure the integrity of location calculation against common adversarial threats. The table shows where the location verification mechanisms contributed by this thesis (two right-most columns) stand with respect to current state-of-the-art mechanisms. A check mark ($\checkmark$) means the solution is sufficient to ensure location integrity against the respective adversarial threat. Note that the threat at row $i$ means the adversary is capable of imposing this threat and all previous threats in upper table rows. Accordingly, a check mark at row $i$ means the respective solution (column) is sufficient to ensure location integrity against an adversary capable of imposing all threats from 1 to $i$ inclusive, or any combination thereof.

\newcommand{\localwidth}{0.4cm}

\begin{table}
\caption[Solutions designed to ensure location integrity, given the respective adversarial threat.]{Solutions designed to ensure the integrity of location calculation against common adversarial threats. The location verification mechanisms contributed by this thesis are in the two right-most columns.}\scalebox{1}{
\begin{tabular}{ll|cccc|ll}

&\multicolumn{1}{c|}{}&  \multicolumn{6}{c}{\multirow{2}{*}{Solutions}}\\
\multicolumn{2}{c|}{Adversarial threat}&\rott{\footnotesize User-declared location}&
\rott{\footnotesize Client self-geolocation}&
\rott{\footnotesize Inference-based or measure.-based IP geoloc.}&
\rott{\footnotesize App-layer measurement-based geoloc.$^{*}$\ \ \ \ \ }&
\rott{\footnotesize CPV}&
\rott{\footnotesize CPV + Proof-of-Work}\\\cline{1-8}
1&\emph{Absence of threats}			&$\checkmark$&$\checkmark$&$\checkmark$&$\checkmark$&$\checkmark$&$\checkmark$\\
2&Falsifying declared location			&&$\checkmark$&$\checkmark$&$\checkmark$&$\checkmark$&$\checkmark$\\
3&Forging transmitted coordinates		&&&$\checkmark$&$\checkmark$&$\checkmark$&$\checkmark$	\\
4&Modifying location hints				&&&&$\checkmark$&$\checkmark$&$\checkmark$	\\
5&Manipulating delays 					&&&&&$\checkmark$&$\checkmark$	\\
6&Colluding with a public MB			&&&&&&$\checkmark$	\\
\Xcline{1-8}{2\arrayrulewidth}
\end{tabular}
}\\[5pt]
\scriptsize
$^{*}$This class of solutions encompasses all measurement-based geolocation techniques when measurements are performed on the application layer of the TCP/IP protocol stack.
\label{table:contributionsummary}
\end{table}

The table categorizes these solutions by their susceptibility to evasion rather than, e.g., by their cost of operation or the magnitude of their accuracy. The ``user-declared location" (column 1) is the mechanism by which the \ac{LSP} simply asks the human user about his/her location. In the absence of any adversarial threats, including the absence of the threat that the user falsifies (or lies about) their location (row 2 in the table), this mechanism is sufficient to ensure location integrity (hence the checkmark in row 1 column 1).

As discussed in Chapter \ref{ch:background}, the ``client self-geolocation" category (column 2 in the table) includes any means by which the client geolocates itself and informs the \acf{LSP}, e.g., using \ac{GPS} or \ac{WPS}. Also recall, from Section \ref{Inferencebasedapproaches} on page \pageref{Inferencebasedapproaches}, that the client's location could be ``inferred" from its IP address (column 3). This is different from measurement-based IP geolocation, which is when the delay-measurement probes are destined to the client's observed IP address. However, we place both, inference-based geolocation (including IP-address based inference) and measurement-based IP geolocation techniques, together under a single solution category (column 3) since they are affected by the same threat: modifying location hints. This threat includes not only modifying browser-based hints (e.g., preferred language---see Section \ref{vulnsinbackgroundall} on page \pageref{vulnsinbackgroundall}), but also using a \ac{MB} to modify the IP address observed by the geolocating party. If measurement-based geolocation is to be used, with delay measurements performed over the application layer (column 4), e.g., through the browser \cite{MuirPaul,shroud} or using websockets \cite{rfc6455,appraising}, it would be sufficient to ensure the integrity of location calculation against the threats in rows 1 to 4 in the table.

The table shows that \ac{CPV} combined with a \acf{PoW} mechanism, as we explain in Chapter \ref{ch:puzzles}, is sufficient to ensure location integrity against all the listed adversarial threats in the table. The absence of \ac{CPV} leaves two threats (rows 5 and 6) unaddressed by current state-of-the-art geolocation techniques. Note however that this list is not exhaustive. For example, there is also the threat of the adversary colluding with a \ac{MB} customized to evade \ac{CPV}, and that \ac{MB} is not colluding with other adversaries. This is the case when, for example, the adversary has its own private \ac{MB} physically located where it wants to fraudulently appear to be. However, it may not be scaleable for an adversary to own a \ac{MB} at every possible geographic location it intends to forge its location to. To that end, we believe the mechanisms for location verification of Internet clients contributed to the literature by this thesis are of practical value to many location-sensitive applications.

\section{Future Research Directions}

We now discuss possible future extensions to the work conducted in this thesis.

{\bf Enhancing the accuracy of delay-based geolocation techniques.} The advantages provided by the \emph{minimum pairs} protocol of Chapter \ref{ch:owd} can be leveraged to enhance the accuracy of delay-based geolocation techniques. The protocol requires three cooperating servers to exchange messages among themselves and the client; in delay-based techniques, sets of three landmarks can cooperate to implement the \emph{minimum pairs} protocol, thus estimating \acp{OWD} instead of \acp{RTT} across \emph{all} the links between the landmarks and the client. No further cooperation would be required from the client beyond echoing the messages, which is similar to what the client does when the landmarks estimate \acp{RTT}. For example, the client would not be required to synchronize its clock with the landmarks, nor to calculate and report its view of the delays.

{\bf Server location verification.} Verifying the geographic locations of servers, e.g., webservers, may provide security benefits to mitigate server impersonation, typically done through phishing, pharming or \ac{MitM} attacks. We believe that some of the ideas in this thesis, including the heuristics used to enhance the delay-measurement process, can be adapted to address the problem of verifying the geographic locations of servers.

\setcounter{tocdepth}{3}

\renewcommand{\bibname}{References}

\pagebreak
\appendix

\definecolor{light-gray}{gray}{0.95}
\lstset{   backgroundcolor=\color{light-gray},  basicstyle=\scriptsize\ttfamily,  breakatwhitespace=false,         	  breaklines=true,                 	  captionpos=b,                    	  commentstyle=\color{ForestGreen},    	  extendedchars=true,             	  frame=single,                    	  keepspaces=true,                 	  keywordstyle=\color{blue},       	  language=C,						  numbersep=5pt,                   	  numbers=left,                      numberstyle=\tiny\color{gray},   stepnumber=1,                    	  rulecolor=\color{black},         	  showspaces=false,                	  showstringspaces=false,          	  showtabs=false,                  	  tabsize=2,                       	  title=\lstname,                 	  escapeinside={(*@}{@*)},
  stringstyle=\color{orange},
  showstringspaces=false,
}

\chapter{RTT Measuring Tools}
\label{app:snippets}

To illustrate the ease of manipulating delays as measured by common network utilities, we show code snippets of example utilities lacking delay-measurement integrity.

Recall from Section \ref{ICMPattack} that the sender in the stateless implementation places the timestamp $s_i$ (packet-creation time) in the \textsc{data} field of the \ac{ICMP} packet.\footnote{\ac{ICMP} types 13 and 14 (timestamp, and timestamp reply), can also be used to measure \acp{RTT}; \ac{RFC} 792 specifies recording sending and receiving timestamps in their \textsc{data} field \cite{rfc792}. However, we did not notice many implementations of these types.} From \ac{GNU}'s \textsl{ping} (\textsl{ping.c}) \cite{GNUping}:
\begin{lstlisting}[firstnumber=502,xleftmargin=13pt]
if (PING_TIMING (data_length))
{
	struct timeval tv;
	gettimeofday (&tv, NULL);
	ping_set_data (ping, &tv, 0, sizeof (tv), USE_IPV6);
}
\end{lstlisting}\vspace{-20pt}
The variable \texttt{tv} represents our $s_i$. When the echo-reply is received, the sender observes the receiving time $r_i$, reads $s_i$ from the echoed packet, and uses them to calculate the \ac{RTT} using (\ref{mainOne}). From \ac{GNU}'s \textsl{ping} (\textsl{ping\_echo.c}) \cite{GNUping}, when the sender receives the echo-reply:
\begin{lstlisting}[firstnumber=181,xleftmargin=13pt]
struct timeval tv;
int timing = 0;
double triptime = 0.0;

gettimeofday (&tv, NULL);
\end{lstlisting}\vspace{-20pt}
\centerline{\raisebox{-3.5pt}[0pt][0pt]{$\vdots$}}
\begin{lstlisting}[firstnumber=196,xleftmargin=13pt]
struct timeval tv1, *tp;

timing++;
tp = (struct timeval *) icmp->icmp_data;

/* Avoid unaligned data: */
memcpy (&tv1, tp, sizeof (tv1));
tvsub (&tv, &tv1);
triptime = ((double) tv.tv_sec) * 1000.0 + (double) tv.tv_usec) / 1000.0;
\end{lstlisting}\vspace{-20pt}
\centerline{\raisebox{-3.5pt}[0pt][0pt]{$\vdots$}}
\begin{lstlisting}[firstnumber=227,xleftmargin=13pt]
if (timing)
	printf (" time=\end{lstlisting}\vspace{-20pt}
The variable \texttt{timing} is true if \texttt{datalen - PING\_HEADER\_LEN >= sizeof (struct timeval)}. Thus, from line 502 in \textsl{ping.c} and 227-228 in \textsl{ping\_echo.c} above, such implementation of \textsl{ping} fails to calculate the \ac{RTT} if the packet size was less than the size of the \textsl{timeval} struct.\footnote{The \textsl{timeval} struct could either be 8 or 16 bytes depending on the platform. The packet size is commonly set by the \textsl{-s} option.}

For the stateful echo-request/reply implementation, recall that the sender records $s_i$ in its local memory. These stateful utilities commonly fill the \textsc{data} field using a fixed predefined pattern; e.g., from \ac{GNU}'s \textsl{traceroute} \cite{GNUping} (\textsl{src/traceroute.c}):
\begin{lstlisting}[firstnumber=664,xleftmargin=13pt]
char data[] = "SUPERMAN";

len = sendto (t->udpfd, (char *) data, sizeof (data),0, (struct sockaddr *) &t->to, sizeof (t->to));
\end{lstlisting}\vspace{-20pt}
\centerline{\raisebox{-3.5pt}[0pt][0pt]{$\vdots$}}
\begin{lstlisting}[firstnumber=679,xleftmargin=13pt]
if (gettimeofday (&t->tsent, NULL) < 0)
	error (EXIT_FAILURE, errno, "gettimeofday");
\end{lstlisting}\vspace{-20pt}
where \texttt{t} is a struct (locally) holding information about an issued \textsl{traceroute} packet. When an echo-reply is received \cite{GNUping} (\textsl{src/traceroute.c}):
\begin{lstlisting}[firstnumber=383,xleftmargin=13pt]
gettimeofday (&now, NULL);

now.tv_usec -= trace->tsent.tv_usec;
now.tv_sec -= trace->tsent.tv_sec;
\end{lstlisting}\vspace{-20pt}
\centerline{\raisebox{-3.5pt}[0pt][0pt]{$\vdots$}}
\begin{lstlisting}[firstnumber=417,xleftmargin=13pt]
triptime = ((double) now.tv_sec) * 1000.0 + ((double) now.tv_usec) / 1000.0;
\end{lstlisting}\vspace{-20pt}
\centerline{\raisebox{-3.5pt}[0pt][0pt]{$\vdots$}}
\begin{lstlisting}[firstnumber=438,xleftmargin=13pt]
printf (" \end{lstlisting}\vspace{-20pt}

The snippets provided herein are only examples of a wide range of utilities adopting similar behaviors. They show how predictable packet contents of commonly-used utilities could be, and provide evidence of lack of integrity in delay measurement. We assert that, at their current state, none of these tools are ready for use in security-sensitive systems. Unfortunately, many such systems either rely on these tools \cite{Dong201285}, or fail to propose integrity-preserving alternatives.

\chapter{Proofs}
\label{app:proofs}

In this appendix, the three claims made in Chapter \ref{ch:cpv} are proved.

\begin{figure}
\centering
\includegraphics[scale=0.5]{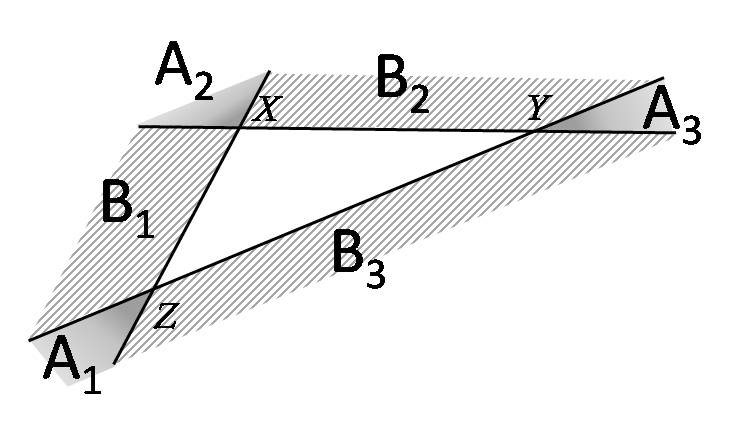}
\caption{Regions $A=A_1\cup  A_2\cup A_3$ and $B=B_1\cup B_2\cup B_3$ outside $\bigtriangleup XYZ$.}
\label{regions}
\end{figure}

\noindent{\bf Notation. }The notation $\bigcirc XY(k)$ refers to the ellipse determined by the foci $X$ and $Y$ whose major axis is $k$ meters long; {\footnotesize $\overline{AB}$} for the \emph{length} of line segment $AB$; and {\footnotesize $\overleftrightarrow{XY}$} refers to the straight line passing by the points $X$ and $Y$. Consider $\bigtriangleup XYZ$ in Fig.~\ref{regions}. Regions $A_1$, $A_2$ and $A_3$ are those outside $\bigtriangleup XYZ$ delimited by the pairs {\footnotesize $(\overleftrightarrow{XZ},\overleftrightarrow{YZ})$}, {{\footnotesize $(\overleftrightarrow{XY},\overleftrightarrow{XZ})$} and {{\footnotesize $(\overleftrightarrow{XY},\overleftrightarrow{YZ})$} respectively, such that none of $\bigtriangleup XYZ$'s exterior angles belong to $A_1$, $A_2$ or $A_3$. Regions $B_1$, $B_2$ and $B_3$ are those outside $\bigtriangleup XYZ$ delimited by the region pairs $(A_1,A_2)$, $(A_2,A_3)$ and $(A_3,A_1)$ respectively. A point $P$ outside $\bigtriangleup XYZ$ will either fall in region $A=A_1\cup  A_2\cup A_3$ or $B=B_1\cup B_2\cup B_3$.\\

{\noindent\bf Proof of Claim \ref{claimarea}}\\
Recall Claim:
\textit{
Let $P$ be a point in the Cartesian plane, and let $\bigtriangleup XYZ$ be the triangle determined by the points $X$, $Y$ and $Z$. If $P$ is strictly outside $\bigtriangleup XYZ$, then the sum of the areas of $\bigtriangleup XYP$, $\bigtriangleup XPZ$ and $\bigtriangleup PYZ$ is greater than the area of $\bigtriangleup XYZ$.\\
}

First, assume that $P$ is in region $A$; then:\newline

\begin{claim}
\label{claimarea1}
If $P$ is in region $A$, then the area of one of the triangles $\bigtriangleup XYP$, $\bigtriangleup XPZ$ or $\bigtriangleup PYZ$ will be larger than the area of $\bigtriangleup XYZ$.
\end{claim}

\begin{figure}
\centering
\subfigure[]
{
    \label{inregionA}
    \includegraphics[scale=0.5]{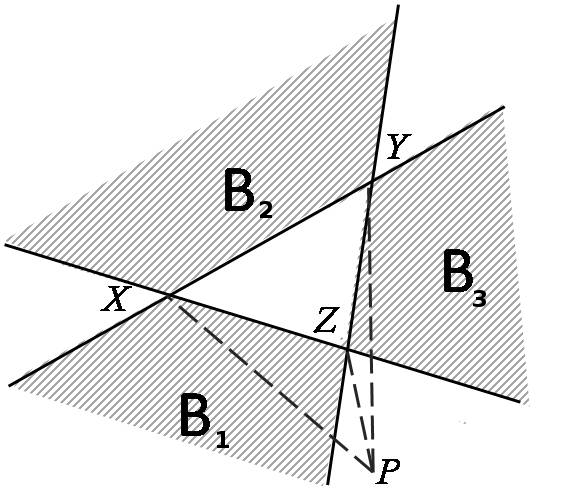}
}
\hspace{20pt}
\subfigure[]
{
    \label{inregionB}
    \includegraphics[scale=0.5]{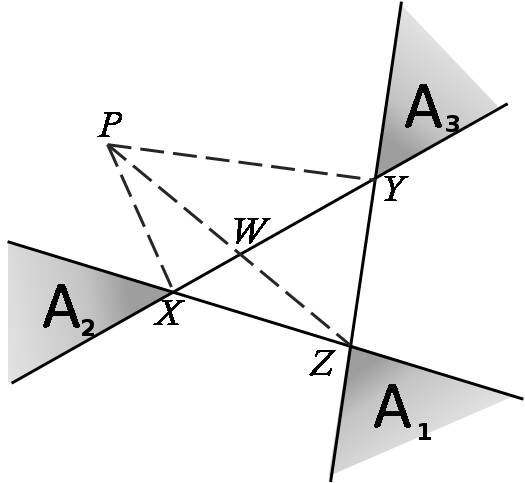}
}
\caption{If $P$ is outside $\bigtriangleup XYZ$, the sum of the areas of $\bigtriangleup XYP$, $\bigtriangleup XPZ$ and $\bigtriangleup ZPY$ will be larger than the area of $\bigtriangleup XYZ$.}
\label{AreaProof}
\end{figure}

Proving claim \ref{claimarea1} suffices to prove claim \ref{claimarea} for region $A$ because if the area of only one triangle by itself exceeds the area $\bigtriangleup XYZ$, then the sum of the areas of the three triangles ($\bigtriangleup XYP$, $\bigtriangleup XPZ$ and $\bigtriangleup PYZ$) will definitely exceed the area of $\bigtriangleup XYZ$. To prove claim \ref{claimarea1}, assume that $P$ is in region $A_1$, as shown in Fig.~\ref{inregionA}. In this case, the one triangle (referred to in claim \ref{claimarea1}) whose area is larger than that of $\bigtriangleup XYZ$ is $\bigtriangleup XYP$. The proof follows.

\begin{framed}
\begin{IEEEproof}\\
Since region $A_1$ is bound by the straight line pair $(\overleftrightarrow{XZ}, \overleftrightarrow{YZ})$.\newline
Therefore $\angle YXP > \angle YXZ$ and $\angle XYP > \angle XYZ$.\newline
Therefore $Z$ is inside $\bigtriangleup XYP$.\newline
Since line segment $XY$ is shared between $\bigtriangleup XYZ$ and $\bigtriangleup XYP$.\newline
Therefore $\bigtriangleup XYZ\subset \bigtriangleup XYP$.\newline
Therefore $area(\bigtriangleup XYZ)<area(\bigtriangleup XYP)$.
\end{IEEEproof}
\end{framed}Note that an analogous proof holds if $P$ is in $A_2$ or $A_3$. For region $B$:

\begin{claim}
\label{claimarea2}
If $P$ is in region $B$, then the sum of the areas of two of the three triangles $\bigtriangleup XYP$, $\bigtriangleup XPZ$ or $\bigtriangleup PYZ$ will be larger than the area of $\bigtriangleup XYZ$.
\end{claim}

Again, proving claim \ref{claimarea2} suffices to prove claim \ref{claimarea} for region $B$ because the sum of the areas of the three triangles ($\bigtriangleup XYP$, $\bigtriangleup XPZ$ and $\bigtriangleup PYZ$) will definitely exceed the area of $\bigtriangleup XYZ$ if the areas of two of the three triangles together exceed the area $\bigtriangleup XYZ$. To prove claim \ref{claimarea2}, assume that $P$ is in region $B_2$, as shown in Fig.~\ref{inregionB}; line segment $PZ$ intersects $XY$ in $W$. In this case, the two triangles (referred to in claim \ref{claimarea2}) are $\bigtriangleup XPZ$ and $\bigtriangleup ZPY$. The proof follows.
\begin{framed}
\begin{IEEEproof}\\
Since $P$, $W$ and $Z$ are collinear, $W$ is between $P$ and $Z$, and\newline
Since line segment $XZ$ is shared between $\bigtriangleup XWZ$ and $\bigtriangleup XPZ$\newline
Therefore $\bigtriangleup XWZ\subset\bigtriangleup XPZ$\newline
Similarly, $\bigtriangleup ZWY\subset\bigtriangleup ZPY$\newline
Therefore $(\bigtriangleup XWZ\cup\bigtriangleup ZWY)\subset(\bigtriangleup XPZ\cup\bigtriangleup ZPY)$\newline
Therefore $\bigtriangleup XYZ\subset(\bigtriangleup XPZ\cup\bigtriangleup ZPY)$.\newline
Therefore $area(\bigtriangleup XYZ)<area(\bigtriangleup XPZ)+area(\bigtriangleup ZPY)$.
\end{IEEEproof}
\end{framed}
Analogous proof holds if $P$ is in $B_1$ or $B_3$. This concludes the proof to Claim \ref{claimarea}.\\

{\noindent\bf Proof of Claim \ref{claimexcessive1}}\\
Recall Claim:
\textit{
Let $W$ be a point in the Cartesian plane, and let $\bigtriangleup XYZ$ be the triangle determined by the points $X$, $Y$ and $Z$ such that $\overline{XZ}\leq\overline{XY}$. If $\overline{XW}>\overline{XY}$, then $W$ is strictly outside of $\bigtriangleup XYZ$.\\
}

\begin{figure}
\centering
\includegraphics[scale=0.65]{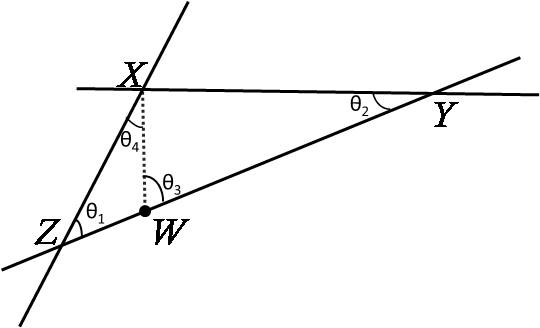}
\caption{If {\footnotesize $\overline{XZ}\leq \overline{XY}$} and $W$ is inside $\bigtriangleup XYZ$, then {\footnotesize $\overline{XW}\leq \overline{XY}$}.}
\label{ExcessiveProof}
\end{figure}

This Claim can be rewritten as:

\begin{claim}
\label{claimexcessive2}
Let $W$ be a point in the Cartesian plane, and let $\bigtriangleup XYZ$ be the triangle determined by the points $X$, $Y$ and $Z$; $\overline{XZ}\leq\overline{XY}$. If $W$ is inside $\bigtriangleup XYZ$, then $\overline{XW}\leq\overline{XY}$.\\
\end{claim}
which is the logical transposition $(P\rightarrow Q) \vdash (\neg Q\rightarrow\neg P)$ of Claim \ref{claimexcessive1}, where $P$ is the event ``$\overline{XW}>\overline{XY}$", and $Q$ is the event ``$W$ is strictly outside of $\bigtriangleup XYZ$". The following proves that $\overline{XW}\leq\overline{XY}$ holds when $\overline{XW}$ is the maximum that maintains $W$ inside $\bigtriangleup XYZ$, which is when $W$ lies on line segment $YZ$ (see Fig.~\ref{ExcessiveProof}).
\begin{framed}
\begin{IEEEproof}\\
Since $\overline{XZ}\leq\overline{XY}$\newline
Therefore $\theta_2\leq\theta_1$.\newline
Since $W$ lies on line segment $YZ$\newline
Therefore $\theta_1+\theta_4=\theta_3$.\newline
Therefore $\theta_1\leq\theta_3$.\newline
Therefore $\theta_2\leq\theta_3$.\newline
Therefore $\overline{XW}\leq\overline{XY}$.
\end{IEEEproof}
\end{framed}

{\noindent\bf Proof of Claim \ref{claimlines}}\\
Recall Claim:
\textit{
Let $P$ be a point in the Cartesian plane, and let $\bigtriangleup XYZ$ be the triangle determined by the points $X$, $Y$ and $Z$. If $P$ is strictly outside $\bigtriangleup XYZ$, then increasing the sums {\footnotesize $\overline{XP}+\overline{PZ}$, $\overline{XP}+\overline{PY}$} or {\footnotesize $\overline{YP}+\overline{PZ}$} without reducing at least one of the other sums cannot place $P$ inside $\bigtriangleup XYZ$.\\
}

\begin{figure}
\centering
\subfigure[]
{
    \label{inregionBxy}
    \includegraphics[scale=0.5]{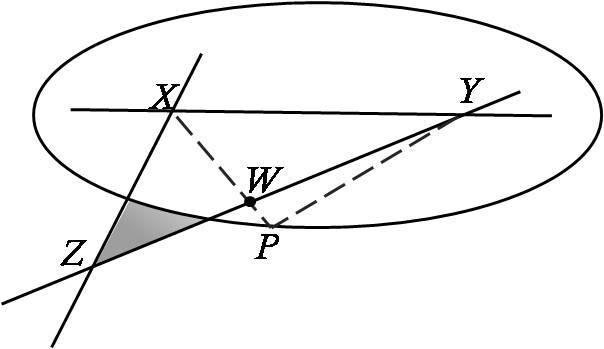}
}
\hspace{20pt}
\subfigure[]
{
    \label{inregionBxz}
   \includegraphics[scale=0.5]{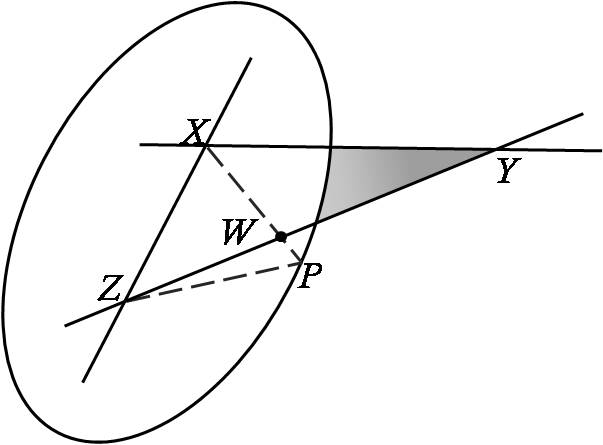}
}
\caption{When $P\in B_3$, then $\bigtriangleup XYZ \subset \{\bigcirc XY(${\footnotesize $\overline{XP}+\overline{PY}$}$)\ \cup\ \bigcirc XZ(${\footnotesize $\overline{XP}+\overline{PZ}$}$)\}$.}
\label{LineProof}
\end{figure}

Similar to the proof of Claim \ref{claimarea}, the proof of Claim \ref{claimlines} is split into two parts: when $P\in A$ and when $P\in B$. For part one, first assume that $P\in A_1$. In this case, according to the isoperimetric inequality, {{\footnotesize $\overline{XP}+\overline{PY}$} must be greater than {{\footnotesize $\overline{XZ}+\overline{ZY}$} because they both have the same starting and ending points, $X$ and $Y$. Therefore, it is impossible to move $P$ inside $\bigtriangleup XYZ$ without decreasing {\footnotesize $\overline{XP}+\overline{PY}$}. Analogous argument applies for regions $A_2$ and $A_3$.

Now to the case where $P\in B$. First assume that $P\in B_3$ as shown in Fig~\ref{LineProof}. If $\bigtriangleup XYZ \subset \{\bigcirc XY(${\footnotesize $\overline{XP}+\overline{PY}$}$)\ \cup\ \bigcirc XZ(${\footnotesize $\overline{XP}+\overline{PZ}$}$)\}$,\footnote{Note that $\bigtriangleup \subset \bigcirc$ if $\forall p\in\bigtriangleup$, $p\in\bigcirc$.} is proved, then $P$ cannot move to inside $\bigtriangleup XYZ$ without reducing {\footnotesize $\overline{XP}+\overline{PY}$} or {\footnotesize $\overline{XP}+\overline{PZ}$} because the sum of the lengths from any point on the ellipse to its pair of foci is constant; hence, the sum of the lengths from any point inside the ellipse to its pair of foci is less than that to any point on the ellipse.

Assume that $\bigtriangleup XYZ$ is split into two: $\bigtriangleup XYW$ and $\bigtriangleup XWZ$, where $W$ is the intersection of line segments $XP$ and $YZ$. Then, proving that $\bigtriangleup XYW\subset \bigcirc XY(${\footnotesize $\overline{XP}+\overline{PY}$}$)$ is as follows (see Fig.~\ref{inregionBxy}).
\begin{framed}
\begin{IEEEproof}\\
Since $X$ is a focus of the ellipse; $P$ is a point on the ellipse; $X$, $W$ and $P$ are collinear; and $P\notin\bigtriangleup XYZ$\newline
Therefore $W$ is inside the ellipse.\newline
Since $Y$ is a focus of the ellipse\newline
Therefore line segments $XW$, $WY$ and $XY$ are inside the ellipse.\newline
Therefore $\bigtriangleup XYW\subset \bigcirc XY(${\footnotesize $\overline{XP}+\overline{PY}$}$)$.
\end{IEEEproof}
\end{framed}
\noindent Analogous proof applies to $\bigtriangleup XWZ\subset \bigcirc XZ(${\footnotesize $\overline{XP}+\overline{PZ}$}$)$ (Fig.~\ref{inregionBxz}). Therefore, when $P\in B_3$, it is impossible to move $P$ inside $\bigtriangleup XYZ$ without reducing the summation {\footnotesize $\overline{XP}+\overline{PY}$} or {\footnotesize $\overline{XP}+\overline{PZ}$}. The remaining regions of $B$ can be proved in the same manner. Therefore, whenever $P\in B$, then  $\bigtriangleup XYZ \subset \{\bigcirc XY(${\footnotesize $\overline{XP}+\overline{PY}$}$)\ \cup\ \bigcirc YZ(${\footnotesize $\overline{YP}+\overline{PZ}$}$)\ \cup\ \bigcirc XZ(${\footnotesize $\overline{XP}+\overline{PZ}$}$)\}$. This concludes the proof.

\end{document}